\documentclass[longauth]{aa}  

\usepackage{amsmath}
\usepackage{graphicx}
\usepackage{graphics}
\usepackage{subfigure}
\usepackage{txfonts}
\usepackage{natbib}
\usepackage{xcolor}
\usepackage{soul}
\usepackage{tablefootnote}
\usepackage{pdflscape}
\usepackage{tabularx}
\usepackage{threeparttable}
\usepackage{rotating}
\usepackage{longtable}
\usepackage{wrapfig}
\usepackage{soul}
\usepackage{booktabs}
\usepackage{hyperref}
\hypersetup{colorlinks=true, citecolor=blue}
\usepackage{lastpage}
\usepackage{placeins}

\usepackage{hyperref}

\bibpunct{(}{)}{;}{a}{}{,}

\newcommand{\hi}{H~{\sc i}}
\newcommand{\hii}{H~{\sc ii}}
\newcommand{\ha}{\ifmmode {\rm H}\alpha \else H$\alpha$\fi}
\newcommand{\hb}{\ifmmode {\rm H}\beta \else H$\beta$\fi}
\newcommand{\lya}{\ifmmode {\rm Ly}\alpha \else Ly$\alpha$\fi}

\newcommand{\ebv}{\ifmmode E_{\rm B-V} \else $E_{\rm B-V}$\fi}
\newcommand{\av}{\ifmmode A_{\rm V} \else $A_{\rm V}$\fi}
\newcommand{\alphaCO}{\ifmmode \alpha_{\rm CO} \else $\alpha_{\rm CO}$\fi}

\def\micron{$\mu$m}

\def\msun{\ifmmode M_{\odot} \else M$_{\odot}$\fi}
\def\msunyr{\ifmmode M_{\odot} {\rm yr}^{-1} \else M$_{\odot}$ yr$^{-1}$\fi}
\def\zsun{\ifmmode Z_{\odot} \else Z$_{\odot}$\fi}
\def\lsun{\ifmmode L_{\odot} \else L$_{\odot}$\fi}
\def\mup{\ifmmode M_{\rm up} \else M$_{\rm up}$\fi}
\def\mlow{\ifmmode M_{\rm low} \else M$_{\rm low}$\fi}

\def\aap{A\&A}

\def\apjl{ApJL}
\def\apjs{ApJS}

\newcommand{\oh}{\ifmmode 12 + \log({\rm O/H}) \else$12 + \log({\rm O/H})$\fi}
\newcommand{\nii}{[N~{\sc ii}]}

\def\fesc{\ifmmode f_{\rm esc} \else $f_{\rm esc}$\fi}
\def\feschii{\ifmmode f_{\rm esc,HII} \else $f_{\rm esc,HII}$\fi}

\defcitealias{Kim_2022_environmental_dep}{K22}

\begin{document}

  \title{Duration and properties of the embedded phase of star formation in 37 nearby galaxies from PHANGS-JWST}
  \subtitle{}
  
\date{Received 02 July 2025 / Accepted 03 December 2025}

\author{Lise~Ramambason \inst{1}\fnmsep\thanks{\email: \href{mailto:lise.ramambason@uni-heidelberg.de}{lise.ramambason@uni-heidelberg.de}} 
\and Mélanie~Chevance\inst{1,2} 
\and Jaeyeon~Kim\inst{3}\fnmsep\thanks{Kavli Postdoctoral Fellow}
\and Francesco~Belfiore\inst{4, 5} 
\and J.~M.~Diederik~Kruijssen\inst{2}
\and Andrea~Romanelli\inst{1}
\and Amirnezam~Amiri\inst{6}
\and Médéric~Boquien\inst{7}
\and Ryan~Chown\inst{8}
\and Daniel~A.~Dale\inst{9}
\and Simthembile~Dlamini\inst{10}
\and Oleg~V.~Egorov\inst{11}
\and Ivan~Gerasimov\inst{7}
\and Simon~C.~O.~Glover\inst{1}
\and Kathryn~Grasha\inst{12,13}\fnmsep\thanks{ARC DECRA Fellow}
\and Hamid~Hassani\inst{14}
\and Hwihyun~Kim\inst{15}
\and Kathryn Kreckel\inst{11}
\and Hannah~Koziol\inst{16}
\and Adam~K.~Leroy\inst{8,17}
\and José Eduardo M\'endez-Delgado\inst{18}
\and Justus~Neumann\inst{19}
\and Lukas~Neumann\inst{5}
\and Hsi-An~Pan\inst{20}
\and Debosmita~Pathak\inst{8, 17}
\and Karin~Sandstrom\inst{16}
\and Sumit~K.~Sarbadhicary\inst{21}
\and Eva~Schinnerer\inst{19}
\and Jiayi~Sun\inst{22}\fnmsep\thanks{NASA Hubble Fellow}
\and Jessica Sutter\inst{23}
\and David A. Thilker\inst{21}
\and Leonardo~Ubeda\inst{24}
\and Tony~D.~Weinbeck\inst{9}
\and Bradley C. ~Whitmore\inst{24}
\and Thomas~G.~Williams\inst{25}}

\institute{
Zentrum für Astronomie der Universit\"{a}t Heidelberg, Institut für Theoretische Astrophysik, Albert-Ueberle-Str. 2, 69120 Heidelberg
\and 
Cosmic Origins Of Life (COOL) Research DAO, \href{https://coolresearch.io}{https://coolresearch.io}
\and
Kavli Institute for Particle Astrophysics \& Cosmology, Stanford University, CA 94305, USA
\and
INAF — Osservatorio Astrofisico di Arcetri, Largo E. Fermi 5, I-50125, Florence, Italy
\and European Southern Observatory, Karl-Schwarzschild Straße 2, D-85748 Garching bei München, Germany
\and 
Department of Physics, University of Arkansas, 226 Physics Building, 825 West Dickson Street, Fayetteville, AR 72701, USA
\and 
Université Côte d'Azur, Observatoire de la Côte d'Azur, CNRS, Laboratoire Lagrange, 06000, Nice, France
\and 
Department of Astronomy, The Ohio State University, 140 West 18th Avenue, Columbus, OH 43210, USA
\and Department of Physics and Astronomy, University of Wyoming, Laramie, WY 82071, USA
\and Department of Astronomy, University of Cape Town, Rondebosch 7701, Cape Town, South Africa
\and Astronomisches Rechen-Institut, Zentrum f\"{u}r Astronomie der Universit\"{a}t Heidelberg, M\"{o}nchhofstra\ss e 12-14, D-69120 Heidelberg, Germany
\and Research School of Astronomy and Astrophysics, Australian National University, Canberra, ACT 2611, Australia
\and ARC Centre of Excellence for All Sky Astrophysics in 3 Dimensions (ASTRO 3D), Australia
\and Dept. of Physics, University of Alberta, 4-183 CCIS, Edmonton, Alberta, T6G 2E1, Canada
\and Gemini Observatory/NSF NOIRLab, 950 N Cherry Avenue, Tucson, AZ 85719, USA
\and Department of Astronomy \& Astrophysics, University of California, San Diego, 9500 Gilman Dr., La Jolla, CA 92093, USA
\and Center for Cosmology and Astroparticle Physics, 191 West Woodruff Avenue, Columbus, OH 43210, USA
\and Instituto de Astronom\'{\i}a, Universidad Nacional Aut\'onoma de M\'exico, Ap. 70-264, 04510 CDMX, Mexico
\and Max-Planck-Institut f\"{u}r Astronomie, K\"{o}nigstuhl 17, D-69117, Heidelberg, Germany
\and Department of Physics, Tamkang University, No.151, Yingzhuan Road, Tamsui District, New Taipei City 251301, Taiwan
\and Department of Physics and Astronomy, The Johns Hopkins University, Baltimore, MD 21218 USA
\and Department of Astrophysical Sciences, Princeton University, 4 Ivy Lane, Princeton, NJ 08544, USA
\and Whitman College, 345 Boyer Avenue, Walla Walla, WA 99362, USA
\and Space Telescope Science Institute, 3700 San Martin Drive, Baltimore, MD 21218, USA
\and Sub-department of Astrophysics, Department of Physics, University of Oxford, Keble Road, Oxford OX1 3RH, UK
}

\abstract{
Light reprocessed by dust grains emitting in the infrared enables the study of the physics at play in dusty embedded regions, where ultraviolet and optical wavelengths are attenuated. Infrared telescopes such as JWST have made it possible to study the earliest feedback phases, when stars are shielded by cocoons of gas and dust. Comprehending this phase is crucial for unravelling the effects of feedback from young stars that leads to their emergence and the dispersal of their host molecular clouds. Here we show that the transition from the embedded to the exposed phase of star formation is short ($< 4$~Myr) and sometimes almost absent ($< 1$~Myr) across a sample of 37 nearby star-forming galaxies covering a wide range of morphologies, from massive barred spirals to irregular dwarfs. The short duration of the dust-clearing timescales suggests a predominant role of pre-supernova feedback mechanisms in revealing newborn stars, confirming previous results on smaller samples and allowing, for the first time, a statistical analysis of their dependencies. We find that the timescales associated with mid-infrared emission at 21~$\mu$m, tracing a dust-embedded feedback phase, are controlled by a complex interplay between giant molecular cloud properties (masses and velocity dispersions) and galaxy morphology. We report relatively longer durations of the embedded phase of star formation in barred spiral galaxies, while this phase is significantly reduced in low-mass irregular dwarf galaxies. We discuss tentative trends with gas-phase metallicity, which may favor faster cloud dispersal at low metallicities. }

\keywords{Galaxies: star formation -- infrared: ISM -- methods: statistical -- ISM : clouds - ISM: dust, extinction}

\authorrunning{L. Ramambason et al.}
\titlerunning{Embedded star formation with 21\micron}

\maketitle

\section{Introduction}
\label{section_intro}

Feedback from massive stars on the surrounding interstellar medium (ISM) is a key element regulating the flow of matter and energy within galaxies and, in turn, their global evolution. Importantly, the various feedback mechanisms at play at different stages of stellar evolution may prevent the accretion of new gas onto giant molecular clouds (GMCs) and may be responsible for the low star-formation efficiencies observed in both Galactic and extra-galactic GMCs \citep[e.g.,][]{Evans_sf_efficiencies_2009, Schruba_sf_efficiencies_2019, Kruijssen_nat_2019}. Understanding which feedback mechanisms (e.g., winds, shocks, radiative feedback, supernovae) dominate in the different evolutionary stages of GMCs represents an important step in deepening our understanding of the cycle of matter in galaxies and how star formation influences galactic evolution \citep[e.g.,][]{Chevance_pp7_review_2023}.

Until recently, the studies able to reach the spatial resolutions necessary to distinguish individual GMCs ($\sim$50-100\,pc) were restricted to a handful of nearby galaxies \citep[e.g.,][]{Wong_lmc_2011, Nieten_m31_2006, Engargiola_m33_2003,Gratier_m33_2010, Leroy_ic10_2006, Schinnerer_paws_2013, Pety_paws_2013, Bigiel_2008, Leroy_m51_2017, Kruijssen_nat_2019, Querejeta_ic342_2023}. However, recent efforts focused on building representative statistical samples of star-forming regions in nearby star-forming galaxies have enabled the assembly of unprecedentedly large catalogs of molecular clouds (e.g., \citealt{Colombo_2014, Rosolowsky_gmc_2021}, Hughes et al. in prep), \hii\ region catalogs \citep[e.g.,][]{Espinosa-Ponce_califa_2020, Congiu_hii_2023, Groves_hii_2023, Lugo-Aranda_HII_catalog_2024}, and stellar clusters \citep[e.g.,][]{Adamo_legus_2017, Cook_legus_dwarf_2019, Maschmann_cluster_2024}, which can be crossmatched \citep[e.g.,][]{Zakardjian_2023}. In particular, the Physics at High Angular resolution in Nearby GalaxieS (PHANGS)\footnote{\url{https://sites.google.com/view/phangs/home}} survey contains the ideal sample of galaxies to perform a detailed comparison of star-forming regions probed through the emission of multiple ISM tracers observed at high-angular resolution, with a continuous wavelength coverage from radio to ultraviolet (UV) with telescopes such as ALMA \citep{Leroy_2021_phang_alma}, MUSE \citep{Emsellem_phangs_muse_2022}, HST \citep{Lee_phangs_hst_2022}, and JWST \citep{Lee_2023_phangs_jwst, Chown_CO_IR_2025}. Current PHANGS catalogs include several tens of thousands of GMCs extracted out of 88 galaxies characterized at $\sim$50-100\,pc scales \citep{Rosolowsky_gmc_2021}, a similarly large number of individual \hii{} regions extracted from a subsample of 19 galaxies at a median spatial scale of $\sim$70\,pc \citep{Santoro_2022_PHANGS_LF}, and $\sim$100~000 stellar clusters and compact associations identified in a subsample of 38 galaxies \citep{Maschmann_cluster_2024}.

While several techniques can be used to age-date young stellar associations \citep[e.g.,][]{Miret-Roig_2024} and the expanding shells of gas around them \citep[e.g.,][]{Bonne_cii_2023, Watkins_bubbles_2023}, deriving direct measurements of the evolutionary timescales is challenging at galactic scales. Nevertheless, the access to high spatial resolution data has enabled the development of different methods to indirectly reconstruct the temporal evolution of star-forming regions (see, e.g., the review in \citealt{Schinnerer_Leroy_2024}). One approach consists of defining subregions within galaxies in an agnostic way by calculating the luminosity-weighted properties of the ISM within geometrical patterns of a fixed size, including the timescales associated with the dynamical state of molecular clouds \citep[such as the gravitational free fall time and the turbulent crossing time;][]{Sun_prop_clouds_2022}. Other avenues that have been followed include either focusing on the kinematics of large-scale gas flow \citep{Meidt_timescales_2015} or leveraging the spatial distribution of different tracers of the stars and gas, assumed to be emitted in a continuous time sequence (see, e.g., the review from \citealt{Chevance_pp7_review_2023}). The above perspective enables a quantitative interpretation of the deviations from the Schmidt-Kennicutt relation \citep[e.g.,][]{SK_law_1998, Reyes_Kennicutt_2019, Kennicutt_Reyes_2021}, which are revealed by the observation of clear spatial offsets between star and gas tracers when focusing on resolutions smaller than the typical GMC separation ($\leq$ 200~pc) or on individual lines of sight \citep[see e.g.,][]{Bigiel_2008, blanc_resolved_sk_2009, Onodera_sk_breakdown_2010, Gratier_2012, Schruba_m33_2010,Leroy_alphaCO_2013, Kruijssen_Longmore_2014, Kreckel_ngc628_2018, Kruijssen_nat_2019, Schinnerer_2019, Pan_morpht_2022}. A statistical analysis of these spatial correlations and decorrelations has been formalized in \cite{Kruijssen_Longmore_2014} and \cite{Kruijssen_tf_2018}, and we describe this analysis in Section \ref{subsec_measuring_timescales}. 

This statistical method has been extensively applied to both observations of local galaxies \citep[e.g.,][]{Kruijssen_nat_2019, Chevance_GMC_spiral_2020, Chevance_lifecycle_2020, Chevance_2022, Zabel_2020, Kim_2021_embedded24_mic, Kim_2022_environmental_dep, Kim_2023_ngc628, Ward_lmc_2020, Ward_lmc_2022, Lu_wisdom_2022, Kruijssen_drift_2024, Romanelli_2025, Kim_pah_2025} and to simulations \citep[e.g.,][]{Haydon_2020, Fujimoto_feedback_recipes_2019, Semenov_ngc300_2021, Keller_early_feedback_2022}. These results depict a baryon cycle in which typical molecular clouds are short-lived (5--30\,Myr) and form stars with low integrated star-formation efficiencies (1--8\%; see, e.g., \citealt{Chevance_GMC_spiral_2020, Kim_2022_environmental_dep}). What is more, the feedback timescales, measured using H$\alpha$ as a star-formation tracer, are short and typically between $1-6$\,Myr, hinting at a predominant role of the pre-supernova feedback mechanisms (e.g., photoionization, stellar winds, radiation pressure) since molecular clouds are dispersed on timescales shorter than (or comparable to) the first supernova explosion \citep[e.g.,][]{Chevance_GMC_spiral_2020, Chevance_2022, Kim_2022_environmental_dep}. Disentangling the impact of winds from young stars and radiative stellar feedback requires probing the earliest stages of stellar feedback, including the dust-embedded phase of star formation. In particular, star-formation rate (SFR) tracers emitting in the mid-infrared (IR) domain are critical to solving long-standing questions regarding the exact nature of the feedback mechanisms responsible for this fast GMC dispersion while also accounting for extinction effects related to the presence of dust grains in star-forming regions. 

The advent of JWST has granted access to high-resolution mid-IR tracers that offer invaluable insights into the earliest stages of star-formation.
Such analyses are challenging due to the complex origin of the mid-IR emission, which correlates both with classical molecular gas tracers such as CO \citep{Leroy_2023_midIR_CO_1kpc, Leroy_2023_midIR_CO_Ha, Chown_CO_IR_2025}, indicating that the mid-IR can be considered as a tracer of the gas column density, and with the attenuation-corrected H$\alpha$ emission \citep{Leroy_2023_midIR_CO_Ha}, because it is sensitive to local heating of the dust grains in star-forming regions. As such, the mid-IR emission arises from the combined emission of diffuse gas reservoirs and compact star-forming regions, which contribute in roughly similar proportions \citep{Leroy_2023_midIR_CO_Ha, Pathak_2024_midIR_pdf}. Among the different photometric bands observable in the mid-IR, the continuum-dominated bands around $\sim$20~$\mu$m (e.g., WISE/22~$\mu$m, MIPS/24~$\mu$m, MIRI/21~$\mu$m) have been extensively used to correct the SFR estimates for dust-attenuation in \hii\ regions \citep[e.g.,][]{Calzetti_midIR_2007, Leroy_sfr_calibration_2012, Kennicutt_resolved_SF_m51_2007, Whitcomb_sfr_tracers_2023, Belfiore_2023b_sfr_midIR} and to locate the sites of embedded star formation \citep[e.g.,][]{Gratier_2012, Corbelli_2017, Kim_2021_embedded24_mic, Kim_2023_ngc628}. 
Focusing on the latter, \cite{Kim_2021_embedded24_mic} used the compact \textit{Spitzer}/24~$\mu$m emission in five nearby galaxies to measure the duration of the heavily obscured star-formation phase ($t_{\rm obscured}$), finding values ranging from 1.4 to 3.8\,Myr. The first analysis carried out with the PHANGS-JWST sample \citep{Lee_2023_phangs_jwst, Chown_CO_IR_2025} suggested that the deeply embedded phase of star formation may indeed be short-lived, but it was restricted to a single measurement obtained for NGC\,628 \citep[\mbox{$t_{\rm obscured}$ = 2.3$^{+2.7}_{-1.4}$\,Myr};][]{Kim_2023_ngc628}. In this paper, we extend the analysis of the embedded phase of star formation to the largest sample to date, consisting of 37 galaxies drawn from the PHANGS surveys, for which sufficiently high spatial resolution can be probed in CO, H$\alpha$, and 21~$\mu$m. For the first time, this allows for analysis of the statistical properties of the timescales associated with the dust-emitted 21~$\mu$m band and their dependence on various physical parameters that may control the earliest phases of feedback.

These longer mid-IR wavelengths, which are associated with dust-continuum emission that is powered by stochastic heating of very small grains, are preferred to other mid-IR bands that reflect the processing of carbonaceous dust through the emission of aromatic and aliphatic features that are attributed to sub-nanometer polycyclic aromatic hydrocarbon-like molecules \cite[PAHs;][]{Leger_puget_pah_1984} or hydrogenated amorphous carbons \citep{Jones_hac_1990}. We refer to \cite{Kim_pah_2025} for a detailed analysis of the timescales associated with PAH-tracing bands. The authors follow a similar framework as the one we used in this study. We also note that the emission associated with PAH can be used to pinpoint the location of embedded young stellar clusters at the highest angular resolution achievable with JWST, providing complementary and independent timescale measurements, which we compare to our approach.

The paper is organized as follows: In Sect.~\ref{section_observations} we describe the multiwavelength data used in this analysis. Section~\ref{section_models} presents the methods used to measure timescales. Our main results are presented in Section~\ref{section_results} and discussed in Section~\ref{section_discussion}. In Sect.~\ref{section_conclusion} we summarize our findings and future prospects.

\section{Data}
\label{section_observations}

\subsection{Sample selection}
\label{section_sample}

\begin{figure*}
\centering
\includegraphics[height=0.9\textheight]{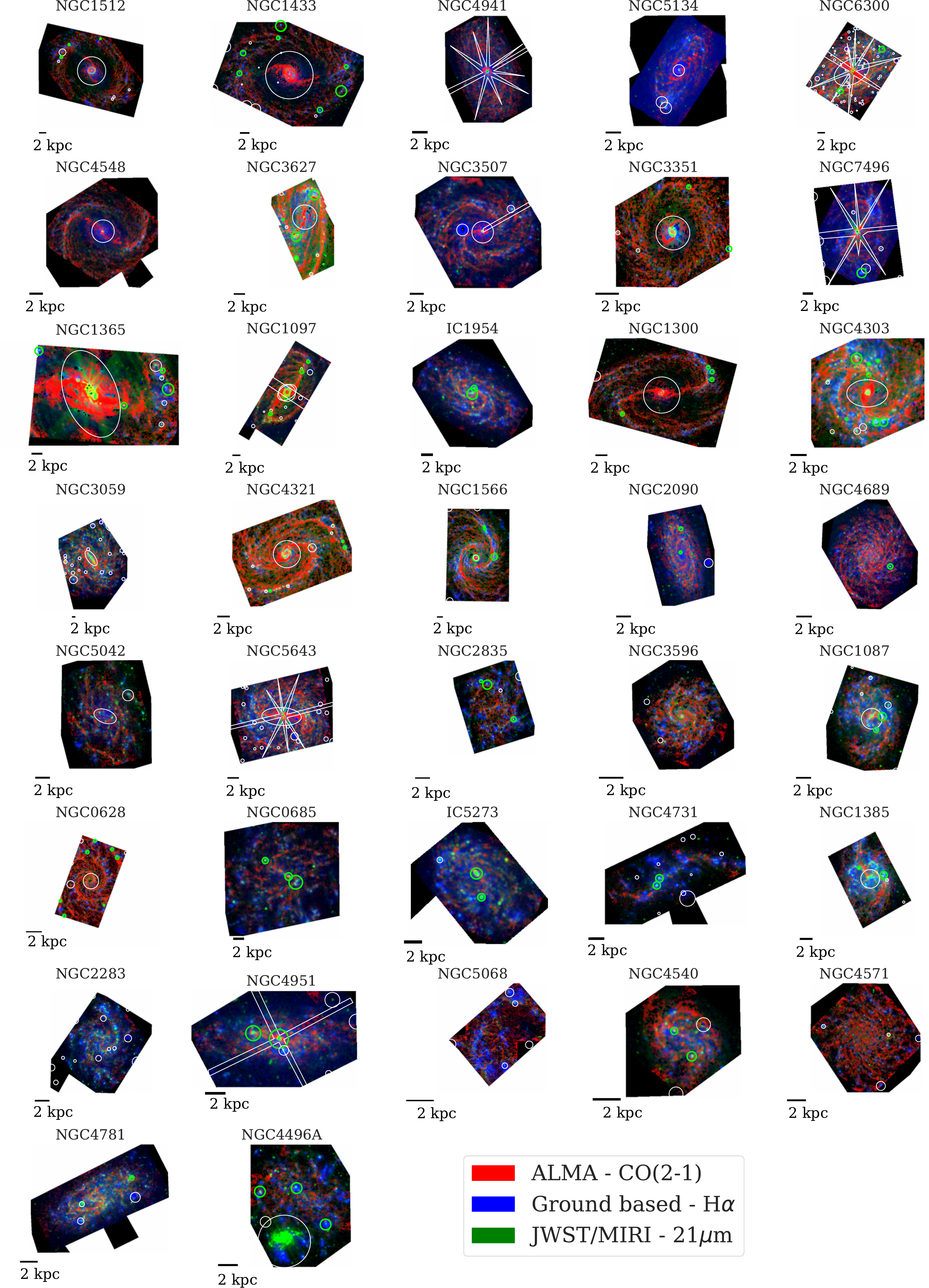}
\caption[]{Composite image of the 37 galaxies drawn from the PHANGS surveys, ordered following their Hubble morphological T-type from the LEDA database, from massive spirals to low-mass irregular galaxies. All the maps have been convolved to the resolution of the ALMA/CO(2-1) maps, corresponding to the minimal aperture size reported in Table~\ref{table_input_param}. The white masks show regions that are excluded from our analysis (H$\alpha$ and CO bright peaks identified in \citetalias{Kim_2022_environmental_dep} as well as artifacts and diffraction spikes in any of the three tracers). The green circles show the brightest peaks in 21~$\mu$m, which are masked in our analysis, as described in Section~\ref{subsect_data_homogenization}. The number of masked peaks and corresponding surface brightness thresholds are reported in the Table~\ref{table_input_param}.}
\label{rgb_with_mask}
\end{figure*}

Our sample consists of 37 galaxies, drawn from the PHANGS surveys, which gathers multiwavelength observations of nearby star forming disk galaxies obtained at high spatial resolution, probing physical scales of 30-250\,pc. These surveys have targeted relatively massive galaxies, with our final sample covering stellar masses between $10^{9.4}-10^{11}$\,M$_{\odot}$, specific star-formation rates (sSFR) between $10^{-11}-10^{-9.6}$\,yr$^{-1}$, having moderate inclination (i $\leq$ 70\,$\deg$) and located within 23.5\,Mpc \citep{Leroy_2021_phang_alma}. They cover a wide range of morphologies, ranging from irregular to flocculent, barred, and grand-design spirals, corresponding to Hubble morphological T-type from 1.2 to 7.4 in the HyperLEDA database \citep{Paturel_hyperleda1_2003, Paturel_hyperleda2_2003, Makarov_HyperLEDA_2014}.

In the current study we aim at characterizing the evolutionary timeline of molecular clouds, including their embedded star-forming phase during which young stars are not yet visible in the UV/optical domain. Our analysis focusing on timescale measurement based on the spatial decorrelation between SFR and gas tracers requires the combination of tracers associated with the emission from different ISM phases, as illustrated in Figure~\ref{rgb_with_mask}: the ALMA/CO(2-1) (shown in red) from the PHANGS-ALMA survey \citep{Leroy_2021_phang_alma}, the H$\alpha$ emission (shown in blue) from ground-based observations (Razza et al. subm.), and the dust-emitted 21~$\mu$m band (shown in green) from the PHANGS-JWST Cycle 1 Treasury \citep{Lee_2023_phangs_jwst} and PHANGS-JWST Cycle 2 Treasury \citep{Chown_CO_IR_2025}. More information about the different surveys and the data reduction is provided in Section \ref{subsec_multiwavelength_tracer}.

Our sample is drawn from the \cite{Kim_2022_environmental_dep} sample, hereafter \citetalias{Kim_2022_environmental_dep}, composed of 54 galaxies, observed in CO and H$\alpha$, 52 of which have been observed at 21~$\mu$m as part of the PHANGS-JWST surveys. We then excluded 15 galaxies for which the physical resolution is insufficient to robustly characterize the decorrelation between CO and 21~$\mu$m peaks of emission, yielding a final sample of 37 galaxies, which are listed in Table~\ref{table_input_param}. Our precise selection criteria are further detailed in Section \ref{section_quant_decorrelations}. We further discuss the reasons why a decorrelation is not observed in a few galaxies from our initial sample and how the resolution of the gas tracers limits the applicability of the method used in this paper in Section \ref{subsect_limitations}. 

\subsection{Multiwavelength tracers of the GMC evolution}
\label{subsec_multiwavelength_tracer}

The observations used in the current study consist in the full-set multiwavelength observations, gathered as part of different PHANGS surveys obtained with a variety of telescope such as ALMA, ground-based facilities targeting H$\alpha$, and JWST. We also use ISM properties derived based on the MUSE observation, available for 20/37 galaxies in our sample, as part of the PHANGS-MUSE survey \citep{Emsellem_phangs_muse_2022} and its recent extension to lower-mass galaxies (PHANGS-dwarf survey, Egorov et al. in prep). 

\subsubsection{Tracers of the molecular gas and of the exposed phase of star formation}
Our work builds upon the analysis from \citetalias{Kim_2022_environmental_dep} and uses the exact same datasets to characterize the timescales associated with the evolution of GMCs based on their CO and H$\alpha$ observations.
As a proxy to trace the cold molecular gas, we use the CO(2-1) maps which were observed as part of the PHANGS-ALMA survey \citep{Leroy_2021_phang_alma}. Data reduction is described in detail in \cite{Leroy_CO_phangs_2021}. In this study, we used the maps resulting from a combination of 12-m, 7-m, and total power antennas of ALMA. The maps have a resolution of $\sim$1\arcsec, resulting in a physical scale of $\sim$25-180~pc for the galaxies considered here. We use the first public release version of moment-0 maps generated with an inclusive signal masking scheme to ensure a high detection completeness (the `broad' masking scheme; see \citealt{Leroy_2021_phang_alma}), at native resolution.

Throughout the paper, the exposed phase of star formation refers to the timescale associated with H$\alpha$ visibility, while the exposed feedback phase traces the overlap time between the CO and H$\alpha$ tracers. We use the continuum-subtracted narrowband H$\alpha$ imaging from PHANGS–H$\alpha$ (Razza et al. subm.). We note that maps of H$\alpha$ obtained with MUSE at higher sensitivity, and probing more of the diffuse emission, are available for a subsample of 20/37 galaxies, which are covered with MUSE (\citealt{Emsellem_phangs_muse_2022}, Egorov et al. in prep). However, such data are not available for the entire sample of galaxies considered here and updating to H$\alpha$ maps observed at higher sensitivity is not crucial in our study, since our analysis focuses only on the compact part of the emission and filters out the diffuse contribution, as further described in Section \ref{subsec_filtering_emission}. We favor homogeneity within our sample and consistency with the previous study from \citetalias{Kim_2022_environmental_dep} and decide to keep the same set of H$\alpha$ observations. 

\subsubsection{21\micron\ as tracer of the embedded star formation}
To extend the previous analysis of \citetalias{Kim_2022_environmental_dep}, we include an additional proxy of the SFR available in the mid-IR. Specifically, we use the 21~$\mu$m associated with dust-continuum emission as a proxy for dust-embedded star formation and refer to the embedded feedback phase as the time when both 21$\mu$m and CO are simultaneously visible, regardless of H$\alpha$ being detected. We further refer to the “obscured” phase of star formation when only 21$\mu$m is visible without H$\alpha$. The MIRI/21~$\mu$m observations are drawn from the JWST-PHANGS surveys (Cycle 1; GO \#2107, \citealt{Lee_2023_phangs_jwst} and Cycle 2, GO \#3707, PI: Leroy, \citealt{Chown_CO_IR_2025}), that observed a subsample of the PHANGS-ALMA survey in the near-IR with MIRI and NIRCam. We focus on the MIRI/21~$\mu$m maps obtained at a spatial resolution of 0.67\arcsec\;\citep{Lee_2023_phangs_jwst}. 
A detailed overview of the data reduction can be found in \cite{Williams_2024_jwstphangs_cycle1}, and the entire PHANGS-JWST sample is presented in \cite{Chown_CO_IR_2025}.

\subsection{Updated maps of ISM properties}
\label{subsect_other_observables}

A few key physical quantities were updated to more recent measurements compared to \citetalias{Kim_2022_environmental_dep}. We briefly describe these updates in this section.

\subsubsection{Total SFR}
\label{subsec_tot_sfr}
In order to derive the total SFR within the field of view considered in our analysis and compare it to the previous analysis from \citetalias{Kim_2022_environmental_dep}, we use the SFR maps from the z0mgs atlas \citep{Leroy_2019_z0mgs}, that combine maps from the Galaxy Evolution Explorer (GALEX) far ultraviolet band (155~nm) and the Wide-field Infrared Survey Explorer (WISE) W4 band (22~$\mu$m), convolved to 15\arcsec\; angular resolution.  For the galaxies which have no GALEX observations, we use instead the SFR maps (Kreckel and Blanc, private communication) derived from PHANGS-H$\alpha$ ground-based imaging (Razza et al. subm), corrected for extinction using WISE W4 band. The derived SFRs include global corrections for H$\alpha$ transmission losses and \nii\ contamination.  We note that these maps were previously used in \citetalias{Kim_2022_environmental_dep} and that using the same prescription to compute the total SFR and total SFR surface density enable a meaningful comparison of the differences between both study (as described in Section \ref{subsect_data_homogenization}). Nevertheless, we also consider updated prescriptions for the SFR surface density measured at kiloparsec-scales ($\Sigma_{\rm SFR}^{\rm kpc}$) based on \cite{Belfiore_2023a_sfr_phangsMUSE}.
We refer to \cite{Sun_2023_calibration} and \cite{Belfiore_2023a_sfr_phangsMUSE} for a comparison of the latter calibration method with other SFR proxies. 
We note that a constant scaling difference in the total SFR and $\Sigma_{\rm SFR}$ does not affect the timescale measurements presented in Section \ref{section_results}.

\subsubsection{Metallicity}
\label{subsec_muse_met_maps}

For most of the galaxies in our sample (20/37), we have access to gas-phase metallicity estimates based on MUSE observations from the PHANGS-MUSE survey \citep{Emsellem_phangs_muse_2022}, covering 15 galaxies from our sample (NGC\,1512, NGC\,1433, NGC\,3351, NGC\,3627, NGC\,7496, NGC\,1365, NGC\,4321,  NGC\,1566, NGC\,4303, NGC\,1300, NGC\,2835, NGC\,1087, NGC\,0628, NGC\,1385, NGC\,5068) and from Egorov et al. (in prep), covering five additional galaxies (IC\,1954, NGC\,3596, NGC\,4496A, NGC\,4731, and NGC\,4781). For the
15 galaxies from PHANGS-MUSE, we estimate the metallicity based on the \hii\ region catalogs from \cite{Groves_hii_2023}, which report metallicity estimates based on the strong line ``S" calibration from \cite{Pilyugin_Grebel_scal_2016}. The latter estimates are available in the megatables v4p0,\footnote{\url{https://www.canfar.net/storage/vault/list/phangs/RELEASES/Sun_etal_2022}} following the data aggregation and table creation schemes described in \cite{Sun_prop_clouds_2022} and \cite{Sun_2023_calibration}. We then calculate the average metallicity as a $\Sigma_{\rm SFR}^{\rm kpc}$-weighted average, excluding the hexagons corresponding to the galactic centers. The resulting metallicity is in excellent agreement with the values obtained by masking and averaging the metallicity from \cite{Williams_2022_mixing_met} (smoothed using a Gaussian Process Regression) and the metallicities at the median galactocentric radius from \cite{Santoro_2022_PHANGS_LF}, both derived using the same MUSE data (see Appendix \ref{appendix_updates}, Figure~\ref{metallicity_checks}). We use a similar method to derive the metallicities for the five galaxies from Egorov et al. (in prep).

For the remaining galaxies (17/37) for which no MUSE observations are available, we derived a first-order metallicity estimate based on the stellar-mass metallicity relation from \cite{Sanchez_2019_metallicity_grad} at the effective radius R$_{\rm eff}$, adopting a metallicity gradient of $-0.1$\,dex/R$_{\rm eff}$ \citep{sanchez_gradient_2014}. A comparison of the latter metallicity estimate, based on the stellar mass-metallicity relations, with the direct metallicity measurements from MUSE is provided in Figure~\ref{metallicity_checks}. Considering the large uncertainties on these estimates, the latter metallicities are only used as luminosity-weighted galactic averages in order to derive the CO-to-H$_2$ conversion factors (from Section \ref{subsec_CO_to_H2}) and the reference timescales (from Section \ref{subsec_measuring_timescales}), but are discarded in the analysis of tentative trends observed between timescales and metallicity from Section \ref{section_results} and \ref{section_discussion}.

\subsubsection{CO-to-\texorpdfstring{H$_2$}{H₂} conversion factor}
\label{subsec_CO_to_H2}

For the PHANGS-MUSE subsample, we used metallicity-dependent $\alphaCO$ maps, based on MUSE metallicity maps from \cite{Williams_2022_mixing_met} and the conversion factor prescription from \cite{Bolatto_2013}. These maps are obtained by reprojecting the metallicity maps to match the coordinate grid corresponding to the PHANGS-ALMA CO maps. 
A local conversion factor was then calculated based on Equation~31 from \cite{Bolatto_2013}, assuming \mbox{$\Sigma_{\rm GMC}$ = 100\,M$_{\odot}$~pc$^{-2}$}. The total surface density term includes molecular gas (CO), atomic gas (\hi), and stellar mass (IRAC 3.6~$\mu$m or WISE 3.4~$\mu$m). The conversion factor is then iteratively solved for as described in \cite{Sun_prop_clouds_2022} (also J.~Sun et al., in prep.).

For the remaining galaxies, we followed the same approach as previously used in \citetalias{Kim_2022_environmental_dep} and estimated a galactic-averaged CO-to-H$_2$ conversion factor based on the prescription from \cite{Sun_2020_cloud_scale} and previously used in \citetalias{Kim_2022_environmental_dep}: 
\begin{equation}
    \label{eq_alphaCO}
    \alphaCO = 4.35 Z'^{-1.6} /R_{21} \rm M_\odot (K\,km\,s^{-1}\,pc^2)^{-1} ,
\end{equation}
where R$_{21}$ is the CO(2–1)-to-CO(1–0) line ratio \citep[0.65;][]{Leroy_alphaCO_2013, den_Brok_2021, Leroy_co_2022} and Z$'$ is the CO luminosity-weighted metallicity in units of the solar value.\footnote{We adopted a solar metallicity of $\log$(O/H)$_{\odot}$=8.69 following \cite{Asplund_2009}.}
This relation is applied to the global metallicity estimates derived using the mass-metallicity relation and assuming a negative metallicity gradient of $-0.1$ dex/kpc. Despite not accounting precisely for the local metallicity variations within each galaxy and assuming a fixed R$_{21}$ ratio \citep[which incorporates systematic uncertainties, see e.g.,][]{Schinnerer_Leroy_2024}, this formula provides a reasonable estimate of the global CO-to-H$_2$ conversion factor, accounting for the metallicity dependencies with a slope corresponding to \cite{Accurso_2017}. 

Comparing the H$_2$ mass estimates derived for galaxies where metallicity-dependent CO-to-H$_2$ maps are available to the predictions obtained with a single galactic-average CO-to-H$_2$ conversion factor, we find that taking the average $\alphaCO$ value from Equation \ref{eq_alphaCO} may overestimate the local CO-to-H$_2$ conversion factor by up to a factor of three, resulting in an overestimation of the total H$_2$ masses (up to a factor of three) and H$_2$ surface densities (up to a factor of two, as shown in the Appendix \ref{appendix_updates}, Figure~\ref{kim_vs_lise_sfr}). We note that, while the CO-to-H$_2$ prescription is important to derive the total H$_2$ mass and H$_2$ surface density, the adopted value does not affect our timescale measurements because the conversion factors cancel each other when we take the ratio of depletion timescales on local and large scales.

\section{Methodology}
\label{section_models}

\subsection{Data homogenization}
\label{subsect_data_homogenization}

Our analysis takes advantage of the rich data available for nearby galaxies by combining maps obtained with different instruments and beam sizes. In order to allow for a meaningful comparison of the peaks of emission identified in different maps, it is then necessary to homogenize the datasets used in the analysis. To do so, we follow the same procedure as detailed in \citetalias{Kim_2022_environmental_dep}, by convolving all the maps to the coarsest resolution, following \cite{Aniano_2011_convolution}, and applying a regridding to ensure that both maps share the same pixel size. The JWST point spread function at 21~$\mu$m is computed following the procedure described in \cite{Williams_2024_jwstphangs_cycle1} and publicly available as a python package\footnote{\url{https://github.com/francbelf/jwst kernels}}, which uses the \texttt{WebbPSF} simulation tool \citep{Perrin_2014_webbpsf} version 1.2.1, taking the detector effects into account. The kernels obtained at 21~$\mu$m are used to convolve to the ALMA point-spread-function, which is assumed to be Gaussian. The two maps are then reprojected to a common pixel grid using the \texttt{reproject\_interp} function from \texttt{astropy}.

The analysis is restricted to regions where all three CO, H$\alpha$, and 21~$\mu$m were observed. We note that, effectively, smaller field of views are considered compared to \citetalias{Kim_2022_environmental_dep}. In addition, the statistical method used in this study aims at providing averaged parameters over typical star-forming regions; hence, bright peaks of emission in any of the three tracers are systematically masked. To do so, we follow the same approach as presented in \citetalias{Kim_2022_environmental_dep}, by ordering the peaks from the brightest to dimmest and looking for a sharp break in the intensity distribution of peaks. This typically results in a maximum of three masked peaks in H$\alpha$ or CO in any galaxy, which are shown as white circles in Figure~\ref{rgb_with_mask}. We updated these masks in order to also exclude bright peaks in the 21~$\mu$m maps, following the same method, which results on average in masking two additional peaks in 21~$\mu$m (shown as green circles in Figure~\ref{rgb_with_mask}). The exact number of masked peaks and the corresponding thresholds in surface brightness are reported in Table~\ref{table_input_param}. Although such peaks account for only a small fraction of the total number of peaks (typically $\leq 1$\%, and $\leq 3$\% for all galaxies), they contribute disproportionately to the total compact 21$\mu$m flux.
Since the derived timescales are flux-weighted, including them biases the  measurements, which become less representative of the averaged cloud population. We further detail the biases introduced by the inclusion of overly bright 21$\mu$m peaks in Appendix \ref{appendix_mask_peaks}.

Following \citetalias{Kim_2022_environmental_dep}, we also mask the galactic centers, which are affected by blending. We note that masking the central regions of galaxies exclude a significant fraction of the total flux of the galaxies, making our analysis only representative of the star-formation timeline in the outer parts of galaxies. The final masks, that include masks of the galactic centers affected by blending, as well as masks of bright peaks and regions affected by instrumental effects such as diffraction spikes, are shown in Figure~\ref{rgb_with_mask}. These slight modifications of the masked regions, as well as the new prescriptions adopted for metallicity and the CO-to-H$_2$ conversion factors (see Section \ref{subsect_other_observables}) involve moderate variations of the predicted parameters compared to the previous analysis from \citetalias{Kim_2022_environmental_dep}, which are further discussed in the Appendix \ref{appendix_material}.

\subsection{Filtering diffuse emission}
\label{subsec_filtering_emission}

In order to characterize the evolution timeline of molecular clouds, we want to focus on the emission arising from molecular clouds and \hii\ regions only. As a result, we consider only the spatially compact emission concentrated around the peaks, identified using the \textsc{Clumpfind} \citep{clumpfind_Williams_1994} algorithm ($\sim 60-750$ peaks identified in each tracer per galaxy), while removing possible contribution from more diffuse neutral and ionized gas components. The peak identification is performed by \textsc{Clumpfind} by calculating flux contour levels in each map, assuming a given range and interval for each tracer in logarithmic scales. The logarithmic ranges and logarithmic intervals below flux maximum covered by flux contour levels for the identification of CO and 21~$\mu$m peaks are reported in Table~\ref{table_input_param}. 

To contrast the H$\alpha$ and 21~$\mu$m emission with the CO emitting molecular gas, we are only interested in the fraction of this emission that is directly associated with the ongoing star-formation cycle, i.e., the compact emission. Nevertheless, diffuse emission can contaminate CO, H$\alpha$, and 21~$\mu$m. Focusing  on the recent MUSE/H$\alpha$ observations available for 19 galaxies, \cite{Belfiore_dig_2022} report that diffuse ionized gas contributes to $19-55$\% of the total H$\alpha$ emission. The diffuse emission in 21~$\mu$m is even larger, with the emission powered by diffuse interstellar radiation field contributing up to $\sim$50-60\% to the total emission \citep[e.g.,][]{Belfiore_2023b_sfr_midIR, Pathak_2024_midIR_pdf}.

Several strategies have been proposed to correct for the diffuse emission. In this work, we follow the methodology developed in \cite{Hygate_2019_diffuse}, by filtering the diffuse component in Fourier space using a Gaussian high-pass filter, characterized by a critical distance in frequency space. Following the recommendations from \cite{Hygate_2019_diffuse}, we ensure that the relative flux losses in compact regions due to the Fourier filtering are less than 10\% in all three tracers. In the current study, we use a softening parameter, n$_\lambda$, (see values reported in Appendix \ref{appendix_material}, Table~\ref{table_input_param}), that filters out the emission on scales n$_\lambda$ times larger than the typical distance between peaks. Effectively, we filter out the emission on scales larger than $\sim 1-3$\,kpc. This filtering is carried out iteratively while updating $\lambda$, until it converges toward a fixed value. This procedure allows us to separate the compact component from the diffuse component in the different emission maps. We discuss the variations in the fractions of diffuse 21~$\mu$m emission (f$_{\rm diffuse}^{21~\mu m}$) in Section \ref{subsec_diffuse_21}. An illustration of the compact emission maps, obtained after this filtering, is presented for four galaxies in Figure~\ref{peak_id_clumpfind}.

\subsection{Measuring evolutionary timescales}
\label{subsec_measuring_timescales}

In the following, we briefly recall the main steps implemented within the \textsc{Heisenberg} code \citep{Kruijssen_tf_2018}, which is publicly available on GitHub.\footnote{\url{https://github.com/mustang-project/Heisenberg}} In the current study, we combined updated results from the analysis of \citetalias{Kim_2022_environmental_dep} using H$\alpha$ as SFR tracer to new runs and leveraging 21$\mu$m as an SFR tracer, which we further describe in this section. The relative timescales were derived based on the spatial distribution of the peaks of emission, identified with the \textsc{Clumpfind} algorithm \citep{clumpfind_Williams_1994}, in both the gas tracer (here, CO) and the SFR tracer (here, 21~$\mu$m). The peak identification in different tracers is shown in Figure~\ref{peak_id_clumpfind} for a subsample of galaxies for which robust timescale measurements are derived (no upper limits). We then placed apertures of various sizes around these peaks and measured the CO-to-21~$\mu$m flux ratio focusing respectively on CO/21~$\mu$m peaks. This enabled the measurement of the deviation of the CO-to-21~$\mu$m flux ratio from the galactic average, as a function of the aperture size, as represented in Figure~\ref{tuning_fork_ha_21}. 

The resulting “tuning-fork” shape is then fitted using two functional forms (equations 81 and 82 from \citealt{Kruijssen_tf_2018}), to mathematically describe the deviation of the CO-to-21~$\mu$m flux ratio from the galactic average, as a function of the aperture size, focusing on the branch associated respectively with the CO and 21$\mu$m peaks. One specificity of the tuning-fork method lies in the fact that the latter functional form depends only on flux-weighted quantities (namely, the flux contrast between peaks and galactic averages, $\mathcal{E}_{\rm 21~\mu m}$ and $\mathcal{E}_{\rm CO}$, the flux ratios between overlapping peaks and isolated peaks in both tracers, $\beta_{\rm 21~\mu m}$ and $\beta_{\rm CO}$ and the enclosed flux fraction of each tracer within a given aperture size, $f_{\rm 21~\mu m}$ and $f_{\rm CO}$), making the method less sensitive to the peak identification process compared to similar approaches relying solely on the counting of emission peaks in different phases. Assuming that the emission follows a Gaussian profile centered around each of the peaks, the spacing between the two branches, their asymmetry and the inflection point, uniquely characterize a function depending on flux-related quantities, that are directly measurable on the emission maps, and of three independent parameters: the separation length between emission peaks $\lambda$, the relative duration of the isolated phase of emission ($t_{\rm 21~\mu m}$/$t_{\rm CO}$), and the relative duration of the overlap or feedback phase ($t_{\rm fb, 21~\mu m}$/$t_{\rm CO}$). Other physical parameters of interest can be derived based on combinations of the latter quantities, for example the star-formation efficiency or feedback velocity, which have been previously analyzed in \citetalias{Kim_2022_environmental_dep}.

\subsection{Reference timescale}
\label{subsec_measuring_timescales}
The timescales derived following this methodology can be considered as a luminosity-weighted average of the evolutionary timeline associated with each star-forming region within a galaxy. Since we exclude galactic centers affected by blending and excessively bright peaks (see Section \ref{subsect_data_homogenization}), these timelines represent the evolution of typical star-forming regions, located in the outer part of galaxies. Converting the relative timescales into absolute ones requires the definition of a reference timescale. In this study, we use the total cloud lifetime traced by CO(2-1) ($t_{\rm CO}$) as the reference timescale for each galaxy. The latter $t_{\rm CO}$ correspond to the measurements obtained by following the exact same methodology, but contrasting maps of CO (as a tracer of the gas peaks) versus H$\alpha$ as a tracer of SFR. These runs adopt the exact same set-up as the ones detailed in \citetalias{Kim_2022_environmental_dep}, but were updated to account for the new masks that match the field of view covered by MIRI observations (see Section \ref{subsect_data_homogenization}). The latter runs are used to estimate the total cloud lifetime $t_{\rm CO}$, as well as the feedback timescale based on H$\alpha$ ($t_{\rm fb,H\alpha}$). 

We note that the $t_{\rm CO}$ measurements are tied to a metallicity-dependent reference timescale (duration of the isolated H$\alpha$ emission) calibrated in \cite{Haydon_2020}. This study uses the output of a hydrodynamical disk galaxy simulation combined with the stellar population synthesis model SLUG \citep{Krumholz_slug_2015} to estimate the reference timescale associated with different observational tracers of stars, assuming a well-sample initial mass function and based on the stellar evolutionary tracks of metallicities $Z/Z_{\odot}$ =
0.05, 0.20, 0.40, 2.00 \citep{Schaller_1992, Charbonnel_1993, Schaerer_1993b, Schaerer_1993a}, where Z$_{\odot}$ corresponds to the solar-metallicity. The authors report that the reference timescale corresponding to the continuum-substracted H$\alpha$ varies slightly with metallicity following
\begin{equation}
    t_{\rm ref}(H\alpha)  [\rm Myr] = (4.32^{+0.09}_{-0.23}) \Big(\frac{Z}{Z_{\odot}} \Big)^{(-0.086^{+0.010}_{-0.023})}.
\end{equation}
Within our sample spanning metallicities of $0.4 Z_\odot -1 Z_\odot$, the H$\alpha$ reference timescales vary by a factor 1.08, from 4.32 up to 4.67\,Myr. This correction therefore enables accounting for a known, but moderate, metallicity-trend of the H$\alpha$ visibility. The corrections introduced to account for metallicity variations are negligible compared to the uncertainties derived on our measurements. The resulting cloud lifetimes, which incorporate those metallicity variations and serve as reference timescales in the current analysis, are reported in Appendix \ref{appendix_material}, Table~\ref{table_outputs_21um}, together with the input parameters used in our analysis.

\section{Results}
\label{section_results}

\subsection{Overview of star-formation cycle}
\label{section_overview}
\subsubsection{Quantifying spatial decorrelations}
\label{section_quant_decorrelations}

\begin{figure*}
\centering
\includegraphics[width=0.8\textwidth]{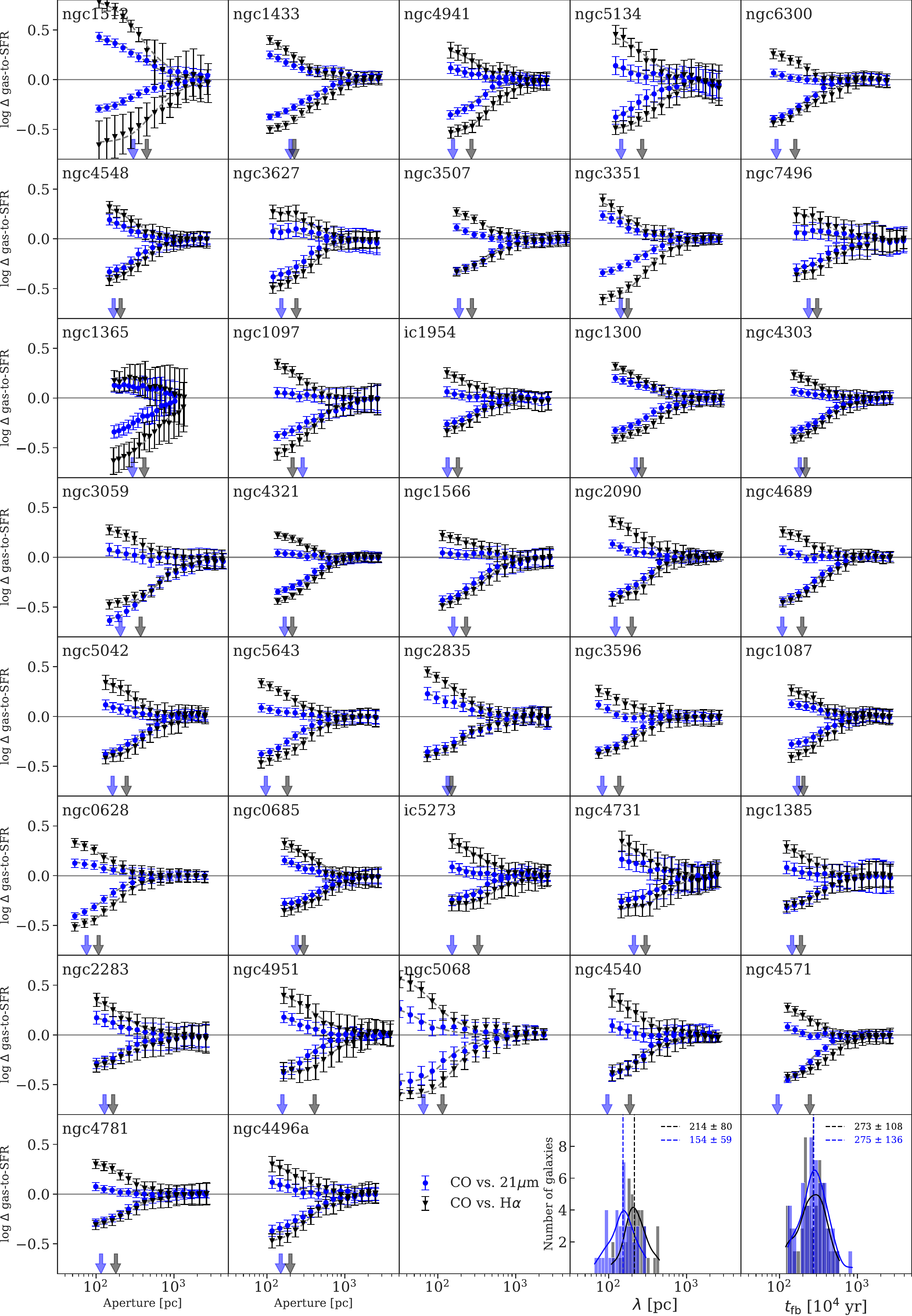}
\caption[]{Measured deviation of the gas-to-stellar flux ratio with respect to the galactic average as a function of the aperture sizes for each galaxy obtained by contrasting CO emission as a gas tracer respectively with H$\alpha$ (black triangles) and 21~$\mu$m (blue circles) as an SFR tracer. The positive deviations correspond to measurements focusing on gas peaks (traced by CO), while the negative deviations were obtained by focusing on stellar peaks (traced respectively by H$\alpha$ or 21~$\mu$m). For each data point, we also show the effective 1$\sigma$ error, after the covariance between data points is taken into account. The horizontal plain line corresponding to a deviation of zero in log represents the galactic average. The dotted gray lines connecting the measurements correspond to a polynomial fit of the tuning-fork branches, following \cite{Kruijssen_tf_2018}. The arrows indicate the typical separation length, $\lambda$, at which the two tracers decorrelate. The two last panels show the histograms of $\lambda$ and inferred feedback timescales ($t_{\rm fb}$, defined in Section \ref{subsec_measuring_timescales}), derived for the whole sample using either H$\alpha$ or 21~$\mu$m as a proxy for SFR, as well as the median and 1$\sigma$ standard deviations associated with these distributions.}
\label{tuning_fork_ha_21}
\end{figure*}

Figure~\ref{tuning_fork_ha_21} shows in blue the deviation of the CO-to-21~$\mu$m flux ratio from the galactic average ($\Delta$~CO-to-21~$\mu$m), when focusing either on gas peaks in the CO maps (leading to positive $\Delta$~CO-to-21~$\mu$m) or on the 21~$\mu$m peaks (leading to negative $\Delta$ CO-to-21~$\mu$m). For comparison, we also show in black the deviations obtained when focusing on CO versus H$\alpha$ peaks ($\Delta$ CO-to-H$\alpha$). We also show the analytical fits constraining the relative duration of the isolated phases and overlap phases for each pair of tracers, as well as the peaks separation lengths calculated for CO and 21~$\mu$m peaks (blue arrows) and for CO and H$\alpha$ peaks (black arrows). The tuning-fork shapes associated with 21~$\mu$m are strikingly flatter than the ones derived using H$\alpha$ as a tracer of SFR, in particular the top branch focusing on CO peaks. This flattening of the upper branch corresponds to a decrease of the \mbox{$t_{21~\mu m}$/$t_{\rm CO}$} ratio. Most of the tuning-fork shapes also display a shift of the inflection point, corresponding to a decrease of the typical separation length $\lambda$. This indicates that the peaks of emission are typically closer in the runs involving 21~$\mu$m and CO, than when using H$\alpha$.

As described in Section \ref{section_intro}, 21~$\mu$m and CO(2-1) both trace well molecular gas column density (see also \citealt{Leroy_2023_midIR_CO_Ha}), leading to stronger correlation between CO and 21~$\mu$m. This makes the application of the statistical method described in Section \ref{subsec_measuring_timescales} challenging, as robust timescale measurements can only be derived for galaxies where the resolution is small enough to resolve the decorrelation scale. Deriving a strict resolution threshold for all galaxies is impossible, since the decorrelation scale itself varies from galaxy to galaxy. As described in Section \ref{section_sample}, our selection criterion combines both a strict cut in resolution (180\,pc), corresponding to the 75\% percentile of the decorrelation scale between CO and 21~$\mu$m measured in our sample, and a criterion excluding galaxies displaying flat tuning-fork branches (see Appendix \ref{appendix_material}; Figure~\ref{tuning_fork_ha_21_all47}). We note that the resolution threshold adopted in the current study (180\,pc) is smaller than the one previously used in \citetalias{Kim_2022_environmental_dep} (200\,pc); this is motivated by the fact that 21$\mu$m and CO peaks are typically closer than H$\alpha$ and CO peaks, as shown by the histogram of decorrelation scales ($\lambda$) in Figure~\ref{tuning_fork_ha_21}.

Physically, the absence of decorrelation (flat upper branch of the tuning-fork) indicates that the brightest peaks of CO emission, which dominate the flux-weighted average, are all associated with 21~$\mu$m emission. We note that this does not necessarily correspond to an absence of isolated peaks of CO emission, but only translates the fact that these isolated peaks are faint and do not contribute much to the total CO emission, compared to the ones overlapping with 21~$\mu$m. For the latter galaxies, the decorrelation between 21~$\mu$m and CO - if present - occurs at scales smaller than our working resolution. In terms of derived quantities, this corresponds to $\lambda \leq l_{\rm ap, min}$, which cannot be properly fitted within our statistical framework (see criterion (ii) from  Appendix \ref{appendix_accurary}). We note that \cite{Kim_pah_2025} reported similar cases of nearly flat tuning-fork branches when using PAH-tracing bands, indicating that this effect is not limited to 21$\mu$m wavelength but might occur systematically when leveraging mid-IR tracer observed at high sensitivity. This limitation, inherent to our statistical method, is further discussed in Section \ref{subsect_limitations}.

\subsubsection{Timescales associated with \texorpdfstring{21~$\mu$m}{21 μm} emission}
\label{subsec_results_embedded_sf}

\begin{figure}
\centering
\includegraphics[width=\columnwidth]{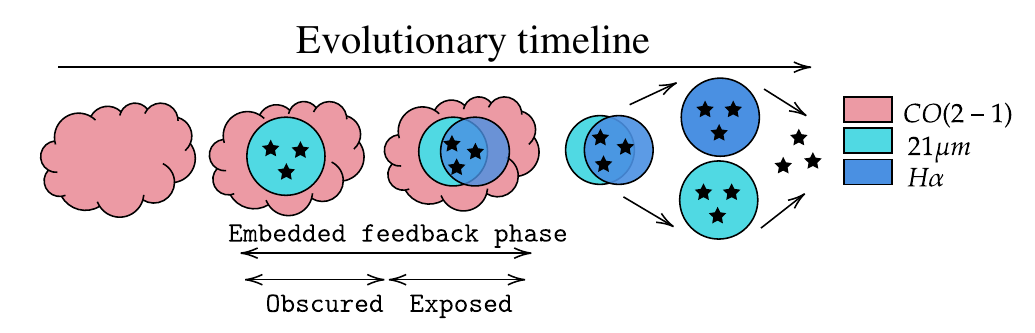}
\caption[]{Schematic evolutionary timeline of a star-forming region from the molecular cloud assembly until it becomes invisible in any of the three tracers we consider. In this study, we focus on the embedded feedback phase, during which stars are embedded within their host GMC. This phase is composed of an exposed phase, during which CO, 21$\mu$m, and H$\alpha$ emit simultaneously, and of an obscured phase, during which H$\alpha$ emission is not detected. After the GMC dispersal, we measured an isolated phase of emission for both SFR tracers before they fade away. For most galaxies, 21$\mu$m fades faster than H$\alpha$ (upper arrow), but a few galaxies display a longer 21$\mu$m emission (bottom arrow).}
\label{gmc_timeline}
\end{figure}

\begin{figure}
\centering
\includegraphics[height=\textheight - 87pt]{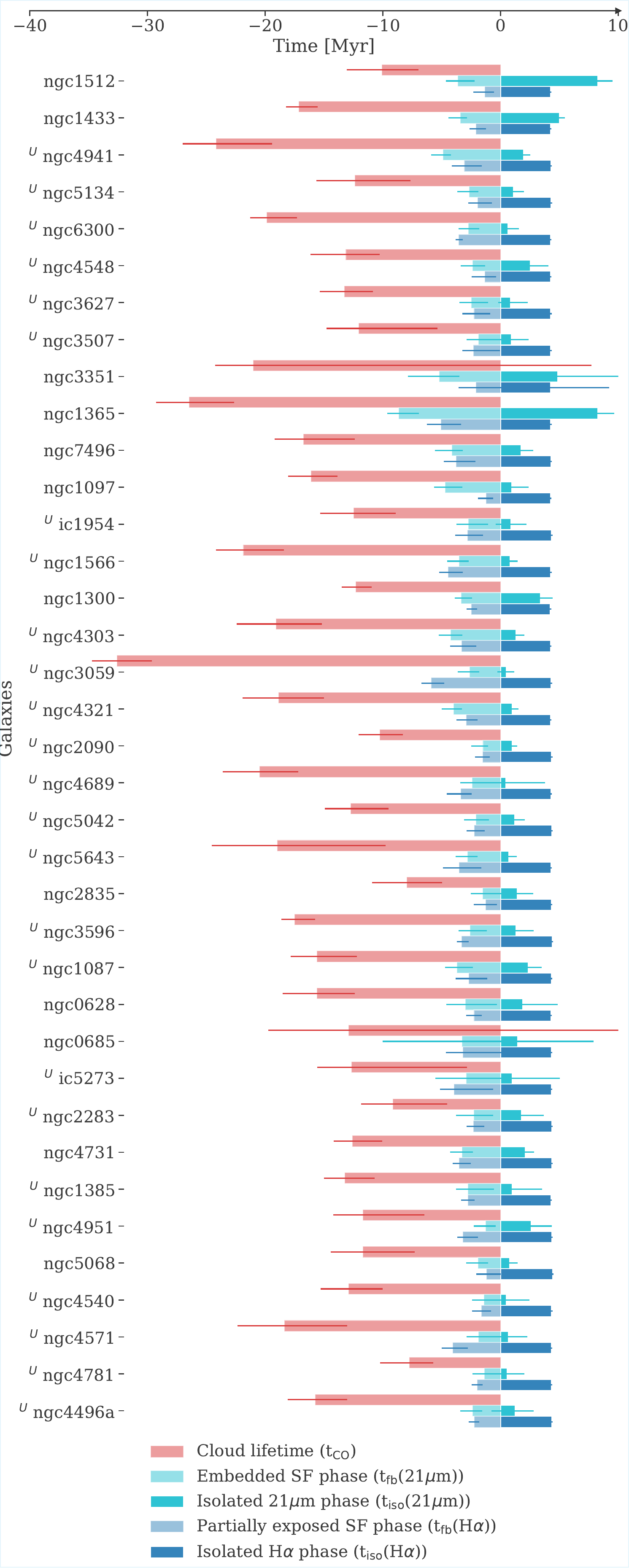}
\caption[]{Typical evolutionary timescales in our sample evolving from an inert CO-emitting phase to an embedded star-forming phase traced with 21~$\mu$m and eventually to an exposed star-forming phase traced with H$\alpha$ emission. The feedback timescale $t_{\rm fb,~21~\mu m}$ is an upper limit in 25 out of 37 galaxies, which are flagged with $U$. The upper and lower errorbars associated with each phase of the evolutionary cycle are shown as thin lines, with the same color. Galaxies are ordered by their Hubble morphological type.} 
\label{timescales}
\end{figure}

\begin{figure}[h!]
\centering
\includegraphics[width=\columnwidth]{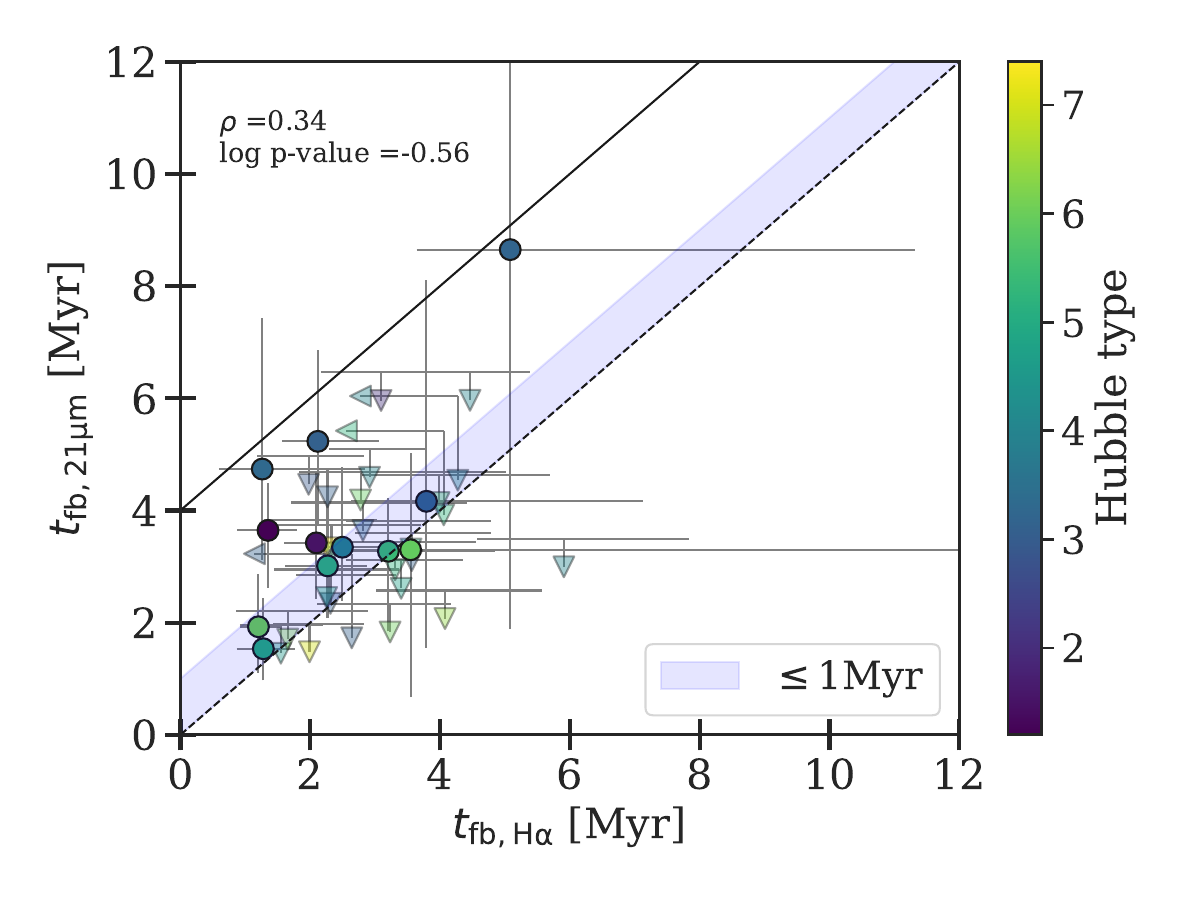}
\caption[]{Feedback timescale measured based on 21~$\mu$m emission versus feedback timescale measured based on H$\alpha$. The color bar shows the Hubble morphological type from the HyperLEDA database \citep{Paturel_hyperleda1_2003, Paturel_hyperleda2_2003, Makarov_HyperLEDA_2014}. The dotted line shows the 1:1 relation. The obscured feedback phase traced by 21~$\mu$m ($t_{\rm 21~\mu m}$ - $t_{\rm H\alpha}$) is less than 4\,Myr in all the galaxies from our sample, as shown by the plain black line, and less than 1\,Myr for 28 out of 37 galaxies, as shown by the shaded area. We display the Spearman correlation coefficient as well as the log p-value of the data, excluding upper limits.
}
\label{tfb_tha}
\end{figure}

In this section, we report the first measurements of the timescales associated with mid-IR 21~$\mu$m emission in a large subsample of the PHANGS-JWST survey. Similar measurements were previously accessible for six galaxies only, based on the observations of Spitzer/24~$\mu$m \citep{Kim_2021_embedded24_mic} and the analysis of MIRI/21~$\mu$m for NGC\,628 \citep{Kim_2023_ngc628}. It is now possible to carry out such analysis for a sample five times larger, for which we measure the timescales associated with the evolutionary cycle of star-forming regions, focusing in particular on the embedded feedback phase (see schematics, Figure~\ref{gmc_timeline}). In the Appendix \ref{appendix_material}, Table~\ref{table_outputs_21um}, we report the new measurements associated with the different phases for which 21~$\mu$m emission is observable, as represented in Figure~\ref{gmc_timeline}. The total duration associated with 21~$\mu$m emission ($t_{\rm 21~\mu m}$), the duration of the overlap phase between CO and 21~$\mu$m (i.e., the embedded feedback associated with 21~$\mu$m emission, $t_{\rm fb, 21~\mu m}$), and the duration of the obscured phase of star formation is defined by
\begin{equation}
    \rm t_{\rm obscured} [\rm Myr] = t_{\rm fb, 21~\mu m} - t_{\rm fb, H\alpha}, 
\label{t_obs}
\end{equation}
where $t_{\rm fb, H\alpha}$ corresponds to the exposed feedback phase, which is visible in optical tracers.

While we observed a decorrelation between CO and 21~$\mu$m in the 37 galaxies of our final sample, this decorrelation remains small for most of them (see Figure~\ref{tuning_fork_ha_21}). Specifically, for 25 out of the 37 galaxies, we derived $\lambda \leq 1.4 \times l_{\rm ap, min}$, which only enabled us to put an upper limit on the separation length, feedback timescale, and duration of the obscured star-formation phase. The temporal evolution of the assembly and dispersal of the clouds, when relying on CO and H$\alpha$ observations, has been characterized previously in \citetalias{Kim_2022_environmental_dep}. Focusing on a subset of their sample and considering the slightly smaller field of view corresponding to the JWST coverage, we repeat the same analysis and find medians for the cloud lifetime ($t_{\rm CO}$=$13.3^{+3.3}_{-2.7}$\,Myr) and the feedback phase associated with H$\alpha$ emission ($t_{\rm fb, H\alpha}$=$2.6^{+1.0}_{-0.8}$\,Myr), which are in good agreement with the previous measurements for the entire sample. In Figure~\ref{timescales}, we then combine these results with the ones presented in Table~\ref{table_outputs_21um} to show an overview of the entire star-formation cycle traced by CO, 21~$\mu$m and H$\alpha$. On average, we measure a total duration of the 21~$\mu$m emission of $t_{\rm 21~\mu m}$= $3.9^{+1.4}_{-0.9}$\,Myr (median of the measurements and of their errorbars across our sample, accounting only for the galaxies which are not affected by blending), a duration of the overlap phase between CO and 21~$\mu$m, $t_{\rm fb, 21~\mu m}$=$3.4^{+1.5}_{-1.2}$\, Myr, and a duration of the obscured phase of star formation, $t_{\rm obscured}$=$0.8^{+1.5}_{-0.8}$\, Myr.

In Figure~\ref{tfb_tha}, we show the evolution of the feedback time derived using the 21~$\mu$m band, as a function of the feedback time measured using H$\alpha$. We confirm that the feedback time measured using a proxy sensitive to dust-embedded star formation tends to be larger than the one measured based on H$\alpha$ tracing only exposed star formation. Nevertheless, we find the obscured feedback phase to be very short ($\leq$ 4\, Myr) in all the galaxies in our sample, with 28/37 galaxies displaying an almost absent ($\leq$ 1\,Myr) obscured star-formation phase. In practice, the only five galaxies (NGC\,1512, NGC1433, NGC\,1097, NGC\,3351, and NGC\,1365) where we clearly resolve a non-zero obscured phase are among the more massive, high-metallicity barred spirals (see Section \ref{subsect_discuss_morpht}, whereas in lower-mass systems the obscured phase is shorter than 1\,Myr. 

\subsection{Environmental dependencies of the timescales}
\label{subsec_statistics}

\subsubsection{Selected environmental properties}
\label{subsec_selected_params}

\begin{figure*}[h!]
\includegraphics[width=\textwidth]{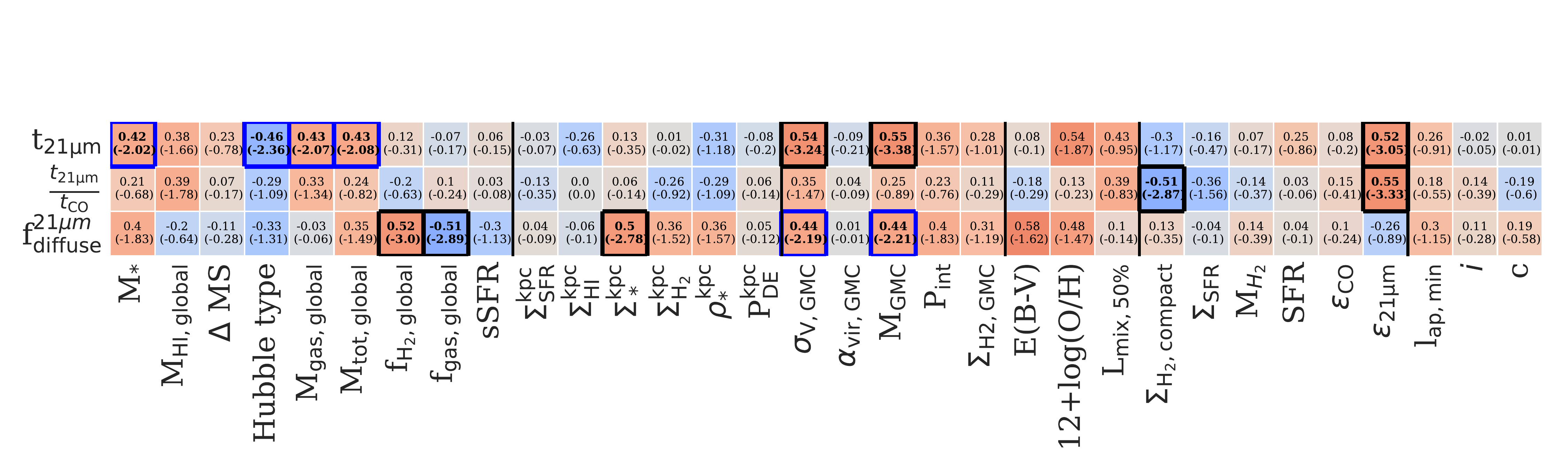}
\caption[]{Spearman's rank correlation coefficients and associated p-values measured between galaxy properties (columns) and our measurements (rows). Statistically significant correlations according to the Holm-Bonferroni method (described in Section~\ref{subsec_significance}) are highlighted as black squares, and marginally significant correlations ($\log p$-values < -2) are shown as blue squares. Our measurements are the total timescale of 21~$\mu$m emission ($t_{\rm 21~\mu m}$), the ratio between timescales of SFR and gas($t_{\rm 21~\mu m}$/$t_{\rm CO}$), and the diffuse emission fractions of 21~$\mu$m (f$_{\rm diffuse}^{21~\mu m}$). We correlate these measurements with various parameters grouped in six categories, described in \ref{subsec_selected_params}, along with the corresponding references.}
\label{correlation}
\end{figure*}

We now examine how various quantities predicted by our models correlate with different galactic parameters, listed in Figure~\ref{correlation}. In this section, we restrict our analysis to the parameters that are well constrained for all 37 galaxies from our sample (i.e., not just upper limits). We derived robust measurements in the whole sample of the total duration of the 21~$\mu$m phase, $t_{\rm 21~\mu m}$ and of the diffuse fraction of 21~$\mu$m emission, which we discuss in sections \ref{subsubsec_total_21} and \ref{subsec_diffuse_21}. We defer the description of potential trends of $\lambda_{\rm fb, 21~\mu m}$, $t_{\rm fb, 21~\mu m}$, and $t_{\rm obscured}$, to the discussion section (Section \ref{section_discussion}) due to the large number of upper limits. 

In Figure~\ref{correlation}, we show the Spearman rank coefficients ($\rho$) and the associated $p$-values between our predictions and numerous parameters. For the parameters that we considered, we report no strong ($\rho \geq 0.6$) correlation, but there are several moderately strong ($\rho \geq 0.4$) correlations, whose significance we examined based on their p-value (see Section \ref{subsec_significance}). In the following, we briefly introduce the parameters considered in our analysis, which are organized in six categories.

\begin{itemize}

\item Global galactic parameters.
We considered global galactic properties of our sample gathered from different studies. We include the stellar masses (M$_{*}$) taken from \cite{Leroy_CO_phangs_2021}, the global \hi\ mass (M$_{\rm HI, global}$) from the Lyon-Meudon Extragalactic Database (HyperLEDA; \citealt{Paturel_hyperleda1_2003, Paturel_hyperleda2_2003, Makarov_HyperLEDA_2014}), the deviation from the main sequence \citep{Leroy_CO_phangs_2021}, and the Hubble morphological T-type from the HyperLEDA. Following \citetalias{Kim_2022_environmental_dep}, we also consider combinations of the later parameters, namely the global gas mass (M$_{\rm gas, global}$ = M$_{\rm HI, global}$ + M$_{\rm H_2, global}$), the total baryonic mass (M$_{\rm tot, global}$ = M$_{\rm gas, global}$ + M$_{*}$), the molecular gas fraction (f$_{\rm H_2, global}$ = M$_{\rm H_2, global}$/M$_{\rm gas, global}$), the global gas mass fraction (f$_{\rm gas, global}$ = M$_{\rm gas, global}$/M$_{\rm tot, global}$), and sSFR = SFR/M$_{*}$.

\item Average kiloparsec-scale galaxy properties.
We also considered kiloparsec-scale galaxy properties from \cite{Sun_prop_clouds_2022} and \cite{Sun_2023_calibration}. Specifically, we consider the surface densities of the SFR, atomic gas, molecular gas, and stellar mass ($\Sigma^{\rm kpc}_{\rm SFR}$, $\Sigma^{\rm kpc}_{\rm HI}$, $\Sigma^{\rm kpc}_{\rm H_2}$, and $\Sigma^{\rm kpc}_{*}$ respectively). We also include the stellar mass volume density near the galaxy midplane ($\rho^{\rm kpc}_{*}$) and the dynamical equilibrium pressure (P$^{\rm kpc}_{\rm DE}$), which is the pressure required to support the ISM weight (per unit area) within the gravitational potential of the galaxy \citep[see e.g.,][]{Sun_dyn_eq_2020}. For each of the aforementioned quantity, we calculate a $\Sigma^{\rm kpc}_{\rm H_2}$-weighted average per galaxy.

\item Averaged properties of GMCs.
To examine potential variations of our predictions with properties of molecular clouds, we also consider properties derived at the GMC scale for the full PHANGS–ALMA sample using the \textsc{Cprops} algorithm in \cite{Rosolowsky_gmc_2021} and Hughes et al. (in prep.). The latter include the velocity dispersion ($\sigma_{\rm V,GMC}$), the virial parameter ($\alpha_{\rm vir, GMC}$), the GMC mass (M$_{\rm GMC}$), the internal pressure (P$_{\rm int}$), and the H$_2$ surface density of GMCs ($\Sigma_{\rm H2, GMC}$). For each of these quantities, we follow the same methodology as in \citetalias{Kim_2022_environmental_dep} and calculate a CO-luminosity-weighted average over the population of clouds in each galaxy.

\item Dust and ionized gas parameters.
Due to the complex nature of the emission at 21~$\mu$m, we may expect dependencies not only with the parameters related to molecular gas, but also with the properties of dust and ionizing sources. For galaxies for which MUSE observations are available (20/37), we also include parameters probing the properties of the ionized gas in \hii\ regions. For these galaxies, we consider the dust attenuation and gas-phase metallicity provided in the \hii\ regions catalog from \cite{Groves_hii_2023}. For both quantities, we consider $\Sigma_{\rm SFR}$-weighted average, excluding the galactic centers (as described in Section \ref{subsec_muse_met_maps}). Finally, we include the 50\% metal-mixing scale (L$_{\rm mix, 50\%}$) from \cite{Kreckel_mixing_2020} and \cite{Williams_2022_mixing_met}, a parameter that quantifies the physical scale at which the metal-mixing in the ISM is effective, using a two-point correlation function.

\item Parameters derived with \textsc{Heisenberg}.
We also include average properties derived by \textsc{Heisenberg} within the masks described in Section \ref{subsect_data_homogenization}. Our analysis includes the molecular gas surface density derived after the filtering of the CO maps ($\Sigma_{\rm H_2, compact}$), the SFR surface density ($\Sigma_{\rm SFR}$; see Section \ref{subsec_tot_sfr}), the total molecular mass associated with compact CO emission and the total SFR recalculated within the field of view of the observations. We also consider the surface density contrasts, which correspond to the flux-ratio between the average emission of the peaks and the galactic average, measured on the filtered map for CO and 21~$\mu$m respectively: $\mathcal{E}_{\rm CO}$ and $\mathcal{E}_{\rm 21~\mu m}$. 

\item Systematic biases.
To check for the potential effects of systematic biases, we complemented the list of physical properties with some important observational features: the minimal aperture size, corresponding to the working resolution of CO observations ($l_{\rm ap, min}$), which is used to convolve all maps (see Section~\ref{subsect_data_homogenization}); the inclination angle of galaxies ($i$) taken from \cite{Lang_rot_curve_2020}; and the mass-weighted average of CO flux completeness on kiloparsec scale (c; \citealt{Sun_prop_clouds_2022}), providing a metric of how deep the CO observation is. A deeper observation leads to a more complete CO flux recovery when a strict signal identification criteria is adopted.
\end{itemize}

\subsubsection{Statistical framework}
\label{subsec_significance}

Following the methodology developed in \citet{Kruijssen_EMOSAICS_2019} (and also used by \citetalias{Kim_2022_environmental_dep}), we ranked each correlation according to their $p$-values for each row in Figure~\ref{correlation}. We note that several of these correlations have $p$-values below the usual significance threshold (hereafter p$_{\rm ref}$) of 0.05. Due to the large number of parameters considered in this study, one might expect a certain number of spurious correlations which should not be considered as significant. Hence, we further determine whether these correlations are statistically significant according to the Holm-Bonferroni method \citep{holm_1979}. This method penalizes the second-order correlations by selecting only correlations which have a $p$-value below a rank-dependent threshold computed as
\begin{equation}
    p_{\rm eff} = \frac{p_{\rm ref}}{N_{\rm corr} + 1 - i},
\end{equation}
where $i$ is the rank of correlation and N$_{\rm corr}$ the number of independent variables being evaluated. Following \citetalias{Kim_2022_environmental_dep}, we treat parameters with a correlation coefficient higher than 0.7 as one single variable, resulting in an estimate of N$_{\rm corr}= 17$.

We note that the Holm-Bonferroni method is by design relatively conservative, since it aims at narrowing the number of apparent correlations to the most significant ones. If we were to relax our significance threshold, other correlations would become significant. To account for this effect, we include in our analysis correlations which we define as “marginally significant”; the latter being associated with $p$-values lower than 0.01, despite not being selected through the Holm-Bonferroni method. 

\subsubsection{Duration of \texorpdfstring{21~$\mu$m}{21 μm} emission}
\label{subsubsec_total_21}

\begin{figure*}[h!]
\centering
\includegraphics[width=\columnwidth]{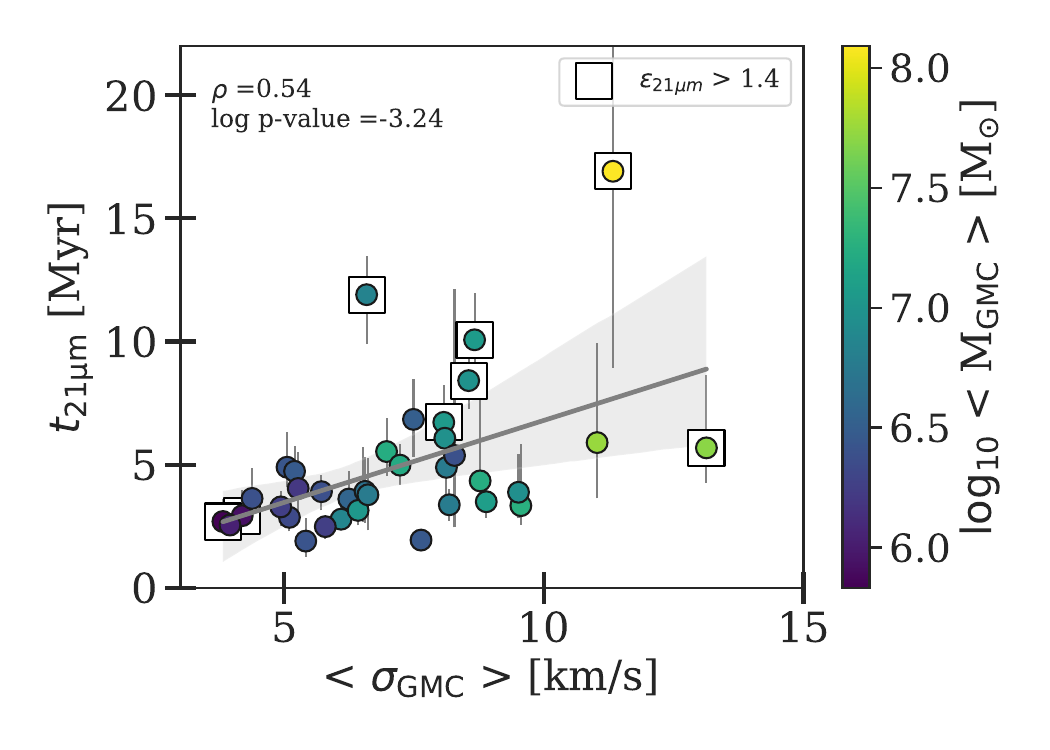}
\includegraphics[width=\columnwidth]{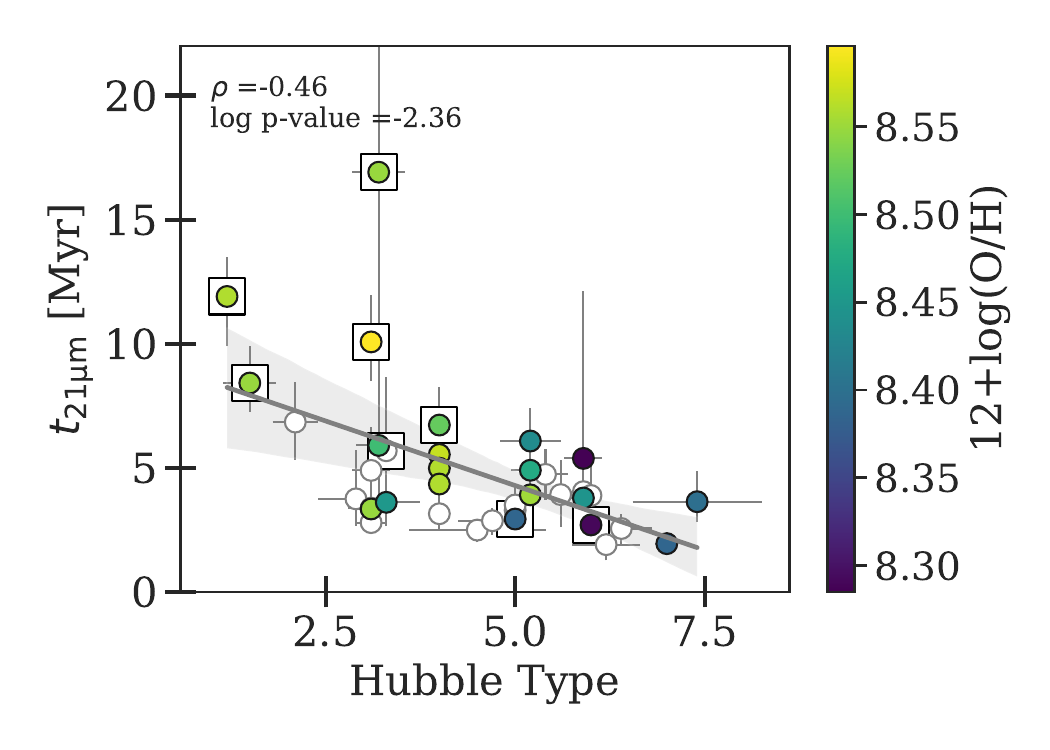}
\caption[]{Dependencies of the total duration of the 21~$\mu$m emitting phase ($t_{\rm 21~\mu m}$). \textbf{Left}: Total duration of the 21~$\mu$m vs. CO-luminosity-weighted average velocity dispersion of GMCs. The color bar shows the CO luminosity-weighted average mass of GMCs. Galaxies with high surface density contrasts are identified with squares. We show in gray a linear regression fitted to the data and the gray-shaded area represents the 95\% confidence interval on the regression, obtained with bootstrapping data. \textbf{Right}: Total duration of the 21~$\mu$m vs. the Hubble morphological type. The color bar shows the metallicity measurements for the galaxies observed with MUSE.}
\label{plot_corr_t21}
\end{figure*}

\begin{figure}[h!]
\centering
\includegraphics[width=\columnwidth]{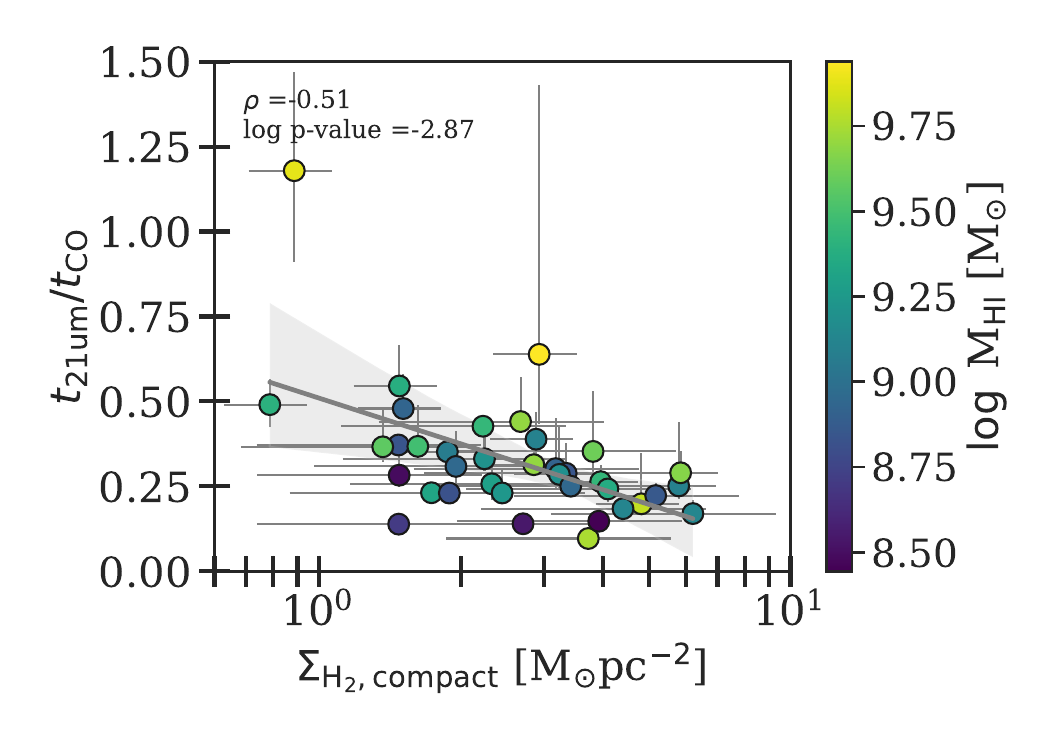}
\caption[]{Ratio of the total duration of the 21~$\mu$m emitting phase ($t_{21~\mu m}$) over the total cloud lifetime ($t_{\rm CO}$) vs. H$_{2}$ surface density estimated after filtering the diffuse CO emission. The color bar shows the total \hi\ mass.}
\label{plot_corr_t21_co}
\end{figure}

In Figure~\ref{plot_corr_t21}, we show the most significant correlations derived for the total duration of 21~$\mu$m emission, $t_{\rm 21~\mu m}$. The left-hand panel of Figure~\ref{plot_corr_t21} displays the evolution of $t_{\rm 21~\mu m}$ as a function of the CO luminosity-weighted average of the velocity dispersion ($\langle \sigma_{\rm GMC} \rangle$) and molecular mass of molecular clouds ($\langle {\rm M}_{\rm GMC} \rangle$). Both parameters correlate with $t_{\rm 21~\mu m}$ (moderately strong correlations with $\rho \sim 0.5$) and are significant according to the Holm-Bonferroni method. These correlations indicate that the duration of the 21~$\mu$m emission is longer in galaxies hosting populations of massive CO-bright molecular clouds, associated with large velocity dispersion. We note that $\langle \sigma_{\rm GMC} \rangle$ and $\langle {\rm M}_{\rm GMC}\rangle$ are correlated parameters (Spearman coefficient \mbox{$\rho = 0.84$}, \mbox{$\log p$-value = $-8.48$}), an effect which cannot be disentangled within our statistical framework. We highlight in Figure~\ref{plot_corr_t21} the galaxies showing high surface density contrast $\mathcal{E}_{\rm 21~\mu m}$, defined as the peak to galactic average flux ratio. Despite moderate variations of this parameter ($1 \leq \mathcal{E}_{\rm 21~\mu m} \leq 2.7$), we find that the galaxies with the highest flux density contrasts tend to show enhanced duration of the 21~$\mu$m emission. We further discuss the physical mechanisms that could be responsible for this enhanced duration of 21~$\mu$m emission in Section \ref{subsect_discuss_gmc}.

While the Holm-Bonferroni method only selects parameters related with local properties of GMCs, we note that the total duration of 21~$\mu$m emission may also be linked to global galactic properties. In particular, the Hubble type is identified as marginally significant correlation ($p$-value $\leq$ 0.01). In the right hand-side panel of Figure~\ref{plot_corr_t21}, we show that the duration of 21~$\mu$m emission anti-correlates with the Hubble morphological type. This anti-correlation includes some significant scatter, and its analysis is complicated by the fact that standard statistical metrics do not strictly apply for discrete variables, such as the Hubble type. Nevertheless, qualitatively, we find that earlier-type galaxies hosting a bar and well-defined spiral arms, are associated with $t_{21~\mu m}$ spanning a large range of durations, up to $\sim$15\, Myr, while later-type galaxies (less massive spirals and irregular dwarf galaxies) are associated with reduced $t_{\rm 21~\mu m}$. We speculate that this trend may be linked to the reduced dust-content in late-type, low-metallicity environments, as we discuss further in Section \ref{subsect_discuss_morpht}.

In Figure~\ref{plot_corr_t21_co}, we examine the dependencies of the ratio of $t_{\rm 21~\mu m}$ to $t_{\rm CO}$. We note that the $t_{\rm 21~\mu m}$/$t_{\rm CO}$ is smaller than $0.5$ in all galaxies in our sample apart from three, meaning that molecular clouds spend a significantly longer time (at least a factor of two) emitting CO than emitting 21~$\mu$m. As demonstrated in \citetalias{Kim_2022_environmental_dep}, galaxies with high molecular gas surface densities are associated with typically higher $t_{\rm CO}$. We find on the other hand that $t_{21~\mu m}$ is relatively insensitive to H$_2$ gas surface density (weak anti-correlation with a Spearman rank coefficient of $-0.3$, see Figure~\ref{correlation}). As a result, we report a significant anti-correlation of $t_{\rm 21~\mu m}$/$t_{\rm CO}$ ratios with $\Sigma_{\rm H_2, compact}$ (molecular surface density calculated after filtering out the diffuse emission), shown in Figure~\ref{plot_corr_t21_co}. Part of the scatter of this relation may be explained by differences in terms of atomic gas reservoir, with larger \hi\ masses at fixed $\Sigma_{\rm H_2,  compact}$ being associated with large $t_{\rm 21~\mu m}$/$t_{\rm CO}$ ratios (as shown by the color bar). 

\subsubsection{Fraction of diffuse \texorpdfstring{21~$\mu$m}{21 μm} emission}
\label{subsec_diffuse_21}

\begin{figure*}[h!]
\centering
\includegraphics[width=\columnwidth]{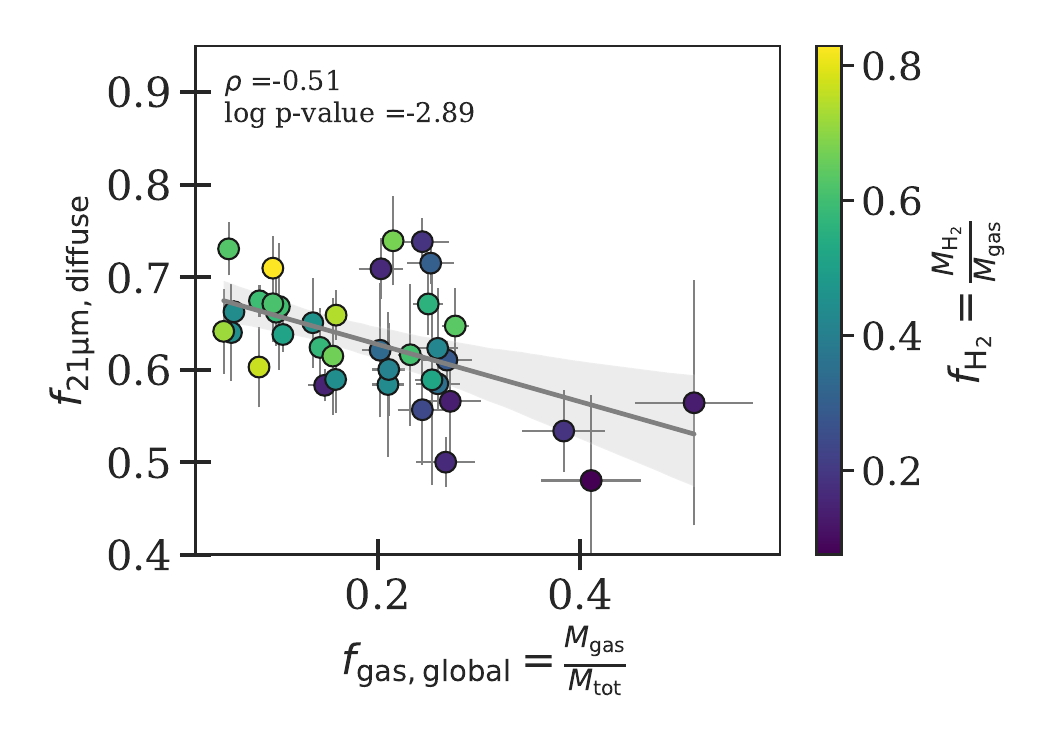}
\includegraphics[width=\columnwidth]{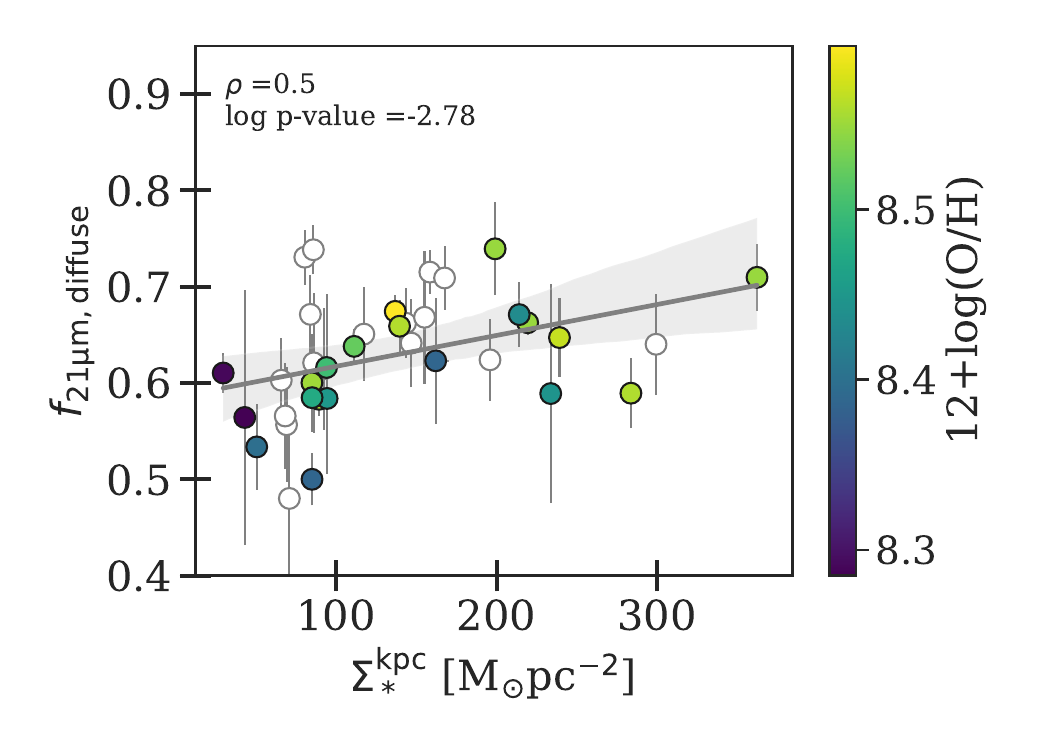}
\caption[]{Dependencies of the diffuse fraction of 21~$\mu$m emission ($f_{21~\mu m, \rm diffuse}$). \textbf{Left}: Diffuse fraction of 21~$\mu$m emission as a function of the global gas mass fraction. The color code shows the H$_2$ gas mass fraction for each galaxy. \textbf{Right}: Diffuse fraction of 21~$\mu$m emission as a function of the kiloparsec-scale stellar surface density. The color bar shows the metallicity.}
\label{plot_corr_fdig21}
\end{figure*}

In Figure~\ref{plot_corr_fdig21}, we show that the fraction of diffuse emission of 21~$\mu$m (f$_{21~\mu m, \rm diffuse}$) is relatively constant and comprised between 48\% and 74\% in our sample. These values are globally in good agreement with previous determinations reported for PHANGS-JWST Cycle 1 galaxies \citep{Belfiore_2023b_sfr_midIR, Leroy_2023_midIR_CO_Ha, Pathak_2024_midIR_pdf, Kim_pah_2025}. In particular, they are similar to the ones reported in \cite{Kim_pah_2025} using the same method as in the current study for the other mid-IR bands studied in a subsample of 17 galaxies, in which they find the diffuse fraction of emission to range between $32-79$\%, $34-81$\%, and, $28-80$\% for 7.7, 10, and 11.3~$\mu$m respectively, with an average and 1$\sigma$ range for $f_{21~\mu m, \rm diffuse}$=62$\pm$4\%, very close to the value of 60$\pm$10\% reported in the latter study as an average for the three PAH-tracing bands.

As shown in Figure~\ref{plot_corr_fdig21} (left panel), we find a significant anti-correlation of $f_{\rm 21~\mu m, diffuse}$ with the global gas fraction, mostly driven by the three galaxies with the highest gas fraction in our sample, for which $f_{\rm 21~\mu m, diffuse}$ appears clearly reduced. Focusing on the color bar, we note that the most gas-rich galaxies in our sample associated with lower $f_{\rm H_2}$, may reach smaller $f_{21~\mu m, \rm diffuse}$ values as low as $\sim$50\%. This suggests that the contribution of diffuse 21$\mu$m dust emission is reduced in galaxies with high gas fractions, for which the inner regions covered in our analysis are atomic-dominated (as shown by the color bar). In the right-hand side panel of Figure~\ref{plot_corr_fdig21}, we show that the latter gas-rich galaxies, correspond to relatively smaller stellar surface densities and lower-metallicity environments. In such galaxies, a greater share of dust heating is localized in \hii\ regions, while more massive, dust-rich galaxies are associated with a stronger diffuse IR component. 

\section{Discussion}
\label{section_discussion}

\subsection{The short duration of the obscured phase of star formation}
\label{subsec_discussion_tobs}

\subsubsection{Comparison with previous studies using \texorpdfstring{21~$\mu$m}{21 μm}}
As described in Section \ref{subsec_results_embedded_sf}, we found that the total duration of the 21~$\mu$m emission ranges between 1.9 and 17\,Myr with a median across our sample of $3.9^{+1.4}_{-0.9}$\,Myr. We measure embedded-feedback timescales, traced by the overlap between CO and 21~$\mu$m, which are always smaller than 9\,Myr, with a median of $3.4^{+1.5}_{-1.2}$\,Myr. We find that the feedback timescales measured with 21~$\mu$m are close to the ones measured based on H$\alpha$ and report a duration of the obscured star-formation phase which is $\leq 4$\,Myr in all the galaxies of our sample, with a median of $0.8^{+1.4}_{-0.8}$\,Myr. These short timescales reinforce the idea that star formation proceeds over a relatively small fraction of the total GMC lifetime (typically, $t_{\rm fb,21~\mu m}$/$t_{\rm CO} \sim 1/5$), even when including the obscured phase of star-formation, which may limit the star-formation efficiencies in GMCs down to a few percent only (e.g., $1-8$\%, \citetalias{Kim_2022_environmental_dep}, \citealt{Chevance_pp7_review_2023}).

This short duration of the obscured star-formation phase is consistent with previous studies focusing on the emission at 21~$\mu$m in the PHANGS-JWST sample. In particular, \cite{Hassani_2023} and \cite{Hassani_2025} classified $>20\,000$ compact 21~$\mu$m sources in the 19 first PHANGS-JWST galaxies using MIRI/21~$\mu$m data. Their results indicate that deeply embedded star-forming regions, that are faint or undetected in H$\alpha$, account for a limited fraction ($\lesssim$10\%) of the total 21~$\mu$m point-source population, while most of the sources are clearly associated with optical counterpart of \hii\ regions. Our results are also consistent with the findings from \cite{Belfiore_2023a_sfr_phangsMUSE} reporting that the mid-IR contamination from deeply embedded star forming region is negligible. Within the framework of our analysis, we suggest that the low number of such deeply embedded regions is representative of their very transient nature, lasting on average less than 4\,Myr. We note that the value previously reported for NGC\,628 in \cite{Kim_2023_ngc628}, following the exact same methodology, falls within the range of new values reported in the current paper ($t_{\rm obscured}$ = 2.3$^{+2.7}_{-1.4}$\,Myr). 

These new findings made possible by JWST capabilities are consistent with previous results based on mid-IR observations performed with \textsc{Spitzer}/24~$\mu$m which reported small fractions of highly obscured regions without H$\alpha$ counterpart \citep[e.g.,][]{Kennicutt_sings_2003, Prescott_obscured_2007, Corbelli_2017}. The same statistical method as the one presented in this study was also applied using 24~$\mu$m as a tracer of embedded star formation in a distinct sample of six nearby galaxies in \cite{Kim_2021_embedded24_mic}. The authors estimated obscured timescales of $1.4 - 3.8$\,Myr, in excellent agreement with the current study ($0 - 3.5$\,Myr). We note, however, that they report a median value of $t_{\rm obscured}$ = 3.0$\pm$0.9\,Myr, which is significantly higher than the one reported in our study, in which most galaxies are consistent with $t_{\rm obscured} \leq 1$\,Myr. Similarly, the values derived for $t_{\rm 24~\mu m}$ and $t_{\rm fb,24~\mu m}$ are compatible in terms of range, but with larger median values than what we measure with 21~$\mu$m. This difference between both studies is surprising, considering that the methodologies adopted in both studies are similar regarding all the aspects that could affect timescale measurements (e.g., masking schemes, filtering of diffuse emission, peak identifications, and reference timescales). Nevertheless, they may arise from biases due to different sample selection, since \cite{Kim_2021_embedded24_mic} is restricted to six galaxies which are mostly lower masses (e.g., M\,33, NGC\,300, and LMC) and located within a maximum of 10\,Mpc. We speculate that these differences could also be attributed to the significant differences in terms of sensitivity, with MIRI being up to a factor of 50 higher in terms of point-source sensitivity compared to MIPS.  

\subsubsection{Link with young embedded stellar populations}
\label{sect_cluster_studies}
Short durations of the embedded phase of star formation are also measured in studies that focus on inferring molecular gas clearing timescales based on the distribution of young stellar populations. Using visual inspection of young stellar cluster (YSCs) in nearby galaxies, several studies reported that the median ages of cluster associated with dusty molecular clouds are short, of the order of $1-4$\,Myr \citep[e.g.,][]{Whitmore_antennae_2014, Calzetti_2015, Hollyhead_2015, Grasha_2018, Grasha_2019}. Similarly short gas-clearing timescales are reported by studies classifying YSCs based on the H$\alpha$ morphology of their surrounding \hii\ regions \citep{Hannon_2019, Hannon_2022}. These gas- and dust-clearing timescales are sensitive to the tracers used to select YSCs; in particular it has been suggested that accounting for the deeply embedded clusters which are only visible in the near-IR could enhance the duration of the embedded phase with respect to measurements based on optical and UV tracers only. The population of embedded clusters, which are missed in the optical, have been shown to represent up to $10-15$\% of the cluster populations in well-studied nearby galaxies \citep[e.g.,][]{Whitmore_antennae_2002, Whitmore_ngc1365_2023, Whitmore_2025}.

Whether this population of previously missed IR clusters significantly enhance the measured embedded timescales remain to be elucidated. Focusing on IR-selected clusters based on Hubble/F336W in M83, \cite{Deshmukh_clearing_2024} measured equally short clearing timescales as in previous studies with 75\% of their IR-selected sources having a dust attenuation $A_V < 1$ by 2\,Myr, and 82\% by 3\,Myr. On the other hand, \cite{Messa_2021} report clearing timescales that are $\sim$1\,Myr longer than previously estimated In NGC\,1313 based on near-UV-optical cluster studies. Similarly, in the dwarf galaxy NGC\,4449, \cite{McQuaid_2024} recently measured slightly higher clearing timescales (5-6\,Myr) and smaller median cluster mass (\mbox{$\sim 7\times 10^3\,M_{\odot}$}) than the values reported for UV/optical catalogs (5\,Myr, and \mbox{$\sim 3.5 \times 10^4\,M_{\odot}$}). They interpret these differences as evidence that clearing timescales depend on the cluster mass, due to reduced efficiency of both pre-supernova and supernovae feedback in lower mass clusters, that can only be probed in the IR.

Complementing these studies, the recent JWST observations enable measurements of the emergence timescale of young stellar clusters via numerous independent techniques, including statistics on different classes of objects, age estimates derived by fitting the spectral energy distributions (SEDs), and age-dating with Pa$\alpha$ equivalent width. Despite differences in the methodologies adopted in different studies, the reported embedded timescales appear consistently short ($\lesssim$ 5\, Myr) across different types of environments (e.g., \citealt{Sun_ring_2024, Whitmore_ngc1365_2023, Linden_ngc3256_2024, Pedrini_feast_2024, Rodriguez_3p3_2024, Knutas_2025, Graham_pah_2025}). These age estimates are in globally good agreement with the embedded feedback timescales reported in our study (with a median across our sample of $t_{\rm fb,21~\mu m}=3.4^{+1.5}_{-1.2}$\,Myr). We note that our luminosity-weighted method bias our results toward the more massive star-forming regions, that may not be representative of the evolution timeline associated with clusters of smaller masses. Recently, \cite{Knutas_2025} reported evidence that the emergence of massive star cluster ($ >5 \times 10^3$M$_{\odot}$) is $\sim$2~Myr shorter than for cluster of masses $\sim 10^3$~M$_{\odot}$ in M\,83.

The mid-IR 3.3\,$\mu$m emission has also been used to trace the survival timescale of PAHs in the earliest star-formation phase, corresponding to our dust-obscured feedback phase. Recent studies report that PAH emission fades within $\sim$3-4\,Myr, consistent with the picture of a very transient obscured star-formation phase \citep{Linden_ngc3256_2024,Rodriguez_3p3_2024, Knutas_2025, Whitmore_2025}. Focusing on NGC\,628, \cite{Whitmore_2025} compared the SEDs of nearly embedded cluster candidates (strong PAH and continuum emitters with a faint optical counterpart) from \cite{Rodriguez_3p3_2024} and \cite{Hassani_2023} to empirical based SED template based on optically identified clusters, and found that their IR SED closely resemble that of optically selected cluster with age 1 to 3\,Myr. Focusing on the 12 nearly embedded clusters from \cite{Hassani_2023} identified with large-aperture photometry at 21$\mu$m, the authors report an age of about 1\,Myr, younger than most of their optically selected clusters and in good agreement with the dust-obscured timescales that we derive in the current study. 

\subsection{Parameters affecting the emission of \texorpdfstring{21~$\mu$m}{21 μm}}
\label{subs_pressure}
\subsubsection{Impact of the average GMC properties}
\label{subsect_discuss_gmc}

\begin{figure}[t]
\centering
\includegraphics[width=\columnwidth]{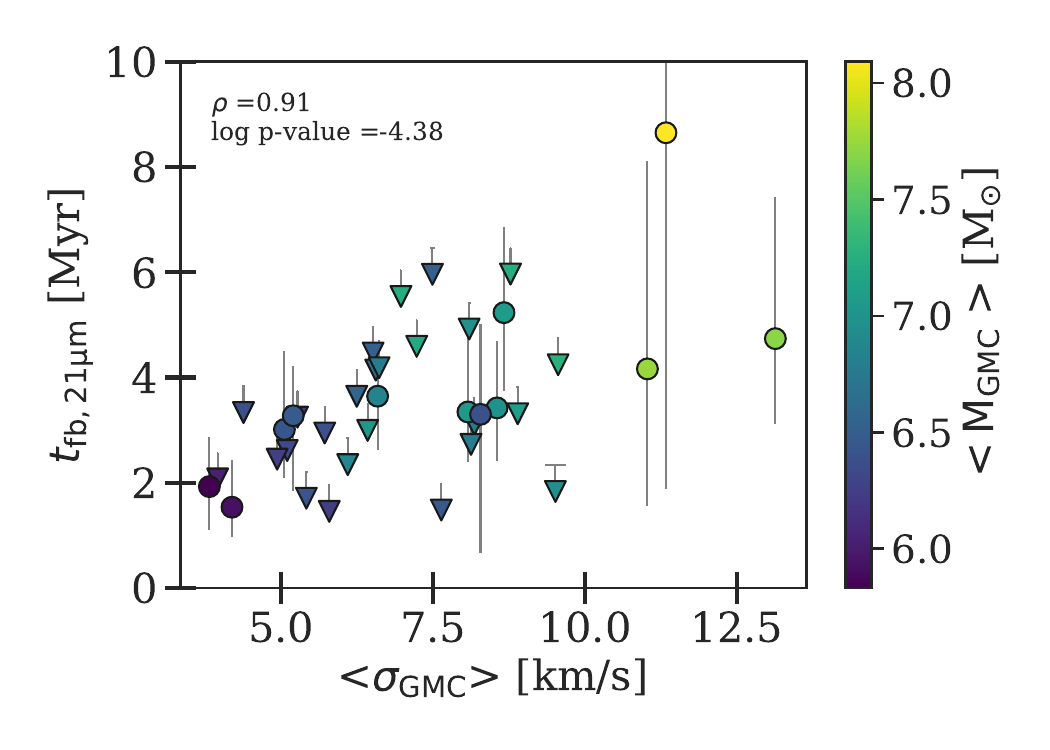}
\caption[]{Duration of the feedback time ($t_{\rm fb,21~\mu m}$) as a function of the CO luminosity-weighted average velocity dispersion of molecular clouds. The color bar shows the CO luminosity-weighted average mass of GMCs.} 
\label{tfb21_vs_gmc}
\end{figure}

As previously discussed in Section \ref{subsec_statistics}, the total duration of the 21~$\mu$m emission, $t_{\rm 21~\mu m}$ is relatively insensitive to global galactic parameters (e.g., SFR, gas and star masses, and surface densities). On the other hand, we find that among the moderately strong correlations of identified in Section \ref{subsubsec_total_21}, the most significant correlation is with the CO luminosity-weighted average of local properties of GMCs measured at 150\,pc, in particular the velocity dispersion and mass of GMCs (see Table~\ref{correlation}). Despite the numerous upper limits, we show on  Figure~\ref{tfb21_vs_gmc} that the feedback timescale measured based on 21~$\mu$m also seem to increase with increasing velocity dispersion and mass of GMCs (the table of correlations derived for these 12 measurements is provided in Appendix~\ref{appendix_material}; Figure~\ref{correlation_fb}). 

Although we find that $t_{\rm 21~\mu m}$, and tentatively $t_{\rm fb, 21~\mu m}$, correlate best with GMC mass and velocity dispersion, these GMC properties themselves correlate with galaxy-scale quantities (stellar mass, molecular gas mass, SFR gas surface density, as well the mid-plane dynamical equilibrium pressure; \citealt{Sun_prop_clouds_2022}). Thus, what we are really seeing could be a reflection of high-pressure environments (found in massive galaxies or galaxy centers) yielding both more massive GMCs and slightly longer embedded phases.

The latter correlations with physical properties of GMCs have important implications for high-redshift galaxies, which host populations of higher-pressure GMCs with higher velocity dispersion \cite[e.g.,][]{Claeyssens_jwstclouds_2023} and masses up to $\sim$100 times larger at $z\sim1$ \citep[e.g.,][]{Dessauges-Zavadsky_snake_2019, Dessauges-Zavadsky_z1_clouds_2023}. Such GMCs are the likely progenitors of globular clusters \citep[e.g.,][]{Kruijssen_2025}. Our results suggest that in such environments, both the total duration of the 21~$\mu$m emission and the feedback timescales could be increased, as star formation may proceed over a longer period before the molecular clouds are disrupted. This result also indicates that the short GMC surviving timescale and rapid cycling of matter derived for nearby galaxies may only hold for nearby galaxies but should be revised for any galaxies in which we expect a significantly different population of molecular clouds. We note, however, that the timescales associated with 21$\mu$m emission do not directly reflect the presence/disruption of gas and dust, but that the observed 21$\mu$m reflects the coupling of dust grains with UV radiation fields. As a result, it may be difficult to make predictions for environments where we expect both the dust properties and heating mechanisms to vary.

Disentangling the underlying physical mechanisms responsible for the emission of a given tracer is not easy with observations. State-of-the-art simulations, such as
STARFORGE \citep{starforge_grudic_2021}, that implement various stellar feedback mechanisms can provide a complementary view on the evolutionary cycle at GMC scales. Using STARFORGE simulation, post-processed in the mid-IR (e.g., 24$\mu$m), \cite{Wainer_embedded_2025} examine the embedded phase of star formation and its observational signatures. Their predicted timescales are in good agreement with the measurements of a few Myr from the current study, indicating that the physical mechanisms implemented in their simulations are compatible with what is probed observationally in nearby galaxies. Interestingly, the authors also report an increased duration of the embedded phase with the cloud free fall time. They attribute this scaling with free fall time to the longer build-up timescales for massive stars in higher-mass GMCs. While a detailed comparison between our observationally based approach and GMC-scale simulations remain to be further explored, this effect provides a possible physical interpretation for the t$_{\rm obscured}$ versus $\langle {\rm M}_{\rm GMC} \rangle$ reported in the current study.

\subsubsection{Impact of morphological type and metallicity}
\label{subsect_discuss_morpht}

\begin{figure*}[t]
\centering
\includegraphics[width=\columnwidth]{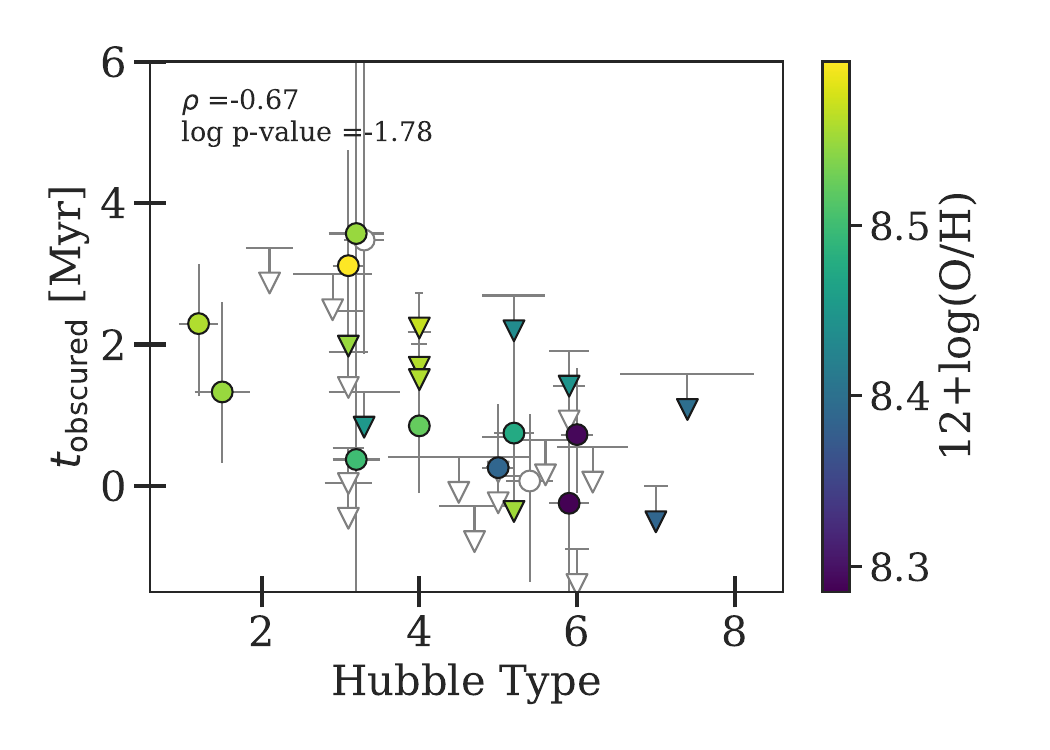}
\includegraphics[width=\columnwidth]{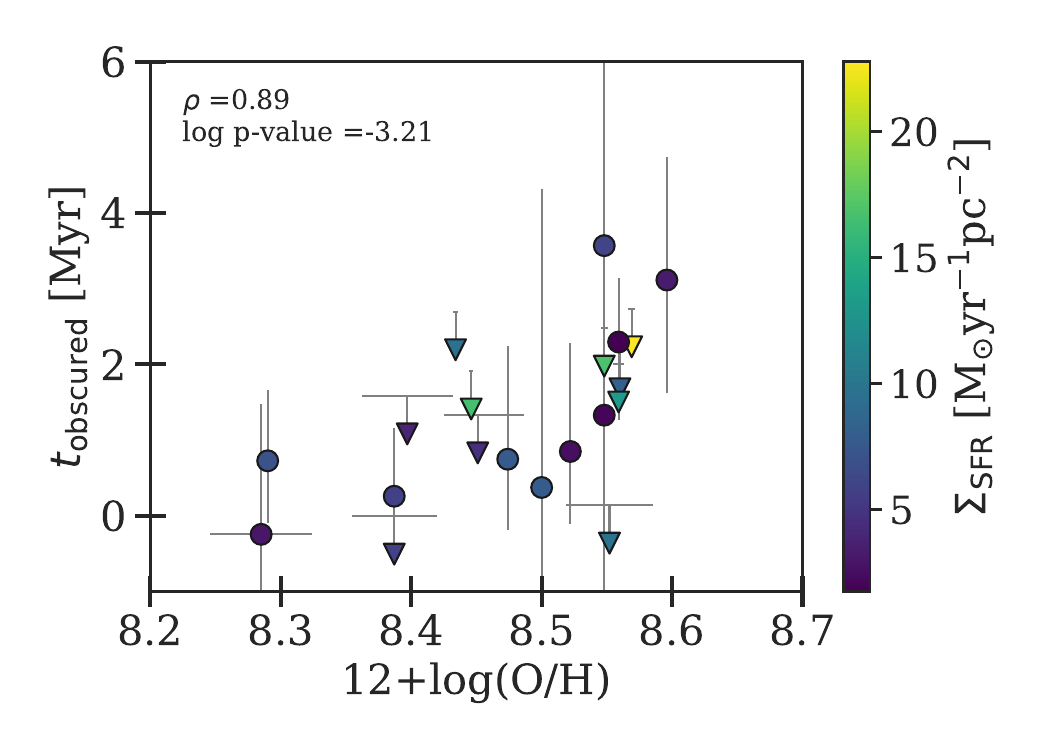}
\caption[]{Dependencies of the duration of the obscured star-formation time ($t_{\rm obscured}$). \textbf{Left:} Duration of the obscured star-formation time as a function of the morphological type of galaxies from the HyperLEDA database \citep{Paturel_hyperleda1_2003, Paturel_hyperleda2_2003, Makarov_HyperLEDA_2014}. The color bar shows the gas-phase metallicity measured with MUSE. \textbf{Right:} Duration of the obscured star-formation phase (bottom) as a function of gas-phase metallicity. The color bar shows the evolution of the SFR surface density.}
\label{tobs_vs_hubbleT}
\end{figure*}

Previous studies have shown that the frequency of regions in different evolutionary stages vary with the morphological type of galaxies \citep[e.g.,][]{Pan_morpht_2022}. Understanding the physical drivers of such differences is complicated by the fact that the morphological type of galaxies strongly correlates with other global parameters, and in particular with the stellar mass and the metallicity, which both decrease in later-type objects. \cite{Pan_morpht_2022} report that the barred spiral galaxies, corresponding the earliest Hubble T-type in our sample, are associated with a large reservoir of CO-emitting gas, unassociated with H$\alpha$ emission, while later type galaxies exhibit more \hii\ regions, associated with low-mass or partially dispersed molecular clouds. The authors report a correlation of the Hubble type with the fraction of sight-lines showing H$\alpha$ only emission and an anti-correlation with the fraction of sight-lines showing CO emission, at 150\,pc. They also report a moderate increase of the fraction of regions where both CO and H$\alpha$ overlap, for later type galaxies. While the exact numerical details differ from our statistical approach, their results should translate into correlations in terms of CO and H$\alpha$ timescales. Specifically, one might expect a longer $t_{\rm CO}$, as well a shorter $t_{\rm fb, H\alpha}$ in early-type galaxies. This anti-correlation is indeed reported in \citetalias{Kim_2022_environmental_dep}, although it is not identified as significant. Nevertheless, both studies remain in good agreement, indicating that longer $t_{\rm CO}$ timescales are expected in earlier-type and higher-mass galaxies.

In the current study, we find a moderate anti-correlation (Spearman coefficient = 0.46, log p-value= $-2.36$) between the Hubble type and the total duration of the 21~$\mu$m emission (see Figure~\ref{plot_corr_t21}). Although this anti-correlation is classified as marginally significant within our statistical framework, it is one of the most significant among the parameters we examined and has a lower p-value than the correlations with the total gas mass and stellar mass. We suggest that the increase of $t_{\rm 21~\mu m}$ observed in massive, early-type galaxies may be driven by metallicity effects, as shown by the color bars from Figure~\ref{plot_corr_t21}. This result indicates that the total duration of the 21$\mu$m emission, seen before and after GMC dispersal, is longer in early type galaxies associated with higher dust-content. In particular, we note that the five barred spiral galaxies showing clear non-zero obscured star-formation phase are associated with significantly longer isolated phase of 21$\mu$m emission, that remain visible for more than 4\,Myr after GMC dispersal (see Figure \ref{timescales}). 

In Figure~\ref{tobs_vs_hubbleT} (left), we further investigate tentative correlations between the duration of the obscured star-formation phase and the Hubble type and metallicity. Despite a small dynamical range in the y-axis, and numerous upper limits, we report a tentative anti-correlation between t$_{\rm obscured}$ and the Hubble type, indicating that in earlier type barred galaxies, associated with higher metal and dust content the newly born stars may stay enshrouded in dusty molecular clouds for a relatively longer time. While a robust confirmation of such claims would require a larger number of galaxies, we further examine possible trends of the timescales associated with 21~$\mu$m as a function of metallicity, for 20/37 galaxies for which metallicity measurements with MUSE are available. Focusing on the outer part of the galaxies (i.e., masking galactic centers) leads to average gas-phase metallicities that are below the solar-value for all galaxies, going from 12+log(O/H)=8.60$\pm$0.01 in NGC\,3351 down to 8.29$\pm$0.003 in NGC\,5068. The latter metallicities also cover a slightly larger range of values than when using a simple mass-metallicity relation, but still somewhat limited ($\sim$0.4\,dex). We stress that the range of metallicities probed within the PHANGS sample remains limited, with only two data points at low-metallicity (NGC\,5068 with 12+log(O/H)=8.29$\pm$0.003 and NGC\,4731 with 12+log(O/H)=8.29$\pm$0.04). 

For ten out of 20 of these galaxies, we derive only upper limits for $t_{\rm fb, 21~\mu m}$, $t_{\rm obscured}$, and $\lambda_{\rm fb,21~\mu m}$. As shown in Figure~\ref{tobs_vs_hubbleT} (right), for this subsample of galaxies we find a strong correlation ($\rho=0.89$, p-value=-3.21) between $t_{\rm obscured}$ and the gas phase metallicity. Similarly, we found correlations of t$_{21\mu m}$ (($\rho=0.54$, p-value=-1.87) and $t_{\rm fb, 21~\mu m}$ ($\rho=0.77$, p-value= -2.05) with metallicity. These correlations with metallicity may reflect a change in the dust-to-gas mass ratio, which is known to evolve with metallicity \citep[e.g.,][]{remy-ruyer_gas--dust_2014, galliano_nearby_2021, MendezDelgado:2024b}, with potentially a strong impact on the population of small grains which power 21~$\mu$m emission. The increased feedback timescales as a function of metallicity may also indicate that feedback by massive stellar winds, which becomes stronger with metallicity, is not responsible for the emergence of young stellar population from CO clouds, which could instead be driven by photoionization feedback mechanisms. In the low-metallicity regime probed by dwarf galaxies (1/2\,Z$_{\odot}$-1/10\,Z$_{\odot}$), changes in terms of gas morphology are expected under the effect of photoionization, leading to an increased porosity of the interstellar medium to UV photons \citep[e.g.,][]{cormier_herschel_2019, polles_modeling_2019, madden_tracing_2020, Chevance_lifecycle_2020, LebouteillerRamambason2022, Ramambason2022}. While the metallicity range probed by our current sample is higher, an increased porosity among the lowest metallicity object would be qualitatively consistent with a reduced duration of the obscured timescale when metallicity decreases. 

Theoretical studies have also shown that the physical mechanisms regulating the cloud assembly and destruction may significantly be affected by metallicity. In \cite{Fukushima_2020}, the authors perform 3D radiation hydrodynamical (RHD) simulations with varying metallicity (1, 1/10, and 1/100~Z\,$_{\odot}$) and report that the star-formation efficiency is systematically reduced with decreasing metallicity (by a factor $\sim$2 at a metallicity of 1/10~Z$_{\odot}$ and $\sim$3 at a metallicity of 1/100~Z$_{\odot}$, compared to the solar metallicity case), regardless of the initial surface density of the clouds. In addition, the authors report a faster expansion of \hii\ region that disrupt the molecular clouds in shorter timescales. Similarly, \cite{yoo_origin_2020} perform RHD simulation of molecular clouds with varying metallicity and calculated that the “enshrouded” time during which the stars remained fully embedded in their birth cloud is on average reduced in their low metallicity runs, going from $\sim 4$\,Myr at solar metallicity down to 2.4\,Myr on average at $\sim$1/10\,Z$_{\odot}$. Both the values and the tentative trends that we report in this section are qualitatively consistent with these predictions.

\subsection{Limitations and future improvements}
\label{subsect_limitations}

One of the main limitation of the current study is the limited resolution and sensitivity of the molecular gas tracer (ALMA/CO(2-1)) compared to the one of the MIRI/21~$\mu$m. The high sensitivity of the mid-IR observations leads to the detection of more peaks of emission than when using ground-based H$\alpha$ as a tracer of star formation (e.g., in \citetalias{Kim_2022_environmental_dep}). The average decorrelation scale between CO and 21~$\mu$m, $\lambda_{21~\mu m}$, is systematically smaller than the average separation length measured between CO and H$\alpha$, as shown by the histogram in Figure~\ref{tuning_fork_ha_21}. While this higher sensitivity is not a limitation in itself, it causes most of the galaxies in our sample to break one of the validity criteria of our statistical method (see item (ii) from Appendix \ref{appendix_accurary}), specifically the requirement that the decorrelation scale must be at least 1.4 times larger than the working resolution for the derived estimates of $\lambda$ and $t_{\rm fb}$ to be robust. Since 25 out of 37 galaxies in our sample are best fitted with a $\lambda$ value smaller than $\sim$1.4 times the minimum aperture (fixed by the resolution of CO maps), only upper limits can be derived for $\lambda_{\rm fb, 21~\mu m}$ and $t_{\rm fb, 21~\mu m}$ \citep{Kruijssen_tf_2018}.

In addition to this first observational limitation, there is a mismatch of a factor of $\sim$2 between 21~$\mu$m and CO(2-1) sensitivities (see criterion number (vi) from Appendix \ref{appendix_accurary}). Indeed, using the 5$\sigma$ sensitivity of the CO maps ($\sim$ 10$^5$~M$_{\odot}$; \citealt{Leroy_CO_phangs_2021}) and the integrated star-formation efficiencies calculated in \citetalias{Kim_2022_environmental_dep}, we estimate a minimal young stellar mass probed by CO observations ranging from 800~M$_{\odot}$ to 8000~M$_{\odot}$, with a median value of $\sim$3600~M$_{\odot}$. Instead, the minimal young stellar mass which is probed with the 21~$\mu$m observation is $\sim$1500~M$_{\odot}$, two times smaller than the values obtained with CO maps (see Appendix \ref{appendix_accurary} for the detailed calculation). This indicates that the young stellar populations probed in the 21~$\mu$m maps may not be detectable with the current sensitivity of our molecular gas tracer. Effectively, this could bias our results toward measuring relatively longer timescales associated with 21~$\mu$m emission. We note that this effect does not affect the conclusions from our study, since most measurements are already upper limits due to the resolution issue mentioned previously. However, it is important to stress that converting the upper limits from the current study into actual measurements would require matching both the resolution and the sensitivity of the SFR and molecular gas tracers. 

One possible avenue to solve this issue would be to use another proxy to probe the neutral gas reservoir, obtained at a higher angular resolution and sensitivity. Falling into the range of detection of JWST imagers, PAH bands have been shown to be good tracers of cold neutral gas distribution \citep[e.g.,][]{Whitcomb_sfr_tracers_2023, Chown_CO_IR_2025}, making them interesting proxies of cold gas at higher spatial resolution than CO. Nevertheless, their complex origin influenced by the ISM distribution, radiation, and abundances of dust and PAHs, complicate the interpretation of their spatial variations, in particular near \hii\ regions where PAHs may be destroyed \citep[e.g.,][]{Chastenet_2023, Egorov_pah_2023, Sutter_qpah_2024}. Other spectroscopic tracers, such as the vibrationally excited H$_2$ lines, only probe a partial, warmer, H$_2$ gas reservoir and their observability is strongly limited by the small field of view of JWST spectrometers. As a result, no tracer other than CO is currently available to trace the bulk molecular gas distribution.

A detailed analysis contrasting the mid-IR PAHs bands with MUSE/H$\alpha$ observations has recently been carried out in \cite{Kim_pah_2025}. Nevertheless, using mid-IR bands to trace both the gas reservoirs (via PAH-dominated bands or H$_2$) and the SFR (via 21~$\mu$m) would bring additional challenges regarding the timeline calibration and the interpretation of the timescales. A first difficulty would be to define a reference timescale associated with the emission of at least one tracer, which is a necessary step to convert the relative timescales derived in \textsc{Heisenberg} into absolute timescales (see Section \ref{subsec_measuring_timescales}). Another challenge is to interpret the timescales associated with tracers which may be only probing part of the star-forming molecular gas reservoirs. Indeed, one important assumption of the statistical approach used here is that the compact emission of the selective tracers used in the analysis traces well the gas that has formed or will form stars.

While the analysis presented in this study relies on classical tracers of molecular gas and SFR, the timescales we derive are also sensitive to how well our selected tracers spatially agree with the underlying distribution of gas and stars. In particular, the interpretation of the timescales associated with the total cloud lifetime and feedback time would be complicated if mechanisms other than stellar feedback are causing the decorrelation between the gas and SFR tracers (e.g., drift of GMCs, ionizing photons leaking out of star-formation sites, runaway stars; \citealt{Koda_tf_limits_2023, Hu_limitation_2024}). If present, dynamical drift effects and H$\alpha$ diffusion effects would both likely lead to an underestimation of the cloud lifetime. Disentangling the signatures associated with these various scenarios remains challenging and requires access to observations at higher spatial resolution than currently accessible in the PHANGS survey (below $\sim$50\,pc in both H$\alpha$ and CO(1-0), only achieved in NGC\,628 and NGC\,5068). In a recent analysis including the two latter galaxies, \cite{Kruijssen_drift_2024} have shown that a decorrelation arising solely from the dynamical drift of GMCs is statistically unlikely with respect to the feedback-driven hypothesis, in all the galaxies with sufficient spatial resolution to allow this test to be performed. 

In addition, the presence of CO-dark H$_2$ gas, undetected in CO(2-1) may also bias our timescale estimates toward smaller cloud lifetimes. This effect has recently been quantified in \citet{Kim_pah_2025} and was found to be negligible for most galaxies from the PHANGS sample, which have a metallicity close to the solar value. Significant underestimations of the cloud lifetimes traced by CO are only expected in galaxies associated with low-metallicity (\citealt{Ward_lmc_2020, Ward_lmc_2022, Kim_pah_2025}). As a result, extending our analysis to a sample including low-metallicity dwarf galaxies may become challenging, since CO molecules do not self-shield as effectively as H$_2$, making it notoriously difficult to detect at low metallicity \citep[e.g.,][]{madden_tracing_2020}. Nevertheless, in order to confirm the potential trends of $t_{\rm 21~\mu m}$, $t_{\rm fb,21~\mu m}$, and $t_{\rm obscured}$ with metallicity discussed in the current paper (see Section \ref{subsect_discuss_morpht}), including galaxies that probe the lower metallicity regime is crucially needed.

The timescales we derive here, similar to all the previous work relying on a similar statistical method, are representative of the average timescale of molecular clouds. While the measured uncertainties account for both internal variations within galaxies and systematic effects, the measurements themselves represent luminosity-weighted averages, integrated on galactic scales. In the current paper, we find that evolutionary timescales of GMCs depend on the interplay between the chemical properties of galaxies (metallicity and dust content), the gas distribution of star-forming gas reservoirs (morphological type and masses of GMCs), as well as on the physical mechanisms injecting energy at the scale of GMCs and controlling the velocity dispersion of GMCs. To disentangle these complex multiscale effects, studying the internal variations of GMCs within the same galaxy is essential. In particular, a promising avenue consists in examining how local environmental variations may affect the derived timelines \citep[e.g., arm versus\ interarm regions;][]{Romanelli_2025} and how these timescales may evolve radially, from the centers to the outer edges of galaxies (e.g., \citealt{Kruijssen_nat_2019, Chevance_lifecycle_2020, Ward_lmc_2020, Ward_lmc_2022}, Romanelli et al.\ in prep.). Such studies are necessary to probe the impact of the local ISM properties on the derived timescales and assess the potential role of galactic dynamics in GMCs evolution.
\section{Conclusion}
\label{section_conclusion}

In this paper, we have analyzed a sample of 37 nearby star-forming galaxies observed in a wide range of wavelengths as part of the PHANGS surveys. We combined ALMA/CO(2-1), ground-based H$\alpha$, and JWST/21~$\mu$m observations in order to constrain the evolutionary cycle of molecular clouds, including the dust-embedded star-formation phase. We analyzed the spatial correlations and decorrelations between CO and 21~$\mu$m maps after filtering the diffuse emission present in both maps and masking the central regions affected by blending. We then used the \textsc{Heisenberg} code to convert these spatial offsets into timescales associated with 21~$\mu$m emission. Our main findings are summarized below:

\begin{itemize}
    \item Across the sample considered
    in our analysis and excluding galaxies affected by blending, we measured a median total duration of 21~$\mu$m emission of $3.9^{+1.4}_{-0.9}$\,Myr, an embedded feedback timescale associated with 21~$\mu$m emission of $3.4^{+1.5}_{-1.2}$\, Myr, and an obscured phase of star formation of $0.8^{+1.5}_{-0.8}$\, Myr. We find that the obscured phase of star formation, which is not visible in H$\alpha$, is short in all the galaxies from our sample, lasting a maximum of 4\,Myr, and it less than 1\,Myr in 28 out of 37 galaxies.
    
    \item We examined the correlations of the timescales we derived with numerous parameters that include galaxy-integrated properties, average kiloparsec-scale properties, average GMC properties, properties of the ionized gas and dust, measurements with \textsc{Heisenberg}, and systematic biases. We find moderately strong but statistically significant correlations of the total emission time of $t_{\rm 21~\mu m}$ with the mass and velocity dispersion of GMCs. We also identified marginally significant correlations of $t_{\rm 21~\mu m}$ with the Hubble morphological type, which may reflect an impact of the gas-phase metallicity.
    
    \item We measured a fraction of diffuse emission contribution to the emission of 21~$\mu$m, which is relatively constant throughout our sample, in the range of $48-74$\%, with a median value of $62\pm4$\%. This fraction of diffuse emission correlates best with the global gas mass fraction, the molecular gas mass fraction, and the kiloparsec-scale stellar surface density, and to a lesser extent it correlates with the properties of GMCs (mass and velocity dispersion).
    
    \item We discussed the potential trend of the feedback timescale ($t_{\rm fb, 21~\mu m}$) and the embedded phase of star formation ($t_{\rm obscured}$) with various parameters. We find that $t_{\rm fb, 21~\mu m}$ may be sensitive to changes in the GMCs properties, in particular variations of the mass and velocity dispersion of the population of clouds that dominate the emission. On the other hand, $t_{\rm obscured}$ appears to be sensitive to the morphological type of galaxies, with barred spiral galaxies showing a longer obscured phase of star formation($\sim$4\,Myr), while low-mass flocculent spiral and irregular galaxies show a reduced duration of the embedded phase of star formation.
    
    \item Finally, we examined potential trends of the timescales associated with 21~$\mu$m with metallicity and find that $t_{\rm 21~\mu m}$, $t_{\rm fb, 21~\mu m}$, and $t_{\rm obscured}$ tend to increase with metallicity, which could be due to the effect of a higher dust-to-gas mass ratio. If this trend is confirmed, it would favor the dispersal of dust and CO clouds through ionization feedback mechanisms rather than feedback from stellar winds, which is instead expected to increase with metallicity. Extending the parameter space probed in this study to the low-metallicity regime is necessary to confirm the presence of such trends.
\end{itemize}

\begin{acknowledgements} We thank the anonymous referee for constructive comments that significantly improved the clarity of our manuscript.
This work was carried out as part of the PHANGS collaboration. LR, MC, and AR gratefully acknowledge funding from the DFG through an Emmy Noether Research Group (grant number CH2137/1-1). COOL Research DAO \citep{cool_whitepaper} is a Decentralized Autonomous Organization supporting research in astrophysics aimed at uncovering our cosmic origins. J.K. is supported by a Kavli Fellowship at the Kavli Institute for Particle Astrophysics and Cosmology (KIPAC). HAP acknowledges support from the National Science and Technology Council of Taiwan under grant 113-2112-M-032-014-MY3. JS acknowledges support by the National Aeronautics and Space Administration (NASA) through the NASA Hubble Fellowship grant HST-HF2-51544 awarded by the Space Telescope Science Institute (STScI), which is operated by the Association of Universities for Research in Astronomy, Inc., under contract NAS~5-26555. MB acknowledges support by the ANID BASAL project FB210003. This work was supported by the French government through the France 2030 investment plan managed by the National Research Agency (ANR), as part of the Initiative of Excellence of Université Côte d’Azur under reference No. ANR-15-IDEX-01. This research was funded, in whole or in part, by the French National Research Agency (ANR), grant ANR-24-CE92-0044 (project STARCLUSTERS).  We thank the German Science Foundation DFG for financial support in the project STARCLUSTERS (funding ID KL 1358/22-1 and SCHI 536/13-1).. OE acknowledges funding from the Deutsche Forschungsgemeinschaft (DFG, German Research Foundation) -- project-ID 541068876. SCOG acknowledges financial support from the European Research Council via the ERC Synergy Grant ``ECOGAL'' (project ID 855130) and from the German Excellence Strategy via the Heidelberg Cluster of Excellence (EXC 2181 - 390900948) ``STRUCTURES''. KK gratefully acknowledges funding from the Deutsche Forschungsgemeinschaft (DFG, German Research Foundation) in the form of an Emmy Noether Research Group (grant number KR4598/2-1, PI Kreckel) and the European Research Council’s starting grant ERC StG-101077573 (“ISM-METALS"). 

This work is based on observations made with the NASA/ESA/CSA JWST. The data were obtained from the Mikulski Archive for Space Telescopes at the Space Telescope Science Institute, which is operated by the Association of Universities for Research in Astronomy, Inc., under NASA contract NAS5-03127. The observations are associated with JWST programs 2107 and 3707. The specific observations analyzed can be accessed via \url{https://doi.org/10.17909/ew88-jt15}.

This paper includes data gathered with the 2.5 meter du Pont located at Las Campanas Observatory, Chile, and data based on observations carried out at the MPG 2.2m telescope on La Silla, Chile.

Based on observations collected at the European Southern Observatory under ESO programs 0111.C-2109 (PI: Egorov), 0108.B-0249 (PI: Kreckel), 094.C-0623 (PI: Kreckel), 095.C-0473,  098.C-0484 (PI: Blanc), 1100.B-0651 (PHANGS-MUSE; PI: Schinnerer), as well as 094.B-0321 (MAGNUM; PI: Marconi), 099.B-0242, 0100.B-0116, 098.B-0551 (MAD; PI: Carollo) and 097.B-0640 (TIMER; PI: Gadotti). 

This paper makes use of the following ALMA data,
which have been processed as part of the PHANGS–ALMA survey: \\
ADS/JAO.ALMA\#2012.1.00650.S,\\
ADS/JAO.ALMA\#2013.1.01161.S,\\
ADS/JAO.ALMA\#2015.1.00925.S,\\
ADS/JAO.ALMA\#2015.1.00956.S,\\
ADS/JAO.ALMA\#2017.1.00392.S,\\
ADS/JAO.ALMA\#2017.1.00886.L,\\
ADS/JAO.ALMA\#2018.1.01651.S.\\
ALMA is a partnership of ESO (representing its member states), NSF (USA) and NINS (Japan), together with NRC (Canada), MOST and ASIAA (Taiwan), and KASI (Republic of Korea), in cooperation with the Republic of Chile. The Joint ALMA Observatory is operated by ESO, AUI/NRAO and NAOJ. This paper includes data gathered with the 2.5 meter du Pont located at Las Campanas Observatory, Chile, and data based on observations carried out at the MPG 2.2m telescope on La Silla, Chile.

This paper used the following software packages : \texttt{Astropy} \citep{astropy:2013, astropy:2018, astropy:2022}, 
\texttt{Clumpfind} \citep{clumpfind_Williams_1994}, 
\texttt{Heisenberg} \citep{Kruijssen_tf_2018}, 
\texttt{Matplotlib} \citep{hunter2007}, 
\texttt{SciPy} \citep{virtanen2020}, 
\texttt{seaborn} \citep{waskom2021}, 
\texttt{pandas} \citep{mckinney2010}
\end{acknowledgements}

\bibliographystyle{aa}
\bibliography{MyLibrary.bib}

@ARTICLE{cool_whitepaper,
       author = {{Chevance}, M\'elanie and {Kruijssen}, J.~M. Diederik and {Longmore}, Steven~N.},
        title = "{COOL Research DAO Whitepaper - Towards community-owned astrophysics for everyone}",
      journal = {arXiv e-prints},
         year = 2025,
        month = dec,
          eid = {arXiv:2501.13160},
        pages = {arXiv:2501.13160},
          doi = {10.48550/arXiv.2501.13160},
archivePrefix = {arXiv},
       eprint = {2501.13160},
 primaryClass = {astro-ph.IM},
       adsurl = {https://ui.adsabs.harvard.edu/abs/2025arXiv250113160C}
}

@article{cormier_herschel_2019,
	title = {The \textit{{Herschel}} {Dwarf} {Galaxy} {Survey}: {II}. {Physical} conditions, origin of [{C} {II}] emission, and porosity of the multiphase low-metallicity {ISM}⋆},
	volume = {626},
	issn = {0004-6361, 1432-0746},
	shorttitle = {The \textit{{Herschel}} {Dwarf} {Galaxy} {Survey}},
	url = {https://www.aanda.org/10.1051/0004-6361/201834457},
	doi = {10.1051/0004-6361/201834457},
	language = {en},
	urldate = {2020-01-13},
	journal = {A\&A},
	author = {Cormier, D. and Abel, N. P. and Hony, S. and Lebouteiller, V. and Madden, S. C. and Polles, F. L. and Galliano, F. and De Looze, I. and Galametz, M. and Lambert-Huyghe, A.},
	month = jun,
	year = {2019},
	pages = {A23}
}

@article{polles_modeling_2019,
	title = {Modeling ionized gas in low-metallicity environments: the {Local} {Group} dwarf galaxy {IC} 10},
	volume = {622},
	issn = {0004-6361, 1432-0746},
	shorttitle = {Modeling ionized gas in low-metallicity environments},
	url = {https://www.aanda.org/10.1051/0004-6361/201833776},
	doi = {10.1051/0004-6361/201833776},
	language = {en},
	urldate = {2020-01-13},
	journal = {A\&A},
	author = {Polles, F. L. and Madden, S. C. and Lebouteiller, V. and Cormier, D. and Abel, N. and Galliano, F. and Hony, S. and Karczewski, O. Ł. and Lee, M.-Y. and Chevance, M. and Galametz, M. and Lianou, S.},
	month = feb,
	year = {2019},
	pages = {A119}
}

@ARTICLE{yoo_origin_2020,
       author = {{Yoo}, Taehwa and {Kimm}, Taysun and {Rosdahl}, Joakim},
        title = "{On the origin of low escape fractions of ionizing radiation from massive star-forming galaxies at high redshift}",
      journal = {\mnras},
         year = 2020,
        month = dec,
       volume = {499},
       number = {4},
        pages = {5175-5193},
          doi = {10.1093/mnras/staa3187},
archivePrefix = {arXiv},
       eprint = {2001.05508},
 primaryClass = {astro-ph.GA},
       adsurl = {https://ui.adsabs.harvard.edu/abs/2020MNRAS.499.5175Y}
}

@article{remy-ruyer_gas--dust_2014,
	title = {Gas-to-dust mass ratios in local galaxies over a 2 dex metallicity range},
	volume = {563},
	issn = {0004-6361, 1432-0746},
	url = {http://www.aanda.org/10.1051/0004-6361/201322803},
	doi = {10.1051/0004-6361/201322803},
	language = {en},
	urldate = {2020-12-09},
	journal = {A\&A},
	author = {Rémy-Ruyer, A. and Madden, S. C. and Galliano, F. and Galametz, M. and Takeuchi, T. T. and Asano, R. S. and Zhukovska, S. and Lebouteiller, V. and Cormier, D. and Jones, A. and Bocchio, M. and Baes, M. and Bendo, G. J. and Boquien, M. and Boselli, A. and DeLooze, I. and Doublier-Pritchard, V. and Hughes, T. and Karczewski, O. Ł. and Spinoglio, L.},
	month = mar,
	year = {2014},
	pages = {A31}
}

@ARTICLE{galliano_nearby_2021,
       author = {{Galliano}, Fr{\'e}d{\'e}ric and {Nersesian}, Angelos and {Bianchi}, Simone and {De Looze}, Ilse and {Roychowdhury}, Sambit and {Baes}, Maarten and {Casasola}, Viviana and {Cassar{\'a}}, Letizia P. and {Dobbels}, Wouter and {Fritz}, Jacopo and {Galametz}, Maud and {Jones}, Anthony P. and {Madden}, Suzanne C. and {Mosenkov}, Aleksandr and {Xilouris}, Emmanuel M. and {Ysard}, Nathalie},
        title = "{A nearby galaxy perspective on dust evolution. Scaling relations and constraints on the dust build-up in galaxies with the DustPedia and DGS samples}",
      journal = {\aap},
         year = 2021,
        month = may,
       volume = {649},
          eid = {A18},
        pages = {A18},
          doi = {10.1051/0004-6361/202039701},
archivePrefix = {arXiv},
       eprint = {2101.00456},
 primaryClass = {astro-ph.GA},
       adsurl = {https://ui.adsabs.harvard.edu/abs/2021A&A...649A..18G}
}

@ARTICLE{madden_tracing_2020,
       author = {{Madden}, S.~C. and {Cormier}, D. and {Hony}, S. and {Lebouteiller}, V. and {Abel}, N. and {Galametz}, M. and {De Looze}, I. and {Chevance}, M. and {Polles}, F.~L. and {Lee}, M. -Y. and {Galliano}, F. and {Lambert-Huyghe}, A. and {Hu}, D. and {Ramambason}, L.},
        title = "{Tracing the total molecular gas in galaxies: [CII] and the CO-dark gas}",
      journal = {\aap},
         year = 2020,
        month = nov,
       volume = {643},
          eid = {A141},
        pages = {A141},
          doi = {10.1051/0004-6361/202038860},
archivePrefix = {arXiv},
       eprint = {2009.00649},
 primaryClass = {astro-ph.GA},
       adsurl = {https://ui.adsabs.harvard.edu/abs/2020A&A...643A.141M}
}

@ARTICLE{Calzetti_midIR_2007,
       author = {{Calzetti}, D. and {Kennicutt}, R.~C. and {Engelbracht}, C.~W. and {Leitherer}, C. and {Draine}, B.~T. and {Kewley}, L. and {Moustakas}, J. and {Sosey}, M. and {Dale}, D.~A. and {Gordon}, K.~D. and {Helou}, G.~X. and {Hollenbach}, D.~J. and {Armus}, L. and {Bendo}, G. and {Bot}, C. and {Buckalew}, B. and {Jarrett}, T. and {Li}, A. and {Meyer}, M. and {Murphy}, E.~J. and {Prescott}, M. and {Regan}, M.~W. and {Rieke}, G.~H. and {Roussel}, H. and {Sheth}, K. and {Smith}, J.~D.~T. and {Thornley}, M.~D. and {Walter}, F.},
        title = "{The Calibration of Mid-Infrared Star Formation Rate Indicators}",
      journal = {\apj},
         year = 2007,
        month = sep,
       volume = {666},
       number = {2},
        pages = {870-895},
          doi = {10.1086/520082},
archivePrefix = {arXiv},
       eprint = {0705.3377},
 primaryClass = {astro-ph},
       adsurl = {https://ui.adsabs.harvard.edu/abs/2007ApJ...666..870C}
}

@ARTICLE{Asplund_2009,
       author = {{Asplund}, Martin and {Grevesse}, Nicolas and {Sauval}, A. Jacques and {Scott}, Pat},
        title = "{The Chemical Composition of the Sun}",
      journal = {\araa},
         year = 2009,
        month = sep,
       volume = {47},
       number = {1},
        pages = {481-522},
          doi = {10.1146/annurev.astro.46.060407.145222},
archivePrefix = {arXiv},
       eprint = {0909.0948},
 primaryClass = {astro-ph.SR},
       adsurl = {https://ui.adsabs.harvard.edu/abs/2009ARA&A..47..481A}
}

@ARTICLE{Starburst99_Leitherer_1999,
       author = {{Leitherer}, Claus and {Schaerer}, Daniel and {Goldader}, Jeffrey D. and {Delgado}, Rosa M. Gonz{\'a}lez and {Robert}, Carmelle and {Kune}, Denis Foo and {de Mello}, Du{\'\i}lia F. and {Devost}, Daniel and {Heckman}, Timothy M.},
        title = "{Starburst99: Synthesis Models for Galaxies with Active Star Formation}",
      journal = {\apjs},
         year = 1999,
        month = jul,
       volume = {123},
       number = {1},
        pages = {3-40},
          doi = {10.1086/313233},
archivePrefix = {arXiv},
       eprint = {astro-ph/9902334},
 primaryClass = {astro-ph},
       adsurl = {https://ui.adsabs.harvard.edu/abs/1999ApJS..123....3L}
}

@ARTICLE{Chevance_2022,
       author = {{Chevance}, M{\'e}lanie and {Kruijssen}, J.~M. Diederik and {Krumholz}, Mark R. and {Groves}, Brent and {Keller}, Benjamin W. and {Hughes}, Annie and {Glover}, Simon C.~O. and {Henshaw}, Jonathan D. and {Herrera}, Cinthya N. and {Kim}, Jaeyeon and {Leroy}, Adam K. and {Pety}, J{\'e}r{\^o}me and {Razza}, Alessandro and {Rosolowsky}, Erik and {Schinnerer}, Eva and {Schruba}, Andreas and {Barnes}, Ashley T. and {Bigiel}, Frank and {Blanc}, Guillermo A. and {Dale}, Daniel A. and {Emsellem}, Eric and {Faesi}, Christopher M. and {Grasha}, Kathryn and {Klessen}, Ralf S. and {Kreckel}, Kathryn and {Liu}, Daizhong and {Longmore}, Steven N. and {Meidt}, Sharon E. and {Querejeta}, Miguel and {Saito}, Toshiki and {Sun}, Jiayi and {Usero}, Antonio},
        title = "{Pre-supernova feedback mechanisms drive the destruction of molecular clouds in nearby star-forming disc galaxies}",
      journal = {\mnras},
         year = 2022,
        month = jan,
       volume = {509},
       number = {1},
        pages = {272-288},
          doi = {10.1093/mnras/stab2938},
archivePrefix = {arXiv},
       eprint = {2010.13788},
 primaryClass = {astro-ph.GA},
       adsurl = {https://ui.adsabs.harvard.edu/abs/2022MNRAS.509..272C}
}

@ARTICLE{Belfiore_dig_2022,
       author = {{Belfiore}, F. and {Santoro}, F. and {Groves}, B. and {Schinnerer}, E. and {Kreckel}, K. and {Glover}, S.~C.~O. and {Klessen}, R.~S. and {Emsellem}, E. and {Blanc}, G.~A. and {Congiu}, E. and {Barnes}, A.~T. and {Boquien}, M. and {Chevance}, M. and {Dale}, D.~A. and {Kruijssen}, J.~M. Diederik and {Leroy}, A.~K. and {Pan}, H. -A. and {Pessa}, I. and {Schruba}, A. and {Williams}, T.~G.},
        title = "{A tale of two DIGs: The relative role of H II regions and low-mass hot evolved stars in powering the diffuse ionised gas (DIG) in PHANGS-MUSE galaxies}",
      journal = {\aap},
         year = 2022,
        month = mar,
       volume = {659},
          eid = {A26},
        pages = {A26},
          doi = {10.1051/0004-6361/202141859},
archivePrefix = {arXiv},
       eprint = {2111.14876},
 primaryClass = {astro-ph.GA},
       adsurl = {https://ui.adsabs.harvard.edu/abs/2022A&A...659A..26B}
}

@ARTICLE{Chevance_lifecycle_2020,
       author = {{Chevance}, M{\'e}lanie and {Kruijssen}, J.~M. Diederik and {Vazquez-Semadeni}, Enrique and {Nakamura}, Fumitaka and {Klessen}, Ralf and {Ballesteros-Paredes}, Javier and {Inutsuka}, Shu-ichiro and {Adamo}, Angela and {Hennebelle}, Patrick},
        title = "{The Molecular Cloud Lifecycle}",
      journal = {\ssr},
         year = 2020,
        month = apr,
       volume = {216},
       number = {4},
          eid = {50},
        pages = {50},
          doi = {10.1007/s11214-020-00674-x},
archivePrefix = {arXiv},
       eprint = {2004.06113},
 primaryClass = {astro-ph.GA},
       adsurl = {https://ui.adsabs.harvard.edu/abs/2020SSRv..216...50C}
}

@ARTICLE{Chevance_GMC_spiral_2020,
       author = {{Chevance}, M{\'e}lanie and {Kruijssen}, J.~M. Diederik and {Hygate}, Alexander P.~S. and {Schruba}, Andreas and {Longmore}, Steven N. and {Groves}, Brent and {Henshaw}, Jonathan D. and {Herrera}, Cinthya N. and {Hughes}, Annie and {Jeffreson}, Sarah M.~R. and {Lang}, Philipp and {Leroy}, Adam K. and {Meidt}, Sharon E. and {Pety}, J{\'e}r{\^o}me and {Razza}, Alessandro and {Rosolowsky}, Erik and {Schinnerer}, Eva and {Bigiel}, Frank and {Blanc}, Guillermo A. and {Emsellem}, Eric and {Faesi}, Christopher M. and {Glover}, Simon C.~O. and {Haydon}, Daniel T. and {Ho}, I. -Ting and {Kreckel}, Kathryn and {Lee}, Janice C. and {Liu}, Daizhong and {Querejeta}, Miguel and {Saito}, Toshiki and {Sun}, Jiayi and {Usero}, Antonio and {Utomo}, Dyas},
        title = "{The lifecycle of molecular clouds in nearby star-forming disc galaxies}",
      journal = {\mnras},
         year = 2020,
        month = apr,
       volume = {493},
       number = {2},
        pages = {2872-2909},
          doi = {10.1093/mnras/stz3525},
archivePrefix = {arXiv},
       eprint = {1911.03479},
 primaryClass = {astro-ph.GA},
       adsurl = {https://ui.adsabs.harvard.edu/abs/2020MNRAS.493.2872C}
}

@ARTICLE{Adamo_legus_2017,
       author = {{Adamo}, A. and {Ryon}, J.~E. and {Messa}, M. and {Kim}, H. and {Grasha}, K. and {Cook}, D.~O. and {Calzetti}, D. and {Lee}, J.~C. and {Whitmore}, B.~C. and {Elmegreen}, B.~G. and {Ubeda}, L. and {Smith}, L.~J. and {Bright}, S.~N. and {Runnholm}, A. and {Andrews}, J.~E. and {Fumagalli}, M. and {Gouliermis}, D.~A. and {Kahre}, L. and {Nair}, P. and {Thilker}, D. and {Walterbos}, R. and {Wofford}, A. and {Aloisi}, A. and {Ashworth}, G. and {Brown}, T.~M. and {Chandar}, R. and {Christian}, C. and {Cignoni}, M. and {Clayton}, G.~C. and {Dale}, D.~A. and {de Mink}, S.~E. and {Dobbs}, C. and {Elmegreen}, D.~M. and {Evans}, A.~S. and {Gallagher}, III, J.~S. and {Grebel}, E.~K. and {Herrero}, A. and {Hunter}, D.~A. and {Johnson}, K.~E. and {Kennicutt}, R.~C. and {Krumholz}, M.~R. and {Lennon}, D. and {Levay}, K. and {Martin}, C. and {Nota}, A. and {{\"O}stlin}, G. and {Pellerin}, A. and {Prieto}, J. and {Regan}, M.~W. and {Sabbi}, E. and {Sacchi}, E. and {Schaerer}, D. and {Schiminovich}, D. and {Shabani}, F. and {Tosi}, M. and {Van Dyk}, S.~D. and {Zackrisson}, E.},
        title = "{Legacy ExtraGalactic UV Survey with The Hubble Space Telescope: Stellar Cluster Catalogs and First Insights Into Cluster Formation and Evolution in NGC 628}",
      journal = {\apj},
         year = 2017,
        month = jun,
       volume = {841},
       number = {2},
          eid = {131},
        pages = {131},
          doi = {10.3847/1538-4357/aa7132},
archivePrefix = {arXiv},
       eprint = {1705.01588},
 primaryClass = {astro-ph.GA},
       adsurl = {https://ui.adsabs.harvard.edu/abs/2017ApJ...841..131A}
}

@ARTICLE{Cook_legus_dwarf_2019,
       author = {{Cook}, D.~O. and {Lee}, J.~C. and {Adamo}, A. and {Kim}, H. and {Chandar}, R. and {Whitmore}, B.~C. and {Mok}, A. and {Ryon}, J.~E. and {Dale}, D.~A. and {Calzetti}, D. and {Andrews}, J.~E. and {Aloisi}, A. and {Ashworth}, G. and {Bright}, S.~N. and {Brown}, T.~M. and {Christian}, C. and {Cignoni}, M. and {Clayton}, G.~C. and {da Silva}, R. and {de Mink}, S.~E. and {Dobbs}, C.~L. and {Elmegreen}, B.~G. and {Elmegreen}, D.~M. and {Evans}, A.~S. and {Fumagalli}, M. and {Gallagher}, J.~S. and {Gouliermis}, D.~A. and {Grasha}, K. and {Grebel}, E.~K. and {Herrero}, A. and {Hunter}, D.~A. and {Jensen}, E.~I. and {Johnson}, K.~E. and {Kahre}, L. and {Kennicutt}, R.~C. and {Krumholz}, M.~R. and {Lee}, N.~J. and {Lennon}, D. and {Linden}, S. and {Martin}, C. and {Messa}, M. and {Nair}, P. and {Nota}, A. and {{\"O}stlin}, G. and {Parziale}, R.~C. and {Pellerin}, A. and {Regan}, M.~W. and {Sabbi}, E. and {Sacchi}, E. and {Schaerer}, D. and {Schiminovich}, D. and {Shabani}, F. and {Slane}, F.~A. and {Small}, J. and {Smith}, C.~L. and {Smith}, L.~J. and {Taibi}, S. and {Thilker}, D.~A. and {de la Torre}, I.~C. and {Tosi}, M. and {Turner}, J.~A. and {Ubeda}, L. and {Van Dyk}, S.~D. and {Walterbos}, R. AM and {Wofford}, A.},
        title = "{Star cluster catalogues for the LEGUS dwarf galaxies}",
      journal = {\mnras},
         year = 2019,
        month = apr,
       volume = {484},
       number = {4},
        pages = {4897-4919},
          doi = {10.1093/mnras/stz331},
archivePrefix = {arXiv},
       eprint = {1902.00082},
 primaryClass = {astro-ph.GA},
       adsurl = {https://ui.adsabs.harvard.edu/abs/2019MNRAS.484.4897C}
}

@ARTICLE{Calzetti_2015,
       author = {{Calzetti}, D. and {Johnson}, K.~E. and {Adamo}, A. and {Gallagher}, J.~S., III and {Andrews}, J.~E. and {Smith}, L.~J. and {Clayton}, G.~C. and {Lee}, J.~C. and {Sabbi}, E. and {Ubeda}, L. and {Kim}, H. and {Ryon}, J.~E. and {Thilker}, D. and {Bright}, S.~N. and {Zackrisson}, E. and {Kennicutt}, R.~C. and {de Mink}, S.~E. and {Whitmore}, B.~C. and {Aloisi}, A. and {Chandar}, R. and {Cignoni}, M. and {Cook}, D. and {Dale}, D.~A. and {Elmegreen}, B.~G. and {Elmegreen}, D.~M. and {Evans}, A.~S. and {Fumagalli}, M. and {Gouliermis}, D.~A. and {Grasha}, K. and {Grebel}, E.~K. and {Krumholz}, M.~R. and {Walterbos}, R. and {Wofford}, A. and {Brown}, T.~M. and {Christian}, C. and {Dobbs}, C. and {Herrero}, A. and {Kahre}, L. and {Messa}, M. and {Nair}, P. and {Nota}, A. and {{\"O}stlin}, G. and {Pellerin}, A. and {Sacchi}, E. and {Schaerer}, D. and {Tosi}, M.},
        title = "{The Brightest Young Star Clusters in NGC 5253.}",
      journal = {\apj},
         year = 2015,
        month = oct,
       volume = {811},
       number = {2},
          eid = {75},
        pages = {75},
          doi = {10.1088/0004-637X/811/2/75},
archivePrefix = {arXiv},
       eprint = {1508.04476},
 primaryClass = {astro-ph.GA},
       adsurl = {https://ui.adsabs.harvard.edu/abs/2015ApJ...811...75C}
}

@ARTICLE{LebouteillerRamambason2022,
       author = {{Lebouteiller}, V. and {Ramambason}, L.},
        title = "{Topological models to infer multiphase interstellar medium properties}",
      journal = {\aap},
         year = 2022,
        month = nov,
       volume = {667},
          eid = {A34},
        pages = {A34},
          doi = {10.1051/0004-6361/202243865},
archivePrefix = {arXiv},
       eprint = {2207.05657},
 primaryClass = {astro-ph.GA},
       adsurl = {https://ui.adsabs.harvard.edu/abs/2022A&A...667A..34L}
}

@ARTICLE{Ramambason2022,
       author = {{Ramambason}, L. and {Lebouteiller}, V. and {Bik}, A. and {Richardson}, C.~T. and {Galliano}, F. and {Schaerer}, D. and {Morisset}, C. and {Polles}, F.~L. and {Madden}, S.~C. and {Chevance}, M. and {De Looze}, I.},
        title = "{Inferring the HII region escape fraction of ionizing photons from infrared emission lines in metal-poor star-forming dwarf galaxies}",
      journal = {\aap},
         year = 2022,
        month = nov,
       volume = {667},
          eid = {A35},
        pages = {A35},
          doi = {10.1051/0004-6361/202243866},
archivePrefix = {arXiv},
       eprint = {2207.06146},
 primaryClass = {astro-ph.GA},
       adsurl = {https://ui.adsabs.harvard.edu/abs/2022A&A...667A..35R}
}

@ARTICLE{Bolatto_2013,
       author = {{Bolatto}, Alberto D. and {Wolfire}, Mark and {Leroy}, Adam K.},
        title = "{The CO-to-H$_{2}$ Conversion Factor}",
      journal = {\araa},
         year = 2013,
        month = aug,
       volume = {51},
       number = {1},
        pages = {207-268},
          doi = {10.1146/annurev-astro-082812-140944},
archivePrefix = {arXiv},
       eprint = {1301.3498},
 primaryClass = {astro-ph.GA},
       adsurl = {https://ui.adsabs.harvard.edu/abs/2013ARA&A..51..207B}
}

@ARTICLE{Accurso_2017,
       author = {{Accurso}, G. and {Saintonge}, A. and {Catinella}, B. and {Cortese}, L. and {Dav{\'e}}, R. and {Dunsheath}, S.~H. and {Genzel}, R. and {Gracia-Carpio}, J. and {Heckman}, T.~M. and {Jimmy} and {Kramer}, C. and {Li}, Cheng and {Lutz}, K. and {Schiminovich}, D. and {Schuster}, K. and {Sternberg}, A. and {Sturm}, E. and {Tacconi}, L.~J. and {Tran}, K.~V. and {Wang}, J.},
        title = "{Deriving a multivariate {\ensuremath{\alpha}}$_{CO}$ conversion function using the [C II]/CO (1-0) ratio and its application to molecular gas scaling relations}",
      journal = {\mnras},
         year = 2017,
        month = oct,
       volume = {470},
       number = {4},
        pages = {4750-4766},
          doi = {10.1093/mnras/stx1556},
archivePrefix = {arXiv},
       eprint = {1702.03888},
 primaryClass = {astro-ph.GA},
       adsurl = {https://ui.adsabs.harvard.edu/abs/2017MNRAS.470.4750A}
}

@ARTICLE{Bigiel_2008,
       author = {{Bigiel}, F. and {Leroy}, A. and {Walter}, F. and {Brinks}, E. and {de Blok}, W.~J.~G. and {Madore}, B. and {Thornley}, M.~D.},
        title = "{The Star Formation Law in Nearby Galaxies on Sub-Kpc Scales}",
      journal = {\aj},
         year = 2008,
        month = dec,
       volume = {136},
       number = {6},
        pages = {2846-2871},
          doi = {10.1088/0004-6256/136/6/2846},
archivePrefix = {arXiv},
       eprint = {0810.2541},
 primaryClass = {astro-ph},
       adsurl = {https://ui.adsabs.harvard.edu/abs/2008AJ....136.2846B}
}

@ARTICLE{Leroy_CO_phangs_2021,
       author = {{Leroy}, Adam K. and {Schinnerer}, Eva and {Hughes}, Annie and {Rosolowsky}, Erik and {Pety}, J{\'e}r{\^o}me and {Schruba}, Andreas and {Usero}, Antonio and {Blanc}, Guillermo A. and {Chevance}, M{\'e}lanie and {Emsellem}, Eric and {Faesi}, Christopher M. and {Herrera}, Cinthya N. and {Liu}, Daizhong and {Meidt}, Sharon E. and {Querejeta}, Miguel and {Saito}, Toshiki and {Sandstrom}, Karin M. and {Sun}, Jiayi and {Williams}, Thomas G. and {Anand}, Gagandeep S. and {Barnes}, Ashley T. and {Behrens}, Erica A. and {Belfiore}, Francesco and {Benincasa}, Samantha M. and {Be{\v{s}}li{\'c}}, Ivana and {Bigiel}, Frank and {Bolatto}, Alberto D. and {den Brok}, Jakob S. and {Cao}, Yixian and {Chandar}, Rupali and {Chastenet}, J{\'e}r{\'e}my and {Chiang}, I-Da and {Congiu}, Enrico and {Dale}, Daniel A. and {Deger}, Sinan and {Eibensteiner}, Cosima and {Egorov}, Oleg V. and {Garc{\'\i}a-Rodr{\'\i}guez}, Axel and {Glover}, Simon C.~O. and {Grasha}, Kathryn and {Henshaw}, Jonathan D. and {Ho}, I. -Ting and {Kepley}, Amanda A. and {Kim}, Jaeyeon and {Klessen}, Ralf S. and {Kreckel}, Kathryn and {Koch}, Eric W. and {Kruijssen}, J.~M. Diederik and {Larson}, Kirsten L. and {Lee}, Janice C. and {Lopez}, Laura A. and {Machado}, Josh and {Mayker}, Ness and {McElroy}, Rebecca and {Murphy}, Eric J. and {Ostriker}, Eve C. and {Pan}, Hsi-An and {Pessa}, Ismael and {Puschnig}, Johannes and {Razza}, Alessandro and {S{\'a}nchez-Bl{\'a}zquez}, Patricia and {Santoro}, Francesco and {Sardone}, Amy and {Scheuermann}, Fabian and {Sliwa}, Kazimierz and {Sormani}, Mattia C. and {Stuber}, Sophia K. and {Thilker}, David A. and {Turner}, Jordan A. and {Utomo}, Dyas and {Watkins}, Elizabeth J. and {Whitmore}, Bradley},
        title = "{PHANGS-ALMA: Arcsecond CO(2-1) Imaging of Nearby Star-forming Galaxies}",
      journal = {\apjs},
         year = 2021,
        month = dec,
       volume = {257},
       number = {2},
          eid = {43},
        pages = {43},
          doi = {10.3847/1538-4365/ac17f3},
archivePrefix = {arXiv},
       eprint = {2104.07739},
 primaryClass = {astro-ph.GA},
       adsurl = {https://ui.adsabs.harvard.edu/abs/2021ApJS..257...43L}
}

@ARTICLE{Sun_prop_clouds_2022,
       author = {{Sun}, Jiayi and {Leroy}, Adam K. and {Rosolowsky}, Erik and {Hughes}, Annie and {Schinnerer}, Eva and {Schruba}, Andreas and {Koch}, Eric W. and {Blanc}, Guillermo A. and {Chiang}, I-Da and {Groves}, Brent and {Liu}, Daizhong and {Meidt}, Sharon and {Pan}, Hsi-An and {Pety}, J{\'e}r{\^o}me and {Querejeta}, Miguel and {Saito}, Toshiki and {Sandstrom}, Karin and {Sardone}, Amy and {Usero}, Antonio and {Utomo}, Dyas and {Williams}, Thomas G. and {Barnes}, Ashley T. and {Benincasa}, Samantha M. and {Bigiel}, Frank and {Bolatto}, Alberto D. and {Boquien}, M{\'e}d{\'e}ric and {Chevance}, M{\'e}lanie and {Dale}, Daniel A. and {Deger}, Sinan and {Emsellem}, Eric and {Glover}, Simon C.~O. and {Grasha}, Kathryn and {Henshaw}, Jonathan D. and {Klessen}, Ralf S. and {Kreckel}, Kathryn and {Kruijssen}, J.~M. Diederik and {Ostriker}, Eve C. and {Thilker}, David A.},
        title = "{Molecular Cloud Populations in the Context of Their Host Galaxy Environments: A Multiwavelength Perspective}",
      journal = {\aj},
         year = 2022,
        month = aug,
       volume = {164},
       number = {2},
          eid = {43},
        pages = {43},
          doi = {10.3847/1538-3881/ac74bd},
archivePrefix = {arXiv},
       eprint = {2206.07055},
 primaryClass = {astro-ph.GA},
       adsurl = {https://ui.adsabs.harvard.edu/abs/2022AJ....164...43S}
}

@ARTICLE{Reyes_Kennicutt_2019,
       author = {{de los Reyes}, Mithi A.~C. and {Kennicutt}, Robert C., Jr.},
        title = "{Revisiting the Integrated Star Formation Law. I. Non-starbursting Galaxies}",
      journal = {\apj},
         year = 2019,
        month = feb,
       volume = {872},
       number = {1},
          eid = {16},
        pages = {16},
          doi = {10.3847/1538-4357/aafa82},
archivePrefix = {arXiv},
       eprint = {1901.01283},
 primaryClass = {astro-ph.GA},
       adsurl = {https://ui.adsabs.harvard.edu/abs/2019ApJ...872...16D}
}

@ARTICLE{Kennicutt_Reyes_2021,
       author = {{Kennicutt}, Robert C., Jr. and {De Los Reyes}, Mithi A.~C.},
        title = "{Revisiting the Integrated Star Formation Law. II. Starbursts and the Combined Global Schmidt Law}",
      journal = {\apj},
         year = 2021,
        month = feb,
       volume = {908},
       number = {1},
          eid = {61},
        pages = {61},
          doi = {10.3847/1538-4357/abd3a2},
archivePrefix = {arXiv},
       eprint = {2012.05363},
 primaryClass = {astro-ph.GA},
       adsurl = {https://ui.adsabs.harvard.edu/abs/2021ApJ...908...61K}
}

@ARTICLE{Kim_2022_environmental_dep,
       author = {{Kim}, Jaeyeon and {Chevance}, M{\'e}lanie and {Kruijssen}, J.~M. Diederik and {Leroy}, Adam K. and {Schruba}, Andreas and {Barnes}, Ashley T. and {Bigiel}, Frank and {Blanc}, Guillermo A. and {Cao}, Yixian and {Congiu}, Enrico and {Dale}, Daniel A. and {Faesi}, Christopher M. and {Glover}, Simon C.~O. and {Grasha}, Kathryn and {Groves}, Brent and {Hughes}, Annie and {Klessen}, Ralf S. and {Kreckel}, Kathryn and {McElroy}, Rebecca and {Pan}, Hsi-An and {Pety}, J{\'e}r{\^o}me and {Querejeta}, Miguel and {Razza}, Alessandro and {Rosolowsky}, Erik and {Saito}, Toshiki and {Schinnerer}, Eva and {Sun}, Jiayi and {Tomi{\v{c}}i{\'c}}, Neven and {Usero}, Antonio and {Williams}, Thomas G.},
        title = "{Environmental dependence of the molecular cloud lifecycle in 54 main-sequence galaxies}",
      journal = {\mnras},
         year = 2022,
        month = oct,
       volume = {516},
       number = {2},
        pages = {3006-3028},
          doi = {10.1093/mnras/stac2339},
archivePrefix = {arXiv},
       eprint = {2206.09857},
 primaryClass = {astro-ph.GA},
       adsurl = {https://ui.adsabs.harvard.edu/abs/2022MNRAS.516.3006K}
}

@ARTICLE{Kim_2021_embedded24_mic,
       author = {{Kim}, Jaeyeon and {Chevance}, M{\'e}lanie and {Kruijssen}, J.~M. Diederik and {Schruba}, Andreas and {Sandstrom}, Karin and {Barnes}, Ashley T. and {Bigiel}, Frank and {Blanc}, Guillermo A. and {Cao}, Yixian and {Dale}, Daniel A. and {Faesi}, Christopher M. and {Glover}, Simon C.~O. and {Grasha}, Kathryn and {Groves}, Brent and {Herrera}, Cinthya and {Klessen}, Ralf S. and {Kreckel}, Kathryn and {Lee}, Janice C. and {Leroy}, Adam K. and {Pety}, J{\'e}r{\^o}me and {Querejeta}, Miguel and {Schinnerer}, Eva and {Sun}, Jiayi and {Usero}, Antonio and {Ward}, Jacob L. and {Williams}, Thomas G.},
        title = "{On the duration of the embedded phase of star formation}",
      journal = {\mnras},
         year = 2021,
        month = jun,
       volume = {504},
       number = {1},
        pages = {487-509},
          doi = {10.1093/mnras/stab878},
archivePrefix = {arXiv},
       eprint = {2012.00019},
 primaryClass = {astro-ph.GA},
       adsurl = {https://ui.adsabs.harvard.edu/abs/2021MNRAS.504..487K}
}

@ARTICLE{Kim_2023_ngc628,
       author = {{Kim}, Jaeyeon and {Chevance}, M{\'e}lanie and {Kruijssen}, J.~M. Diederik and {Barnes}, Ashley. T. and {Bigiel}, Frank and {Blanc}, Guillermo A. and {Boquien}, M{\'e}d{\'e}ric and {Cao}, Yixian and {Congiu}, Enrico and {Dale}, Daniel A. and {Egorov}, Oleg V. and {Faesi}, Christopher M. and {Glover}, Simon C.~O. and {Grasha}, Kathryn and {Groves}, Brent and {Hassani}, Hamid and {Hughes}, Annie and {Klessen}, Ralf S. and {Kreckel}, Kathryn and {Larson}, Kirsten L. and {Lee}, Janice C. and {Leroy}, Adam K. and {Liu}, Daizhong and {Longmore}, Steven N. and {Meidt}, Sharon E. and {Pan}, Hsi-An and {Pety}, J{\'e}r{\^o}me and {Querejeta}, Miguel and {Rosolowsky}, Erik and {Saito}, Toshiki and {Sandstrom}, Karin and {Schinnerer}, Eva and {Smith}, Rowan J. and {Usero}, Antonio and {Watkins}, Elizabeth J. and {Williams}, Thomas G.},
        title = "{PHANGS-JWST First Results: Duration of the Early Phase of Massive Star Formation in NGC 628}",
      journal = {\apjl},
         year = 2023,
        month = feb,
       volume = {944},
       number = {2},
          eid = {L20},
        pages = {L20},
          doi = {10.3847/2041-8213/aca90a},
archivePrefix = {arXiv},
       eprint = {2211.15698},
 primaryClass = {astro-ph.GA},
       adsurl = {https://ui.adsabs.harvard.edu/abs/2023ApJ...944L..20K}
}

@ARTICLE{Aniano_2011_convolution,
       author = {{Aniano}, G. and {Draine}, B.~T. and {Gordon}, K.~D. and {Sandstrom}, K.},
        title = "{Common-Resolution Convolution Kernels for Space- and Ground-Based Telescopes}",
      journal = {\pasp},
         year = 2011,
        month = oct,
       volume = {123},
       number = {908},
        pages = {1218},
          doi = {10.1086/662219},
archivePrefix = {arXiv},
       eprint = {1106.5065},
 primaryClass = {astro-ph.IM},
       adsurl = {https://ui.adsabs.harvard.edu/abs/2011PASP..123.1218A}
}

@ARTICLE{Williams_2024_jwstphangs_cycle1,
       author = {{Williams}, Thomas G. and {Lee}, Janice C. and {Larson}, Kirsten L. and {Leroy}, Adam K. and {Sandstrom}, Karin and {Schinnerer}, Eva and {Thilker}, David A. and {Belfiore}, Francesco and {Egorov}, Oleg V. and {Rosolowsky}, Erik and {Sutter}, Jessica and {DePasquale}, Joseph and {Pagan}, Alyssa and {Berger}, Travis A. and {Anand}, Gagandeep S. and {Barnes}, Ashley T. and {Bigiel}, Frank and {Boquien}, M{\'e}d{\'e}ric and {Cao}, Yixian and {Chastenet}, J{\'e}r{\'e}my and {Chevance}, M{\'e}lanie and {Chown}, Ryan and {Dale}, Daniel A. and {Deger}, Sinan and {Eibensteiner}, Cosima and {Emsellem}, Eric and {Faesi}, Christopher M. and {Glover}, Simon C.~O. and {Grasha}, Kathryn and {Hannon}, Stephen and {Hassani}, Hamid and {Henshaw}, Jonathan D. and {Jim{\'e}nez-Donaire}, Mar{\'\i}a J. and {Kim}, Jaeyeon and {Klessen}, Ralf S. and {Koch}, Eric W. and {Li}, Jing and {Liu}, Daizhong and {Meidt}, Sharon E. and {M{\'e}ndez-Delgado}, J. Eduardo and {Murphy}, Eric J. and {Neumann}, Justus and {Neumann}, Lukas and {Neumayer}, Nadine and {Oakes}, Elias K. and {Pathak}, Debosmita and {Pety}, J{\'e}r{\^o}me and {Pinna}, Francesca and {Querejeta}, Miguel and {Ramambason}, Lise and {Romanelli}, Andrea and {Sormani}, Mattia C. and {Stuber}, Sophia K. and {Sun}, Jiayi and {Teng}, Yu-Hsuan and {Usero}, Antonio and {Watkins}, Elizabeth J. and {Weinbeck}, Tony D.},
        title = "{PHANGS-JWST: Data-processing Pipeline and First Full Public Data Release}",
      journal = {\apjs},
         year = 2024,
        month = jul,
       volume = {273},
       number = {1},
          eid = {13},
        pages = {13},
          doi = {10.3847/1538-4365/ad4be5},
archivePrefix = {arXiv},
       eprint = {2401.15142},
 primaryClass = {astro-ph.GA},
       adsurl = {https://ui.adsabs.harvard.edu/abs/2024ApJS..273...13W}
}

@INPROCEEDINGS{Perrin_2014_webbpsf,
       author = {{Perrin}, Marshall D. and {Sivaramakrishnan}, Anand and {Lajoie}, Charles-Philippe and {Elliott}, Erin and {Pueyo}, Laurent and {Ravindranath}, Swara and {Albert}, Lo{\"\i}c.},
        title = "{Updated point spread function simulations for JWST with WebbPSF}",
    booktitle = {Space Telescopes and Instrumentation 2014: Optical, Infrared, and Millimeter Wave},
         year = 2014,
       editor = {{Oschmann}, Jacobus M., Jr. and {Clampin}, Mark and {Fazio}, Giovanni G. and {MacEwen}, Howard A.},
       series = {Society of Photo-Optical Instrumentation Engineers (SPIE) Conference Series},
       volume = {9143},
        month = aug,
          eid = {91433X},
        pages = {91433X},
          doi = {10.1117/12.2056689},
       adsurl = {https://ui.adsabs.harvard.edu/abs/2014SPIE.9143E..3XP}
}

@ARTICLE{Leroy_m51_2017,
       author = {{Leroy}, Adam K. and {Schinnerer}, Eva and {Hughes}, Annie and {Kruijssen}, J.~M. Diederik and {Meidt}, Sharon and {Schruba}, Andreas and {Sun}, Jiayi and {Bigiel}, Frank and {Aniano}, Gonzalo and {Blanc}, Guillermo A. and {Bolatto}, Alberto and {Chevance}, M{\'e}lanie and {Colombo}, Dario and {Gallagher}, Molly and {Garcia-Burillo}, Santiago and {Kramer}, Carsten and {Querejeta}, Miguel and {Pety}, Jerome and {Thompson}, Todd A. and {Usero}, Antonio},
        title = "{Cloud-scale ISM Structure and Star Formation in M51}",
      journal = {\apj},
         year = 2017,
        month = sep,
       volume = {846},
       number = {1},
          eid = {71},
        pages = {71},
          doi = {10.3847/1538-4357/aa7fef},
archivePrefix = {arXiv},
       eprint = {1706.08540},
 primaryClass = {astro-ph.GA},
       adsurl = {https://ui.adsabs.harvard.edu/abs/2017ApJ...846...71L}
}

@ARTICLE{Leroy_2019_z0mgs,
       author = {{Leroy}, Adam K. and {Sandstrom}, Karin M. and {Lang}, Dustin and {Lewis}, Alexia and {Salim}, Samir and {Behrens}, Erica A. and {Chastenet}, J{\'e}r{\'e}my and {Chiang}, I-Da and {Gallagher}, Molly J. and {Kessler}, Sarah and {Utomo}, Dyas},
        title = "{A z = 0 Multiwavelength Galaxy Synthesis. I. A WISE and GALEX Atlas of Local Galaxies}",
      journal = {\apjs},
         year = 2019,
        month = oct,
       volume = {244},
       number = {2},
          eid = {24},
        pages = {24},
          doi = {10.3847/1538-4365/ab3925},
archivePrefix = {arXiv},
       eprint = {1910.13470},
 primaryClass = {astro-ph.GA},
       adsurl = {https://ui.adsabs.harvard.edu/abs/2019ApJS..244...24L}
}

@ARTICLE{Leroy_sfr_calibration_2012,
       author = {{Leroy}, Adam K. and {Bigiel}, Frank and {de Blok}, W.~J.~G. and {Boissier}, Samuel and {Bolatto}, Alberto and {Brinks}, Elias and {Madore}, Barry and {Munoz-Mateos}, Juan-Carlos and {Murphy}, Eric and {Sandstrom}, Karin and {Schruba}, Andreas and {Walter}, Fabian},
        title = "{Estimating the Star Formation Rate at 1 kpc Scales in nearby Galaxies}",
      journal = {\aj},
         year = 2012,
        month = jul,
       volume = {144},
       number = {1},
          eid = {3},
        pages = {3},
          doi = {10.1088/0004-6256/144/1/3},
archivePrefix = {arXiv},
       eprint = {1202.2873},
 primaryClass = {astro-ph.CO},
       adsurl = {https://ui.adsabs.harvard.edu/abs/2012AJ....144....3L}
}

@ARTICLE{Williams_2022_mixing_met,
       author = {{Williams}, Thomas G. and {Kreckel}, Kathryn and {Belfiore}, Francesco and {Groves}, Brent and {Sandstrom}, Karin and {Santoro}, Francesco and {Blanc}, Guillermo A. and {Bigiel}, Frank and {Boquien}, M{\'e}d{\'e}ric and {Chevance}, M{\'e}lanie and {Congiu}, Enrico and {Emsellem}, Eric and {Glover}, Simon C.~O. and {Grasha}, Kathryn and {Klessen}, Ralf S. and {Koch}, Eric and {Kruijssen}, J.~M. Diederik and {Leroy}, Adam K. and {Liu}, Daizhong and {Meidt}, Sharon and {Pan}, Hsi-An and {Querejeta}, Miguel and {Rosolowsky}, Erik and {Saito}, Toshiki and {S{\'a}nchez-Bl{\'a}zquez}, Patricia and {Schinnerer}, Eva and {Schruba}, Andreas and {Watkins}, Elizabeth J.},
        title = "{The 2D metallicity distribution and mixing scales of nearby galaxies}",
      journal = {\mnras},
         year = 2022,
        month = jan,
       volume = {509},
       number = {1},
        pages = {1303-1322},
          doi = {10.1093/mnras/stab3082},
archivePrefix = {arXiv},
       eprint = {2110.10697},
 primaryClass = {astro-ph.GA},
       adsurl = {https://ui.adsabs.harvard.edu/abs/2022MNRAS.509.1303W}
}

@ARTICLE{Lee_2023_phangs_jwst,
       author = {{Lee}, Janice C. and {Sandstrom}, Karin M. and {Leroy}, Adam K. and {Thilker}, David A. and {Schinnerer}, Eva and {Rosolowsky}, Erik and {Larson}, Kirsten L. and {Egorov}, Oleg V. and {Williams}, Thomas G. and {Schmidt}, Judy and {Emsellem}, Eric and {Anand}, Gagandeep S. and {Barnes}, Ashley T. and {Belfiore}, Francesco and {Be{\v{s}}li{\'c}}, Ivana and {Bigiel}, Frank and {Blanc}, Guillermo A. and {Bolatto}, Alberto D. and {Boquien}, M{\'e}d{\'e}ric and {den Brok}, Jakob and {Cao}, Yixian and {Chandar}, Rupali and {Chastenet}, J{\'e}r{\'e}my and {Chevance}, M{\'e}lanie and {Chiang}, I-Da and {Congiu}, Enrico and {Dale}, Daniel A. and {Deger}, Sinan and {Eibensteiner}, Cosima and {Faesi}, Christopher M. and {Glover}, Simon C.~O. and {Grasha}, Kathryn and {Groves}, Brent and {Hassani}, Hamid and {Henny}, Kiana F. and {Henshaw}, Jonathan D. and {Hoyer}, Nils and {Hughes}, Annie and {Jeffreson}, Sarah and {Jim{\'e}nez-Donaire}, Mar{\'\i}a J. and {Kim}, Jaeyeon and {Kim}, Hwihyun and {Klessen}, Ralf S. and {Koch}, Eric W. and {Kreckel}, Kathryn and {Kruijssen}, J.~M. Diederik and {Li}, Jing and {Liu}, Daizhong and {Lopez}, Laura A. and {Maschmann}, Daniel and {Chen}, Ness Mayker and {Meidt}, Sharon E. and {Murphy}, Eric J. and {Neumann}, Justus and {Neumayer}, Nadine and {Pan}, Hsi-An and {Pessa}, Ismael and {Pety}, J{\'e}r{\^o}me and {Querejeta}, Miguel and {Pinna}, Francesca and {Rodr{\'\i}guez}, M. Jimena and {Saito}, Toshiki and {S{\'a}nchez-Bl{\'a}zquez}, Patricia and {Santoro}, Francesco and {Sardone}, Amy and {Smith}, Rowan J. and {Sormani}, Mattia C. and {Scheuermann}, Fabian and {Stuber}, Sophia K. and {Sutter}, Jessica and {Sun}, Jiayi and {Teng}, Yu-Hsuan and {Tre{\ss}}, Robin G. and {Usero}, Antonio and {Watkins}, Elizabeth J. and {Whitmore}, Bradley C. and {Razza}, Alessandro},
        title = "{The PHANGS-JWST Treasury Survey: Star Formation, Feedback, and Dust Physics at High Angular Resolution in Nearby GalaxieS}",
      journal = {\apjl},
         year = 2023,
        month = feb,
       volume = {944},
       number = {2},
          eid = {L17},
        pages = {L17},
          doi = {10.3847/2041-8213/acaaae},
archivePrefix = {arXiv},
       eprint = {2212.02667},
 primaryClass = {astro-ph.GA},
       adsurl = {https://ui.adsabs.harvard.edu/abs/2023ApJ...944L..17L}
}

@ARTICLE{Sun_2023_calibration,
       author = {{Sun}, Jiayi and {Leroy}, Adam K. and {Ostriker}, Eve C. and {Meidt}, Sharon and {Rosolowsky}, Erik and {Schinnerer}, Eva and {Wilson}, Christine D. and {Utomo}, Dyas and {Belfiore}, Francesco and {Blanc}, Guillermo A. and {Emsellem}, Eric and {Faesi}, Christopher and {Groves}, Brent and {Hughes}, Annie and {Koch}, Eric W. and {Kreckel}, Kathryn and {Liu}, Daizhong and {Pan}, Hsi-An and {Pety}, J{\'e}r{\^o}me and {Querejeta}, Miguel and {Razza}, Alessandro and {Saito}, Toshiki and {Sardone}, Amy and {Usero}, Antonio and {Williams}, Thomas G. and {Bigiel}, Frank and {Bolatto}, Alberto D. and {Chevance}, M{\'e}lanie and {Dale}, Daniel A. and {Gensior}, Jindra and {Glover}, Simon C.~O. and {Grasha}, Kathryn and {Henshaw}, Jonathan D. and {Jim{\'e}nez-Donaire}, Mar{\'\i}a J. and {Klessen}, Ralf S. and {Kruijssen}, J.~M. Diederik and {Murphy}, Eric J. and {Neumann}, Lukas and {Teng}, Yu-Hsuan and {Thilker}, David A.},
        title = "{Star Formation Laws and Efficiencies across 80 Nearby Galaxies}",
      journal = {\apjl},
         year = 2023,
        month = mar,
       volume = {945},
       number = {2},
          eid = {L19},
        pages = {L19},
          doi = {10.3847/2041-8213/acbd9c},
archivePrefix = {arXiv},
       eprint = {2302.12267},
 primaryClass = {astro-ph.GA},
       adsurl = {https://ui.adsabs.harvard.edu/abs/2023ApJ...945L..19S}
}

@ARTICLE{Sun_2020_cloud_scale,
       author = {{Sun}, Jiayi and {Leroy}, Adam K. and {Schinnerer}, Eva and {Hughes}, Annie and {Rosolowsky}, Erik and {Querejeta}, Miguel and {Schruba}, Andreas and {Liu}, Daizhong and {Saito}, Toshiki and {Herrera}, Cinthya N. and {Faesi}, Christopher and {Usero}, Antonio and {Pety}, J{\'e}r{\^o}me and {Kruijssen}, J.~M. Diederik and {Ostriker}, Eve C. and {Bigiel}, Frank and {Blanc}, Guillermo A. and {Bolatto}, Alberto D. and {Boquien}, M{\'e}d{\'e}ric and {Chevance}, M{\'e}lanie and {Dale}, Daniel A. and {Deger}, Sinan and {Emsellem}, Eric and {Glover}, Simon C.~O. and {Grasha}, Kathryn and {Groves}, Brent and {Henshaw}, Jonathan and {Jimenez-Donaire}, Maria J. and {Kim}, Jenny J. and {Klessen}, Ralf S. and {Kreckel}, Kathryn and {Lee}, Janice C. and {Meidt}, Sharon and {Sandstrom}, Karin and {Sardone}, Amy E. and {Utomo}, Dyas and {Williams}, Thomas G.},
        title = "{Molecular Gas Properties on Cloud Scales across the Local Star-forming Galaxy Population}",
      journal = {\apjl},
         year = 2020,
        month = sep,
       volume = {901},
       number = {1},
          eid = {L8},
        pages = {L8},
          doi = {10.3847/2041-8213/abb3be},
archivePrefix = {arXiv},
       eprint = {2009.01842},
 primaryClass = {astro-ph.GA},
       adsurl = {https://ui.adsabs.harvard.edu/abs/2020ApJ...901L...8S}
}

@ARTICLE{Hygate_2019_diffuse,
       author = {{Hygate}, Alexander P.~S. and {Kruijssen}, J.~M. Diederik and {Chevance}, M{\'e}lanie and {Schruba}, Andreas and {Haydon}, Daniel T. and {Longmore}, Steven N.},
        title = "{An uncertainty principle for star formation - IV. On the nature and filtering of diffuse emission}",
      journal = {\mnras},
         year = 2019,
        month = sep,
       volume = {488},
       number = {2},
        pages = {2800-2824},
          doi = {10.1093/mnras/stz1779},
archivePrefix = {arXiv},
       eprint = {1810.11405},
 primaryClass = {astro-ph.GA},
       adsurl = {https://ui.adsabs.harvard.edu/abs/2019MNRAS.488.2800H}
}

@ARTICLE{Sanchez_2019_metallicity_grad,
       author = {{S{\'a}nchez}, S.~F. and {Barrera-Ballesteros}, J.~K. and {L{\'o}pez-Cob{\'a}}, C. and {Brough}, S. and {Bryant}, J.~J. and {Bland-Hawthorn}, J. and {Croom}, S.~M. and {van de Sande}, J. and {Cortese}, L. and {Goodwin}, M. and {Lawrence}, J.~S. and {L{\'o}pez-S{\'a}nchez}, A.~R. and {Sweet}, S.~M. and {Owers}, M.~S. and {Richards}, S.~N. and {Walcher}, C.~J.},
        title = "{The SAMI galaxy survey: exploring the gas-phase mass-metallicity relation}",
      journal = {\mnras},
         year = 2019,
        month = apr,
       volume = {484},
       number = {3},
        pages = {3042-3070},
          doi = {10.1093/mnras/stz019},
archivePrefix = {arXiv},
       eprint = {1812.11263},
 primaryClass = {astro-ph.GA},
       adsurl = {https://ui.adsabs.harvard.edu/abs/2019MNRAS.484.3042S}
}

@ARTICLE{Santoro_2022_PHANGS_LF,
       author = {{Santoro}, Francesco and {Kreckel}, Kathryn and {Belfiore}, Francesco and {Groves}, Brent and {Congiu}, Enrico and {Thilker}, David A. and {Blanc}, Guillermo A. and {Schinnerer}, Eva and {Ho}, I. -Ting and {Kruijssen}, J.~M. Diederik and {Meidt}, Sharon and {Klessen}, Ralf S. and {Schruba}, Andreas and {Querejeta}, Miguel and {Pessa}, Ismael and {Chevance}, M{\'e}lanie and {Kim}, Jaeyeon and {Emsellem}, Eric and {McElroy}, Rebecca and {Barnes}, Ashley T. and {Bigiel}, Frank and {Boquien}, M{\'e}d{\'e}ric and {Dale}, Daniel A. and {Glover}, Simon C.~O. and {Grasha}, Kathryn and {Lee}, Janice and {Leroy}, Adam K. and {Pan}, Hsi-An and {Rosolowsky}, Erik and {Saito}, Toshiki and {Sanchez-Blazquez}, Patricia and {Watkins}, Elizabeth J. and {Williams}, Thomas G.},
        title = "{PHANGS-MUSE: The H II region luminosity function of local star-forming galaxies}",
      journal = {\aap},
         year = 2022,
        month = feb,
       volume = {658},
          eid = {A188},
        pages = {A188},
          doi = {10.1051/0004-6361/202141907},
archivePrefix = {arXiv},
       eprint = {2111.09362},
 primaryClass = {astro-ph.GA},
       adsurl = {https://ui.adsabs.harvard.edu/abs/2022A&A...658A.188S}
}

@ARTICLE{Leroy_2023_midIR_CO_1kpc,
       author = {{Leroy}, Adam K. and {Bolatto}, Alberto D. and {Sandstrom}, Karin and {Rosolowsky}, Erik and {Barnes}, Ashley. T. and {Bigiel}, F. and {Boquien}, M{\'e}d{\'e}ric and {den Brok}, Jakob S. and {Cao}, Yixian and {Chastenet}, J{\'e}r{\'e}my and {Chevance}, M{\'e}lanie and {Chiang}, I-Da and {Chown}, Ryan and {Colombo}, Dario and {Ellison}, Sara L. and {Emsellem}, Eric and {Grasha}, Kathryn and {Henshaw}, Jonathan D. and {Hughes}, Annie and {Klessen}, Ralf S. and {Koch}, Eric W. and {Kim}, Jaeyeon and {Kreckel}, Kathryn and {Kruijssen}, J.~M. Diederik and {Larson}, Kirsten L. and {Lee}, Janice C. and {Levy}, Rebecca C. and {Lin}, Lihwai and {Liu}, Daizhong and {Meidt}, Sharon E. and {Pety}, J{\'e}r{\^o}me and {Querejeta}, Miguel and {Rubio}, M{\'o}nica and {Saito}, Toshiki and {Salim}, Samir and {Schinnerer}, Eva and {Sormani}, Mattia C. and {Sun}, Jiayi and {Thilker}, David A. and {Usero}, Antonio and {Vogel}, Stuart N. and {Watkins}, Elizabeth J. and {Whitcomb}, Cory M. and {Williams}, Thomas G. and {Wilson}, Christine D.},
        title = "{PHANGS-JWST First Results: A Global and Moderately Resolved View of Mid-infrared and CO Line Emission from Galaxies at the Start of the JWST Era}",
      journal = {\apjl},
         year = 2023,
        month = feb,
       volume = {944},
       number = {2},
          eid = {L10},
        pages = {L10},
          doi = {10.3847/2041-8213/acab01},
archivePrefix = {arXiv},
       eprint = {2212.09774},
 primaryClass = {astro-ph.GA},
       adsurl = {https://ui.adsabs.harvard.edu/abs/2023ApJ...944L..10L}
}

@ARTICLE{Leroy_2023_midIR_CO_Ha,
       author = {{Leroy}, Adam K. and {Sandstrom}, Karin and {Rosolowsky}, Erik and {Belfiore}, Francesco and {Bolatto}, Alberto D. and {Cao}, Yixian and {Koch}, Eric W. and {Schinnerer}, Eva and {Barnes}, Ashley. T. and {Be{\v{s}}li{\'c}}, Ivana and {Bigiel}, F. and {Blanc}, Guillermo A. and {Chastenet}, J{\'e}r{\'e}my and {Chen}, Ness Mayker and {Chevance}, M{\'e}lanie and {Chown}, Ryan and {Congiu}, Enrico and {Dale}, Daniel A. and {Egorov}, Oleg V. and {Emsellem}, Eric and {Eibensteiner}, Cosima and {Faesi}, Christopher M. and {Glover}, Simon C.~O. and {Grasha}, Kathryn and {Groves}, Brent and {Hassani}, Hamid and {Henshaw}, Jonathan D. and {Hughes}, Annie and {Jim{\'e}nez-Donaire}, Mar{\'\i}a J. and {Kim}, Jaeyeon and {Klessen}, Ralf S. and {Kreckel}, Kathryn and {Kruijssen}, J.~M. Diederik and {Larson}, Kirsten L. and {Lee}, Janice C. and {Levy}, Rebecca C. and {Liu}, Daizhong and {Lopez}, Laura A. and {Meidt}, Sharon E. and {Murphy}, Eric J. and {Neumann}, Justus and {Pessa}, Ismael and {Pety}, J{\'e}r{\^o}me and {Saito}, Toshiki and {Sardone}, Amy and {Sun}, Jiayi and {Thilker}, David A. and {Usero}, Antonio and {Watkins}, Elizabeth J. and {Whitcomb}, Cory M. and {Williams}, Thomas G.},
        title = "{PHANGS-JWST First Results: Mid-infrared Emission Traces Both Gas Column Density and Heating at 100 pc Scales}",
      journal = {\apjl},
         year = 2023,
        month = feb,
       volume = {944},
       number = {2},
          eid = {L9},
        pages = {L9},
          doi = {10.3847/2041-8213/acaf85},
archivePrefix = {arXiv},
       eprint = {2212.10574},
 primaryClass = {astro-ph.GA},
       adsurl = {https://ui.adsabs.harvard.edu/abs/2023ApJ...944L...9L}
}

@ARTICLE{Pathak_2024_midIR_pdf,
       author = {{Pathak}, Debosmita and {Leroy}, Adam K. and {Thompson}, Todd A. and {Lopez}, Laura A. and {Belfiore}, Francesco and {Boquien}, M{\'e}d{\'e}ric and {Dale}, Daniel A. and {Glover}, Simon C.~O. and {Klessen}, Ralf S. and {Koch}, Eric W. and {Rosolowsky}, Erik and {Sandstrom}, Karin M. and {Schinnerer}, Eva and {Smith}, Rowan and {Sun}, Jiayi and {Sutter}, Jessica and {Williams}, Thomas G. and {Bigiel}, Frank and {Cao}, Yixian and {Chastenet}, J{\'e}r{\'e}my and {Chevance}, M{\'e}lanie and {Chown}, Ryan and {Emsellem}, Eric and {Faesi}, Christopher M. and {Larson}, Kirsten L. and {Lee}, Janice C. and {Meidt}, Sharon and {Ostriker}, Eve C. and {Ramambason}, Lise and {Sarbadhicary}, Sumit K. and {Thilker}, David A.},
        title = "{A Two-Component Probability Distribution Function Describes the Mid-IR Emission from the Disks of Star-forming Galaxies}",
      journal = {\aj},
         year = 2024,
        month = jan,
       volume = {167},
       number = {1},
          eid = {39},
        pages = {39},
          doi = {10.3847/1538-3881/ad110d},
archivePrefix = {arXiv},
       eprint = {2311.18067},
 primaryClass = {astro-ph.GA},
       adsurl = {https://ui.adsabs.harvard.edu/abs/2024AJ....167...39P}
}

@ARTICLE{Belfiore_2023a_sfr_phangsMUSE,
       author = {{Belfiore}, Francesco and {Leroy}, Adam K. and {Sun}, Jiayi and {Barnes}, Ashley T. and {Boquien}, M{\'e}d{\'e}ric and {Cao}, Yixian and {Congiu}, Enrico and {Dale}, Daniel A. and {Egorov}, Oleg V. and {Eibensteiner}, Cosima and {Glover}, Simon C.~O. and {Grasha}, Kathryn and {Groves}, Brent and {Klessen}, Ralf S. and {Kreckel}, Kathryn and {Neumann}, Lukas and {Querejeta}, Miguel and {Sanchez-Blazquez}, Patricia and {Schinnerer}, Eva and {Williams}, Thomas G.},
        title = "{Calibration of hybrid resolved star formation rate recipes based on PHANGS-MUSE H{\ensuremath{\alpha}} and H{\ensuremath{\beta}} maps}",
      journal = {\aap},
         year = 2023,
        month = feb,
       volume = {670},
          eid = {A67},
        pages = {A67},
          doi = {10.1051/0004-6361/202244863},
archivePrefix = {arXiv},
       eprint = {2211.08487},
 primaryClass = {astro-ph.GA},
       adsurl = {https://ui.adsabs.harvard.edu/abs/2023A&A...670A..67B}
}

@ARTICLE{Belfiore_2023b_sfr_midIR,
       author = {{Belfiore}, Francesco and {Leroy}, Adam K. and {Williams}, Thomas G. and {Barnes}, Ashley T. and {Bigiel}, Frank and {Boquien}, M{\'e}d{\'e}ric and {Cao}, Yixian and {Chastenet}, J{\'e}r{\'e}my and {Congiu}, Enrico and {Dale}, Daniel A. and {Egorov}, Oleg V. and {Eibensteiner}, Cosima and {Emsellem}, Eric and {Glover}, Simon C.~O. and {Groves}, Brent and {Hassani}, Hamid and {Klessen}, Ralf S. and {Kreckel}, Kathryn and {Neumann}, Lukas and {Neumann}, Justus and {Querejeta}, Miguel and {Rosolowsky}, Erik and {Sanchez-Blazquez}, Patricia and {Sandstrom}, Karin and {Schinnerer}, Eva and {Sun}, Jiayi and {Sutter}, Jessica and {Watkins}, Elizabeth J.},
        title = "{Calibrating mid-infrared emission as a tracer of obscured star formation on H II-region scales in the era of JWST}",
      journal = {\aap},
         year = 2023,
        month = oct,
       volume = {678},
          eid = {A129},
        pages = {A129},
          doi = {10.1051/0004-6361/202347175},
archivePrefix = {arXiv},
       eprint = {2306.11811},
 primaryClass = {astro-ph.GA},
       adsurl = {https://ui.adsabs.harvard.edu/abs/2023A&A...678A.129B}
}

@ARTICLE{Leroy_2021_phang_alma,
       author = {{Leroy}, Adam K. and {Schinnerer}, Eva and {Hughes}, Annie and {Rosolowsky}, Erik and {Pety}, J{\'e}r{\^o}me and {Schruba}, Andreas and {Usero}, Antonio and {Blanc}, Guillermo A. and {Chevance}, M{\'e}lanie and {Emsellem}, Eric and {Faesi}, Christopher M. and {Herrera}, Cinthya N. and {Liu}, Daizhong and {Meidt}, Sharon E. and {Querejeta}, Miguel and {Saito}, Toshiki and {Sandstrom}, Karin M. and {Sun}, Jiayi and {Williams}, Thomas G. and {Anand}, Gagandeep S. and {Barnes}, Ashley T. and {Behrens}, Erica A. and {Belfiore}, Francesco and {Benincasa}, Samantha M. and {Be{\v{s}}li{\'c}}, Ivana and {Bigiel}, Frank and {Bolatto}, Alberto D. and {den Brok}, Jakob S. and {Cao}, Yixian and {Chandar}, Rupali and {Chastenet}, J{\'e}r{\'e}my and {Chiang}, I-Da and {Congiu}, Enrico and {Dale}, Daniel A. and {Deger}, Sinan and {Eibensteiner}, Cosima and {Egorov}, Oleg V. and {Garc{\'\i}a-Rodr{\'\i}guez}, Axel and {Glover}, Simon C.~O. and {Grasha}, Kathryn and {Henshaw}, Jonathan D. and {Ho}, I. -Ting and {Kepley}, Amanda A. and {Kim}, Jaeyeon and {Klessen}, Ralf S. and {Kreckel}, Kathryn and {Koch}, Eric W. and {Kruijssen}, J.~M. Diederik and {Larson}, Kirsten L. and {Lee}, Janice C. and {Lopez}, Laura A. and {Machado}, Josh and {Mayker}, Ness and {McElroy}, Rebecca and {Murphy}, Eric J. and {Ostriker}, Eve C. and {Pan}, Hsi-An and {Pessa}, Ismael and {Puschnig}, Johannes and {Razza}, Alessandro and {S{\'a}nchez-Bl{\'a}zquez}, Patricia and {Santoro}, Francesco and {Sardone}, Amy and {Scheuermann}, Fabian and {Sliwa}, Kazimierz and {Sormani}, Mattia C. and {Stuber}, Sophia K. and {Thilker}, David A. and {Turner}, Jordan A. and {Utomo}, Dyas and {Watkins}, Elizabeth J. and {Whitmore}, Bradley},
        title = "{PHANGS-ALMA: Arcsecond CO(2-1) Imaging of Nearby Star-forming Galaxies}",
      journal = {\apjs},
         year = 2021,
        month = dec,
       volume = {257},
       number = {2},
          eid = {43},
        pages = {43},
          doi = {10.3847/1538-4365/ac17f3},
archivePrefix = {arXiv},
       eprint = {2104.07739},
 primaryClass = {astro-ph.GA},
       adsurl = {https://ui.adsabs.harvard.edu/abs/2021ApJS..257...43L}
}

@ARTICLE{Pan_morpht_2022,
       author = {{Pan}, Hsi-An and {Schinnerer}, Eva and {Hughes}, Annie and {Leroy}, Adam and {Groves}, Brent and {Barnes}, Ashley Thomas and {Belfiore}, Francesco and {Bigiel}, Frank and {Blanc}, Guillermo A. and {Cao}, Yixian and {Chevance}, M{\'e}lanie and {Congiu}, Enrico and {Dale}, Daniel A. and {Eibensteiner}, Cosima and {Emsellem}, Eric and {Faesi}, Christopher M. and {Glover}, Simon C.~O. and {Grasha}, Kathryn and {Herrera}, Cinthya N. and {Ho}, I. -Ting and {Klessen}, Ralf S. and {Kruijssen}, J.~M. Diederik and {Lang}, Philipp and {Liu}, Daizhong and {McElroy}, Rebecca and {Meidt}, Sharon E. and {Murphy}, Eric J. and {Pety}, J{\'e}r{\^o}me and {Querejeta}, Miguel and {Razza}, Alessandro and {Rosolowsky}, Erik and {Saito}, Toshiki and {Santoro}, Francesco and {Schruba}, Andreas and {Sun}, Jiayi and {Tomi{\v{c}}i{\'c}}, Neven and {Usero}, Antonio and {Utomo}, Dyas and {Williams}, Thomas G.},
        title = "{The Gas-Star Formation Cycle in Nearby Star-forming Galaxies. II. Resolved Distributions of CO and H{\ensuremath{\alpha}} Emission for 49 PHANGS Galaxies}",
      journal = {\apj},
         year = 2022,
        month = mar,
       volume = {927},
       number = {1},
          eid = {9},
        pages = {9},
          doi = {10.3847/1538-4357/ac474f},
archivePrefix = {arXiv},
       eprint = {2201.01403},
 primaryClass = {astro-ph.GA},
       adsurl = {https://ui.adsabs.harvard.edu/abs/2022ApJ...927....9P}
}

@ARTICLE{Kruijssen_tf_2018,
       author = {{Kruijssen}, J.~M. Diederik and {Schruba}, Andreas and {Hygate}, Alexander P.~S. and {Hu}, Chia-Yu and {Haydon}, Daniel T. and {Longmore}, Steven N.},
        title = "{An uncertainty principle for star formation - II. A new method for characterizing the cloud-scale physics of star formation and feedback across cosmic history}",
      journal = {\mnras},
         year = 2018,
        month = sep,
       volume = {479},
       number = {2},
        pages = {1866-1952},
          doi = {10.1093/mnras/sty1128},
archivePrefix = {arXiv},
       eprint = {1805.00012},
 primaryClass = {astro-ph.GA},
       adsurl = {https://ui.adsabs.harvard.edu/abs/2018MNRAS.479.1866K}
}

@ARTICLE{Kruijssen_nat_2019,
       author = {{Kruijssen}, J.~M. Diederik and {Schruba}, Andreas and {Chevance}, M{\'e}lanie and {Longmore}, Steven N. and {Hygate}, Alexander P.~S. and {Haydon}, Daniel T. and {McLeod}, Anna F. and {Dalcanton}, Julianne J. and {Tacconi}, Linda J. and {van Dishoeck}, Ewine F.},
        title = "{Fast and inefficient star formation due to short-lived molecular clouds and rapid feedback}",
      journal = {\nat},
         year = 2019,
        month = may,
       volume = {569},
       number = {7757},
        pages = {519-522},
          doi = {10.1038/s41586-019-1194-3},
archivePrefix = {arXiv},
       eprint = {1905.08801},
 primaryClass = {astro-ph.GA},
       adsurl = {https://ui.adsabs.harvard.edu/abs/2019Natur.569..519K}
}

@article{holm_1979,
 ISSN = {03036898, 14679469},
 URL = {http://www.jstor.org/stable/4615733},
 author = {Sture Holm},
 journal = {Scandinavian Journal of Statistics},
 number = {2},
 pages = {65--70},
 publisher = {[Board of the Foundation of the Scandinavian Journal of Statistics, Wiley]},
 title = {A Simple Sequentially Rejective Multiple Test Procedure},
 urldate = {2024-11-18},
 volume = {6},
 year = {1979}
}

@ARTICLE{Rosolowsky_gmc_2021,
       author = {{Rosolowsky}, Erik and {Hughes}, Annie and {Leroy}, Adam K. and {Sun}, Jiayi and {Querejeta}, Miguel and {Schruba}, Andreas and {Usero}, Antonio and {Herrera}, Cinthya N. and {Liu}, Daizhong and {Pety}, J{\'e}r{\^o}me and {Saito}, Toshiki and {Be{\v{s}}li{\'c}}, Ivana and {Bigiel}, Frank and {Blanc}, Guillermo and {Chevance}, M{\'e}lanie and {Dale}, Daniel A. and {Deger}, Sinan and {Faesi}, Christopher M. and {Glover}, Simon C.~O. and {Henshaw}, Jonathan D. and {Klessen}, Ralf S. and {Kruijssen}, J.~M. Diederik and {Larson}, Kirsten and {Lee}, Janice and {Meidt}, Sharon and {Mok}, Angus and {Schinnerer}, Eva and {Thilker}, David A. and {Williams}, Thomas G.},
        title = "{Giant molecular cloud catalogues for PHANGS-ALMA: methods and initial results}",
      journal = {\mnras},
         year = 2021,
        month = mar,
       volume = {502},
       number = {1},
        pages = {1218-1245},
          doi = {10.1093/mnras/stab085},
archivePrefix = {arXiv},
       eprint = {2101.04697},
 primaryClass = {astro-ph.GA},
       adsurl = {https://ui.adsabs.harvard.edu/abs/2021MNRAS.502.1218R}
}

@ARTICLE{Lang_rot_curve_2020,
       author = {{Lang}, Philipp and {Meidt}, Sharon E. and {Rosolowsky}, Erik and {Nofech}, Joseph and {Schinnerer}, Eva and {Leroy}, Adam K. and {Emsellem}, Eric and {Pessa}, Ismael and {Glover}, Simon C.~O. and {Groves}, Brent and {Hughes}, Annie and {Kruijssen}, J.~M. Diederik and {Querejeta}, Miguel and {Schruba}, Andreas and {Bigiel}, Frank and {Blanc}, Guillermo A. and {Chevance}, M{\'e}lanie and {Colombo}, Dario and {Faesi}, Christopher and {Henshaw}, Jonathan D. and {Herrera}, Cinthya N. and {Liu}, Daizhong and {Pety}, J{\'e}r{\^o}me and {Puschnig}, Johannes and {Saito}, Toshiki and {Sun}, Jiayi and {Usero}, Antonio},
        title = "{PHANGS CO Kinematics: Disk Orientations and Rotation Curves at 150 pc Resolution}",
      journal = {\apj},
         year = 2020,
        month = jul,
       volume = {897},
       number = {2},
          eid = {122},
        pages = {122},
          doi = {10.3847/1538-4357/ab9953},
archivePrefix = {arXiv},
       eprint = {2005.11709},
 primaryClass = {astro-ph.GA},
       adsurl = {https://ui.adsabs.harvard.edu/abs/2020ApJ...897..122L}
}

@ARTICLE{Hollyhead_2015,
       author = {{Hollyhead}, K. and {Bastian}, N. and {Adamo}, A. and {Silva-Villa}, E. and {Dale}, J. and {Ryon}, J.~E. and {Gazak}, Z.},
        title = "{Studying the YMC population of M83: how long clusters remain embedded, their interaction with the ISM and implications for GC formation theories}",
      journal = {\mnras},
         year = 2015,
        month = may,
       volume = {449},
       number = {1},
        pages = {1106-1117},
          doi = {10.1093/mnras/stv331},
archivePrefix = {arXiv},
       eprint = {1502.03823},
 primaryClass = {astro-ph.GA},
       adsurl = {https://ui.adsabs.harvard.edu/abs/2015MNRAS.449.1106H}
}

@ARTICLE{Corbelli_2017,
       author = {{Corbelli}, Edvige and {Braine}, Jonathan and {Bandiera}, Rino and {Brouillet}, Nathalie and {Combes}, Fran{\c{c}}oise and {Druard}, Cl{\'e}ment and {Gratier}, Pierre and {Mata}, Jimmy and {Schuster}, Karl and {Xilouris}, Manolis and {Palla}, Francesco},
        title = "{From molecules to young stellar clusters: the star formation cycle across the disk of M 33}",
      journal = {\aap},
         year = 2017,
        month = may,
       volume = {601},
          eid = {A146},
        pages = {A146},
          doi = {10.1051/0004-6361/201630034},
archivePrefix = {arXiv},
       eprint = {1703.09183},
 primaryClass = {astro-ph.GA},
       adsurl = {https://ui.adsabs.harvard.edu/abs/2017A&A...601A.146C}
}

@ARTICLE{Hassani_2023,
       author = {{Hassani}, Hamid and {Rosolowsky}, Erik and {Leroy}, Adam K. and {Boquien}, M{\'e}d{\'e}ric and {Lee}, Janice C. and {Barnes}, Ashley T. and {Belfiore}, Francesco and {Bigiel}, F. and {Cao}, Yixian and {Chevance}, M{\'e}lanie and {Dale}, Daniel A. and {Egorov}, Oleg V. and {Emsellem}, Eric and {Faesi}, Christopher M. and {Grasha}, Kathryn and {Kim}, Jaeyeon and {Klessen}, Ralf S. and {Kreckel}, Kathryn and {Kruijssen}, J.~M. Diederik and {Larson}, Kirsten L. and {Meidt}, Sharon E. and {Sandstrom}, Karin M. and {Schinnerer}, Eva and {Thilker}, David A. and {Watkins}, Elizabeth J. and {Whitmore}, Bradley C. and {Williams}, Thomas G.},
        title = "{PHANGS-JWST First Results: The 21 {\ensuremath{\mu}}m Compact Source Population}",
      journal = {\apjl},
         year = 2023,
        month = feb,
       volume = {944},
       number = {2},
          eid = {L21},
        pages = {L21},
          doi = {10.3847/2041-8213/aca8ab},
archivePrefix = {arXiv},
       eprint = {2212.01526},
 primaryClass = {astro-ph.GA},
       adsurl = {https://ui.adsabs.harvard.edu/abs/2023ApJ...944L..21H}
}

@ARTICLE{Gratier_2012,
       author = {{Gratier}, P. and {Braine}, J. and {Rodriguez-Fernandez}, N.~J. and {Schuster}, K.~F. and {Kramer}, C. and {Corbelli}, E. and {Combes}, F. and {Brouillet}, N. and {van der Werf}, P.~P. and {R{\"o}llig}, M.},
        title = "{Giant molecular clouds in the Local Group galaxy M 33{\ensuremath{\star}}}",
      journal = {\aap},
         year = 2012,
        month = jun,
       volume = {542},
          eid = {A108},
        pages = {A108},
          doi = {10.1051/0004-6361/201116612},
archivePrefix = {arXiv},
       eprint = {1111.4320},
 primaryClass = {astro-ph.CO},
       adsurl = {https://ui.adsabs.harvard.edu/abs/2012A&A...542A.108G}
}

@ARTICLE{McQuaid_2024,
       author = {{McQuaid}, Timothy and {Calzetti}, Daniela and {Linden}, Sean T. and {Messa}, Matteo and {Adamo}, Angela and {Elmegreen}, Bruce and {Grasha}, Kathryn and {Johnson}, Kelsey E. and {Smith}, Linda J. and {Bajaj}, Varun},
        title = "{The Timescales of Star Cluster Emergence: The Case of NGC 4449}",
      journal = {\apj},
         year = 2024,
        month = jun,
       volume = {967},
       number = {2},
          eid = {102},
        pages = {102},
          doi = {10.3847/1538-4357/ad3e64},
archivePrefix = {arXiv},
       eprint = {2404.09112},
 primaryClass = {astro-ph.GA},
       adsurl = {https://ui.adsabs.harvard.edu/abs/2024ApJ...967..102M}
}

@ARTICLE{Whitmore_ngc1365_2023,
       author = {{Whitmore}, Bradley C. and {Chandar}, Rupali and {Rodr{\'\i}guez}, M. Jimena and {Lee}, Janice C. and {Emsellem}, Eric and {Floyd}, Matthew and {Kim}, Hwihyun and {Kruijssen}, J.~M. Diederik and {Mok}, Angus and {Sormani}, Mattia C. and {Boquien}, M{\'e}d{\'e}ric and {Dale}, Daniel A. and {Faesi}, Christopher M. and {Henny}, Kiana F. and {Hannon}, Stephen and {Thilker}, David A. and {White}, Richard L. and {Barnes}, Ashley T. and {Bigiel}, F. and {Chevance}, M{\'e}lanie and {Henshaw}, Jonathan D. and {Klessen}, Ralf S. and {Leroy}, Adam K. and {Liu}, Daizhong and {Maschmann}, Daniel and {Meidt}, Sharon E. and {Rosolowsky}, Erik and {Schinnerer}, Eva and {Sun}, Jiayi and {Watkins}, Elizabeth J. and {Williams}, Thomas G.},
        title = "{PHANGS-JWST First Results: Massive Young Star Clusters and New Insights from JWST Observations of NGC 1365}",
      journal = {\apjl},
         year = 2023,
        month = feb,
       volume = {944},
       number = {2},
          eid = {L14},
        pages = {L14},
          doi = {10.3847/2041-8213/acae94},
archivePrefix = {arXiv},
       eprint = {2212.12039},
 primaryClass = {astro-ph.GA},
       adsurl = {https://ui.adsabs.harvard.edu/abs/2023ApJ...944L..14W}
}

@ARTICLE{Sun_ring_2024,
       author = {{Sun}, Jiayi and {He}, Hao and {Batschkun}, Kyle and {Levy}, Rebecca C. and {Emig}, Kimberly and {Rodr{\'\i}guez}, M. Jimena and {Hassani}, Hamid and {Leroy}, Adam K. and {Schinnerer}, Eva and {Ostriker}, Eve C. and {Wilson}, Christine D. and {Bolatto}, Alberto D. and {Mills}, Elisabeth A.~C. and {Rosolowsky}, Erik and {Lee}, Janice C. and {Dale}, Daniel A. and {Larson}, Kirsten L. and {Thilker}, David A. and {Ubeda}, Leonardo and {Whitmore}, Bradley C. and {Williams}, Thomas G. and {Barnes}, Ashley T. and {Bigiel}, Frank and {Chevance}, M{\'e}lanie and {Glover}, Simon C.~O. and {Grasha}, Kathryn and {Groves}, Brent and {Henshaw}, Jonathan D. and {Indebetouw}, R{\'e}my and {Jim{\'e}nez-Donaire}, Mar{\'\i}a J. and {Klessen}, Ralf S. and {Koch}, Eric W. and {Liu}, Daizhong and {Mathur}, Smita and {Meidt}, Sharon and {Menon}, Shyam H. and {Neumann}, Justus and {Pinna}, Francesca and {Querejeta}, Miguel and {Sormani}, Mattia C. and {Tress}, Robin G.},
        title = "{Hidden Gems on a Ring: Infant Massive Clusters and Their Formation Timeline Unveiled by ALMA, HST, and JWST in NGC 3351}",
      journal = {\apj},
         year = 2024,
        month = jun,
       volume = {967},
       number = {2},
          eid = {133},
        pages = {133},
          doi = {10.3847/1538-4357/ad3de6},
archivePrefix = {arXiv},
       eprint = {2401.14453},
 primaryClass = {astro-ph.GA},
       adsurl = {https://ui.adsabs.harvard.edu/abs/2024ApJ...967..133S}
}

@ARTICLE{Linden_ngc3256_2024,
       author = {{Linden}, Sean T. and {Lai}, Thomas and {Evans}, Aaron S. and {Armus}, Lee and {Larson}, Kirsten L. and {Rich}, Jeffrey A. and {U}, Vivian and {Privon}, George C. and {Inami}, Hanae and {Song}, Yiqing and {Bianchin}, Marina and {Bohn}, Thomas and {Buiten}, Victorine A. and {Sanchez-Garc{\'\i}a}, Maria and {Kader}, Justin and {Lenki{\'c}}, Laura and {Medling}, Anne M. and {B{\"o}ker}, Torsten and {D{\'\i}az-Santos}, Tanio and {Charmandaris}, Vassilis and {Barcos-Mu{\~n}oz}, Loreto and {van der Werf}, Paul and {Stierwalt}, Sabrina and {Aalto}, Susanne and {Appleton}, Philip and {Hayward}, Christopher C. and {Howell}, Justin H. and {Malkan}, Matthew A. and {Mazzarella}, Joseph M. and {Murphy}, Eric J. and {Surace}, Jason},
        title = "{GOALS-JWST: Constraining the Emergence Timescale for Massive Star Clusters in NGC 3256}",
      journal = {\apjl},
         year = 2024,
        month = oct,
       volume = {974},
       number = {2},
          eid = {L27},
        pages = {L27},
          doi = {10.3847/2041-8213/ad7eae},
archivePrefix = {arXiv},
       eprint = {2409.16503},
 primaryClass = {astro-ph.GA},
       adsurl = {https://ui.adsabs.harvard.edu/abs/2024ApJ...974L..27L}
}

@ARTICLE{Deshmukh_clearing_2024,
       author = {{Deshmukh}, Suyash and {Linden}, Sean T. and {Calzetti}, Daniela and {Adamo}, Angela and {Messa}, Matteo and {Grasha}, Kathryn and {Sabbi}, Elena and {Smith}, Linda and {Johnson}, Kelsey E.},
        title = "{The Clearing Timescale for Infrared-selected Star Clusters in M83 with HST}",
      journal = {\apjl},
         year = 2024,
        month = oct,
       volume = {974},
       number = {2},
          eid = {L24},
        pages = {L24},
          doi = {10.3847/2041-8213/ad7ba9},
archivePrefix = {arXiv},
       eprint = {2409.08994},
 primaryClass = {astro-ph.GA},
       adsurl = {https://ui.adsabs.harvard.edu/abs/2024ApJ...974L..24D}
}

@ARTICLE{Dessauges-Zavadsky_z1_clouds_2023,
       author = {{Dessauges-Zavadsky}, Miroslava and {Richard}, Johan and {Combes}, Fran{\c{c}}oise and {Messa}, Matteo and {Nagy}, David and {Mayer}, Lucio and {Schaerer}, Daniel and {Egami}, Eiichi and {Adamo}, Angela},
        title = "{Molecular gas cloud properties at z ≃ 1 revealed by the superb angular resolution achieved with ALMA and gravitational lensing}",
      journal = {\mnras},
         year = 2023,
        month = mar,
       volume = {519},
       number = {4},
        pages = {6222-6238},
          doi = {10.1093/mnras/stad113},
archivePrefix = {arXiv},
       eprint = {2301.05715},
 primaryClass = {astro-ph.GA},
       adsurl = {https://ui.adsabs.harvard.edu/abs/2023MNRAS.519.6222D}
}

@ARTICLE{Claeyssens_jwstclouds_2023,
       author = {{Claeyssens}, Ad{\'e}la{\"\i}de and {Adamo}, Angela and {Richard}, Johan and {Mahler}, Guillaume and {Messa}, Matteo and {Dessauges-Zavadsky}, Miroslava},
        title = "{Star formation at the smallest scales: a JWST study of the clump populations in SMACS0723}",
      journal = {\mnras},
         year = 2023,
        month = apr,
       volume = {520},
       number = {2},
        pages = {2180-2203},
          doi = {10.1093/mnras/stac3791},
archivePrefix = {arXiv},
       eprint = {2208.10450},
 primaryClass = {astro-ph.GA},
       adsurl = {https://ui.adsabs.harvard.edu/abs/2023MNRAS.520.2180C}
}

@ARTICLE{Dessauges-Zavadsky_snake_2019,
       author = {{Dessauges-Zavadsky}, Miroslava and {Richard}, Johan and {Combes}, Fran{\c{c}}oise and {Schaerer}, Daniel and {Rujopakarn}, Wiphu and {Mayer}, Lucio and {Cava}, Antonio and {Boone}, Fr{\'e}d{\'e}ric and {Egami}, Eiichi and {Kneib}, Jean-Paul and {P{\'e}rez-Gonz{\'a}lez}, Pablo G. and {Pfenniger}, Daniel and {Rawle}, Tim D. and {Teyssier}, Romain and {van der Werf}, Paul P.},
        title = "{Molecular clouds in the Cosmic Snake normal star-forming galaxy 8 billion years ago}",
      journal = {Nature Astronomy},
         year = 2019,
        month = sep,
       volume = {3},
        pages = {1115-1121},
          doi = {10.1038/s41550-019-0874-0},
archivePrefix = {arXiv},
       eprint = {1909.08010},
 primaryClass = {astro-ph.GA},
       adsurl = {https://ui.adsabs.harvard.edu/abs/2019NatAs...3.1115D}
}

@ARTICLE{Fukushima_2020,
       author = {{Fukushima}, Hajime and {Yajima}, Hidenobu and {Sugimura}, Kazuyuki and {Hosokawa}, Takashi and {Omukai}, Kazuyuki and {Matsumoto}, Tomoaki},
        title = "{Star cluster formation and cloud dispersal by radiative feedback: dependence on metallicity and compactness}",
      journal = {\mnras},
         year = 2020,
        month = sep,
       volume = {497},
       number = {3},
        pages = {3830-3845},
          doi = {10.1093/mnras/staa2062},
archivePrefix = {arXiv},
       eprint = {2005.13401},
 primaryClass = {astro-ph.GA},
       adsurl = {https://ui.adsabs.harvard.edu/abs/2020MNRAS.497.3830F}
}

@ARTICLE{Groves_hii_2023,
       author = {{Groves}, B. and {Kreckel}, K. and {Santoro}, F. and {Belfiore}, F. and {Zavodnik}, E. and {Congiu}, E. and {Egorov}, O.~V. and {Emsellem}, E. and {Grasha}, K. and {Leroy}, A. and {Scheuermann}, F. and {Schinnerer}, E. and {Watkins}, E.~J. and {Barnes}, A.~T. and {Bigiel}, F. and {Dale}, D.~A. and {Glover}, S.~C.~O. and {Pessa}, I. and {Sanchez-Blazquez}, P. and {Williams}, T.~G.},
        title = "{The PHANGS-MUSE nebular catalogue}",
      journal = {\mnras},
         year = 2023,
        month = apr,
       volume = {520},
       number = {4},
        pages = {4902-4952},
          doi = {10.1093/mnras/stad114},
archivePrefix = {arXiv},
       eprint = {2301.03811},
 primaryClass = {astro-ph.GA},
       adsurl = {https://ui.adsabs.harvard.edu/abs/2023MNRAS.520.4902G}
}

@ARTICLE{Zakardjian_2023,
       author = {{Zakardjian}, Antoine and {Pety}, J{\'e}r{\^o}me and {Herrera}, Cinthya N. and {Hughes}, Annie and {Oakes}, Elias and {Kreckel}, Kathryn and {Faesi}, Chris and {Glover}, Simon C.~O. and {Groves}, Brent and {Klessen}, Ralf S. and {Meidt}, Sharon and {Barnes}, Ashley and {Belfiore}, Francesco and {Be{\v{s}}li{\'c}}, Ivana and {Bigiel}, Frank and {Blanc}, Guillermo A. and {Chevance}, M{\'e}lanie and {Dale}, Daniel A. and {den Brok}, Jakob and {Eibensteiner}, Cosima and {Emsellem}, Eric and {Garc{\'\i}a-Rodr{\'\i}guez}, Axel and {Grasha}, Kathryn and {Koch}, Eric W. and {Leroy}, Adam K. and {Liu}, Daizhong and {Mc Elroy}, Rebecca and {Neumann}, Lukas and {Pan}, Hsi-An and {Querejeta}, Miguel and {Razza}, Alessandro and {Rosolowsky}, Erik and {Saito}, Toshiki and {Santoro}, Francesco and {Schinnerer}, Eva and {Sun}, Jiayi and {Usero}, Antonio and {Watkins}, Elizabeth J. and {Williams}, Thomas},
        title = "{The impact of H II regions on giant molecular cloud properties in nearby galaxies sampled by PHANGS ALMA and MUSE}",
      journal = {\aap},
         year = 2023,
        month = oct,
       volume = {678},
          eid = {A171},
        pages = {A171},
          doi = {10.1051/0004-6361/202244520},
archivePrefix = {arXiv},
       eprint = {2305.03650},
 primaryClass = {astro-ph.GA},
       adsurl = {https://ui.adsabs.harvard.edu/abs/2023A&A...678A.171Z}
}

@ARTICLE{Schruba_m33_2010,
       author = {{Schruba}, Andreas and {Leroy}, Adam K. and {Walter}, Fabian and {Sandstrom}, Karin and {Rosolowsky}, Erik},
        title = "{The Scale Dependence of the Molecular Gas Depletion Time in M33}",
      journal = {\apj},
         year = 2010,
        month = oct,
       volume = {722},
       number = {2},
        pages = {1699-1706},
          doi = {10.1088/0004-637X/722/2/1699},
archivePrefix = {arXiv},
       eprint = {1009.1651},
 primaryClass = {astro-ph.CO},
       adsurl = {https://ui.adsabs.harvard.edu/abs/2010ApJ...722.1699S}
}

@ARTICLE{Kruijssen_Longmore_2014,
       author = {{Kruijssen}, J.~M. Diederik and {Longmore}, Steven N.},
        title = "{An uncertainty principle for star formation - I. Why galactic star formation relations break down below a certain spatial scale}",
      journal = {\mnras},
         year = 2014,
        month = apr,
       volume = {439},
       number = {4},
        pages = {3239-3252},
          doi = {10.1093/mnras/stu098},
archivePrefix = {arXiv},
       eprint = {1401.4459},
 primaryClass = {astro-ph.GA},
       adsurl = {https://ui.adsabs.harvard.edu/abs/2014MNRAS.439.3239K}
}

@ARTICLE{Chown_CO_IR_2025,
       author = {{Chown}, Ryan and {Leroy}, Adam K. and {Sandstrom}, Karin and {Chastenet}, J{\'e}r{\'e}my and {Sutter}, Jessica and {Koch}, Eric W. and {Koziol}, Hannah B. and {Neumann}, Lukas and {Sun}, Jiayi and {Williams}, Thomas G. and {Baron}, Dalya and {Anand}, Gagandeep S. and {Barnes}, Ashley. T. and {Bazzi}, Zein and {Belfiore}, Francesco and {Bigiel}, Frank and {Bolatto}, Alberto and {Boquien}, M{\'e}d{\'e}ric and {Cao}, Yixian and {Chevance}, M{\'e}lanie and {Colombo}, Dario and {Dale}, Daniel A. and {den Brok}, Jakob and {Egorov}, Oleg V. and {Eibensteiner}, Cosima and {Emsellem}, Eric and {Hassani}, Hamid and {Henshaw}, Jonathan D. and {He}, Hao and {Kim}, Jaeyeon and {Klessen}, Ralf S. and {Kreckel}, Kathryn and {Larson}, Kirsten L. and {Lee}, Janice C. and {Meidt}, Sharon E. and {Murphy}, Eric J. and {Oakes}, Elias K. and {Ostriker}, Eve C. and {Pan}, Hsi-An and {Pathak}, Debosmita and {Rosolowsky}, Erik and {Sarbadhicary}, Sumit K. and {Schinnerer}, Eva and {Teng}, Yu-Hsuan and {Thilker}, David A. and {Weinbeck}, Tony D. and {Watkins}, Elizabeth J.},
        title = "{Polycyclic Aromatic Hydrocarbon and CO(2{\textendash}1) Emission at 50{\textendash}150 pc Scales in 70 Nearby Galaxies}",
      journal = {\apj},
         year = 2025,
        month = apr,
       volume = {983},
       number = {1},
          eid = {64},
        pages = {64},
          doi = {10.3847/1538-4357/adbd40},
archivePrefix = {arXiv},
       eprint = {2410.05397},
 primaryClass = {astro-ph.GA},
       adsurl = {https://ui.adsabs.harvard.edu/abs/2025ApJ...983...64C}
}

@ARTICLE{Maschmann_cluster_2024,
       author = {{Maschmann}, Daniel and {Lee}, Janice C. and {Thilker}, David A. and {Whitmore}, Bradley C. and {Deger}, Sinan and {Boquien}, M{\'e}d{\'e}ric and {Chandar}, Rupali and {Dale}, Daniel A. and {Wofford}, Aida and {Hannon}, Stephen and {Larson}, Kirsten L. and {Leroy}, Adam K. and {Schinnerer}, Eva and {Rosolowsky}, Erik and {{\'U}beda}, Leonardo and {Barnes}, Ashley T. and {Emsellem}, Eric and {Grasha}, Kathryn and {Groves}, Brent and {Indebetouw}, R{\'e}my and {Kim}, Hwihyun and {Klessen}, Ralf S. and {Kreckel}, Kathryn and {Levy}, Rebecca C. and {Pinna}, Francesca and {Rodr{\'\i}guez}, M. Jimena and {Tian}, Qiushi and {Williams}, Thomas G.},
        title = "{PHANGS-HST Catalogs for {\ensuremath{\sim}}100,000 Star Clusters and Compact Associations in 38 Galaxies. I. Observed Properties}",
      journal = {\apjs},
         year = 2024,
        month = jul,
       volume = {273},
       number = {1},
          eid = {14},
        pages = {14},
          doi = {10.3847/1538-4365/ad3cd3},
archivePrefix = {arXiv},
       eprint = {2403.04901},
 primaryClass = {astro-ph.GA},
       adsurl = {https://ui.adsabs.harvard.edu/abs/2024ApJS..273...14M}
}

@ARTICLE{clumpfind_Williams_1994,
       author = {{Williams}, Jonathan P. and {de Geus}, Eugene J. and {Blitz}, Leo},
        title = "{Determining Structure in Molecular Clouds}",
      journal = {\apj},
         year = 1994,
        month = jun,
       volume = {428},
        pages = {693},
          doi = {10.1086/174279},
       adsurl = {https://ui.adsabs.harvard.edu/abs/1994ApJ...428..693W}
}

@ARTICLE{SK_law_1998,
       author = {{Kennicutt}, Jr., Robert C.},
        title = "{The Global Schmidt Law in Star-forming Galaxies}",
      journal = {\apj},
         year = 1998,
        month = may,
       volume = {498},
       number = {2},
        pages = {541-552},
          doi = {10.1086/305588},
archivePrefix = {arXiv},
       eprint = {astro-ph/9712213},
 primaryClass = {astro-ph},
       adsurl = {https://ui.adsabs.harvard.edu/abs/1998ApJ...498..541K}
}

@ARTICLE{Schinnerer_Leroy_2024,
       author = {{Schinnerer}, E. and {Leroy}, A.~K.},
        title = "{Molecular Gas and the Star-Formation Process on Cloud Scales in Nearby Galaxies}",
      journal = {\araa},
         year = 2024,
        month = sep,
       volume = {62},
       number = {1},
        pages = {369-436},
          doi = {10.1146/annurev-astro-071221-052651},
archivePrefix = {arXiv},
       eprint = {2403.19843},
 primaryClass = {astro-ph.GA},
       adsurl = {https://ui.adsabs.harvard.edu/abs/2024ARA&A..62..369S}
}

@ARTICLE{blanc_resolved_sk_2009,
       author = {{Blanc}, Guillermo A. and {Heiderman}, Amanda and {Gebhardt}, Karl and {Evans}, II, Neal J. and {Adams}, Joshua},
        title = "{The Spatially Resolved Star Formation Law From Integral Field Spectroscopy: VIRUS-P Observations of NGC 5194}",
      journal = {\apj},
         year = 2009,
        month = oct,
       volume = {704},
       number = {1},
        pages = {842-862},
          doi = {10.1088/0004-637X/704/1/842},
archivePrefix = {arXiv},
       eprint = {0908.2810},
 primaryClass = {astro-ph.CO},
       adsurl = {https://ui.adsabs.harvard.edu/abs/2009ApJ...704..842B}
}

@ARTICLE{Onodera_sk_breakdown_2010,
       author = {{Onodera}, Sachiko and {Kuno}, Nario and {Tosaki}, Tomoka and {Kohno}, Kotaro and {Nakanishi}, Kouichiro and {Sawada}, Tsuyoshi and {Muraoka}, Kazuyuki and {Komugi}, Shinya and {Miura}, Rie and {Kaneko}, Hiroyuki and {Hirota}, Akihiko and {Kawabe}, Ryohei},
        title = "{Breakdown of Kennicutt-Schmidt Law at Giant Molecular Cloud Scales in M33}",
      journal = {\apjl},
         year = 2010,
        month = oct,
       volume = {722},
       number = {2},
        pages = {L127-L131},
          doi = {10.1088/2041-8205/722/2/L127},
archivePrefix = {arXiv},
       eprint = {1009.1971},
 primaryClass = {astro-ph.GA},
       adsurl = {https://ui.adsabs.harvard.edu/abs/2010ApJ...722L.127O}
}

@ARTICLE{Kreckel_ngc628_2018,
       author = {{Kreckel}, K. and {Faesi}, C. and {Kruijssen}, J.~M.~D. and {Schruba}, A. and {Groves}, B. and {Leroy}, A.~K. and {Bigiel}, F. and {Blanc}, G.~A. and {Chevance}, M. and {Herrera}, C. and {Hughes}, A. and {McElroy}, R. and {Pety}, J. and {Querejeta}, M. and {Rosolowsky}, E. and {Schinnerer}, E. and {Sun}, J. and {Usero}, A. and {Utomo}, D.},
        title = "{A 50 pc Scale View of Star Formation Efficiency across NGC 628}",
      journal = {\apjl},
         year = 2018,
        month = aug,
       volume = {863},
       number = {2},
          eid = {L21},
        pages = {L21},
          doi = {10.3847/2041-8213/aad77d},
archivePrefix = {arXiv},
       eprint = {1807.11506},
 primaryClass = {astro-ph.GA},
       adsurl = {https://ui.adsabs.harvard.edu/abs/2018ApJ...863L..21K}
}

@ARTICLE{Schinnerer_2019,
       author = {{Schinnerer}, Eva and {Hughes}, Annie and {Leroy}, Adam and {Groves}, Brent and {Blanc}, Guillermo A. and {Kreckel}, Kathryn and {Bigiel}, Frank and {Chevance}, M{\'e}lanie and {Dale}, Daniel and {Emsellem}, Eric and {Faesi}, Christopher and {Glover}, Simon and {Grasha}, Kathryn and {Henshaw}, Jonathan and {Hygate}, Alexander and {Kruijssen}, J.~M. Diederik and {Meidt}, Sharon and {Pety}, Jerome and {Querejeta}, Miguel and {Rosolowsky}, Erik and {Saito}, Toshiki and {Schruba}, Andreas and {Sun}, Jiayi and {Utomo}, Dyas},
        title = "{The Gas-Star Formation Cycle in Nearby Star-forming Galaxies. I. Assessment of Multi-scale Variations}",
      journal = {\apj},
         year = 2019,
        month = dec,
       volume = {887},
       number = {1},
          eid = {49},
        pages = {49},
          doi = {10.3847/1538-4357/ab50c2},
archivePrefix = {arXiv},
       eprint = {1910.10520},
 primaryClass = {astro-ph.GA},
       adsurl = {https://ui.adsabs.harvard.edu/abs/2019ApJ...887...49S}
}

@ARTICLE{Leroy_alphaCO_2013,
       author = {{Leroy}, Adam K. and {Walter}, Fabian and {Sandstrom}, Karin and {Schruba}, Andreas and {Munoz-Mateos}, Juan-Carlos and {Bigiel}, Frank and {Bolatto}, Alberto and {Brinks}, Elias and {de Blok}, W.~J.~G. and {Meidt}, Sharon and {Rix}, Hans-Walter and {Rosolowsky}, Erik and {Schinnerer}, Eva and {Schuster}, Karl-Friedrich and {Usero}, Antonio},
        title = "{Molecular Gas and Star Formation in nearby Disk Galaxies}",
      journal = {\aj},
         year = 2013,
        month = aug,
       volume = {146},
       number = {2},
          eid = {19},
        pages = {19},
          doi = {10.1088/0004-6256/146/2/19},
archivePrefix = {arXiv},
       eprint = {1301.2328},
 primaryClass = {astro-ph.CO},
       adsurl = {https://ui.adsabs.harvard.edu/abs/2013AJ....146...19L}
}

@ARTICLE{den_Brok_2021,
       author = {{den Brok}, J.~S. and {Chatzigiannakis}, D. and {Bigiel}, F. and {Puschnig}, J. and {Barnes}, A.~T. and {Leroy}, A.~K. and {Jim{\'e}nez-Donaire}, M.~J. and {Usero}, A. and {Schinnerer}, E. and {Rosolowsky}, E. and {Faesi}, C.~M. and {Grasha}, K. and {Hughes}, A. and {Kruijssen}, J.~M.~D. and {Liu}, D. and {Neumann}, L. and {Pety}, J. and {Querejeta}, M. and {Saito}, T. and {Schruba}, A. and {Stuber}, S.},
        title = "{New constraints on the $^{12}$CO(2-1)/(1-0) line ratio across nearby disc galaxies}",
      journal = {\mnras},
         year = 2021,
        month = jul,
       volume = {504},
       number = {3},
        pages = {3221-3245},
          doi = {10.1093/mnras/stab859},
archivePrefix = {arXiv},
       eprint = {2103.10442},
 primaryClass = {astro-ph.GA},
       adsurl = {https://ui.adsabs.harvard.edu/abs/2021MNRAS.504.3221D}
}

@ARTICLE{Leroy_co_2022,
       author = {{Leroy}, Adam K. and {Rosolowsky}, Erik and {Usero}, Antonio and {Sandstrom}, Karin and {Schinnerer}, Eva and {Schruba}, Andreas and {Bolatto}, Alberto D. and {Sun}, Jiayi and {Barnes}, Ashley. T. and {Belfiore}, Francesco and {Bigiel}, Frank and {den Brok}, Jakob S. and {Cao}, Yixian and {Chiang}, I-Da and {Chevance}, M{\'e}lanie and {Dale}, Daniel A. and {Eibensteiner}, Cosima and {Faesi}, Christopher M. and {Glover}, Simon C.~O. and {Hughes}, Annie and {Jim{\'e}nez Donaire}, Mar{\'\i}a J. and {Klessen}, Ralf S. and {Koch}, Eric W. and {Kruijssen}, J.~M. Diederik and {Liu}, Daizhong and {Meidt}, Sharon E. and {Pan}, Hsi-An and {Pety}, J{\'e}r{\^o}me and {Puschnig}, Johannes and {Querejeta}, Miguel and {Saito}, Toshiki and {Sardone}, Amy and {Watkins}, Elizabeth J. and {Weiss}, Axel and {Williams}, Thomas G.},
        title = "{Low-J CO Line Ratios from Single-dish CO Mapping Surveys and PHANGS-ALMA}",
      journal = {\apj},
         year = 2022,
        month = mar,
       volume = {927},
       number = {2},
          eid = {149},
        pages = {149},
          doi = {10.3847/1538-4357/ac3490},
archivePrefix = {arXiv},
       eprint = {2109.11583},
 primaryClass = {astro-ph.GA},
       adsurl = {https://ui.adsabs.harvard.edu/abs/2022ApJ...927..149L}
}

@ARTICLE{Ward_lmc_2020,
       author = {{Ward}, Jacob L. and {Chevance}, M{\'e}lanie and {Kruijssen}, J.~M. Diederik and {Hygate}, Alexander P.~S. and {Schruba}, Andreas and {Longmore}, Steven N.},
        title = "{Towards a multitracer timeline of star formation in the LMC - I. Deriving the lifetimes of H I clouds}",
      journal = {\mnras},
         year = 2020,
        month = sep,
       volume = {497},
       number = {2},
        pages = {2286-2301},
          doi = {10.1093/mnras/staa1977},
archivePrefix = {arXiv},
       eprint = {2007.03691},
 primaryClass = {astro-ph.GA},
       adsurl = {https://ui.adsabs.harvard.edu/abs/2020MNRAS.497.2286W}
}

@ARTICLE{Ward_lmc_2022,
       author = {{Ward}, Jacob L. and {Kruijssen}, J.~M. Diederik and {Chevance}, M{\'e}lanie and {Kim}, Jaeyeon and {Longmore}, Steven N.},
        title = "{Towards a multitracer timeline of star formation in the LMC - II. The formation and destruction of molecular clouds}",
      journal = {\mnras},
         year = 2022,
        month = nov,
       volume = {516},
       number = {3},
        pages = {4025-4042},
          doi = {10.1093/mnras/stac2467},
archivePrefix = {arXiv},
       eprint = {2209.05541},
 primaryClass = {astro-ph.GA},
       adsurl = {https://ui.adsabs.harvard.edu/abs/2022MNRAS.516.4025W}
}

@ARTICLE{Lu_wisdom_2022,
       author = {{Lu}, Anan and {Boyce}, Hope and {Haggard}, Daryl and {Bureau}, Martin and {Liang}, Fu-Heng and {Liu}, Lijie and {Choi}, Woorak and {Cappellari}, Michele and {Chemin}, Laurent and {Chevance}, M{\'e}lanie and {Davis}, Timothy A. and {Drissen}, Laurent and {Elford}, Jacob S. and {Gensior}, Jindra and {Kruijssen}, J.~M. Diederik and {Martin}, Thomas and {Mass{\'e}}, Etienne and {Robert}, Carmelle and {Ruffa}, Ilaria and {Rousseau-Nepton}, Laurie and {Sarzi}, Marc and {Savard}, Gabriel and {Williams}, Thomas G.},
        title = "{WISDOM project - XI. Star formation efficiency in the bulge of the AGN-host Galaxy NGC 3169 with SITELLE and ALMA}",
      journal = {\mnras},
         year = 2022,
        month = aug,
       volume = {514},
       number = {4},
        pages = {5035-5055},
          doi = {10.1093/mnras/stac1583},
archivePrefix = {arXiv},
       eprint = {2206.03316},
 primaryClass = {astro-ph.GA},
       adsurl = {https://ui.adsabs.harvard.edu/abs/2022MNRAS.514.5035L}
}

@ARTICLE{Kruijssen_EMOSAICS_2019,
       author = {{Kruijssen}, J.~M. Diederik and {Pfeffer}, Joel L. and {Crain}, Robert A. and {Bastian}, Nate},
        title = "{The E-MOSAICS project: tracing galaxy formation and assembly with the age-metallicity distribution of globular clusters}",
      journal = {\mnras},
         year = 2019,
        month = jul,
       volume = {486},
       number = {3},
        pages = {3134-3179},
          doi = {10.1093/mnras/stz968},
archivePrefix = {arXiv},
       eprint = {1904.04261},
 primaryClass = {astro-ph.GA},
       adsurl = {https://ui.adsabs.harvard.edu/abs/2019MNRAS.486.3134K}
}

@ARTICLE{Kruijssen_drift_2024,
       author = {{Kruijssen}, J.~M. Diederik and {Chevance}, M{\'e}lanie and {Longmore}, Steven N. and {Ginsburg}, Adam and {Ramambason}, Lise and {Romanelli}, Andrea},
        title = "{Death of the Immortal Molecular Cloud: Resolution Dependence of the Gas-Star Formation Relation Rules out Decoupling by Stellar Drift}",
      journal = {submitted to The Open Journal of Astrophysics, arXiv e-prints},
         year = 2024,
        month = apr,
          eid = {arXiv:2404.14495},
        pages = {arXiv:2404.14495},
          doi = {10.48550/arXiv.2404.14495},
archivePrefix = {arXiv},
       eprint = {2404.14495},
 primaryClass = {astro-ph.GA},
       adsurl = {https://ui.adsabs.harvard.edu/abs/2024arXiv240414495K}
}

@ARTICLE{Kruijssen_2025,
       author = {{Kruijssen}, J.~M. Diederik},
        title = "{The formation of globular clusters}",
    journal = {Encyclopedia of Astrophysics, Volume 4, ISBN 978-0-443-21439-4, Elsevier},
         year = 2026,
       volume = {4},
        month = jan,
        pages = {500-534},
          doi = {10.1016/B978-0-443-21439-4.00078-X},
archivePrefix = {arXiv},
       eprint = {2501.16438},
 primaryClass = {astro-ph.GA},
       adsurl = {https://ui.adsabs.harvard.edu/abs/2026enap....4..500K}
}

@ARTICLE{Semenov_ngc300_2021,
       author = {{Semenov}, Vadim A. and {Kravtsov}, Andrey V. and {Gnedin}, Nickolay Y.},
        title = "{Spatial Decorrelation of Young Stars and Dense Gas as a Probe of the Star Formation-Feedback Cycle in Galaxies}",
      journal = {\apj},
         year = 2021,
        month = sep,
       volume = {918},
       number = {1},
          eid = {13},
        pages = {13},
          doi = {10.3847/1538-4357/ac0a77},
archivePrefix = {arXiv},
       eprint = {2103.13406},
 primaryClass = {astro-ph.GA},
       adsurl = {https://ui.adsabs.harvard.edu/abs/2021ApJ...918...13S}
}

@ARTICLE{Fujimoto_feedback_recipes_2019,
       author = {{Fujimoto}, Yusuke and {Chevance}, M{\'e}lanie and {Haydon}, Daniel T. and {Krumholz}, Mark R. and {Kruijssen}, J.~M. Diederik},
        title = "{A fundamental test for stellar feedback recipes in galaxy simulations}",
      journal = {\mnras},
         year = 2019,
        month = aug,
       volume = {487},
       number = {2},
        pages = {1717-1728},
          doi = {10.1093/mnras/stz641},
archivePrefix = {arXiv},
       eprint = {1905.09839},
 primaryClass = {astro-ph.GA},
       adsurl = {https://ui.adsabs.harvard.edu/abs/2019MNRAS.487.1717F}
}

@ARTICLE{Keller_early_feedback_2022,
       author = {{Keller}, Benjamin W. and {Kruijssen}, J.~M. Diederik and {Chevance}, M{\'e}lanie},
        title = "{Empirically motivated early feedback: momentum input by stellar feedback in galaxy simulations inferred through observations}",
      journal = {\mnras},
         year = 2022,
        month = aug,
       volume = {514},
       number = {4},
        pages = {5355-5374},
          doi = {10.1093/mnras/stac1607},
archivePrefix = {arXiv},
       eprint = {2206.06391},
 primaryClass = {astro-ph.GA},
       adsurl = {https://ui.adsabs.harvard.edu/abs/2022MNRAS.514.5355K}
}

@ARTICLE{Grasha_2018,
       author = {{Grasha}, K. and {Calzetti}, D. and {Bittle}, L. and {Johnson}, K.~E. and {Donovan Meyer}, J. and {Kennicutt}, R.~C. and {Elmegreen}, B.~G. and {Adamo}, A. and {Krumholz}, M.~R. and {Fumagalli}, M. and {Grebel}, E.~K. and {Gouliermis}, D.~A. and {Cook}, D.~O. and {Gallagher}, J.~S. and {Aloisi}, A. and {Dale}, D.~A. and {Linden}, S. and {Sacchi}, E. and {Thilker}, D.~A. and {Walterbos}, R.~A.~M. and {Messa}, M. and {Wofford}, A. and {Smith}, L.~J.},
        title = "{Connecting young star clusters to CO molecular gas in NGC 7793 with ALMA-LEGUS}",
      journal = {\mnras},
         year = 2018,
        month = nov,
       volume = {481},
       number = {1},
        pages = {1016-1027},
          doi = {10.1093/mnras/sty2154},
archivePrefix = {arXiv},
       eprint = {1808.02496},
 primaryClass = {astro-ph.GA},
       adsurl = {https://ui.adsabs.harvard.edu/abs/2018MNRAS.481.1016G}
}

@ARTICLE{Grasha_2019,
       author = {{Grasha}, K. and {Calzetti}, D. and {Adamo}, A. and {Kennicutt}, R.~C. and {Elmegreen}, B.~G. and {Messa}, M. and {Dale}, D.~A. and {Fedorenko}, K. and {Mahadevan}, S. and {Grebel}, E.~K. and {Fumagalli}, M. and {Kim}, H. and {Dobbs}, C.~L. and {Gouliermis}, D.~A. and {Ashworth}, G. and {Gallagher}, J.~S. and {Smith}, L.~J. and {Tosi}, M. and {Whitmore}, B.~C. and {Schinnerer}, E. and {Colombo}, D. and {Hughes}, A. and {Leroy}, A.~K. and {Meidt}, S.~E.},
        title = "{The spatial relation between young star clusters and molecular clouds in M51 with LEGUS}",
      journal = {\mnras},
         year = 2019,
        month = mar,
       volume = {483},
       number = {4},
        pages = {4707-4723},
          doi = {10.1093/mnras/sty3424},
archivePrefix = {arXiv},
       eprint = {1812.06109},
 primaryClass = {astro-ph.GA},
       adsurl = {https://ui.adsabs.harvard.edu/abs/2019MNRAS.483.4707G}
}

@ARTICLE{Whitmore_antennae_2014,
       author = {{Whitmore}, Bradley C. and {Brogan}, Crystal and {Chandar}, Rupali and {Evans}, Aaron and {Hibbard}, John and {Johnson}, Kelsey and {Leroy}, Adam and {Privon}, George and {Remijan}, Anthony and {Sheth}, Kartik},
        title = "{ALMA Observations of the Antennae Galaxies. I. A New Window on a Prototypical Merger}",
      journal = {\apj},
         year = 2014,
        month = nov,
       volume = {795},
       number = {2},
          eid = {156},
        pages = {156},
          doi = {10.1088/0004-637X/795/2/156},
archivePrefix = {arXiv},
       eprint = {1410.4473},
 primaryClass = {astro-ph.GA},
       adsurl = {https://ui.adsabs.harvard.edu/abs/2014ApJ...795..156W}
}

@ARTICLE{Messa_2021,
       author = {{Messa}, Matteo and {Calzetti}, Daniela and {Adamo}, Angela and {Grasha}, Kathryn and {Johnson}, Kelsey E. and {Sabbi}, Elena and {Smith}, Linda J. and {Bajaj}, Varun and {Finn}, Molly K. and {Lin}, Zesen},
        title = "{Looking for Obscured Young Star Clusters in NGC 1313}",
      journal = {\apj},
         year = 2021,
        month = mar,
       volume = {909},
       number = {2},
          eid = {121},
        pages = {121},
          doi = {10.3847/1538-4357/abe0b5},
archivePrefix = {arXiv},
       eprint = {2011.09392},
 primaryClass = {astro-ph.GA},
       adsurl = {https://ui.adsabs.harvard.edu/abs/2021ApJ...909..121M}
}

@ARTICLE{Kreckel_mixing_2020,
       author = {{Kreckel}, Kathryn and {Ho}, I. -Ting and {Blanc}, Guillermo A. and {Glover}, Simon C.~O. and {Groves}, Brent and {Rosolowsky}, Erik and {Bigiel}, Frank and {Boqu{\'\i}en}, M{\'e}d{\'e}ric and {Chevance}, M{\'e}lanie and {Dale}, Daniel A. and {Deger}, Sinan and {Emsellem}, Eric and {Grasha}, Kathryn and {Kim}, Jenny J. and {Klessen}, Ralf S. and {Kruijssen}, J.~M. Diederik and {Lee}, Janice C. and {Leroy}, Adam K. and {Liu}, Daizhong and {McElroy}, Rebecca and {Meidt}, Sharon E. and {Pessa}, Ismael and {Sanchez-Blazquez}, Patricia and {Sandstrom}, Karin and {Santoro}, Francesco and {Scheuermann}, Fabian and {Schinnerer}, Eva and {Schruba}, Andreas and {Utomo}, Dyas and {Watkins}, Elizabeth J. and {Williams}, Thomas G.},
        title = "{Measuring the mixing scale of the ISM within nearby spiral galaxies}",
      journal = {\mnras},
         year = 2020,
        month = nov,
       volume = {499},
       number = {1},
        pages = {193-209},
          doi = {10.1093/mnras/staa2743},
archivePrefix = {arXiv},
       eprint = {2009.02342},
 primaryClass = {astro-ph.GA},
       adsurl = {https://ui.adsabs.harvard.edu/abs/2020MNRAS.499..193K}
}

@ARTICLE{Zabel_2020,
       author = {{Zabel}, N. and {Davis}, T.~A. and {Sarzi}, M. and {Nedelchev}, Boris and {Chevance}, M. and {Kruijssen}, J.~M. Diederik and {Iodice}, E. and {Baes}, M. and {Bendo}, G.~J. and {Corsini}, E. Maria and {De Looze}, I. and {de Zeeuw}, P. Tim and {Gadotti}, D.~A. and {Grossi}, M. and {Peletier}, R. and {Pinna}, F. and {Serra}, Paolo and {van de Voort}, F. and {Venhola}, A. and {Viaene}, S. and {Vlahakis}, C.},
        title = "{AlFoCS + Fornax3D: resolved star formation in the Fornax cluster with ALMA and MUSE}",
      journal = {\mnras},
         year = 2020,
        month = aug,
       volume = {496},
       number = {2},
        pages = {2155-2182},
          doi = {10.1093/mnras/staa1513},
archivePrefix = {arXiv},
       eprint = {2005.13454},
 primaryClass = {astro-ph.GA},
       adsurl = {https://ui.adsabs.harvard.edu/abs/2020MNRAS.496.2155Z}
}

@ARTICLE{Haydon_2020,
       author = {{Haydon}, Daniel T. and {Kruijssen}, J.~M. Diederik and {Chevance}, M{\'e}lanie and {Hygate}, Alexander P.~S. and {Krumholz}, Mark R. and {Schruba}, Andreas and {Longmore}, Steven N.},
        title = "{An uncertainty principle for star formation - III. The characteristic emission time-scales of star formation rate tracers}",
      journal = {\mnras},
         year = 2020,
        month = oct,
       volume = {498},
       number = {1},
        pages = {235-257},
          doi = {10.1093/mnras/staa2430},
archivePrefix = {arXiv},
       eprint = {1810.10897},
 primaryClass = {astro-ph.GA},
       adsurl = {https://ui.adsabs.harvard.edu/abs/2020MNRAS.498..235H}
}

@ARTICLE{Egorov_pah_2023,
       author = {{Egorov}, Oleg V. and {Kreckel}, Kathryn and {Sandstrom}, Karin M. and {Leroy}, Adam K. and {Glover}, Simon C.~O. and {Groves}, Brent and {Kruijssen}, J.~M. Diederik and {Barnes}, Ashley. T. and {Belfiore}, Francesco and {Bigiel}, F. and {Blanc}, Guillermo A. and {Boquien}, M{\'e}d{\'e}ric and {Cao}, Yixian and {Chastenet}, J{\'e}r{\'e}my and {Chevance}, M{\'e}lanie and {Congiu}, Enrico and {Dale}, Daniel A. and {Emsellem}, Eric and {Grasha}, Kathryn and {Klessen}, Ralf S. and {Larson}, Kirsten L. and {Liu}, Daizhong and {Murphy}, Eric J. and {Pan}, Hsi-An and {Pessa}, Ismael and {Pety}, J{\'e}r{\^o}me and {Rosolowsky}, Erik and {Scheuermann}, Fabian and {Schinnerer}, Eva and {Sutter}, Jessica and {Thilker}, David A. and {Watkins}, Elizabeth J. and {Williams}, Thomas G.},
        title = "{PHANGS-JWST First Results: Destruction of the PAH Molecules in H II Regions Probed by JWST and MUSE}",
      journal = {\apjl},
         year = 2023,
        month = feb,
       volume = {944},
       number = {2},
          eid = {L16},
        pages = {L16},
          doi = {10.3847/2041-8213/acac92},
archivePrefix = {arXiv},
       eprint = {2212.09159},
 primaryClass = {astro-ph.GA},
       adsurl = {https://ui.adsabs.harvard.edu/abs/2023ApJ...944L..16E}
}

@ARTICLE{Sutter_qpah_2024,
       author = {{Sutter}, Jessica and {Sandstrom}, Karin and {Chastenet}, J{\'e}r{\'e}my and {Leroy}, Adam K. and {Koch}, Eric W. and {Williams}, Thomas G. and {Chown}, Ryan and {Belfiore}, Francesco and {Bigiel}, Frank and {Boquien}, M{\'e}d{\'e}ric and {Cao}, Yixian and {Chevance}, M{\'e}lanie and {Dale}, Daniel A. and {Egorov}, Oleg V. and {Glover}, Simon C.~O. and {Groves}, Brent and {Klessen}, Ralf S. and {Kreckel}, Kathryn and {Larson}, Kirsten L. and {Oakes}, Elias K. and {Pathak}, Debosmita and {Ramambason}, Lise and {Rosolowsky}, Erik and {Watkins}, Elizabeth J.},
        title = "{The Fraction of Dust Mass in the Form of Polycyclic Aromatic Hydrocarbons on 10{\textendash}50 pc Scales in Nearby Galaxies}",
      journal = {\apj},
         year = 2024,
        month = aug,
       volume = {971},
       number = {2},
          eid = {178},
        pages = {178},
          doi = {10.3847/1538-4357/ad54bd},
archivePrefix = {arXiv},
       eprint = {2405.15102},
 primaryClass = {astro-ph.GA},
       adsurl = {https://ui.adsabs.harvard.edu/abs/2024ApJ...971..178S}
}

@ARTICLE{Chastenet_2023,
       author = {{Chastenet}, J{\'e}r{\'e}my and {Sutter}, Jessica and {Sandstrom}, Karin and {Belfiore}, Francesco and {Egorov}, Oleg V. and {Larson}, Kirsten L. and {Leroy}, Adam K. and {Liu}, Daizhong and {Rosolowsky}, Erik and {Thilker}, David A. and {Watkins}, Elizabeth J. and {Williams}, Thomas G. and {Barnes}, Ashley. T. and {Bigiel}, Frank and {Boquien}, M{\'e}d{\'e}ric and {Chevance}, M{\'e}lanie and {Chiang}, I-Da and {Dale}, Daniel A. and {Kruijssen}, J.~M. Diederik and {Emsellem}, Eric and {Grasha}, Kathryn and {Groves}, Brent and {Hassani}, Hamid and {Hughes}, Annie and {Kreckel}, Kathryn and {Meidt}, Sharon E. and {Rickards Vaught}, Ryan J. and {Sardone}, Amy and {Schinnerer}, Eva},
        title = "{PHANGS-JWST First Results: Variations in PAH Fraction as a Function of ISM Phase and Metallicity}",
      journal = {\apjl},
         year = 2023,
        month = feb,
       volume = {944},
       number = {2},
          eid = {L11},
        pages = {L11},
          doi = {10.3847/2041-8213/acadd7},
archivePrefix = {arXiv},
       eprint = {2301.00578},
 primaryClass = {astro-ph.GA},
       adsurl = {https://ui.adsabs.harvard.edu/abs/2023ApJ...944L..11C}
}

@ARTICLE{Pessa_SFH_2023,
       author = {{Pessa}, I. and {Schinnerer}, E. and {Sanchez-Blazquez}, P. and {Belfiore}, F. and {Groves}, B. and {Emsellem}, E. and {Neumann}, J. and {Leroy}, A.~K. and {Bigiel}, F. and {Chevance}, M. and {Dale}, D.~A. and {Glover}, S.~C.~O. and {Grasha}, K. and {Klessen}, R.~S. and {Kreckel}, K. and {Kruijssen}, J.~M.~D. and {Pinna}, F. and {Querejeta}, M. and {Rosolowsky}, E. and {Williams}, T.~G.},
        title = "{Resolved stellar population properties of PHANGS-MUSE galaxies}",
      journal = {\aap},
         year = 2023,
        month = may,
       volume = {673},
          eid = {A147},
        pages = {A147},
          doi = {10.1051/0004-6361/202245673},
archivePrefix = {arXiv},
       eprint = {2303.13676},
 primaryClass = {astro-ph.GA},
       adsurl = {https://ui.adsabs.harvard.edu/abs/2023A&A...673A.147P}
}

@ARTICLE{sanchez_gradient_2014,
author = {{S{\'a}nchez}, S.~F. and {Rosales-Ortega}, F.~F. and {Iglesias-P{\'a}ramo}, J. and {Moll{\'a}}, M. and {Barrera-Ballesteros}, J. and {Marino}, R.~A. and {P{\'e}rez}, E. and {S{\'a}nchez-Blazquez}, P. and {Gonz{\'a}lez Delgado}, R. and {Cid Fernandes}, R. and {de Lorenzo-C{\'a}ceres}, A. and {Mendez-Abreu}, J. and {Galbany}, L. and {Falcon-Barroso}, J. and {Miralles-Caballero}, D. and {Husemann}, B. and {Garc{\'\i}a-Benito}, R. and {Mast}, D. and {Walcher}, C.~J. and {Gil de Paz}, A. and {Garc{\'\i}a-Lorenzo}, B. and {Jungwiert}, B. and {V{\'\i}lchez}, J.~M. and {J{\'\i}lkov{\'a}}, Lucie and {Lyubenova}, M. and {Cortijo-Ferrero}, C. and {D{\'\i}az}, A.~I. and {Wisotzki}, L. and {M{\'a}rquez}, I. and {Bland-Hawthorn}, J. and {Ellis}, S. and {van de Ven}, G. and {Jahnke}, K. and {Papaderos}, P. and {Gomes}, J.~M. and {Mendoza}, M.~A. and {L{\'o}pez-S{\'a}nchez}, {\'A}. R.},
        title = "{A characteristic oxygen abundance gradient in galaxy disks unveiled with CALIFA}",
      journal = {\aap},
         year = 2014,
        month = mar,
       volume = {563},
          eid = {A49},
        pages = {A49},
          doi = {10.1051/0004-6361/201322343},
archivePrefix = {arXiv},
       eprint = {1311.7052},
 primaryClass = {astro-ph.CO},
       adsurl = {https://ui.adsabs.harvard.edu/abs/2014A&A...563A..49S}
}

@ARTICLE{Pilyugin_Grebel_scal_2016,
       author = {{Pilyugin}, L.~S. and {Grebel}, E.~K.},
        title = "{New calibrations for abundance determinations in H II regions}",
      journal = {\mnras},
         year = 2016,
        month = apr,
       volume = {457},
       number = {4},
        pages = {3678-3692},
          doi = {10.1093/mnras/stw238},
archivePrefix = {arXiv},
       eprint = {1601.08217},
 primaryClass = {astro-ph.GA},
       adsurl = {https://ui.adsabs.harvard.edu/abs/2016MNRAS.457.3678P}
}

@ARTICLE{Makarov_HyperLEDA_2014,
       author = {{Makarov}, Dmitry and {Prugniel}, Philippe and {Terekhova}, Nataliya and {Courtois}, H{\'e}l{\`e}ne and {Vauglin}, Isabelle},
        title = "{HyperLEDA. III. The catalogue of extragalactic distances}",
      journal = {\aap},
         year = 2014,
        month = oct,
       volume = {570},
          eid = {A13},
        pages = {A13},
          doi = {10.1051/0004-6361/201423496},
archivePrefix = {arXiv},
       eprint = {1408.3476},
 primaryClass = {astro-ph.GA},
       adsurl = {https://ui.adsabs.harvard.edu/abs/2014A&A...570A..13M}
}

@ARTICLE{Emsellem_phangs_muse_2022,
       author = {{Emsellem}, Eric and {Schinnerer}, Eva and {Santoro}, Francesco and {Belfiore}, Francesco and {Pessa}, Ismael and {McElroy}, Rebecca and {Blanc}, Guillermo A. and {Congiu}, Enrico and {Groves}, Brent and {Ho}, I. -Ting and {Kreckel}, Kathryn and {Razza}, Alessandro and {Sanchez-Blazquez}, Patricia and {Egorov}, Oleg and {Faesi}, Chris and {Klessen}, Ralf S. and {Leroy}, Adam K. and {Meidt}, Sharon and {Querejeta}, Miguel and {Rosolowsky}, Erik and {Scheuermann}, Fabian and {Anand}, Gagandeep S. and {Barnes}, Ashley T. and {Be{\v{s}}li{\'c}}, Ivana and {Bigiel}, Frank and {Boquien}, M{\'e}d{\'e}ric and {Cao}, Yixian and {Chevance}, M{\'e}lanie and {Dale}, Daniel A. and {Eibensteiner}, Cosima and {Glover}, Simon C.~O. and {Grasha}, Kathryn and {Henshaw}, Jonathan D. and {Hughes}, Annie and {Koch}, Eric W. and {Kruijssen}, J.~M. Diederik and {Lee}, Janice and {Liu}, Daizhong and {Pan}, Hsi-An and {Pety}, J{\'e}r{\^o}me and {Saito}, Toshiki and {Sandstrom}, Karin M. and {Schruba}, Andreas and {Sun}, Jiayi and {Thilker}, David A. and {Usero}, Antonio and {Watkins}, Elizabeth J. and {Williams}, Thomas G.},
        title = "{The PHANGS-MUSE survey. Probing the chemo-dynamical evolution of disc galaxies}",
      journal = {\aap},
         year = 2022,
        month = mar,
       volume = {659},
          eid = {A191},
        pages = {A191},
          doi = {10.1051/0004-6361/202141727},
archivePrefix = {arXiv},
       eprint = {2110.03708},
 primaryClass = {astro-ph.GA},
       adsurl = {https://ui.adsabs.harvard.edu/abs/2022A&A...659A.191E}
}

@ARTICLE{Hu_limitation_2024,
       author = {{Hu}, Zipeng and {Wibking}, Benjamin D. and {Krumholz}, Mark R. and {Federrath}, Christoph},
        title = "{On the limitations of H{\ensuremath{\alpha}} luminosity as a star formation tracer in spatially resolved observations}",
      journal = {\mnras},
         year = 2024,
        month = nov,
       volume = {534},
       number = {3},
        pages = {2426-2437},
          doi = {10.1093/mnras/stae2241},
archivePrefix = {arXiv},
       eprint = {2406.11155},
 primaryClass = {astro-ph.GA},
       adsurl = {https://ui.adsabs.harvard.edu/abs/2024MNRAS.534.2426H}
}

@ARTICLE{MendezDelgado:2024b,
       author = {{M{\'e}ndez-Delgado}, J.~E. and {Kreckel}, K. and {Esteban}, C. and {Garc{\'\i}a-Rojas}, J. and {Carigi}, L. and {Sander}, A.~A.~C. and {Palla}, M. and {Chru{\'s}li{\'n}ska}, M. and {De Looze}, I. and {Rela{\~n}o}, M. and {van der Giessen}, S.~A. and {Reyes-Rodr{\'\i}guez}, E. and {S{\'a}nchez}, S.~F.},
        title = "{Gas-phase Fe/O and Fe/N abundances in star-forming regions: Relations between nucleosynthesis, metallicity, and dust}",
      journal = {\aap},
         year = 2024,
        month = oct,
       volume = {690},
          eid = {A248},
        pages = {A248},
          doi = {10.1051/0004-6361/202450928},
archivePrefix = {arXiv},
       eprint = {2408.06215},
 primaryClass = {astro-ph.GA},
       adsurl = {https://ui.adsabs.harvard.edu/abs/2024A&A...690A.248M}
}

@ARTICLE{Koda_tf_limits_2023,
       author = {{Koda}, Jin and {Tan}, Jonathan C.},
        title = "{On the Lifetime of Molecular Clouds with the ``Tuning-fork'' Analysis}",
      journal = {\apj},
         year = 2023,
        month = dec,
       volume = {959},
       number = {1},
          eid = {1},
        pages = {1},
          doi = {10.3847/1538-4357/ad05c6},
archivePrefix = {arXiv},
       eprint = {2308.11717},
 primaryClass = {astro-ph.GA},
       adsurl = {https://ui.adsabs.harvard.edu/abs/2023ApJ...959....1K}
}

@ARTICLE{Paturel_hyperleda1_2003,
       author = {{Paturel}, G. and {Petit}, C. and {Prugniel}, Ph. and {Theureau}, G. and {Rousseau}, J. and {Brouty}, M. and {Dubois}, P. and {Cambr{\'e}sy}, L.},
        title = "{HYPERLEDA.  I. Identification and designation of galaxies}",
      journal = {\aap},
         year = 2003,
        month = dec,
       volume = {412},
        pages = {45-55},
          doi = {10.1051/0004-6361:20031411},
       adsurl = {https://ui.adsabs.harvard.edu/abs/2003A&A...412...45P}
}

@ARTICLE{Paturel_hyperleda2_2003,
       author = {{Paturel}, G. and {Theureau}, G. and {Bottinelli}, L. and {Gouguenheim}, L. and {Coudreau-Durand}, N. and {Hallet}, N. and {Petit}, C.},
        title = "{HYPERLEDA. II. The homogenized HI data}",
      journal = {\aap},
         year = 2003,
        month = dec,
       volume = {412},
        pages = {57-67},
          doi = {10.1051/0004-6361:20031412},
       adsurl = {https://ui.adsabs.harvard.edu/abs/2003A&A...412...57P}
}

@ARTICLE{Bonne_cii_2023,
       author = {{Bonne}, L. and {Kabanovic}, S. and {Schneider}, N. and {Zavagno}, A. and {Keilmann}, E. and {Simon}, R. and {Buchbender}, C. and {G{\"u}sten}, R. and {Jacob}, A.~M. and {Jacobs}, K. and {Kavak}, U. and {Polles}, F.~L. and {Tiwari}, M. and {Wyrowski}, F. and {Tielens}, A.~G.~G.~M.},
        title = "{The SOFIA FEEDBACK [CII] Legacy Survey: Rapid molecular cloud dispersal in RCW 79}",
      journal = {\aap},
         year = 2023,
        month = nov,
       volume = {679},
          eid = {L5},
        pages = {L5},
          doi = {10.1051/0004-6361/202347721},
archivePrefix = {arXiv},
       eprint = {2310.01657},
 primaryClass = {astro-ph.GA},
       adsurl = {https://ui.adsabs.harvard.edu/abs/2023A&A...679L...5B}
}

@INPROCEEDINGS{Chevance_pp7_review_2023,
       author = {{Chevance}, M. and {Krumholz}, M.~R. and {McLeod}, A.~F. and {Ostriker}, E.~C. and {Rosolowsky}, E.~W. and {Sternberg}, A.},
        title = "{The Life and Times of Giant Molecular Clouds}",
    booktitle = {Protostars and Planets VII},
         year = 2023,
       editor = {{Inutsuka}, S. and {Aikawa}, Y. and {Muto}, T. and {Tomida}, K. and {Tamura}, M.},
       series = {Astronomical Society of the Pacific Conference Series},
       volume = {534},
        month = jul,
        pages = {1},
          doi = {10.48550/arXiv.2203.09570},
archivePrefix = {arXiv},
       eprint = {2203.09570},
 primaryClass = {astro-ph.GA},
       adsurl = {https://ui.adsabs.harvard.edu/abs/2023ASPC..534....1C}
}

@ARTICLE{Wong_lmc_2011,
       author = {{Wong}, Tony and {Hughes}, Annie and {Ott}, J{\"u}rgen and {Muller}, Erik and {Pineda}, Jorge L. and {Bernard}, Jean-Philippe and {Chu}, You-Hua and {Fukui}, Yasuo and {Gruendl}, Robert A. and {Henkel}, Christian and {Kawamura}, Akiko and {Klein}, Ulrich and {Looney}, Leslie W. and {Maddison}, Sarah and {Mizuno}, Yoji and {Paradis}, Deborah and {Seale}, Jonathan and {Welty}, Daniel E.},
        title = "{The Magellanic Mopra Assessment (MAGMA). I. The Molecular Cloud Population of the Large Magellanic Cloud}",
      journal = {\apjs},
         year = 2011,
        month = dec,
       volume = {197},
       number = {2},
          eid = {16},
        pages = {16},
          doi = {10.1088/0067-0049/197/2/16},
archivePrefix = {arXiv},
       eprint = {1108.5715},
 primaryClass = {astro-ph.GA},
       adsurl = {https://ui.adsabs.harvard.edu/abs/2011ApJS..197...16W}
}

@ARTICLE{Nieten_m31_2006,
       author = {{Nieten}, Ch. and {Neininger}, N. and {Gu{\'e}lin}, M. and {Ungerechts}, H. and {Lucas}, R. and {Berkhuijsen}, E.~M. and {Beck}, R. and {Wielebinski}, R.},
        title = "{Molecular gas in the Andromeda galaxy}",
      journal = {\aap},
         year = 2006,
        month = jul,
       volume = {453},
       number = {2},
        pages = {459-475},
          doi = {10.1051/0004-6361:20035672},
archivePrefix = {arXiv},
       eprint = {astro-ph/0512563},
 primaryClass = {astro-ph},
       adsurl = {https://ui.adsabs.harvard.edu/abs/2006A&A...453..459N}
}

@ARTICLE{Engargiola_m33_2003,
       author = {{Engargiola}, G. and {Plambeck}, R.~L. and {Rosolowsky}, E. and {Blitz}, L.},
        title = "{Giant Molecular Clouds in M33. I. BIMA All-Disk Survey}",
      journal = {\apjs},
         year = 2003,
        month = dec,
       volume = {149},
       number = {2},
        pages = {343-363},
          doi = {10.1086/379165},
archivePrefix = {arXiv},
       eprint = {astro-ph/0308388},
 primaryClass = {astro-ph},
       adsurl = {https://ui.adsabs.harvard.edu/abs/2003ApJS..149..343E}
}

@ARTICLE{Schruba_sf_efficiencies_2019,
       author = {{Schruba}, Andreas and {Kruijssen}, J.~M. Diederik and {Leroy}, Adam K.},
        title = "{How Galactic Environment Affects the Dynamical State of Molecular Clouds and Their Star Formation Efficiency}",
      journal = {\apj},
         year = 2019,
        month = sep,
       volume = {883},
       number = {1},
          eid = {2},
        pages = {2},
          doi = {10.3847/1538-4357/ab3a43},
archivePrefix = {arXiv},
       eprint = {1908.04306},
 primaryClass = {astro-ph.GA},
       adsurl = {https://ui.adsabs.harvard.edu/abs/2019ApJ...883....2S}
}

@ARTICLE{Leroy_ic10_2006,
       author = {{Leroy}, A. and {Bolatto}, A. and {Walter}, F. and {Blitz}, L.},
        title = "{Molecular Gas in the Low-Metallicity, Star-forming Dwarf IC 10}",
      journal = {\apj},
         year = 2006,
        month = jun,
       volume = {643},
       number = {2},
        pages = {825-843},
          doi = {10.1086/503024},
archivePrefix = {arXiv},
       eprint = {astro-ph/0602056},
 primaryClass = {astro-ph},
       adsurl = {https://ui.adsabs.harvard.edu/abs/2006ApJ...643..825L}
}

@ARTICLE{Schinnerer_paws_2013,
       author = {{Schinnerer}, Eva and {Meidt}, Sharon E. and {Pety}, J{\'e}r{\^o}me and {Hughes}, Annie and {Colombo}, Dario and {Garc{\'\i}a-Burillo}, Santiago and {Schuster}, Karl F. and {Dumas}, Ga{\"e}lle and {Dobbs}, Clare L. and {Leroy}, Adam K. and {Kramer}, Carsten and {Thompson}, Todd A. and {Regan}, Michael W.},
        title = "{The PdBI Arcsecond Whirlpool Survey (PAWS). I. A Cloud-scale/Multi-wavelength View of the Interstellar Medium in a Grand-design Spiral Galaxy}",
      journal = {\apj},
         year = 2013,
        month = dec,
       volume = {779},
       number = {1},
          eid = {42},
        pages = {42},
          doi = {10.1088/0004-637X/779/1/42},
archivePrefix = {arXiv},
       eprint = {1304.1801},
 primaryClass = {astro-ph.CO},
       adsurl = {https://ui.adsabs.harvard.edu/abs/2013ApJ...779...42S}
}

@ARTICLE{Pety_paws_2013,
       author = {{Pety}, J{\'e}r{\^o}me and {Schinnerer}, Eva and {Leroy}, Adam K. and {Hughes}, Annie and {Meidt}, Sharon E. and {Colombo}, Dario and {Dumas}, Gaelle and {Garc{\'\i}a-Burillo}, Santiago and {Schuster}, Karl F. and {Kramer}, Carsten and {Dobbs}, Clare L. and {Thompson}, Todd A.},
        title = "{The Plateau de Bure + 30 m Arcsecond Whirlpool Survey Reveals a Thick Disk of Diffuse Molecular Gas in the M51 Galaxy}",
      journal = {\apj},
         year = 2013,
        month = dec,
       volume = {779},
       number = {1},
          eid = {43},
        pages = {43},
          doi = {10.1088/0004-637X/779/1/43},
archivePrefix = {arXiv},
       eprint = {1304.1396},
 primaryClass = {astro-ph.GA},
       adsurl = {https://ui.adsabs.harvard.edu/abs/2013ApJ...779...43P}
}

@ARTICLE{Gratier_m33_2010,
       author = {{Gratier}, P. and {Braine}, J. and {Rodriguez-Fernandez}, N.~J. and {Schuster}, K.~F. and {Kramer}, C. and {Xilouris}, E.~M. and {Tabatabaei}, F.~S. and {Henkel}, C. and {Corbelli}, E. and {Israel}, F. and {van der Werf}, P.~P. and {Calzetti}, D. and {Garcia-Burillo}, S. and {Sievers}, A. and {Combes}, F. and {Wiklind}, T. and {Brouillet}, N. and {Herpin}, F. and {Bontemps}, S. and {Aalto}, S. and {Koribalski}, B. and {van der Tak}, F. and {Wiedner}, M.~C. and {R{\"o}llig}, M. and {Mookerjea}, B.},
        title = "{Molecular and atomic gas in the Local Group galaxy M 33}",
      journal = {\aap},
         year = 2010,
        month = nov,
       volume = {522},
          eid = {A3},
        pages = {A3},
          doi = {10.1051/0004-6361/201014441},
archivePrefix = {arXiv},
       eprint = {1003.3222},
 primaryClass = {astro-ph.CO},
       adsurl = {https://ui.adsabs.harvard.edu/abs/2010A&A...522A...3G}
}

@ARTICLE{Querejeta_ic342_2023,
       author = {{Querejeta}, Miguel and {Pety}, J{\'e}r{\^o}me and {Schruba}, Andreas and {Leroy}, Adam K. and {Herrera}, Cinthya N. and {Chiang}, I-Da and {Meidt}, Sharon E. and {Rosolowsky}, Erik and {Schinnerer}, Eva and {Schuster}, Karl and {Sun}, Jiayi and {Herrmann}, Kimberly A. and {Barnes}, Ashley T. and {Be{\v{s}}li{\'c}}, Ivana and {Bigiel}, Frank and {Cao}, Yixian and {Chevance}, M{\'e}lanie and {Eibensteiner}, Cosima and {Emsellem}, Eric and {Faesi}, Christopher M. and {Hughes}, Annie and {Kim}, Jaeyeon and {Klessen}, Ralf S. and {Kreckel}, Kathryn and {Kruijssen}, J.~M. Diederik and {Liu}, Daizhong and {Neumayer}, Nadine and {Pan}, Hsi-An and {Saito}, Toshiki and {Sandstrom}, Karin and {Teng}, Yu-Hsuan and {Usero}, Antonio and {Williams}, Thomas G. and {Zakardjian}, Antoine},
        title = "{A sensitive, high-resolution, wide-field IRAM NOEMA CO(1-0) survey of the very nearby spiral galaxy IC 342}",
      journal = {\aap},
         year = 2023,
        month = dec,
       volume = {680},
          eid = {A4},
        pages = {A4},
          doi = {10.1051/0004-6361/202143023},
archivePrefix = {arXiv},
       eprint = {2310.06501},
 primaryClass = {astro-ph.GA},
       adsurl = {https://ui.adsabs.harvard.edu/abs/2023A&A...680A...4Q}
}

@ARTICLE{Meidt_timescales_2015,
       author = {{Meidt}, Sharon E. and {Hughes}, Annie and {Dobbs}, Clare L. and {Pety}, J{\'e}r{\^o}me and {Thompson}, Todd A. and {Garc{\'\i}a-Burillo}, Santiago and {Leroy}, Adam K. and {Schinnerer}, Eva and {Colombo}, Dario and {Querejeta}, Miguel and {Kramer}, Carsten and {Schuster}, Karl F. and {Dumas}, Ga{\"e}lle},
        title = "{Short GMC Lifetimes: An Observational Estimate with the PdBI Arcsecond Whirlpool Survey (PAWS)}",
      journal = {\apj},
         year = 2015,
        month = jun,
       volume = {806},
       number = {1},
          eid = {72},
        pages = {72},
          doi = {10.1088/0004-637X/806/1/72},
archivePrefix = {arXiv},
       eprint = {1504.04528},
 primaryClass = {astro-ph.GA},
       adsurl = {https://ui.adsabs.harvard.edu/abs/2015ApJ...806...72M}
}

@ARTICLE{Evans_sf_efficiencies_2009,
       author = {{Evans}, II, Neal J. and {Dunham}, Michael M. and {J{\o}rgensen}, Jes K. and {Enoch}, Melissa L. and {Mer{\'\i}n}, Bruno and {van Dishoeck}, Ewine F. and {Alcal{\'a}}, Juan M. and {Myers}, Philip C. and {Stapelfeldt}, Karl R. and {Huard}, Tracy L. and {Allen}, Lori E. and {Harvey}, Paul M. and {van Kempen}, Tim and {Blake}, Geoffrey A. and {Koerner}, David W. and {Mundy}, Lee G. and {Padgett}, Deborah L. and {Sargent}, Anneila I.},
        title = "{The Spitzer c2d Legacy Results: Star-Formation Rates and Efficiencies; Evolution and Lifetimes}",
      journal = {\apjs},
         year = 2009,
        month = apr,
       volume = {181},
       number = {2},
        pages = {321-350},
          doi = {10.1088/0067-0049/181/2/321},
archivePrefix = {arXiv},
       eprint = {0811.1059},
 primaryClass = {astro-ph},
       adsurl = {https://ui.adsabs.harvard.edu/abs/2009ApJS..181..321E}
}

@ARTICLE{Colombo_2014,
       author = {{Colombo}, Dario and {Hughes}, Annie and {Schinnerer}, Eva and {Meidt}, Sharon E. and {Leroy}, Adam K. and {Pety}, J{\'e}r{\^o}me and {Dobbs}, Clare L. and {Garc{\'\i}a-Burillo}, Santiago and {Dumas}, Ga{\"e}lle and {Thompson}, Todd A. and {Schuster}, Karl F. and {Kramer}, Carsten},
        title = "{The PdBI Arcsecond Whirlpool Survey (PAWS): Environmental Dependence of Giant Molecular Cloud Properties in M51}",
      journal = {\apj},
         year = 2014,
        month = mar,
       volume = {784},
       number = {1},
          eid = {3},
        pages = {3},
          doi = {10.1088/0004-637X/784/1/3},
archivePrefix = {arXiv},
       eprint = {1401.1505},
 primaryClass = {astro-ph.GA},
       adsurl = {https://ui.adsabs.harvard.edu/abs/2014ApJ...784....3C}
}

@ARTICLE{Lugo-Aranda_HII_catalog_2024,
       author = {{Lugo-Aranda}, A.~Z. and {S{\'a}nchez}, S.~F. and {Barrera-Ballesteros}, J.~K. and {L{\'o}pez-Cob{\'a}}, C. and {Espinosa-Ponce}, C. and {Galbany}, L. and {Anderson}, Joseph P.},
        title = "{H II regions and diffuse ionized gas in the AMUSING++ Compilation - I. Catalogue presentation}",
      journal = {\mnras},
         year = 2024,
        month = mar,
       volume = {528},
       number = {4},
        pages = {6099-6118},
          doi = {10.1093/mnras/stae345},
archivePrefix = {arXiv},
       eprint = {2401.15807},
 primaryClass = {astro-ph.GA},
       adsurl = {https://ui.adsabs.harvard.edu/abs/2024MNRAS.528.6099L}
}

@ARTICLE{Espinosa-Ponce_califa_2020,
       author = {{Espinosa-Ponce}, C. and {S{\'a}nchez}, S.~F. and {Morisset}, C. and {Barrera-Ballesteros}, J.~K. and {Galbany}, L. and {Garc{\'\i}a-Benito}, R. and {Lacerda}, E.~A.~D. and {Mast}, D.},
        title = "{H II regions in the CALIFA survey: I. catalogue presentation}",
      journal = {\mnras},
         year = 2020,
        month = may,
       volume = {494},
       number = {2},
        pages = {1622-1646},
          doi = {10.1093/mnras/staa782},
archivePrefix = {arXiv},
       eprint = {2003.07865},
 primaryClass = {astro-ph.GA},
       adsurl = {https://ui.adsabs.harvard.edu/abs/2020MNRAS.494.1622E}
}

@ARTICLE{Lee_phangs_hst_2022,
       author = {{Lee}, Janice C. and {Whitmore}, Bradley C. and {Thilker}, David A. and {Deger}, Sinan and {Larson}, Kirsten L. and {Ubeda}, Leonardo and {Anand}, Gagandeep S. and {Boquien}, M{\'e}d{\'e}ric and {Chandar}, Rupali and {Dale}, Daniel A. and {Emsellem}, Eric and {Leroy}, Adam K. and {Rosolowsky}, Erik and {Schinnerer}, Eva and {Schmidt}, Judy and {Lilly}, James and {Turner}, Jordan and {Van Dyk}, Schuyler and {White}, Richard L. and {Barnes}, Ashley T. and {Belfiore}, Francesco and {Bigiel}, Frank and {Blanc}, Guillermo A. and {Cao}, Yixian and {Chevance}, Melanie and {Congiu}, Enrico and {Egorov}, Oleg V. and {Glover}, Simon C.~O. and {Grasha}, Kathryn and {Groves}, Brent and {Henshaw}, Jonathan D. and {Hughes}, Annie and {Klessen}, Ralf S. and {Koch}, Eric and {Kreckel}, Kathryn and {Kruijssen}, J.~M. Diederik and {Liu}, Daizhong and {Lopez}, Laura A. and {Mayker}, Ness and {Meidt}, Sharon E. and {Murphy}, Eric J. and {Pan}, Hsi-An and {Pety}, J{\'e}r{\^o}me and {Querejeta}, Miguel and {Razza}, Alessandro and {Saito}, Toshiki and {S{\'a}nchez-Bl{\'a}zquez}, Patricia and {Santoro}, Francesco and {Sardone}, Amy and {Scheuermann}, Fabian and {Schruba}, Andreas and {Sun}, Jiayi and {Usero}, Antonio and {Watkins}, E. and {Williams}, Thomas G.},
        title = "{The PHANGS-HST Survey: Physics at High Angular Resolution in Nearby Galaxies with the Hubble Space Telescope}",
      journal = {\apjs},
         year = 2022,
        month = jan,
       volume = {258},
       number = {1},
          eid = {10},
        pages = {10},
          doi = {10.3847/1538-4365/ac1fe5},
archivePrefix = {arXiv},
       eprint = {2101.02855},
 primaryClass = {astro-ph.GA},
       adsurl = {https://ui.adsabs.harvard.edu/abs/2022ApJS..258...10L}
}

@ARTICLE{Pedrini_feast_2024,
       author = {{Pedrini}, Alex and {Adamo}, Angela and {Calzetti}, Daniela and {Bik}, Arjan and {Gregg}, Benjamin and {Linden}, Sean T. and {Bajaj}, Varun and {Ryon}, Jenna E. and {Ali}, Ahmad A. and {Bortolini}, Giacomo and {Correnti}, Matteo and {Elmegreen}, Bruce G. and {Elmegreen}, Debra Meloy and {Gallagher}, John S. and {Grasha}, Kathryn and {Gutermuth}, Robert A. and {Johnson}, Kelsey E. and {Melinder}, Jens and {Messa}, Matteo and {{\"O}stlin}, G{\"o}ran and {Sabbi}, Elena and {Smith}, Linda J. and {Tosi}, Monica and {Vieira}, Helena Faustino},
        title = "{FEAST: Feedback in Emerging extragAlactic Star ClusTers: JWST Spots Polycyclic Aromatic Hydrocarbon Destruction in NGC 628 during the Emerging Phase of Star Formation}",
      journal = {\apj},
         year = 2024,
        month = aug,
       volume = {971},
       number = {1},
          eid = {32},
        pages = {32},
          doi = {10.3847/1538-4357/ad534d},
archivePrefix = {arXiv},
       eprint = {2406.01666},
 primaryClass = {astro-ph.GA},
       adsurl = {https://ui.adsabs.harvard.edu/abs/2024ApJ...971...32P}
}

@ARTICLE{Rodriguez_3p3_2024,
       author = {{Rodr{\'\i}guez}, M. Jimena and {Lee}, Janice C. and {Indebetouw}, Remy and {Whitmore}, B.~C. and {Maschmann}, Daniel and {Williams}, Thomas G. and {Chandar}, Rupali and {Barnes}, A.~T. and {Gnedin}, Oleg Y. and {Sandstrom}, Karin M. and {Rosolowsky}, Erik and {Leroy}, Adam K. and {Thilker}, David A. and {Kim}, Hwihyun and {Sun}, Jiayi and {Klessen}, Ralf S. and {Groves}, Brent and {Wofford}, Aida and {Boquien}, M{\'e}d{\'e}ric and {Dale}, Daniel A. and {{\'U}beda}, Leonardo and {Larson}, Kirsten L. and {Grasha}, Kathryn and {Johnson}, Kelsey E. and {Levy}, Rebecca C. and {Bigiel}, Frank and {Hassani}, Hamid and {Sarbadhicary}, Sumit K.},
        title = "{Tracing the Earliest Stages of Star and Cluster Formation in 19 Nearby Galaxies with PHANGS-JWST and HST: Compact 3.3 {\ensuremath{\mu}}m Polycyclic Aromatic Hydrocarbon Emitters and Their Relation to the Optical Census of Star Clusters}",
      journal = {\apj},
         year = 2025,
        month = apr,
       volume = {983},
       number = {2},
          eid = {137},
        pages = {137},
          doi = {10.3847/1538-4357/adbb69},
archivePrefix = {arXiv},
       eprint = {2412.07862},
 primaryClass = {astro-ph.GA},
       adsurl = {https://ui.adsabs.harvard.edu/abs/2025ApJ...983..137R}
}

@ARTICLE{Kennicutt_resolved_SF_m51_2007,
       author = {{Kennicutt}, Jr., Robert C. and {Calzetti}, Daniela and {Walter}, Fabian and {Helou}, George and {Hollenbach}, David J. and {Armus}, Lee and {Bendo}, George and {Dale}, Daniel A. and {Draine}, Bruce T. and {Engelbracht}, Charles W. and {Gordon}, Karl D. and {Prescott}, Moire K.~M. and {Regan}, Michael W. and {Thornley}, Michele D. and {Bot}, Caroline and {Brinks}, Elias and {de Blok}, Erwin and {de Mello}, Dulia and {Meyer}, Martin and {Moustakas}, John and {Murphy}, Eric J. and {Sheth}, Kartik and {Smith}, J.~D.~T.},
        title = "{Star Formation in NGC 5194 (M51a). II. The Spatially Resolved Star Formation Law}",
      journal = {\apj},
         year = 2007,
        month = dec,
       volume = {671},
       number = {1},
        pages = {333-348},
          doi = {10.1086/522300},
archivePrefix = {arXiv},
       eprint = {0708.0922},
 primaryClass = {astro-ph},
       adsurl = {https://ui.adsabs.harvard.edu/abs/2007ApJ...671..333K}
}

@ARTICLE{Whitcomb_sfr_tracers_2023,
       author = {{Whitcomb}, C.~M. and {Sandstrom}, K. and {Leroy}, A. and {Smith}, J. -D.~T.},
        title = "{Star Formation and Molecular Gas Diagnostics with Mid- and Far-infrared Emission}",
      journal = {\apj},
         year = 2023,
        month = may,
       volume = {948},
       number = {2},
          eid = {88},
        pages = {88},
          doi = {10.3847/1538-4357/acc316},
archivePrefix = {arXiv},
       eprint = {2212.00180},
 primaryClass = {astro-ph.GA},
       adsurl = {https://ui.adsabs.harvard.edu/abs/2023ApJ...948...88W}
}

@ARTICLE{Leger_puget_pah_1984,
       author = {{Leger}, A. and {Puget}, J.~L.},
        title = "{Identification of the Unidentified Infrared Emission Features of Interstellar Dust}",
      journal = {\aap},
         year = 1984,
        month = aug,
       volume = {137},
        pages = {L5-L8},
       adsurl = {https://ui.adsabs.harvard.edu/abs/1984A&A...137L...5L}
}

@ARTICLE{Jones_hac_1990,
       author = {{Jones}, A.~P. and {Duley}, W.~W. and {Williams}, D.~A.},
        title = "{The structure and evolution of hydrogenated amorphous carbon grains and mantles in the interstellar medium}",
      journal = {\qjras},
         year = 1990,
        month = dec,
       volume = {31},
        pages = {567-582},
       adsurl = {https://ui.adsabs.harvard.edu/abs/1990QJRAS..31..567J}
}

@ARTICLE{Prescott_obscured_2007,
       author = {{Prescott}, Moire K.~M. and {Kennicutt}, Jr., Robert C. and {Bendo}, George J. and {Buckalew}, Brent A. and {Calzetti}, Daniela and {Engelbracht}, Charles W. and {Gordon}, Karl D. and {Hollenbach}, David J. and {Lee}, Janice C. and {Moustakas}, John and {Dale}, Daniel A. and {Helou}, George and {Jarrett}, Thomas H. and {Murphy}, Eric J. and {Smith}, John-David T. and {Akiyama}, Sanae and {Sosey}, Megan L.},
        title = "{The Incidence of Highly Obscured Star-forming Regions in SINGS Galaxies}",
      journal = {\apj},
         year = 2007,
        month = oct,
       volume = {668},
       number = {1},
        pages = {182-202},
          doi = {10.1086/521071},
archivePrefix = {arXiv},
       eprint = {0706.3501},
 primaryClass = {astro-ph},
       adsurl = {https://ui.adsabs.harvard.edu/abs/2007ApJ...668..182P}
}

@ARTICLE{Congiu_hii_2023,
       author = {{Congiu}, Enrico and {Blanc}, Guillermo A. and {Belfiore}, Francesco and {Santoro}, Francesco and {Scheuermann}, Fabian and {Kreckel}, Kathryn and {Emsellem}, Eric and {Groves}, Brent and {Pan}, Hsi-An and {Bigiel}, Frank and {Dale}, Daniel A. and {Glover}, Simon C.~O. and {Grasha}, Kathryn and {Egorov}, Oleg V. and {Leroy}, Adam and {Schinnerer}, Eva and {Watkins}, Elizabeth J. and {Williams}, Thomas G.},
        title = "{PHANGS-MUSE: Detection and Bayesian classification of  40 000 ionised nebulae in nearby spiral galaxies}",
      journal = {\aap},
         year = 2023,
        month = apr,
       volume = {672},
          eid = {A148},
        pages = {A148},
          doi = {10.1051/0004-6361/202245153},
archivePrefix = {arXiv},
       eprint = {2302.03062},
 primaryClass = {astro-ph.GA},
       adsurl = {https://ui.adsabs.harvard.edu/abs/2023A&A...672A.148C}
}

@ARTICLE{Hannon_2022,
       author = {{Hannon}, Stephen and {Lee}, Janice C. and {Whitmore}, B.~C. and {Mobasher}, B. and {Thilker}, D. and {Chandar}, R. and {Adamo}, A. and {Wofford}, A. and {Orozco-Duarte}, R. and {Calzetti}, D. and {Della Bruna}, L. and {Kreckel}, K. and {Groves}, B. and {Barnes}, A.~T. and {Boquien}, M. and {Belfiore}, F. and {Linden}, S.},
        title = "{H {\ensuremath{\alpha}} morphologies of star clusters in 16 LEGUS galaxies: Constraints on H II region evolution time-scales}",
      journal = {\mnras},
         year = 2022,
        month = may,
       volume = {512},
       number = {1},
        pages = {1294-1316},
          doi = {10.1093/mnras/stac550},
archivePrefix = {arXiv},
       eprint = {2203.01339},
 primaryClass = {astro-ph.GA},
       adsurl = {https://ui.adsabs.harvard.edu/abs/2022MNRAS.512.1294H}
}

@ARTICLE{Hannon_2019,
       author = {{Hannon}, Stephen and {Lee}, Janice C. and {Whitmore}, B.~C. and {Chandar}, R. and {Adamo}, A. and {Mobasher}, B. and {Aloisi}, A. and {Calzetti}, D. and {Cignoni}, M. and {Cook}, D.~O. and {Dale}, D. and {Deger}, S. and {Della Bruna}, L. and {Elmegreen}, D.~M. and {Gouliermis}, D.~A. and {Grasha}, K. and {Grebel}, E.~K. and {Herrero}, A. and {Hunter}, D.~A. and {Johnson}, K.~E. and {Kennicutt}, R. and {Kim}, H. and {Sacchi}, E. and {Smith}, L. and {Thilker}, D. and {Turner}, J. and {Walterbos}, R.~A.~M. and {Wofford}, A.},
        title = "{H {\ensuremath{\alpha}} morphologies of star clusters: a LEGUS study of H II region evolution time-scales and stochasticity in low-mass clusters}",
      journal = {\mnras},
         year = 2019,
        month = dec,
       volume = {490},
       number = {4},
        pages = {4648-4665},
          doi = {10.1093/mnras/stz2820},
archivePrefix = {arXiv},
       eprint = {1910.02983},
 primaryClass = {astro-ph.GA},
       adsurl = {https://ui.adsabs.harvard.edu/abs/2019MNRAS.490.4648H}
}

@ARTICLE{Romanelli_2025,
       author = {{Romanelli}, Andrea and {Chevance}, M{\'e}lanie and {Kruijssen}, J.~M. Diederik and {Ramambason}, Lise and {Querejeta}, Miguel and {Boquien}, Mederic and {Dale}, Daniel A. and {den Brok}, Jakob and {Glover}, Simon C.~O. and {Grasha}, Kathryn and {Hughes}, Annie and {Kim}, Jaeyeon and {Longmore}, Steven and {Meidt}, Sharon E. and {Mendez-Delgado}, Jos{\'e} Eduardo and {Neumann}, Lukas and {Pety}, J{\'e}r{\^o}me and {Schinnerer}, Eva and {Smith}, Rowan and {Sun}, Jiayi and {Williams}, Thomas G.},
        title = "{The impact of spiral arms on the star formation life cycle}",
      journal = {\aap},
         year = 2025,
        month = jun,
       volume = {698},
          eid = {A296},
        pages = {A296},
          doi = {10.1051/0004-6361/202553895},
archivePrefix = {arXiv},
       eprint = {2505.10908},
 primaryClass = {astro-ph.GA},
       adsurl = {https://ui.adsabs.harvard.edu/abs/2025A&A...698A.296R}
}

@ARTICLE{Whitmore_2025,
       author = {{Whitmore}, Bradley C. and {Chandar}, Rupali and {Lee}, Janice C. and {Henny}, Kiana F. and {Rodr{\'\i}guez}, M. Jimena and {Baron}, Dalya and {Bigiel}, F. and {Boquien}, M{\'e}d{\'e}ric and {Chevance}, M{\'e}lanie and {Chown}, Ryan and {Dale}, Daniel A. and {Floyd}, Matthew and {Grasha}, Kathryn and {Glover}, Simon C.~O. and {Gnedin}, Oleg and {Hassani}, Hamid and {Indebetouw}, Remy and {Kapoor}, Anand Utsav and {Larson}, Kirsten L. and {Leroy}, Adam K. and {Maschmann}, Daniel and {Scheuermann}, Fabian and {Sutter}, Jessica and {Schinnerer}, Eva and {Sarbadhicary}, Sumit K. and {Thilker}, David A. and {Williams}, Thomas G. and {Wofford}, Aida},
        title = "{Empirical SED Templates for Star Clusters Observed with HST and JWST: No Strong PAH or IR Dust Emission after 5 Myr}",
      journal = {\apj},
         year = 2025,
        month = mar,
       volume = {982},
       number = {1},
          eid = {50},
        pages = {50},
          doi = {10.3847/1538-4357/adb3a2},
archivePrefix = {arXiv},
       eprint = {2503.17921},
 primaryClass = {astro-ph.GA},
       adsurl = {https://ui.adsabs.harvard.edu/abs/2025ApJ...982...50W}
}

@ARTICLE{starforge_grudic_2021,
       author = {{Grudi{\'c}}, Michael Y. and {Guszejnov}, D{\'a}vid and {Hopkins}, Philip F. and {Offner}, Stella S.~R. and {Faucher-Gigu{\`e}re}, Claude-Andr{\'e}},
        title = "{STARFORGE: Towards a comprehensive numerical model of star cluster formation and feedback}",
      journal = {\mnras},
         year = 2021,
        month = sep,
       volume = {506},
       number = {2},
        pages = {2199-2231},
          doi = {10.1093/mnras/stab1347},
archivePrefix = {arXiv},
       eprint = {2010.11254},
 primaryClass = {astro-ph.IM},
       adsurl = {https://ui.adsabs.harvard.edu/abs/2021MNRAS.506.2199G}
}

@ARTICLE{Sun_dyn_eq_2020,
       author = {{Sun}, Jiayi and {Leroy}, Adam K. and {Ostriker}, Eve C. and {Hughes}, Annie and {Rosolowsky}, Erik and {Schruba}, Andreas and {Schinnerer}, Eva and {Blanc}, Guillermo A. and {Faesi}, Christopher and {Kruijssen}, J.~M. Diederik and {Meidt}, Sharon and {Utomo}, Dyas and {Bigiel}, Frank and {Bolatto}, Alberto D. and {Chevance}, M{\'e}lanie and {Chiang}, I-Da and {Dale}, Daniel and {Emsellem}, Eric and {Glover}, Simon C.~O. and {Grasha}, Kathryn and {Henshaw}, Jonathan and {Herrera}, Cinthya N. and {Jimenez-Donaire}, Maria Jesus and {Lee}, Janice C. and {Pety}, J{\'e}r{\^o}me and {Querejeta}, Miguel and {Saito}, Toshiki and {Sandstrom}, Karin and {Usero}, Antonio},
        title = "{Dynamical Equilibrium in the Molecular ISM in 28 Nearby Star-forming Galaxies}",
      journal = {\apj},
         year = 2020,
        month = apr,
       volume = {892},
       number = {2},
          eid = {148},
        pages = {148},
          doi = {10.3847/1538-4357/ab781c},
archivePrefix = {arXiv},
       eprint = {2002.08964},
 primaryClass = {astro-ph.GA},
       adsurl = {https://ui.adsabs.harvard.edu/abs/2020ApJ...892..148S}
}

@ARTICLE{Kennicutt_sings_2003,
       author = {{Kennicutt}, Jr., Robert C. and {Armus}, Lee and {Bendo}, George and {Calzetti}, Daniela and {Dale}, Daniel A. and {Draine}, Bruce T. and {Engelbracht}, Charles W. and {Gordon}, Karl D. and {Grauer}, Albert D. and {Helou}, George and {Hollenbach}, David J. and {Jarrett}, Thomas H. and {Kewley}, Lisa J. and {Leitherer}, Claus and {Li}, Aigen and {Malhotra}, Sangeeta and {Regan}, Michael W. and {Rieke}, George H. and {Rieke}, Marcia J. and {Roussel}, H{\'e}l{\`e}ne and {Smith}, John-David T. and {Thornley}, Michele D. and {Walter}, Fabian},
        title = "{SINGS: The SIRTF Nearby Galaxies Survey}",
      journal = {\pasp},
         year = 2003,
        month = aug,
       volume = {115},
       number = {810},
        pages = {928-952},
          doi = {10.1086/376941},
archivePrefix = {arXiv},
       eprint = {astro-ph/0305437},
 primaryClass = {astro-ph},
       adsurl = {https://ui.adsabs.harvard.edu/abs/2003PASP..115..928K}
}

@ARTICLE{Kim_pah_2025,
       author = {{Kim}, Jaeyeon and {Chevance}, M{\'e}lanie and {Ramambason}, Lise and {Kreckel}, Kathryn and {Klessen}, Ralf S. and {Dale}, Daniel A. and {Leroy}, Adam K. and {Sandstrom}, Karin and {Chown}, Ryan and {Williams}, Thomas G. and {Sarbadhicary}, Sumit K. and {Belfiore}, Francesco and {Bigiel}, Frank and {Congiu}, Enrico and {Egorov}, Oleg V. and {Emsellem}, Eric and {Glover}, Simon C.~O. and {Grasha}, Kathryn and {Hughes}, Annie and {Kruijssen}, J.~M. Diederik and {Lee}, Janice C. and {Pathak}, Debosmita and {Pessa}, Ismael and {Rosolowsky}, Erik and {Sun}, Jiayi and {Sutter}, Jessica and {Thilker}, David A.},
        title = "{Timescales of Polycyclic Aromatic Hydrocarbon and Dust Continuum Emission from Gas Clouds Compared to Molecular Gas Cloud Lifetimes in PHANGS-JWST Galaxies}",
      journal = {\apj},
         year = 2025,
        month = aug,
       volume = {988},
       number = {2},
          eid = {215},
        pages = {215},
          doi = {10.3847/1538-4357/ade443},
archivePrefix = {arXiv},
       eprint = {2506.10063},
 primaryClass = {astro-ph.GA},
       adsurl = {https://ui.adsabs.harvard.edu/abs/2025ApJ...988..215K}
}

@article{astropy:2013,
  author = {{Astropy Collaboration} and Robitaille, T. P. and Tollerud, E. J. and Greenfield, P. and Droettboom, M. and Bray, E. and Aldcroft, T. and Davis, M. and Ginsburg, A. and Price-Whelan, A. M. and others},
  title = {Astropy: A community Python package for astronomy},
  journal = {A\&A},
  volume = {558},
  pages = {A33},
  year = {2013},
  doi = {10.1051/0004-6361/201322068}
}

@article{astropy:2018,
  author = {{Astropy Collaboration} and Price-Whelan, A. M. and Sip\H{o}cz, B. M. and G\"unther, H. M. and Lim, P. L. and Crawford, S. M. and Conseil, S. and Shupe, D. L. and Craig, M. W. and Dencheva, N. and others},
  title = {The Astropy Project: Building an Open-science Project and Status of the v2.0 Core Package},
  journal = {AJ},
  volume = {156},
  number = {3},
  pages = {123},
  year = {2018},
  doi = {10.3847/1538-3881/aabc4f}
}

@article{astropy:2022,
  author = {{Astropy Collaboration} and Pedregosa, F. and Tollerud, E. J. and Greenfield, P. and Robitaille, T. and others},
  title = {The Astropy Project: Sustaining and Growing a Community-oriented Open-source Project and the Transition to v5.0},
  journal = {ApJ},
  volume = {935},
  pages = {167},
  year = {2022},
  doi = {10.3847/1538-4357/ac7c74}
}

@article{hunter2007,
  author = {Hunter, J. D.},
  title = {Matplotlib: A 2D Graphics Environment},
  journal = {Comput. Sci. Eng.},
  volume = {9},
  number = {3},
  pages = {90--95},
  year = {2007},
  doi = {10.1109/MCSE.2007.55}
}

@article{virtanen2020,
  author = {Virtanen, P. and Gommers, R. and Oliphant, T. E. and Haberland, M. and Reddy, T. and Cournapeau, D. and Burovski, E. and Peterson, P. and Weckesser, W. and Bright, J. and others},
  title = {SciPy 1.0: Fundamental Algorithms for Scientific Computing in Python},
  journal = {Nat. Methods},
  volume = {17},
  pages = {261--272},
  year = {2020},
  doi = {10.1038/s41592-019-0686-2}
}

@misc{waskom2021,
  author = {Waskom, M.},
  title = {seaborn: statistical data visualization},
  year = {2021},
  howpublished = {\url{https://seaborn.pydata.org}},
  note = {Version 0.11.1}
}

@incollection{mckinney2010,
  author = {McKinney, Wes},
  title = {Data Structures for Statistical Computing in Python},
  booktitle = {Proceedings of the 9th Python in Science Conference},
  year = {2010},
  editor = {van der Walt, Stéfan and Millman, Jarrod},
  pages = {51--56},
  publisher = {SciPy},
  address = {Austin, Texas},
  doi = {10.25080/Majora-92bf1922-00a}
}

@ARTICLE{Knutas_2025,
       author = {{Knutas}, Alice and {Adamo}, Angela and {Pedrini}, Alex and {Linden}, Sean T. and {Bajaj}, Varun and {Ryon}, Jenna E. and {Gregg}, Benjamin and {Ali}, Ahmad A. and {Andersson}, Eric P. and {Bik}, Arjan and {Bortolini}, Giacomo and {Buckner}, Anne S.~M. and {Calzetti}, Daniela and {Duarte-Cabral}, Ana and {Elmegreen}, Bruce G. and {Faustino Vieira}, Helena and {Gallagher}, John S. and {Grasha}, Kathryn and {Johnson}, Kelsey and {Lai}, Thomas S.-Y. and {Lapeer}, Drew and {Messa}, Matteo and {{\"O}stlin}, G{\"o}ran and {Sabbi}, Elena and {Smith}, Linda J. and {Tosi}, Monica},
        title = "{FEAST: JWST Uncovers the Emerging Timescales of Young Star Clusters in M83}",
      journal = {\apj},
         year = 2025,
        month = nov,
       volume = {993},
       number = {1},
          eid = {13},
        pages = {13},
          doi = {10.3847/1538-4357/ae018c},
archivePrefix = {arXiv},
       eprint = {2505.08874},
 primaryClass = {astro-ph.GA},
       adsurl = {https://ui.adsabs.harvard.edu/abs/2025ApJ...993...13K}
}

@ARTICLE{Watkins_bubbles_2023,
       author = {{Watkins}, E.~J. and {Kreckel}, K. and {Groves}, B. and {Glover}, S.~C.~O. and {Whitmore}, B.~C. and {Leroy}, A.~K. and {Schinnerer}, E. and {Meidt}, S.~E. and {Egorov}, O.~V. and {Barnes}, A.~T. and {Lee}, J.~C. and {Bigiel}, F. and {Boquien}, M. and {Chandar}, R. and {Chevance}, M. and {Dale}, D.~A. and {Grasha}, K. and {Klessen}, R.~S. and {Kruijssen}, J.~M.~D. and {Larson}, K.~L. and {Li}, J. and {M{\'e}ndez-Delgado}, J.~E. and {Pessa}, I. and {Saito}, T. and {Sanchez-Blazquez}, P. and {Sarbadhicary}, S.~K. and {Scheuermann}, F. and {Thilker}, D.~A. and {Williams}, T.~G.},
        title = "{Quantifying the energetics of molecular superbubbles in PHANGS galaxies}",
      journal = {\aap},
         year = 2023,
        month = aug,
       volume = {676},
          eid = {A67},
        pages = {A67},
          doi = {10.1051/0004-6361/202346075},
archivePrefix = {arXiv},
       eprint = {2302.03699},
 primaryClass = {astro-ph.GA},
       adsurl = {https://ui.adsabs.harvard.edu/abs/2023A&A...676A..67W}
}

@ARTICLE{Wainer_embedded_2025,
       author = {{Wainer}, Tobin M. and {Dalcanton}, Julianne J. and {Grudi{\'c}}, Michael Y. and {Offner}, Stella S.~R. and {Smercina}, Adam and {Williams}, Benjamin F. and {Johnson}, L. Clifton and {Peltonen}, J. and {Koch}, Eric W. and {Neralwar}, Kartik R.},
        title = "{The Timescales of Embedded Star Formation as Observed in STARFORGE}",
      journal = {Submitted to AAS Journals, arXiv e-prints},
         year = 2025,
        month = sep,
          eid = {arXiv:2509.18322},
        pages = {arXiv:2509.18322},
          doi = {10.48550/arXiv.2509.18322},
archivePrefix = {arXiv},
       eprint = {2509.18322},
 primaryClass = {astro-ph.SR},
       adsurl = {https://ui.adsabs.harvard.edu/abs/2025arXiv250918322W}
}

@ARTICLE{Hassani_2025,
       author = {{Hassani}, Hamid and {Rosolowsky}, Erik and {Leroy}, Adam K. and {Sandstrom}, Karin and {Boquien}, M{\'e}d{\'e}ric and {Thilker}, David A. and {Whitmore}, Bradley C. and {Anand}, Gagandeep S. and {Barnes}, Ashley T. and {Cao}, Yixian and {Chown}, Ryan and {Congiu}, Enrico and {Dale}, Daniel A. and {Egorov}, Oleg V. and {Gerasimov}, Ivan and {Grasha}, Kathryn and {Indebetouw}, Remy and {Lee}, Janice C. and {Liang}, Fu-Heng and {Maschmann}, Daniel and {Meidt}, Sharon E. and {Oakes}, Elias K. and {Pessa}, Ismael and {Pety}, J{\'e}r{\^o}me and {Querejeta}, Miguel and {Ramambason}, Lise and {Jimena Rodr{\'\i}guez}, M. and {Sarbadhicary}, Sumit K. and {Sutter}, Jessica and {{\'U}beda}, Leonardo and {Williams}, Thomas G.},
        title = "{The Hidden Life of Stars: Embedded Beginnings to AGB Endings in the PHANGS-JWST Sample. I. Catalog of Mid-IR Sources}",
      journal = {Submitted to ApJS, arXiv e-prints},
         year = 2025,
        month = sep,
          eid = {arXiv:2509.16459},
        pages = {arXiv:2509.16459},
          doi = {10.48550/arXiv.2509.16459},
archivePrefix = {arXiv},
       eprint = {2509.16459},
 primaryClass = {astro-ph.GA},
       adsurl = {https://ui.adsabs.harvard.edu/abs/2025arXiv250916459H}
}

@ARTICLE{Krumholz_slug_2015,
       author = {{Krumholz}, Mark R. and {Fumagalli}, Michele and {da Silva}, Robert L. and {Rendahl}, Theodore and {Parra}, Jonathan},
        title = "{SLUG - stochastically lighting up galaxies - III. A suite of tools for simulated photometry, spectroscopy, and Bayesian inference with stochastic stellar populations}",
      journal = {\mnras},
         year = 2015,
        month = sep,
       volume = {452},
       number = {2},
        pages = {1447-1467},
          doi = {10.1093/mnras/stv1374},
archivePrefix = {arXiv},
       eprint = {1502.05408},
 primaryClass = {astro-ph.GA},
       adsurl = {https://ui.adsabs.harvard.edu/abs/2015MNRAS.452.1447K}
}

@ARTICLE{Whitmore_antennae_2002,
       author = {{Whitmore}, Bradley C. and {Zhang}, Qing},
        title = "{What Fraction of the Young Clusters in the Antennae Galaxies Are ``Missing''?}",
      journal = {\aj},
         year = 2002,
        month = sep,
       volume = {124},
       number = {3},
        pages = {1418-1434},
          doi = {10.1086/341822},
archivePrefix = {arXiv},
       eprint = {astro-ph/0207100},
 primaryClass = {astro-ph},
       adsurl = {https://ui.adsabs.harvard.edu/abs/2002AJ....124.1418W}
}

@ARTICLE{Miret-Roig_2024,
       author = {{Miret-Roig}, N{\'u}ria and {Alves}, Jo{\~a}o and {Barrado}, David and {Burkert}, Andreas and {Ratzenb{\"o}ck}, Sebastian and {Konietzka}, Ralf},
        title = "{Insights into star formation and dispersal from the synchronization of stellar clocks}",
      journal = {Nature Astronomy},
         year = 2024,
        month = feb,
       volume = {8},
        pages = {216-222},
          doi = {10.1038/s41550-023-02132-4},
archivePrefix = {arXiv},
       eprint = {2311.13042},
 primaryClass = {astro-ph.SR},
       adsurl = {https://ui.adsabs.harvard.edu/abs/2024NatAs...8..216M}
}

@ARTICLE{Schaller_1992,
       author = {{Schaller}, G. and {Schaerer}, D. and {Meynet}, G. and {Maeder}, A.},
        title = "{New Grids of Stellar Models from 0.8-SOLAR-MASS to 120-SOLAR-MASSES at Z=0.020 and Z=0.001}",
      journal = {\aaps},
         year = 1992,
        month = dec,
       volume = {96},
        pages = {269},
       adsurl = {https://ui.adsabs.harvard.edu/abs/1992A&AS...96..269S}
}

@ARTICLE{Charbonnel_1993,
       author = {{Charbonnel}, C. and {Meynet}, G. and {Maeder}, A. and {Schaller}, G. and {Schaerer}, D.},
        title = "{Grids of stellar models. III. From 0.8 to 120 Msolar at Z=0.004}",
      journal = {\aaps},
         year = 1993,
        month = oct,
       volume = {101},
        pages = {415},
       adsurl = {https://ui.adsabs.harvard.edu/abs/1993A&AS..101..415C}
}

@ARTICLE{Schaerer_1993a,
       author = {{Schaerer}, D. and {Meynet}, G. and {Maeder}, A. and {Schaller}, G.},
        title = "{Grids of stellar models. II. From 0.8 to 120 M\_solar at Z=0.008}",
      journal = {\aaps},
         year = 1993,
        month = may,
       volume = {98},
        pages = {523},
       adsurl = {https://ui.adsabs.harvard.edu/abs/1993A&AS...98..523S}
}

@ARTICLE{Schaerer_1993b,
       author = {{Schaerer}, D. and {Charbonnel}, C. and {Meynet}, G. and {Maeder}, A. and {Schaller}, G.},
        title = "{Grids of stellar models. IV. From 0.8 to 120 M\_solar at Z=0.040}",
      journal = {\aaps},
         year = 1993,
        month = dec,
       volume = {102},
        pages = {339},
       adsurl = {https://ui.adsabs.harvard.edu/abs/1993A&AS..102..339S}
}

@ARTICLE{Graham_pah_2025,
       author = {{Graham}, Gabrielle B. and {Dale}, Daniel A. and {Smith}, Chase L. and {Brann}, Elisabeth and {Conder}, Kaycee D. and {Crowe}, Samuel and {Dhileepkumar}, Sumitra and {Imming}, Nicole A. and {Mendez}, Emilio and {Pleska}, Zachary and {Sako}, Kelsey and {Amiri}, Amirnezam and {Barnes}, Ashley T. and {Boquien}, M{\'e}d{\'e}ric and {Chandar}, Rupali and {Chown}, Ryan and {Gnedin}, Oleg Y. and {Grasha}, Kathryn and {Hannon}, Stephen and {Hassani}, Hamid and {Indebetouw}, R{\'e}my and {Kim}, Hwihyun and {Kim}, Jaeyeon and {Koziol}, Hannah and {Larson}, Kirsten L. and {Lee}, Janice C. and {Leroy}, Adam K. and {Oakes}, Elias K. and {Rodr{\'\i}guez}, M. Jimena and {Rosolowsky}, Erik and {Sandstrom}, Karin and {Schinnerer}, Eva and {Sutter}, Jessica and {Thilker}, David A. and {Ubeda}, Leonardo and {Whitmore}, Bradley C. and {Weinbeck}, Tony D. and {Williams}, Thomas G. and {Wofford}, Aida and {M{\'e}ndez-Delgado}, J. Eduardo and {Tian}, Qiushi Chris and {Phangs Team}},
        title = "{PAH Marks the Spot: Digging for Buried Clusters in Nearby Star-forming Galaxies}",
      journal = {\aj},
         year = 2025,
        month = dec,
       volume = {170},
       number = {6},
          eid = {340},
        pages = {340},
          doi = {10.3847/1538-3881/ae10ad},
archivePrefix = {arXiv},
       eprint = {2511.11920},
 primaryClass = {astro-ph.GA},
       adsurl = {https://ui.adsabs.harvard.edu/abs/2025AJ....170..340G}
}

\begin{appendix}

\section{Accuracy of the results}
\label{appendix_accurary}

\begin{figure}[h!]
\includegraphics[width=0.45\textwidth]{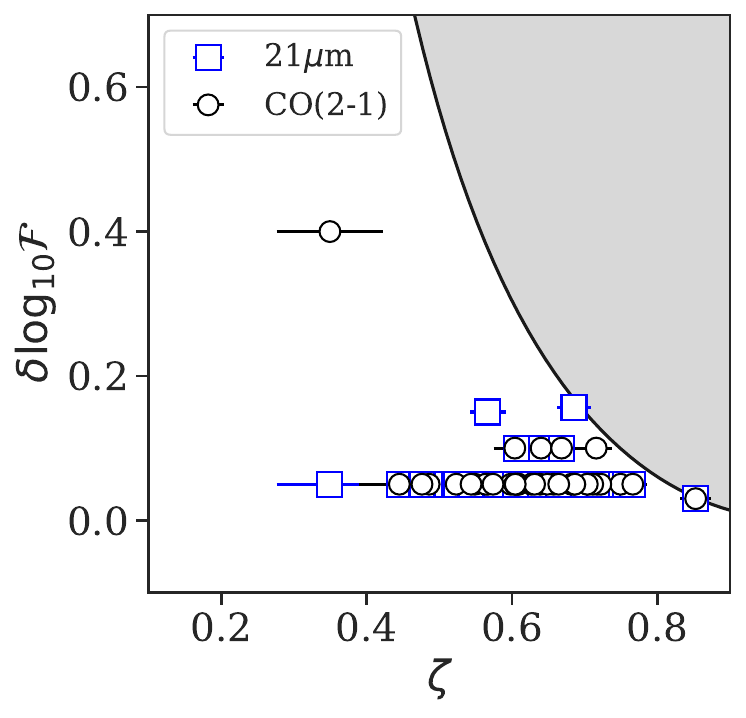}
\includegraphics[width=0.45\textwidth]{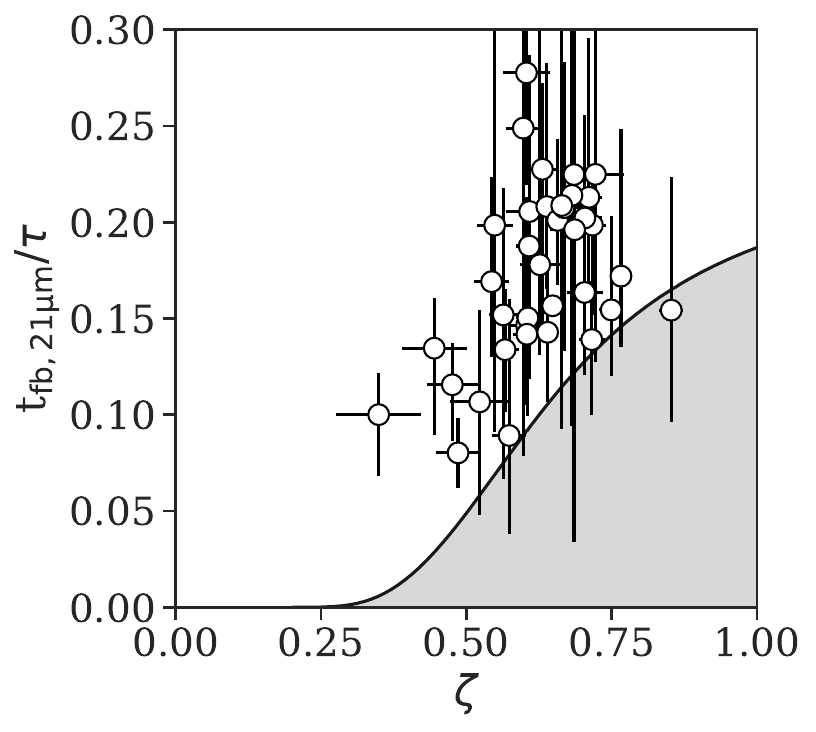}
\caption[]{Accuracy criteria from Heisenberg. Top: Flux contrast ($\delta \log_{10} \mathcal{F}$) adopted during the peak identification for 21~$\mu$m (blue squares) and CO(2-1) (black circles) observations, as a function of the average filling factor ($\zeta$). Bottom: Measured $t_{\rm fb,21~\mu m}/\tau$ as a function of $\zeta$. The gray region indicates the parameter space where high filling factors biase our measurements of the $t_{\rm fb, 21~\mu m}$ from \cite{Kruijssen_tf_2018}.}
\label{check_blending}
\end{figure}

To validate the accuracy of our measurements, we verified that the requirements listed in Sect. 4.4 of \cite{Kruijssen_tf_2018} were fulfilled. The constrained parameters are measured with an accuracy of at least 30\%, upon satisfaction of the following criteria :
 \begin{itemize}
     \item  (i) The duration of $t_{\rm CO}$ and $t_{21~\mu m}$ (resp.\ $t_{\rm CO}$ and $t_{\rm H\alpha}$)  should differ by less than one order of magnitude. This criterion is matched in all of our sample, both for CO versus\ H$\alpha$ and CO versus\ 21~$\mu$m runs. We note that for one galaxy (NGC\,3059), $\log_{10}$($t_{21~\mu m}$/$t_{\rm CO}$) $\sim$ 1.0.
     
     \item (ii) The ratio $\lambda_{\rm fb,21}$/l$_{\rm ap, min}$ (resp.\ $\lambda_{\rm fb,H\alpha}$/l$_{\rm ap, min}$) should be below 1.5. For the H$\alpha$ versus\ CO runs, this criterion is matched for all but two galaxies (NGC\,4303 and NGC\,4548) for which only upper limits are derived. We note the values we derive for $\lambda_{\rm fb,H\alpha}$ may slightly differ from \citetalias{Kim_2022_environmental_dep} due to our updated mask. Hence, we report only an upper limit for NGC\,4303 while a measurement was reported in \citetalias{Kim_2022_environmental_dep}. Conversely, we now provide measurements of $t_{\rm fb,H\alpha}$ for NGC\,1087, NGC\,1385, and NGC\,4540, while only an upper limit was derived in \citetalias{Kim_2022_environmental_dep}. For the runs with 21~$\mu$m, the combination of the high sensitivity of 21~$\mu$m maps tends to increase the number of detected peaks and decrease $\lambda$, while the aperture size is fixed by the worst resolution of all maps, i.e., the resolution of the ALMA/CO(2-1) maps. Hence, 31 out of 37 galaxies do not meet this criterion in the 21~$\mu$m versus CO runs. To enlarge our sample size, we slightly relax the recommended threshold and consider also as measurements galaxies for which $\lambda_{\rm fb,21}$/l$_{\rm ap, min}$ > 1.4, allowing us to include four additional galaxies visually associated with a clear decorrelation between 21~$\mu$m and CO (NGC\,7496, NGC\,0628, NGC\,4731, and NGC\,0685). The 25/37 galaxies with $\lambda_{\rm fb,21}$/l$_{\rm ap, min} \leq 1.4$, are flagged in our analysis (see Figure~\ref{timescales}), as we can only derive an upper limit $t_{\rm fb, 21~\mu m}$, $\lambda_{\rm fb, 21~\mu m}$, and $t_{\rm obs}$ (see discussion in Section \ref{subsec_measuring_timescales}).
     
     \item  (iii) We verify that the number of identified peaks in both CO and H$\alpha$ (respectively in both CO and 21~$\mu$m) emission maps is always above 35. 
     
      \item (iv) The CO-to-SFR flux ratios measured locally, focusing on CO (resp.\ H$\alpha$ or 21~$\mu$m) peaks, is never below (resp.\ above) the galactic average. In the context of this analysis, we used a slightly more stringent criterion than the recommendation from \cite{Kruijssen_tf_2018} by imposing a $\Delta_{\rm min} > 0.05$ (see Figure~\ref{tuning_fork_ha_21_all47}), hence removing galaxies where no decorrelation was obtained between CO and 21~$\mu$m.
     
     \item (v) The global star-formation history of the analyzed regions should not vary by more than 0.2\,dex, during the last evolutionary cycle ($\tau$ = $t_{\rm gas}$ + $t_{\rm star}$ - $t_{\rm fb}$), when averaged over a time period of $t_{\rm gas}$ or $t_{\rm star}$. A relatively stable SFR is expected in disk galaxies following a secular evolution, in particular when masking the central regions. For the subsample for which observations from the PHANGS-MUSE survey are available, we additionally ensure that this condition is verified based on the SFR histories measured in \cite{Pessa_SFH_2023}.
     
     (vi) Each region, independently undergoing evolution from gas to star, should be detectable in both tracers at some point in their life. This implies that sensitivity of the CO and 21~$\mu$m should be matched, allowing the faintest CO peak to evolve into \hii\ regions that are bright enough to be detected in the 21~$\mu$m map. To verify this, we first compute the minimum mass of the young stellar population expected to form within the observed clouds, by multiplying the 5$\sigma$ sensitivity of the CO maps ($\sim$ 10$^5$M$_{\odot}$; \citealt{Leroy_CO_phangs_2021}) by the integrated star-formation efficiencies reported in \citetalias{Kim_2022_environmental_dep}. This results in minimum young stellar population masses ranging from 800~M$_{\odot}$ to 8000~M$_{\odot}$, with a median value of $\sim$3600~M$_{\odot}$. We then compare this value to the mass of a hypothetical young stellar population that emits photons at the 5$\sigma$ sensitivity of the 21~$\mu$m map on the scale of star-forming regions ($\lambda_{21~\mu m}$). We use the \textsc{Starburst99} model \citep{Starburst99_Leitherer_1999} to estimate H$\alpha$ luminosity assuming instantaneous star formation, 5\,Myr ago and a fully sampled initial mass function. The latter luminosity is converted to 21~$\mu$m emission using the conversion factor from \cite{Leroy_2023_midIR_CO_Ha} to estimate the mass, resulting into a minimum  young stellar population mass of $\sim$1500~M$_{\odot}$. As a result, we find that the minimum  young stellar population mass estimated for CO and 21~$\mu$m maps are in good agreement with each other within a factor of two (respectively three) for 13 galaxies (respectively 21) out of the 37 from our sample. For the rest of the sample, the greater sensitivity of the 21~$\mu$m observations compared to the CO ones biases the measured timescales associated with 21~$\mu$m (i.e., $t_{\rm 21~\mu m}$, $t_{\rm fb, 21~\mu m}$, and $t_{\rm obscured}$ toward larger values. We note that for most of the galaxies, the reported $t_{\rm fb, 21~\mu m}$ and $t_{\rm obscured}$ are already upper limits (see discussion Section \ref{subsect_limitations}). Hence, this bias does not affect our conclusions regarding the short duration of the embedded feedback phase (see Section \ref{subsec_discussion_tobs}). 
     
     \item (vii) When peaks are densely distributed, potentially overlapping with each other, the density contrast used for identifying peaks ($\delta \log_{10} \mathcal{F}$) should be small enough to identify adjacent peaks. In the top panel of Figure~\ref{check_blending}, we plot the density contrast used in each map of tracers as a function of the mean filling factor for gas and stellar peaks, defined  as $\zeta = 1/(t_{\rm CO}+t_{\rm 21\mu m}) \times (t_{\rm CO} \times \zeta_{\rm CO} +t_{\rm 21\mu m} \times \zeta_{\rm 21\mu m})$, where $\zeta_{\rm CO}$= 2r$_{\rm CO}$/$\lambda_{\rm fb, 21\mu m}$ and $\zeta_{\rm 21\mu m}$= 2r$_{\rm 21\mu m}$/$\lambda_{\rm fb, 21\mu m}$ with r$_{\rm CO}$ (resp. r$_{\rm 21\mu m}$) is the radii of the gas peaks (resp. SFR peaks). We confirm that our adopted values are small enough for all galaxies, compared to the upper limit identifying regions affected by blending from \cite{Kruijssen_tf_2018}. 
     
     \item (viii) In order to check whether we sufficiently resolve independent regions, in the bottom panel of Figure~\ref{check_blending} we compare the analytical prescription of \cite{Kruijssen_tf_2018} with our measurements of $t_{\rm fb, 21~\mu m}$/$\tau$, where $\tau$ is the total duration of the entire evolutionary cycle (\mbox{$\tau$ = $t_{\rm CO}$ + $t_{\rm 21~\mu m}$ - $t_{\rm fb, 21~\mu m}$}). We confirm that the total duration of the evolutionary cycle is large enough for all galaxies compared to the upper limit identifying regions affected by blending from \cite{Kruijssen_tf_2018}, except for NGC\,5068 which lies slightly below the theoretical limit. 

     \item (ix) As shown in the lower panel of Figure~\ref{check_blending}, we ensure that the conditions $t_{\rm fb,21~\mu m}/\tau$ > 0.05 and $t_{\rm fb,21~\mu m}/\tau$ < 0.95 are verified for all galaxies. 
     
     \item (x) Similarly to condition (v), the global SFR of the analyzed region should not vary more than 0.2\,dex during the entire evolutionary lifecycle when averaged over $t_{\rm fb,21~\mu m}$. This is satisfied using the same argument in (v) stated above. 
     
     \item (xi) After masking obviously crowed regions such as the galaxy center, visual inspection does not reveal abundant region blending. 
     
     In conclusion, we find that our measurements are constrained with high accuracy for 12 out of 37 galaxies. For the remaining 25 galaxies, our analysis is mainly limited by the limited spatial resolution of the CO maps (criterion ii) and the difference in sensitivity (criterion v)), leading to the derivation of upper limits only for $\lambda_{\rm fb, 21~\mu m}$, $t_{\rm fb, 21~\mu m} $, and $t_{\rm obscured}$.
 \end{itemize}

\section{Outputs from the Tuning-fork analysis}
\label{appendix_material}

\begin{figure}
\centering
\includegraphics[width=\columnwidth]{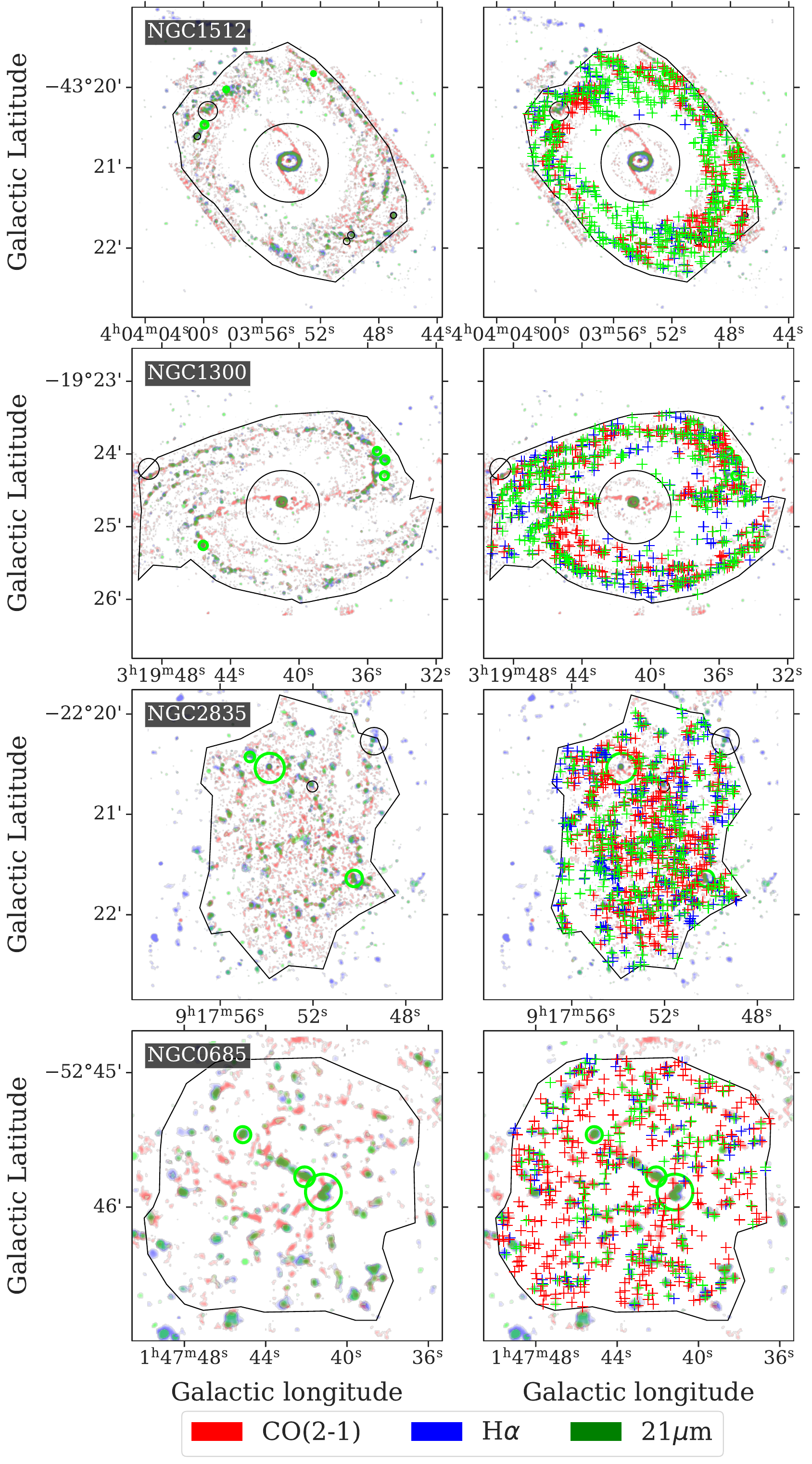}
\caption[]{Composite image of four galaxies for which robust timescale measurements are derived. The galaxies are ordered following their Hubble morphological T-type from the LEDA database, from massive barred spiral galaxies to low-mass irregular galaxies. The left column shows the compact emission maps obtained after filtering the large scale component. On the right column, the superimposed crosses show the peaks of emission selected by \textsc{Clumpfind} \citep{clumpfind_Williams_1994} in each tracer. Isolated 21$\mu$m peaks (green crosses) are identified in the interarm region of barred spiral galaxies, while most 21$\mu$m peaks overlap with CO or H$\alpha$ peaks in flocculent and irregular galaxies. The black masks show regions which are considered in our analysis, excluding galactic centers, bright peaks of emission, and outer regions where CO is undetected.}
\label{peak_id_clumpfind}
\end{figure}

\begin{figure*}[h]
\centering
\includegraphics[width=0.8\textwidth]{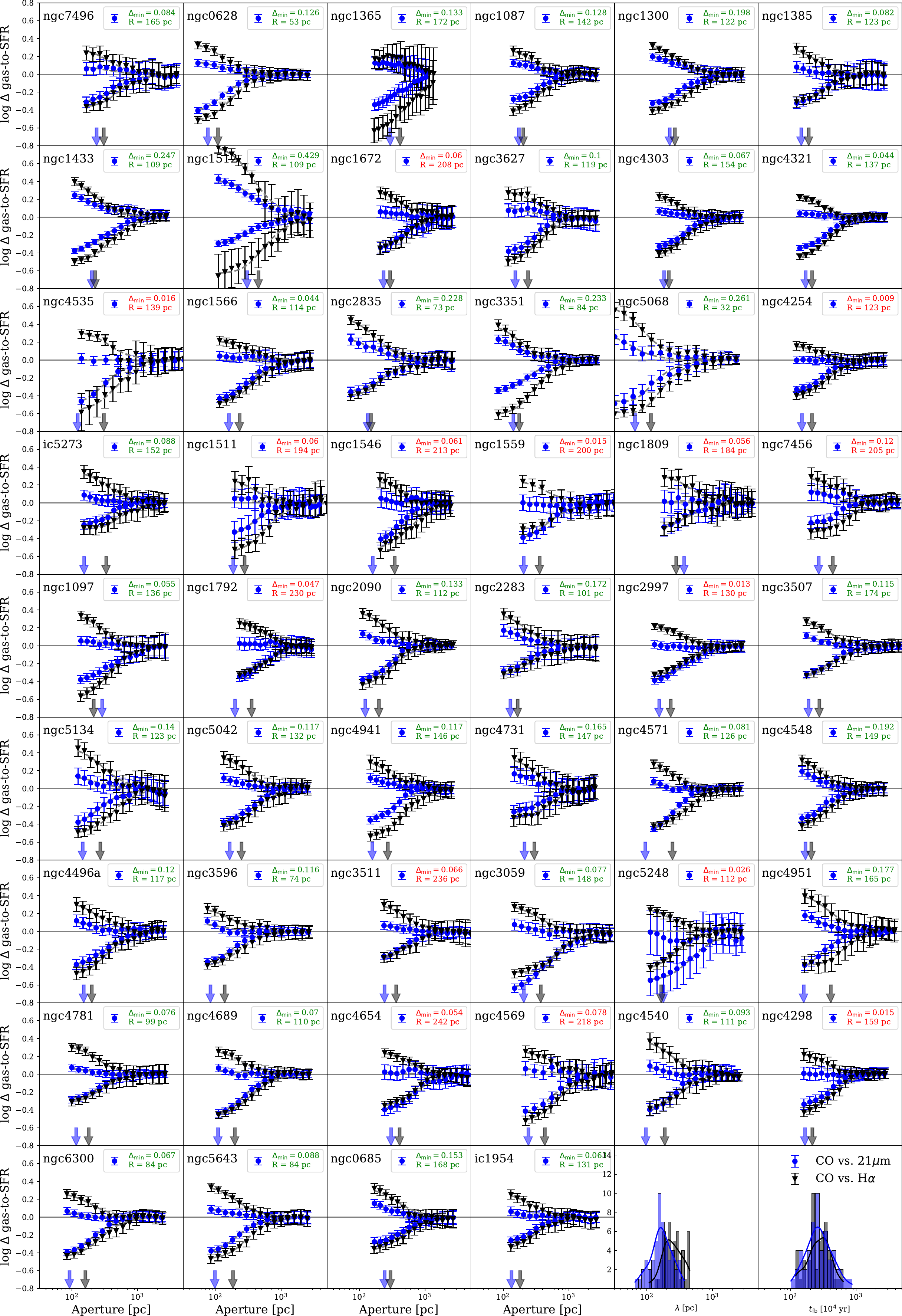}
\caption[]{Deviation of the 21~$\mu$m-to-CO flux ratio (blue) and H$\alpha$-to-CO flux ratio (black) with respect to the galactic average as a function of aperture sizes for each galaxy. The arrows indicate the typical separation length, $\lambda$, at which the two tracers decorrelate. The amplitude of the decorrelation and the separation length are systematically smaller for CO versus 21~$\mu$m runs than for CO versus H$\alpha$ runs, highlighting the greater overlap between CO and 21~$\mu$m. The legends indicate the criteria used to select our final sample: (1) a resolution $R\leq 180$\,pc (2) a clear decorrelation between CO and 21~$\mu$m with $\Delta_{\rm min}$>0.04 dex. 
\label{tuning_fork_ha_21_all47}}
\end{figure*}

In this Section, we provide additional figures illustrating specific aspects of our analysis, which are mentioned throughout the text. In Figure \ref{peak_id_clumpfind}, we illustrate the filtering of the diffuse emission and peak identification (described in Section \ref{subsec_filtering_emission}), that enable us to derive timescale measurements, based on the "tuning-fork" framework, as explained in Section \ref{subsec_measuring_timescales}. In Figure~\ref{tuning_fork_ha_21_all47}, we show the 52 galaxies for which CO, H$\alpha$, and 21$\mu$m maps are available and illustrate our selection criteria, described in Section \ref{section_sample}, leading to a final sample of 37 galaxies. The resulting sample is presented in Table~\ref{table_input_param}, ordered by morphological type, along with the main input parameters used in our analysis. Table~\ref{table_outputs_21um} lists the values of physical quantities derived in our analysis, whose trends are discussed in Section \ref{section_results}. 

\begin{figure}[h]
\includegraphics[width=0.99\textwidth]{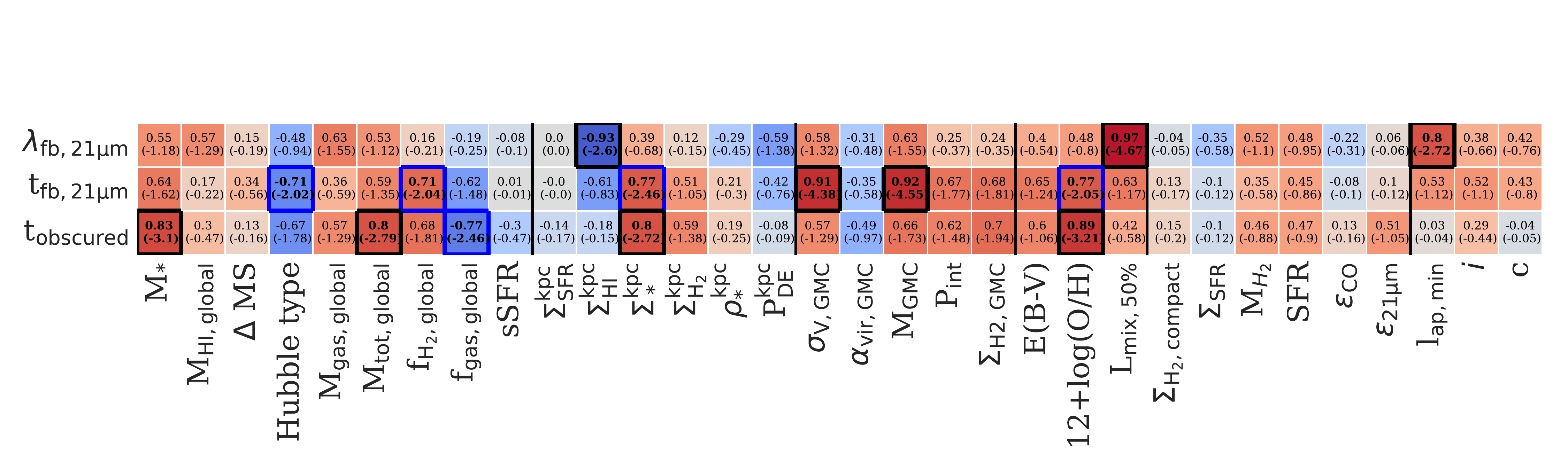}
\caption[]{Spearman’s rank correlation coefficients measured between galaxy properties (columns) and our measurements (rows), adopting a methodology similar to the one described in Figure~\ref{correlation}. The correlations derived for these quantities ($\lambda_{\rm fb, 21~\mu m}$, $t_{\rm fb, 21~\mu m}$, and $t_{\rm obscured}$) are based only on the robust measurements are obtained for 12/37 galaxies.}
\label{correlation_fb}
\end{figure}

For some of these parameters ($t_{\rm fb, 21~\mu m}$, $t_{\rm obscured}$, and $\lambda_{\rm fb, 21~\mu m}$), robust measurements can only be derived in 12/37 galaxies in our sample, while only upper limits are obtained for the others. The correlations obtained for these parameters with various physical quantities are shown in Figure~\ref{correlation_fb}. While these are based on a small number of objects, they are used to guide the discussion in Sections \ref{subsect_discuss_gmc} and \ref{subsect_discuss_morpht}.

\FloatBarrier

\begin{table*}
\centering
\caption[t]{Main input parameters used in the analysis for galaxies ordered by increasing Hubble morphological T-types.}
\label{table_input_param}

\begin{tabular}{|c|c|c|c|c|c|c|c|c|c|c|c|}
\hline
Galaxy & $t_{\rm CO}$\tablefootmark{a} & l$_{\rm ap, min}$\tablefootmark{b} & l$_{\rm ap, max}$\tablefootmark{c} & N$_{\rm pix}$\tablefootmark{d} & $\Delta \rm F_{CO}$\tablefootmark{e} & $\delta \rm F_{CO}$\tablefootmark{f} & $\Delta \rm F_{21}$\tablefootmark{g} & $\delta \rm F_{21}$\tablefootmark{h} & n$_{\lambda}$\tablefootmark{i} & N$_{\rm mask}^{21}$\tablefootmark{j} & $F_{\rm mask}^{21}$\tablefootmark{k} \\
\hline
NGC1512 & 10.1$\pm_{2.3}^{2.7}$ & 110 & 3000 & 15 & 1.2 & 0.05 & 2.0 & 0.05 & 11 & 3 (0.3\%) & 22.3 (12.8\%)\\%
NGC1433 & 17.2$\pm_{2.5}^{2.0}$ & 109 & 3000 & 15 & 1.9 & 0.05 & 2.0 & 0.05 & 12 & 7 (1\%) & 22.8 (17.3\%) \\%
NGC4941 & 24.2$\pm_{4.0}^{5.4}$ & 149 & 3000 & 10 & 2.5 & 0.05 & 1.6 & 0.05 & 12 & 0 & - \\
NGC5134 & 12.4$\pm_{2.4}^{2.9}$ & 123 & 3000 & 10 & 3.4 & 0.05 & 2.0 & 0.05 & 12 & 3 (1.6\%) & 64.8 (40.1\%)\\%
NGC4548 & 13.2$\pm_{2.5}^{5.3}$ & 149 & 3000 & 8 & 2.5 & 0.05 & 2.5 & 0.05 & 14 & 0 & -\\
NGC3507 & 12.1$\pm_{1.6}^{2.5}$ & 176 & 7000 & 7 & 3.0 & 0.1 & 1.8 & 0.05 & 12 & 0& - \\
NGC3351 & 21.0$\pm_{2.7}^{4.6}$ & 84 & 3000 & 15 & 1.8 & 0.05 & 2.0 & 0.05 & 12 & 2 (0.5\%) & 24.8 (11.6\%) \\%
NGC3627 & 13.3$\pm_{1.8}^{2.6}$ & 121 & 3000 & 15 & 3.2 & 0.05 & 2.0 & 0.05 & 13  & 4 (2.8\%) & 371.5 (58.1\%) \\%
NGC6300 & 19.9$\pm_{2.7}^{4.4}$ & 85 & 3000 & 15 & 2.2 & 0.05 & 2.2 & 0.15 & 13 & 2 (0.4\%) & 83.6 (5\%) \\%
NGC7496 & 16.8$\pm_{2.9}^{9.8}$ & 168 & 5000 & 15 & 3.0 & 0.05 & 2.5 & 0.05 & 13 & 1 (0.8\%) & 37.1  (20.2\%)\\%
NGC1365 & 23.8$\pm_{6.1}^{95.2}$ & 173 & 1500 & 20 & 2.7 & 0.05 & 2.0 & 0.05 & 13 & 4 (2.3\%) & 16.3 (61.4\%)  \\%
NGC1097 & 16.1$\pm_{2.9}^{3.2}$ & 137 & 3000 & 7 & 2.1 & 0.05 & 2.1 & 0.05 & 14 & 3  (1.1\%) & 45.7 (16.6\%)\\ %
IC1954 & 12.5$\pm_{2.2}^{3.4}$ & 131 & 3000 & 7 & 2.5 & 0.05 & 2.0 & 0.05 & 14 & 2 (1.4\%) & 46.2 (6.7\%)\\%
NGC4303 & 19.1$\pm_{2.9}^{3.0}$ & 155 & 3000 & 15 & 1.6 & 0.05 & 2.5 & 0.05 & 13 & 7 (2.4\%) & 155.2 (31.6\%)\\%
NGC1566 & 21.9$\pm_{3.1}^{3.3}$ & 115 & 3000 & 15 & 2.5 & 0.05 & 2.5 & 0.05 & 12 & 4  (1.4\%) & 272.5 (34.4\%)\\%
NGC3059 & 32.6$\pm_{5.5}^{9.2}$ & 149 & 5000 & 20 & 3.5 & 0.05 & 2.0 & 0.05 & 10 & 0 & -\\
NGC1300 & 12.3$\pm_{1.1}^{1.8}$ & 122 & 3000 & 15 & 1.5 & 0.05 & 2.0 & 0.05 & 11 & 4 (2.3\%) & 41.3 (28.7\%) \\%
NGC4321 & 18.9$\pm_{2.2}^{3.3}$ & 138 & 3000 & 15 & 1.6 & 0.05 & 2.0 & 0.05 & 12 & 3 (0.5\%) & 116.2 (8.7\%) \\%
NGC2090 & 10.3$\pm_{1.8}^{2.0}$ & 112 & 3000 & 10 & 2.5 & 0.05 & 2.0 & 0.05 & 12 & 2 (1.1\%) & 22.7 (9.5\%)\\%
NGC4689 & 20.5$\pm_{3.0}^{3.9}$ & 110 & 3000 & 15 & 3.0 & 0.05 & 2.5 & 0.05 & 13 & 0 & -\\
NGC5042 & 12.8$\pm_{2.1}^{3.0}$ & 133 & 3000 & 10 & 2.5 & 0.05 & 2.5 & 0.1 & 12 & 0 & -\\
NGC5643 & 19.0$\pm_{3.3}^{3.9}$ & 85 & 3000 & 15 & 2.5 & 0.1 & 2.5 & 0.1 & 11 & 0 & -\\
NGC2835 & 8.0$\pm_{1.2}^{1.4}$ & 74 & 3000 & 15 & 1.2 & 0.05 & 2.0 & 0.05 & 12 & 3 (1.1\%) & 134.5 (17.5\%)\\%
NGC0628 & 15.6$\pm_{1.9}^{2.3}$ & 53 & 3000 & 15 & 1.1 & 0.05 & 2.0 & 0.05 & 13 & 7 (1.4\%) & 71.6 (46.1\%)\\%
NGC1087 & 15.6$\pm_{2.8}^{3.6}$ & 143 & 3000 & 15 & 1.5 & 0.1 & 2.5 & 0.1 & 13 & 2 (1.5\%) & 45.6 (43.2\%) \\
NGC3596 & 17.5$\pm_{2.3}^{3.5}$ & 74 & 3000 & 4 & 3.5 & 0.1 & 2.0 & 0.05 & 14 & 0 & -\\
NGC0685 & 12.9$\pm_{2.4}^{4.4}$ & 169 & 3000 & 10 & 3.0 & 0.05 & 2.0 & 0.05 & 13 & 3 (1.6\%) & 10.5 (15.4\%) \\
IC5273 & 12.7$\pm_{2.8}^{3.8}$ & 153 & 3000 & 6 & 2.5 & 0.05 & 2.0 & 0.05 & 12 & 2 (1.5\%) & 91.0 (43.5\%) \\
NGC4731 & 12.6$\pm_{3.2}^{28.7}$ & 148 & 3000 & 5 & 2.5 & 0.05 & 2.5 & 0.156 & 13 & 2 (2.7\%) & 21.8 (50.2\%) \\
NGC1385 & 13.3$\pm_{2.7}^{6.7}$ & 124 & 3000 & 10 & 1.7 & 0.05 & 3.0 & 0.05 & 14 & 2 (1.1\%) & 230.0\ (31.5\%) \\
NGC2283 & 9.2$\pm_{2.1}^{2.4}$ & 102 & 3000 & 15 & 1.4 & 0.05 & 2.5 & 0.05 & 13 & 0 & -\\
NGC5068 & 11.7$\pm_{1.4}^{2.6}$ & 32 & 3000 & 15 & 1.6 & 0.03 & 2.0 & 0.03 & 10 & 0 & -\\
NGC4951 & 11.7$\pm_{3.0}^{2.9}$ & 167 & 5000 & 10 & 2.4 & 0.05 & 2.0 & 0.05 & 12 & 2 (1.5\%) & 15.5 (9.7\%) \\
NGC4540 & 12.9$\pm_{3.3}^{4.7}$ & 111 & 3000 & 15 & 3.0 & 0.05 & 3.0 & 0.05 & 14 & 2 (1.5\%) & 39.5 (17.6\%) \\
NGC4571 & 18.4$\pm_{2.8}^{4.8}$ & 127 & 3000 & 7 & 3.0 & 0.4 & 2.0 & 0.05 & 12 & 0 & -\\
NGC4781 & 7.8$\pm_{1.1}^{1.6}$ & 99 & 3000 & 15 & 2.5 & 0.05 & 2.3 & 0.05 & 12 & 2 (1.0\%) & 49.4 (14.4\%) \\
NGC4496A & 15.8$\pm_{3.0}^{3.1}$ & 117 & 3000 & 10 & 3.0 & 0.05 & 2.5 & 0.05 & 12 & 3 (1.5\%) & 42.5 (10.0\%)\\
\hline
\end{tabular}
\tablefoot{
\tablefoottext{a}{Total cloud lifetime in Myr (reference timescale in the current study), recalculated with updated masks with respect to \citetalias{Kim_2022_environmental_dep}.}\\
\tablefoottext{b}{Minimal aperture size (in  parsecs ).}\\
\tablefoottext{c}{Maximal aperture size (in  parsecs ).}\\
\tablefoottext{d}{Minimal number of pixels to used in the peak identification with \textsc{Clumpfind}.}\\
\tablefoottext{e}{Logarithmic range below flux maximum covered by flux contour levels for molecular gas peak identification in CO maps.}\\
\tablefoottext{f}{Logarithmic interval between flux contour levels for molecular gas peak identification in CO maps.}\\
\tablefoottext{g}{Logarithmic range below flux maximum covered by flux contour levels for SFR peak identification in $21~\mu m$ maps.}\\
\tablefoottext{h}{Logarithmic interval between flux contour levels for SFR peak identification in $21~\mu m$ maps.}\\
\tablefoottext{i}{Softening parameters used to filter diffuse emission in both CO and 21~$\mu$m maps.}\\
\tablefoottext{j}{Total number of bright 21~$\mu$m peaks masked in our analysis (and fraction with respect to the total number of 21$\mu$m peaks).}\\
\tablefoottext{h}{Surface brightness threshold (in MJy/sr) used to mask bright peaks (and fraction of the total 21$\mu$m compact flux they represent).}\\ 
Other parameters which are not listed here are set to the values adopted in \citetalias{Kim_2022_environmental_dep}\ for the variables adjusted individually for each galaxy, and \cite{Kruijssen_tf_2018}\ for the default quantities chosen to run \textsc{Heisenberg}.}
\end{table*}

\begin{table*}
\centering
\caption[]{Physical quantities associated with 21~$\mu$m describing the evolution timeline of molecular clouds.}
\label{table_outputs_21um}
\begin{tabular}{|c|c|c|c|c|c|c|c|c|}
\hline
Galaxy & Hubble T-type\tablefootmark{a} & $t_{21~\mu m}$\tablefootmark{b} & $t_{\rm fb, 21~\mu m}$\tablefootmark{c} & $t_{\rm obscured}$\tablefootmark{d} & $\lambda_{\rm fb, 21~\mu m}$\tablefootmark{e} & f$_{\rm diff, 21~\mu m}$\tablefootmark{f} & $\mathcal{E}_{21~\mu m}$\tablefootmark{g} \\
\hline
NGC1512 & 1.2$\pm$0.25 & 11.9$\pm_{2.01}^{1.59}$ & 3.64$\pm_{1.02}^{0.85}$ & 2.29$\pm_{1.49}^{1.29}$ & 302.33$\pm_{34.13}^{37.64}$ & 0.58$\pm_{0.02}^{0.02}$ & 2.7$\pm_{0.81}^{0.7}$ \\
NGC1433 & 1.5$\pm$0.35 & 8.42$\pm_{1.16}^{1.49}$ & 3.42$\pm_{1.0}^{1.27}$ & 1.33$\pm_{1.49}^{1.72}$ & 201.2$\pm_{26.66}^{41.78}$ & 0.66$\pm_{0.01}^{0.01}$ & 1.66$\pm_{0.38}^{0.97}$ \\
NGC4941 & 2.1$\pm$0.3 & 6.84$\pm_{1.53}^{1.62}$ & $\leq$6.46 & $\leq$4.64 & $\leq$188.38 & 0.73$\pm_{0.03}^{0.03}$ & 1.06$\pm_{0.25}^{0.22}$ \\
NGC5134 & 2.9$\pm$0.5 & 3.75$\pm_{1.08}^{1.97}$ & $\leq$4.96 & $\leq$3.84 & $\leq$271.38 & 0.64$\pm_{0.05}^{0.05}$ & 1.11$\pm_{0.27}^{1.38}$ \\
NGC3627 & 3.1$\pm$0.2 & 3.34$\pm_{0.77}^{2.52}$ & $\leq$4.74 & $\leq$3.05 & $\leq$376.56 & 0.71$\pm_{0.03}^{0.03}$ & 1.12$\pm_{0.17}^{1.77}$ \\
NGC3507 & 3.1$\pm$0.2 & 2.79$\pm_{0.46}^{0.77}$ & $\leq$2.84 & $\leq$1.54 & $\leq$240.29 & 0.65$\pm_{0.05}^{0.05}$ & 1.05$\pm_{0.25}^{0.52}$ \\
NGC3351 & 3.1$\pm$0.2 & 10.07$\pm_{1.58}^{1.89}$ & 5.23$\pm_{1.49}^{1.63}$ & 3.11$\pm_{2.05}^{2.56}$ & 143.3$\pm_{20.07}^{27.16}$ & 0.67$\pm_{0.02}^{0.02}$ & 1.46$\pm_{0.29}^{0.6}$ \\
NGC4548 & 3.1$\pm$0.25 & 4.89$\pm_{1.33}^{1.74}$ & $\leq$3.23 & $\leq$3.19 & $\leq$208.1 & 0.64$\pm_{0.05}^{0.05}$ & 1.08$\pm_{0.19}^{0.18}$ \\
NGC6300 & 3.1$\pm$0.3 & 3.38$\pm_{0.65}^{0.66}$ & $\leq$3.61 & $\leq$1.27 & $\leq$111.96 & 0.66$\pm_{0.04}^{0.04}$ & 1.06$\pm_{0.19}^{0.25}$ \\
NGC7496 & 3.2$\pm$0.3 & 5.9$\pm_{2.27}^{4.04}$ & 4.16$\pm_{2.61}^{3.94}$ & 0.37$\pm_{3.77}^{7.28}$ & 237.45$\pm_{72.99}^{361.49}$ & 0.62$\pm_{0.08}^{0.08}$ & 1.36$\pm_{0.65}^{4.68}$ \\
NGC1365 & 3.2$\pm$0.35 & 16.91$\pm_{7.98}^{6.43}$ & 8.65$\pm_{6.76}^{6.07}$ & 3.57$\pm_{8.21}^{12.3}$ & 295.8$\pm_{82.68}^{282.53}$ & 0.74$\pm_{0.05}^{0.05}$ & 1.44$\pm_{0.54}^{2.92}$ \\
NGC1097 & 3.3$\pm$0.25 & 5.69$\pm_{1.42}^{2.97}$ & 4.74$\pm_{1.62}^{2.69}$ & 3.48$\pm_{2.28}^{3.37}$ & 288.95$\pm_{71.62}^{184.19}$ & 0.62$\pm_{0.04}^{0.04}$ & 1.85$\pm_{0.66}^{2.16}$ \\
IC1954 & 3.3$\pm$0.45 & 3.61$\pm_{0.96}^{1.14}$ & $\leq$4.14 & $\leq$2.94 & $\leq$180.45 & 0.58$\pm_{0.08}^{0.08}$ & 1.01$\pm_{0.32}^{0.45}$ \\
NGC4303 & 4.0$\pm$0.05 & 5.54$\pm_{1.0}^{1.38}$ & $\leq$6.04 & $\leq$3.7 & $\leq$246.38 & 0.65$\pm_{0.04}^{0.04}$ & 1.11$\pm_{0.22}^{0.55}$ \\
NGC1566 & 4.0$\pm$0.1 & 4.34$\pm_{0.84}^{3.33}$ & $\leq$6.46 & $\leq$2.93 & $\leq$272.22 & 0.59$\pm_{0.04}^{0.04}$ & 1.21$\pm_{0.21}^{1.32}$ \\
NGC1300 & 4.0$\pm$0.1 & 6.72$\pm_{0.88}^{1.52}$ & 3.34$\pm_{0.95}^{1.43}$ & 0.85$\pm_{1.37}^{2.02}$ & 222.9$\pm_{34.93}^{70.51}$ & 0.64$\pm_{0.02}^{0.02}$ & 1.6$\pm_{0.38}^{1.24}$ \\
NGC3059 & 4.0$\pm$0.1 & 3.15$\pm_{0.6}^{0.68}$ & $\leq$3.49 & $\leq$-0.49 & $\leq$258.53 & 0.72$\pm_{0.02}^{0.02}$ & 1.33$\pm_{0.34}^{0.75}$ \\
NGC4321 & 4.0$\pm$0.15 & 4.98$\pm_{0.8}^{0.86}$ & $\leq$5.09 & $\leq$3.06 & $\leq$200.78 & 0.66$\pm_{0.03}^{0.03}$ & 1.13$\pm_{0.17}^{0.36}$ \\
NGC2090 & 4.5$\pm$0.9 & 2.49$\pm_{0.5}^{0.43}$ & $\leq$1.95 & $\leq$1.05 & $\leq$148.24 & 0.71$\pm_{0.03}^{0.03}$ & 1.05$\pm_{0.19}^{0.22}$ \\
NGC4689 & 4.7$\pm$0.45 & 2.87$\pm_{0.56}^{0.52}$ & $\leq$3.11 & $\leq$0.66 & $\leq$131.55 & 0.6$\pm_{0.04}^{0.04}$ & 1.0$\pm_{0.21}^{0.2}$ \\
NGC5643 & 5.0$\pm$0.15 & 3.5$\pm_{0.64}^{0.71}$ & $\leq$3.81 & $\leq$1.5 & $\leq$117.74 & 0.61$\pm_{0.06}^{0.06}$ & 1.08$\pm_{0.22}^{0.36}$ \\
NGC5042 & 5.0$\pm$0.2 & 3.29$\pm_{0.76}^{0.66}$ & $\leq$2.95 & $\leq$1.82 & $\leq$210.99 & 0.56$\pm_{0.06}^{0.06}$ & 1.13$\pm_{0.23}^{0.52}$ \\
NGC2835 & 5.0$\pm$0.2 & 2.93$\pm_{0.52}^{1.03}$ & 1.53$\pm_{0.56}^{0.9}$ & 0.26$\pm_{0.98}^{1.39}$ & 134.05$\pm_{26.66}^{76.09}$ & 0.5$\pm_{0.03}^{0.03}$ & 1.55$\pm_{0.4}^{1.89}$ \\
NGC0628 & 5.2$\pm$0.25 & 4.9$\pm_{0.82}^{1.43}$ & 3.01$\pm_{0.93}^{1.5}$ & 0.75$\pm_{1.58}^{2.11}$ & 75.98$\pm_{12.45}^{24.89}$ & 0.58$\pm_{0.02}^{0.02}$ & 1.28$\pm_{0.24}^{0.63}$ \\
NGC3596 & 5.2$\pm$0.25 & 3.91$\pm_{0.76}^{0.66}$ & $\leq$3.44 & $\leq$1.38 & $\leq$100.47 & 0.6$\pm_{0.05}^{0.05}$ & 1.07$\pm_{0.22}^{0.36}$ \\
NGC1087 & 5.2$\pm$0.4 & 6.08$\pm_{1.3}^{1.33}$ & $\leq$5.42 & $\leq$4.02 & $\leq$223.61 & 0.67$\pm_{0.03}^{0.03}$ & 1.12$\pm_{0.27}^{0.8}$ \\
NGC0685 & 5.4$\pm$0.3 & 4.73$\pm_{1.03}^{1.02}$ & 3.27$\pm_{1.43}^{0.95}$ & 0.07$\pm_{2.46}^{2.6}$ & 241.91$\pm_{56.0}^{76.17}$ & 0.57$\pm_{0.05}^{0.05}$ & 1.23$\pm_{0.29}^{0.64}$ \\
IC5273 & 5.6$\pm$0.5 & 3.92$\pm_{1.29}^{1.38}$ & $\leq$4.64 & $\leq$2.36 & $\leq$225.9 & 0.62$\pm_{0.07}^{0.07}$ & 1.0$\pm_{0.37}^{0.63}$ \\
NGC2283 & 5.9$\pm$0.2 & 4.04$\pm_{1.03}^{1.47}$ & $\leq$3.74 & $\leq$2.8 & $\leq$206.17 & 0.48$\pm_{0.09}^{0.09}$ & 1.14$\pm_{0.28}^{1.56}$ \\
NGC4731 & 5.9$\pm$0.25 & 5.38$\pm_{2.92}^{6.74}$ & 3.3$\pm_{2.63}^{1.72}$ & -0.25$\pm_{4.1}^{13.05}$ & 212.05$\pm_{83.24}^{185.02}$ & 0.56$\pm_{0.13}^{0.13}$ & 1.27$\pm_{0.49}^{3.87}$ \\
1111NGC1385 & 5.9$\pm$0.25 & 3.78$\pm_{1.45}^{1.49}$ & $\leq$4.68 & $\leq$4.15 & $\leq$230.8 & 0.59$\pm_{0.11}^{0.11}$ & 1.11$\pm_{0.43}^{2.37}$ \\
NGC4951 & 6.0$\pm$0.15 & 3.88$\pm_{1.06}^{1.56}$ & $\leq$2.33 & $\leq$0.04 & $\leq$204.49 & 0.74$\pm_{0.03}^{0.03}$ & 1.0$\pm_{0.18}^{0.29}$ \\
NGC5068 & 6.0$\pm$0.2 & 2.7$\pm_{0.55}^{0.93}$ & 1.93$\pm_{0.82}^{0.93}$ & 0.72$\pm_{1.1}^{1.28}$ & 65.86$\pm_{16.79}^{29.17}$ & 0.61$\pm_{0.02}^{0.02}$ & 1.94$\pm_{0.71}^{2.91}$ \\
NGC4540 & 6.2$\pm$0.45 & 1.9$\pm_{0.64}^{0.92}$ & $\leq$2.2 & $\leq$1.8 & $\leq$126.28 & 0.67$\pm_{0.07}^{0.07}$ & 1.0$\pm_{0.25}^{0.19}$ \\
NGC4571 & 6.4$\pm$0.4 & 2.55$\pm_{0.62}^{0.57}$ & $\leq$2.57 & $\leq$-0.01 & $\leq$108.54 & 0.67$\pm_{0.04}^{0.04}$ & 1.0$\pm_{0.4}^{0.07}$ \\
NGC4781 & 7.0$\pm$0.15 & 1.94$\pm_{0.4}^{0.46}$ & $\leq$1.98 & $\leq$0.84 & $\leq$159.88 & 0.62$\pm_{0.07}^{0.07}$ & 1.09$\pm_{0.24}^{0.76}$ \\
NGC4496A & 7.4$\pm$0.85 & 3.63$\pm_{0.82}^{1.24}$ & $\leq$3.83 & $\leq$2.36 & $\leq$208.88 & 0.53$\pm_{0.04}^{0.04}$ & 1.16$\pm_{0.24}^{0.79}$ \\
\hline
Median & - & 3.91$\pm_{0.88}^{1.38}$ & 3.38$\pm_{1.23}^{1.47}$ & 0.80$\pm_{0.80}^{1.47}$ & 217.48$\pm_{34.53}^{73.30}$ & 0.62$\pm_{0.04}^{0.04}$ & 1.12$\pm_{0.27}^{0.64}$ \\
\hline
\end{tabular}
\tablefoot{\tablefoottext{a}{Hubble morphological T-types taken from the HyperLEDA database \citep{Paturel_hyperleda1_2003, Paturel_hyperleda2_2003, Makarov_HyperLEDA_2014}}.\\
\tablefoottext{b}{Total duration of 21~$\mu$m emission (in megayears).}\\ \tablefoottext{c}{Feedback timescale associated with 21~$\mu$m (in megayears).}\\ \tablefoottext{d}{Duration of the obscured star-formation phase (in megayears, defined in Eq. \ref{t_obs}).} \\ \tablefoottext{e}{Characteristic region separation length (in  parsecs ).}\\
\tablefoottext{f}{Diffuse emission fraction of 21~$\mu$m.}\\
\tablefoottext{g}{Peak-to-galactic average flux constrast.} \\
For some galaxies, only a 1$\sigma$ upper limit can be obtained on $t_{\rm fb, 21~\mu m}$ and $\lambda_{\rm fb, 21~\mu m}$ because the independent star-forming regions are not sufficiently resolved (see Appendix \ref{appendix_accurary}).}
\end{table*}

\pagebreak
\onecolumn

\section{Updates compare to \citetalias{Kim_2022_environmental_dep} analysis}
\label{appendix_updates}

\begin{figure*}
\centering
\includegraphics[width=0.35\textwidth]{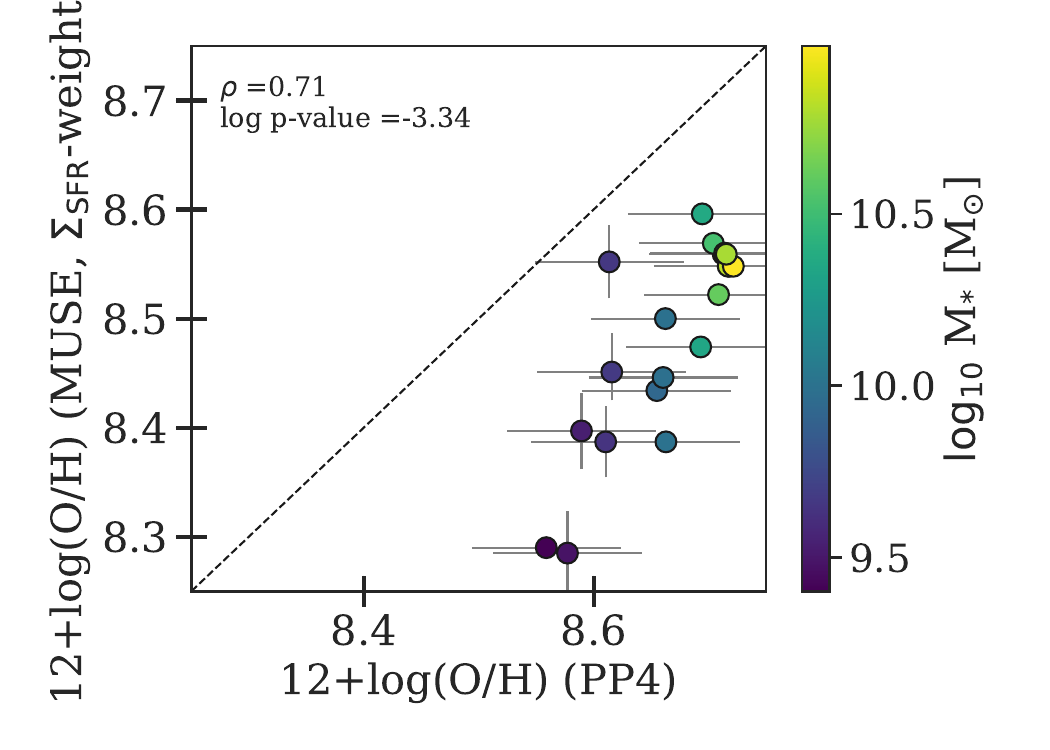}
\includegraphics[width=0.35\textwidth]{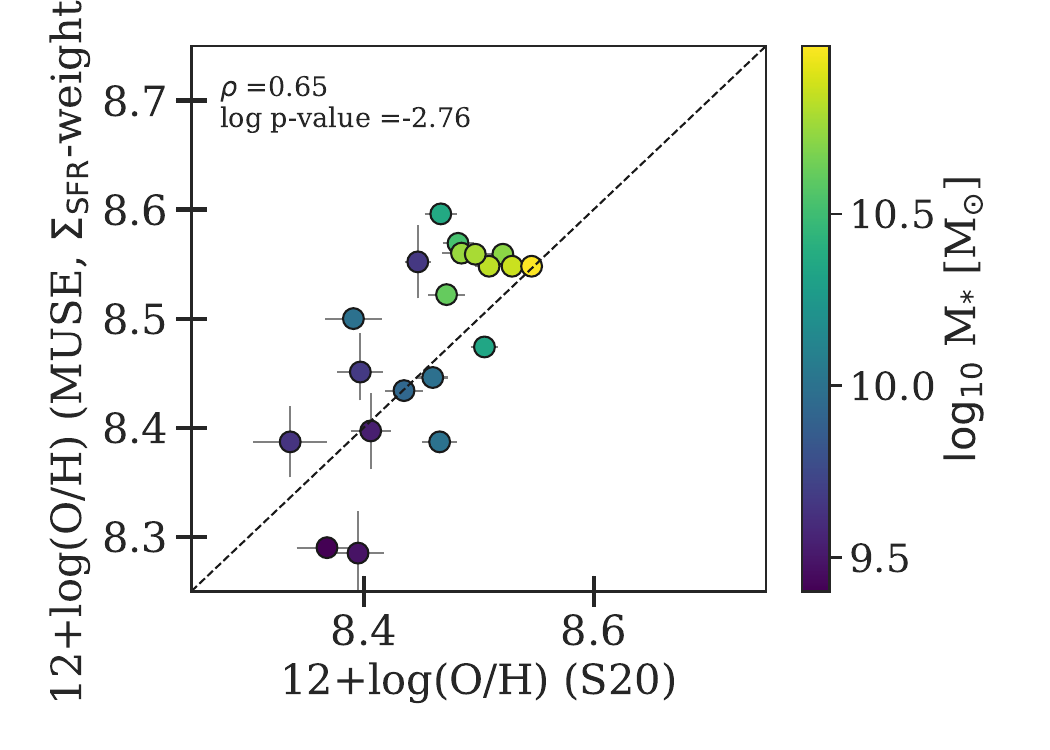}
\includegraphics[width=0.35\textwidth]{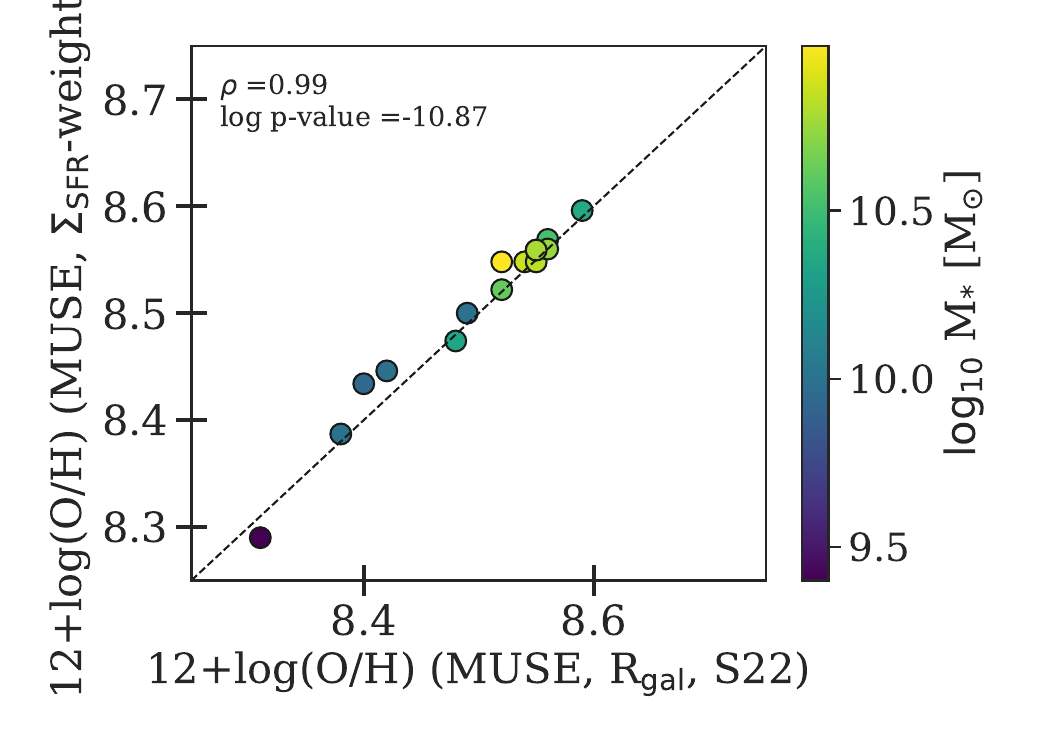}
\includegraphics[width=0.35\textwidth]{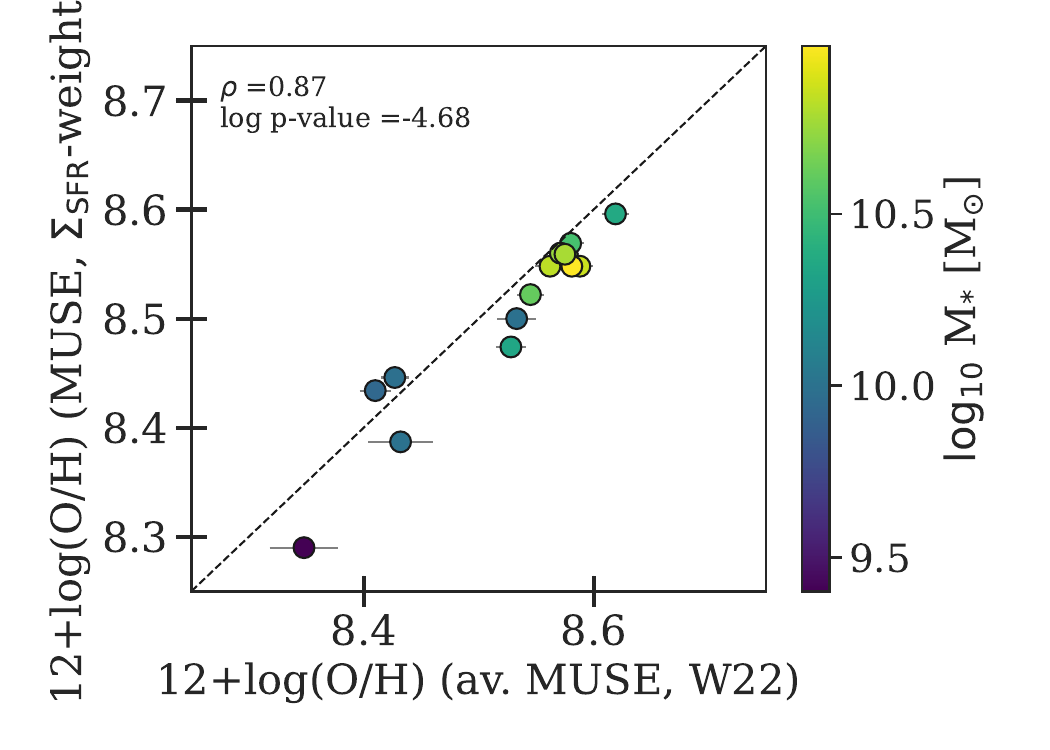}
\caption[]{Comparison of the $\Sigma_{\rm SFR}$-weighted metallicity based on metallicity estimates obtained with MUSE data compared with (i) the integrated metallicity estimated from a direct conversion of the stellar mass based on the MZ relation from \cite{Sanchez_2019_metallicity_grad} at the effective radius R$_{\rm eff}$; (ii) the metallicity estimates from \cite{Sanchez_2019_metallicity_grad} at R$_{\rm eff}$, corrected assuming a fixed radial gradient within the galaxy (-0.1\,dex/,R$_{\rm eff}$; \citealt{sanchez_gradient_2014}); (iii) the metallicity estimated using the S-calibration \citep{Pilyugin_Grebel_scal_2016} at the median galactocentric radius of MUSE observations in \cite{Santoro_2022_PHANGS_LF}; and (iv) 
metallicity estimates based on the maps from \cite{Williams_2022_mixing_met}. To ensure a meaningful comparison, all the estimates are averaged within the masks used in our analysis, excluding galactic centers. The color bar shows the stellar masses in log values.}
\label{metallicity_checks}
\end{figure*}

\begin{figure*}
\centering
\includegraphics[width=0.45\textwidth]{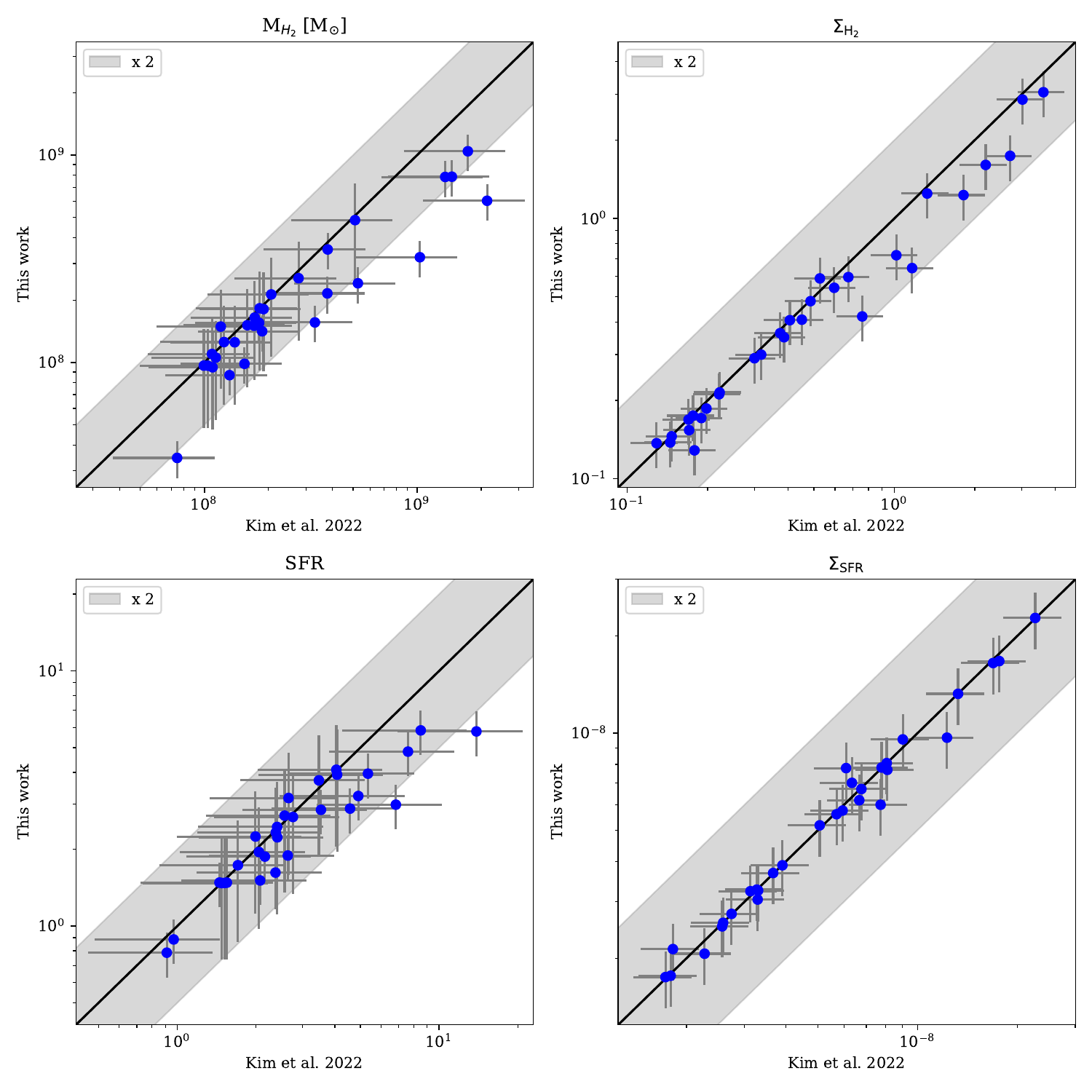}
\includegraphics[width=0.45\textwidth]{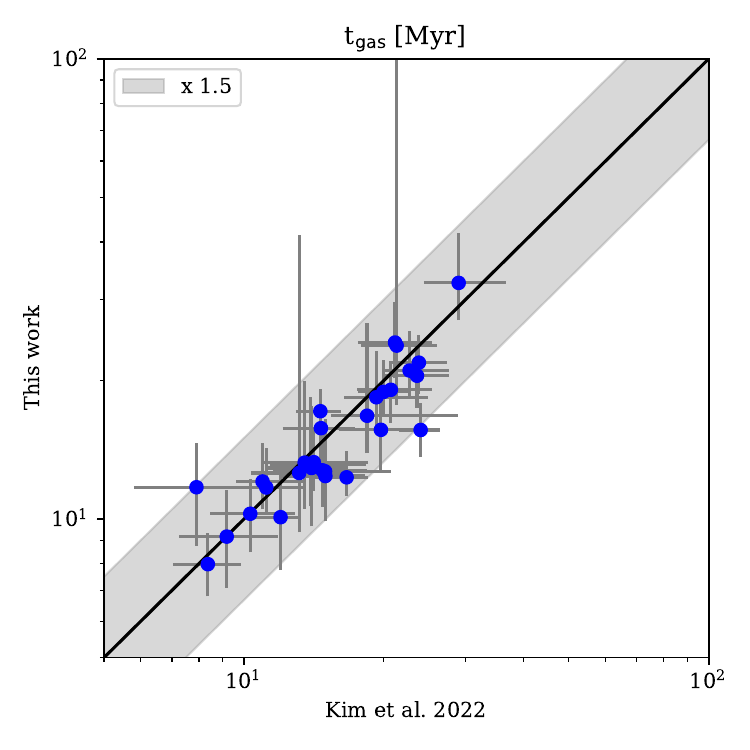}

\caption[]{Comparison between the main physical quantities derived in \citetalias{Kim_2022_environmental_dep} and the ones derived in this study when applying \textsc{Heisenberg} to the same CO and H$\alpha$ maps, within masks that correspond to the MIRI/21~$\mu$m observations, and adopting an updated CO-to-H$_2$ conversion factor (see Section \ref{subsect_data_homogenization}). \textbf{Left:} Comparison of the predicted molecular gas masses (upper left panel), H$_2$ gas surface density (upper-right panel), SFR (bottom left panel), and SFR surface densities (bottom-right panel). Deviations remain moderate between both studies: most of our sample lie on the one-one relation while the few galaxies that deviate from previous measurements remain in good agreement within a factor of 2. \textbf{Right: } Comparison between the cloud lifetimes ($t_{\rm gas}$ = $t_{\rm CO}$) derived in \citetalias{Kim_2022_environmental_dep} and the ones derived in this study when applying \textsc{Heisenberg} to the same CO and H$\alpha$ maps, within masks that correspond to the MIRI/21~$\mu$m observations. Both measurements are in agreement within a factor of 1.5.}
\label{kim_vs_lise_sfr}
\end{figure*}

One major update of our study is to account for the metallicity information available with the MUSE maps. In Figure~\ref{metallicity_checks}, we provide a comparison of the adopted $\Sigma_{\rm SFR}$-weighted metallicity average with other metallicity estimates that were used in previous studies. Finally, in Figure~\ref{kim_vs_lise_sfr} we provide a direct comparison of key parameters previously derived in \citetalias{Kim_2022_environmental_dep}, that we updated using new physical prescriptions and different masks. The modifications of the masked regions, as well as the new prescriptions adopted for metallicity and the CO-to-H$_2$ conversion factors (see Section  \ref{subsect_other_observables}) involve variations of the predicted parameters compared to the previous analysis from \citetalias{Kim_2022_environmental_dep}. 
Nevertheless, the latter variations remain moderate for the predicted SFR, molecular gas masses (M$_{\rm H_2}$) and the associated surface density ($\Sigma_{\rm SFR}$ and $\Sigma_{\rm H_2}$). In particular, while the predicted masses and SFR, which are sensitive to the covered area, tend to be smaller in the current study, we checked that the H$_2$ and SFR surface densities do not vary more than a factor of two compared to the predictions from \citetalias{Kim_2022_environmental_dep}. Similarly, we also checked that all the quantities predicted by our method remain in good agreement with the values obtained in \citetalias{Kim_2022_environmental_dep} within a factor of two to three. Most importantly, we show in Figure~\ref{kim_vs_lise_sfr} that the total cloud lifetime, used as the reference timescale in our study, remains in excellent agreement between both studies, varying by a factor of 1.5 at maximum. 
\FloatBarrier
\pagebreak

\FloatBarrier

\newpage
\onecolumn
\section{Impact of masking extremely bright 21$\mu$m peaks}
\label{appendix_mask_peaks}

The statistical method used in the current study provides flux-weighted measurements, averaged on the population of peaks detected in gas and stellar maps. In order to derive statistically meaningful averages, exceedingly bright peaks, identified as outliers in the peak luminosity function, need to be excluded. They represent a small fraction of the total number of detected peaks (median $\sim$ 1.4\%, and $< 3$\% in all galaxies), but contribute disproportionately to the emission (5-58\%, with a median of 19\%, see Table \ref{table_input_param}). 

In previous studies, the inclusion of such bright peaks has been found to significantly bias the measurements, leading to enlarged error bars on the derived quantities \citep[e.g.,][]{Ward_lmc_2022} or to overestimation of the timescale associated with the phase in which these bright peaks are captured \citep{Chevance_GMC_spiral_2020}. In other more favorable cases, the contribution from bright peaks may average out, and lead to measurements that remain compatible within the errors, regardless of the masking scheme considered \citep[e.g.,][]{Kim_pah_2025}. In Table \ref{table_outputs_21um_mask}, we assess the effect of masking the brightest peaks in 21$\mu$m for the subsample of galaxies for which robust timescales measurements could be derived, both including and excluding bright peaks.

For 8/26 galaxies, including bright peaks, most of the time captured in a phase emitting both CO and 21$\mu$m simultaneously, flattens the tuning-fork upper branch, rendering robust measurements impossible. This effect is specific to the tracers considered in the current study, since the correlation between CO-and-21$\mu$m is  strong and already challenging to resolve (see Section \ref{subsect_limitations}). Based on the subsample where robust measurements could be derived with and without masking bright peaks, we find that including them tend to enlarge the uncertainties associated with $\lambda_{\rm fb, 21\mu m}$, with no systematic increase or decrease. It also tends to moderately decrease all the timescales associated with 21$\mu$m, although the results remain compatible within the errors.

\begin{table*}[h]
\centering
\caption[]{Comparison of physical quantities associated with 21~$\mu$m described in Table \ref{table_outputs_21um} with (default) and without (in parenthesis) masking bright peaks.}
\label{table_outputs_21um_mask}
\begin{tabular}{|c|c|c|c|c|c|}
\hline
Galaxy & t$_{21\mu m}$ & t$_{\rm fb, 21\mu m}$ & t$_{\rm obscured}$ & $\lambda_{\rm fb, 21\mu m}$  \\
\hline
NGC1512 & 11.9$\pm_{2.01}^{1.59}$ (9.19$\pm_{2.13}^{1.24}$) & 3.64$\pm_{1.02}^{0.85}$ (3.28$\pm_{1.55}^{0.93}$) & 2.29$\pm_{1.49}^{1.29}$ (1.93$\pm_{2.02}^{1.38}$) & 302.33$\pm_{34.13}^{37.64}$ (283.02$\pm_{72.67}^{43.91}$) \\
NGC1433 & 8.42$\pm_{1.16}^{1.49}$ (7.29$\pm_{1.01}^{1.5}$) & 3.42$\pm_{1.0}^{1.27}$ (2.98$\pm_{0.87}^{1.18}$) & 1.33$\pm_{1.49}^{1.72}$ (0.89$\pm_{1.37}^{1.63}$) & 201.2$\pm_{26.66}^{41.78}$ (170.78$\pm_{22.51}^{34.27}$) \\
NGC5134 & 3.75$\pm_{1.08}^{1.97}$ (2.58$\pm_{0.83}^{2.9}$) & $\leq$4.96 ($\leq$5.14) & $\leq$3.84 ($\leq$4.01) & $\leq$271.38 ($\leq$412.75) \\
NGC6300 & 3.38$\pm_{0.65}^{0.66}$ (3.5$\pm_{0.7}^{0.69}$) & $\leq$3.61  ($\leq$3.71) & $\leq$1.27 ($\leq$1.38) & $\leq$111.96 ($\leq$104.66) \\
NGC1365 & 16.91$\pm_{7.98}^{6.43}$ (14.55$\pm_{8.66}^{3.9}$) & 8.65$\pm_{6.76}^{6.07}$ (9.33$\pm_{7.54}^{4.53}$) & 3.57$\pm_{8.21}^{12.3}$ (4.25$\pm_{8.99}^{10.76}$) & 295.8$\pm_{82.68}^{282.53}$ (459.93$\pm_{151.73}^{334.53}$) \\
IC1954 & 3.61$\pm_{0.96}^{1.14}$ (3.26$\pm_{0.82}^{1.08}$) & $\leq$4.14 ($\leq$3.93) & $\leq$2.94 ($\leq$2.72) &  $\leq$180.45 ($\leq$210.3) \\
NGC4303 & 5.54$\pm_{1.0}^{1.38}$ (5.08$\pm_{1.13}^{1.52}$) & $\leq$6.04 ($\leq$6.15) & $\leq$3.7 ($\leq$3.81) & $\leq$246.38 ($\leq$292.56) \\
NGC4321 & 4.98$\pm_{0.8}^{0.86}$ (4.16$\pm_{0.78}^{0.73}$) & $\leq$5.09 ($\leq$4.42) & $\leq$3.06 ($\leq$2.39) & $\leq$200.78 ($\leq$194.14) \\
NGC2090 & 2.49$\pm_{0.5}^{0.43}$ (2.2$\pm_{0.44}^{0.44}$) & $\leq$1.95 ($\leq$1.95) & $\leq$1.05 ($\leq$1.05) & $\leq$162.71 \\
NGC2835 & 2.93$\pm_{0.52}^{1.03}$ (2.52$\pm_{0.51}^{1.11}$) & 1.53$\pm_{0.56}^{0.9}$ (1.49$\pm_{0.55}^{0.95}$) & 0.26$\pm_{0.98}^{1.39}$ (0.21$\pm_{0.96}^{1.44}$) & 134.05$\pm_{26.66}^{76.09}$ (142.52$\pm_{36.51}^{100.33}$) \\
NGC0628 & 4.9$\pm_{0.82}^{1.43}$ (2.65$\pm_{0.54}^{1.05}$) & 3.01$\pm_{0.93}^{1.5}$ ($\leq$3.14) & 0.75$\pm_{1.58}^{2.11}$ ($\leq$1.48) & 75.98$\pm_{12.45}^{24.89}$ ($\leq$92.31) \\
NGC1087 & 6.08$\pm_{1.3}^{1.33}$ (4.52$\pm_{1.15}^{1.97}$) & $\leq$5.42 (4.15$\pm_{1.73}^{2.09}$) & $\leq$4.02 (1.42$\pm_{2.78}^{3.42}$) & $\leq$223.61 (257.88$\pm_{90.54}^{295.15}$) \\
NGC0685 & 4.73$\pm_{1.03}^{1.02}$ (4.71$\pm_{0.95}^{1.14}$) & 3.27$\pm_{1.43}^{0.95}$ (3.26$\pm_{1.38}^{1.1}$) & 0.07$\pm_{2.46}^{2.6}$ (0.05$\pm_{2.41}^{2.74}$) & 241.91$\pm_{56.0}^{76.17}$ (247.87$\pm_{60.55}^{74.35}$) \\
NGC1385 & 3.78$\pm_{1.45}^{1.49}$ (3.61$\pm_{1.32}^{1.06}$) & $\leq$4.68 ($\leq$3.91) & $\leq$4.15 ($\leq$3.37) &  $\leq$230.8 ($\leq$219.26) \\
NGC4731 & 5.38$\pm_{2.92}^{6.74}$ (2.96$\pm_{1.82}^{2.83}$) & 3.3$\pm_{2.63}^{1.72}$ (2.42$\pm_{2.16}^{2.56}$) & -0.25$\pm_{4.1}^{13.05}$ (-1.12$\pm_{3.63}^{13.9}$) & 212.05$\pm_{83.24}^{185.02}$ (222.15$\pm_{101.07}^{489.28}$) \\
NGC4951 & 3.88$\pm_{1.06}^{1.56}$ (3.99$\pm_{1.11}^{1.53}$) & $\leq$2.33 ($\leq$2.5) & $\leq$0.04 ($\leq$0.21) & $\leq$204.49 ($\leq$223.13) \\
NGC4540 & 1.9$\pm_{0.64}^{0.92}$ (1.87$\pm_{0.64}^{0.8}$) & $\leq$2.2 ($\leq$2.17) & $\leq$1.8 ($\leq$1.77) & $\leq$126.28 ($\leq$127.11) \\
NGC4496A & 3.63$\pm_{0.82}^{1.24}$ (3.7$\pm_{0.85}^{1.1}$) & $\leq$3.83 ($\leq$3.7) & $\leq$2.36 ($\leq$2.22) & $\leq$208.88 ($\leq$193.45) \\
\hline
Median & $4.31_{-1.01}^{+1.35} (3.66_{-0.90}^{+1.12})$ & $3.42_{-1.43}^{+1.27} (3.26_{-1.55}^{+1.18})$ & $1.00_{-1.00}^{+2.45} (0.89_{-0.89}^{+2.74})$ & $212.05_{-34.1}^{+76.1} (247.89_{72.66}^{+100.33})$\\
\hline
\end{tabular}
\tablefoot{For 18/26 galaxies in which bright 21$\mu$m peaks have been identified, and for which robust measurements could be derived in both cases.}
\end{table*}
\end{appendix}
\end{document}